\documentclass[hidelinks]{article}
\usepackage{PRIMEarxiv}
\usepackage[numbers]{natbib}
\usepackage[utf8]{inputenc} %
\usepackage[T1]{fontenc}    %
\usepackage[hyphens]{url}
\usepackage{tablefootnote}
\usepackage{hyperref}  %
\usepackage{amsmath}
\usepackage{booktabs}       %
\usepackage{amsfonts}       %
\usepackage{nicefrac}       %
\usepackage{microtype}      %
\usepackage{lipsum}
\usepackage{fancyhdr}       %
\usepackage{graphicx}       %
\graphicspath{{Survey_arXiv/}}
\usepackage{mathtools}
\usepackage{bm}
\usepackage{tikz}
\usepackage{esvect}
\usepackage{caption}
\usepackage{qcircuit}
\usepackage{array}
\usepackage{braket}
\usepackage{xcolor}
\usepackage{subfig}
\usepackage{todonotes}
\newcommand{\smallcite}[1]{\textsuperscript{\tiny\cite{#1}}}
\usepackage[normalem]{ulem}
 
\graphicspath{{media/}}     %

\newcommand{\measoz}[1] {\mbox{$\left< \hat{O}\right>$}}
\newcommand{\measham}[1] {\mbox{$\left< H\right>$}}

\newcommand{\citelink}[2]{\hyperlink{cite.#1}{#2}}

\title{ Quantum Machine Learning on Near-Term Quantum Devices: Current State of Supervised and Unsupervised Techniques for Real-World Applications }

\author{
  Yaswitha Gujju \\
  Dept. of Computer Science, The University of Tokyo \\
  \texttt{yaswitha-gujju@g.ecc.u-tokyo.ac.jp} \\
  \And
  Atsushi Matsuo \\
  IBM Quantum, IBM Research - Tokyo \\
\texttt{matsuoa@jp.ibm.com} \\
\And
  Rudy Raymond\thanks{Part of this work was written while RR was with IBM Research -- Tokyo.} \\
  Global Technology and Applied Research, J.P. Morgan Chase \& Co. \\
  Dept. of Computer Science, The University of Tokyo\\
Quantum Computing Center, Keio University\\
  \texttt{raymond.putra@jpmchase.com} \\
}

\begin{document}
\maketitle
\nocite{*}

\begin{abstract}
The past decade has witnessed significant advancements in quantum hardware, encompassing improvements in speed, qubit quantity, and quantum volume—a metric defining the maximum size of a quantum circuit effectively implementable on near-term quantum devices. This progress has led to a surge in Quantum Machine Learning (QML) applications on real hardware, aiming to achieve quantum advantage over classical approaches. This survey focuses on selected supervised and unsupervised learning applications executed on quantum hardware, specifically tailored for real-world scenarios. The exploration includes a thorough analysis of current QML implementation limitations on quantum hardware, covering techniques like encoding, ansatz structure, error mitigation, and gradient methods to address these challenges. Furthermore, the survey evaluates the performance of QML implementations in comparison to classical counterparts. In conclusion, we discuss existing bottlenecks related to applying QML on real quantum devices and propose potential solutions to overcome these challenges in the future.
\end{abstract}

\keywords{Quantum Machine Learning \and Real hardware \and Variational circuits\and Quantum Kernel methods \and Data encoding \and High Energy Physics \and Healthcare \and Finance \and Bottlenecks}

\section{Introduction}

Machine Learning (ML) is ubiquitous, with applications spanning image recognition, healthcare diagnosis, text translation, anomaly detection, and physics. In parallel, near-term quantum devices have shown potential in addressing classically intractable problems, even with the challenges of noise and limited qubit connectivity \cite{preskill2018quantum, preskill2023quantum}. While quantum factoring algorithms, such as Shor's, remain challenging, there have been notable successes, like the factorization of  $N=15$ using nuclear spins as quantum bits with room temperature liquid state nuclear magnetic resonance techniques \cite{vandersypen2001experimental}. \color{black} The combination of quantum computing \cite{feynman2018simulating, nielsen2002quantum} and machine learning, termed, Quantum Machine Learning (QML) \cite{biamonte2017quantum, wittek2014quantum,das2019machine} has become an active research area with great advancements being made in the last decade.\color{black} 
 Within QML, subdomains arise based on the data and algorithm types, whether classical or quantum. In this survey, we delve into different aspects of QML, specifically focusing on algorithms that leverage real quantum hardware, either in supervised or unsupervised contexts. \color{black}
In addition to the paradigms mentioned, reinforcement learning (RL) represents the third paradigm. Although not addressed in our current survey, we direct readers to \cite{meyer2022survey, dong2008quantum, dunjko2017advances} for a comprehensive overview of the literature on quantum reinforcement learning.
\color{black}

\begin{figure*}[t]
\centering  
\subfloat[Applications of QML. These are the different subproblems identified among the papers surveyed for real world domains that include High Energy Physics, Healthcare and Finance.]{\label{fig:sa}
    \includegraphics[width=9cm]{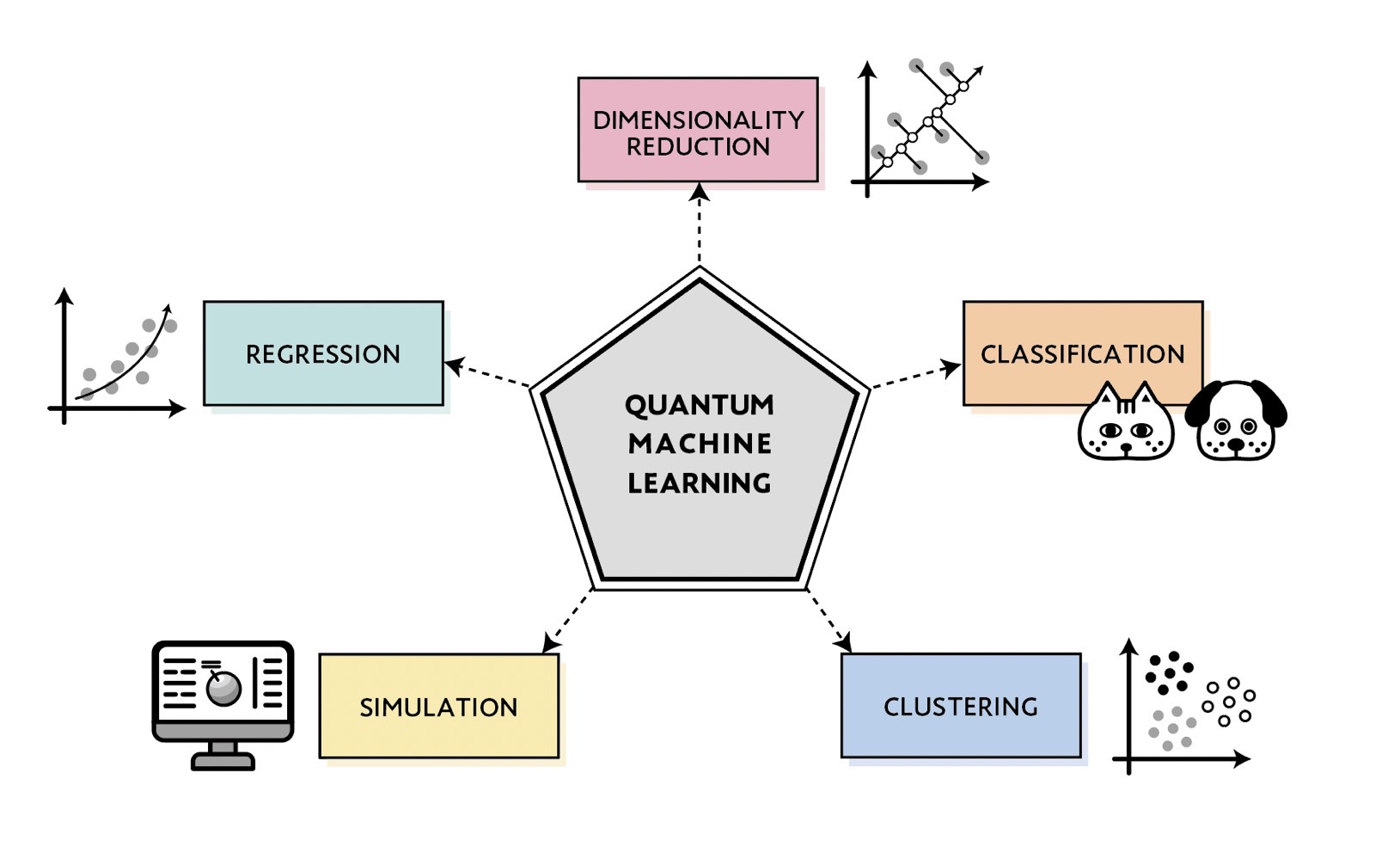}}\\
\subfloat[Outline of QML. The data can be inherently quantum or classical depending on the application. Consequently, we perform quantum state preparation for classical data. Based on the papers reviewed, we study different encoding techniques for classical data. We primarily focus on Variational Quantum Circuit and Kernel models and study the drawbacks and current challenges in the field.]
{\label{fig:apisearch}%
\includegraphics[width=9cm]{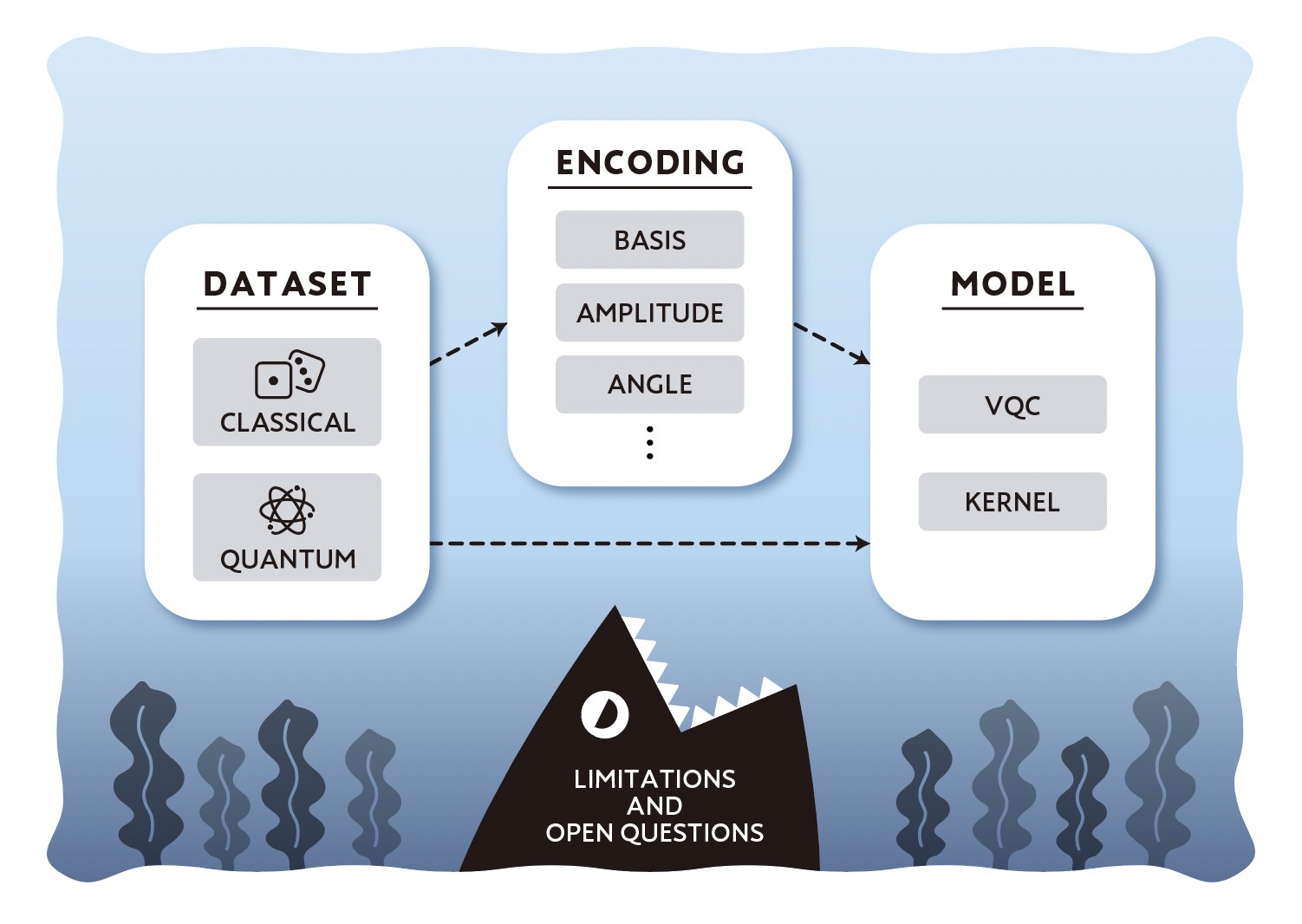}%
}
\caption{Overview of the paper.}
 \label{fig:Distribution_overview}
\end{figure*}

The past decade has witnessed significant advancements in the performance of quantum hardware, including the number of qubits, speed, and quantum volume \cite{volume1}. Consequently, there has been an increase in the number of works implementing Quantum Machine Learning (QML) on real hardware. The common objective of these works is to demonstrate the advantages of utilizing quantum computers, with their unique properties such as entanglement and superposition, for practical machine learning tasks. To gain a comprehensive understanding of the current performance and limitations of near-term quantum devices in QML, it is necessary to conduct a thorough study. In this survey, we aim to consolidate and analyze works that involve the implementation of QML on real hardware to assess their performance. There has been a growing trend in utilizing quantum computing for commercial and industrial applications, as evidenced by several studies 
\cite{wurtz2022industry, doi:10.1080/09537325.2022.2110056, GUPTA2023102544, Preskill1998, Bayerstadler2021, hassija2020present, Bova2021, singh2023contemporary,jadhav2023quantum,fedorov2022quantum}. In light of these recent publications, our focus is specifically directed towards exploring applications and techniques that hold relevance for real-world scenarios. Consequently, we have identified high energy physics \cite{humble2022snowmass,guan2021quantum, sharma2021quantum1}, finance \cite{wilkens2023quantum,ORUS2019100028,herman2022survey,egger2020quantum}, and healthcare \cite{fi15030094,wei2023quantum,flother2023state} as the domains of interest for our survey\color{black}.
Moreover, we recognize quantum chemistry as another promising field where QML holds substantial potential. 
For example, a recent paper by \cite{vakili2024quantum} claims to be the first instance of a quantum-classical generative model, trained on a 16-qubit IBM quantum computer, that yields experimentally confirmed biological hits for designing small molecules in cancer therapy, thereby indicating its practical potential in drug discovery. In particular, introducing innovative methodologies for molecular simulation \cite{sajjan2022quantum,mcardle2020quantum, von2018quantum,von2020exploring,rupp2018guest,von2013first, 
zaspel2018boosting} is a promising avenue. However, due to the similar nature of QML application in quantum chemistry and high energy physics (HEP), both primarily focused on simulating complex quantum systems to understand their properties, we have chosen to exclude quantum chemistry from our current study.
\color{black}

Two main QML frameworks have gained widespread use due to their ability to be implemented with relative ease on quantum hardware, and their demonstrated capacity to work on general datasets. These frameworks are the quantum kernel methods \cite{schuldkernel, havlivcek2019supervised, blank2020quantum, bartkiewicz2020experimental, huang2021power, liu2021rigorous, schuld2019quantum} and the variational quantum algorithms \cite{schuld2014quest, havlivcek2019supervised, qae1, abbas2021power, basisencoding1, killoran2019continuous, beer2020training, cerezo2021variational}. The quantum kernel method involves building a kernel similar to the technique used in Support Vector Machines (SVM) \cite{svm1, noble2006support}. On the other hand, the variational quantum algorithm employs a parameterized quantum circuit (PQC) whose parameters must be optimized.

For a more thorough review of quantum machine learning frameworks designed to solve classification problems such as support vector machines, kernel methods, decision tree classifiers, nearest neighbor algorithms, and annealing-based classifiers, we recommend referring to \cite{li2022recent}. The article also discusses the vulnerability of quantum classifiers in adversarial learning.

The main aim of this study is to comprehend and emphasize the constraints and methods applied in various fields that use different datasets and algorithms to run on the current ion trap and superconducting-based quantum hardware. It is noteworthy that quantum computing includes hardware architectures beyond gate-based systems, such as D-Wave's quantum computer that employs quantum annealing \cite{finnila1994quantum,hauke2020perspectives,das2008colloquium}. However, for this study, we will focus on gate-based architectures as they follow the circuit model paradigm, which makes them different from other approaches.

The current state of Quantum Machine Learning (QML) \cite{schuld2022quantum, ciliberto2018quantum, dunjko2018machine} faces numerous challenges, largely tied to quantum hardware capabilities. These encompass limited qubit connectivity, noise, coherence times, and errors in both state preparation and measurement. Prolonged running times on quantum hardware further affect the execution and outcomes of QML algorithms.
One core challenge is efficiently encoding classical data into quantum features. Alongside this, loading and storing prepared quantum states while resisting decoherence is a significant challenge. After preparing the states, it is crucial to develop efficient quantum algorithms. Challenges vary based on the type of algorithm such as kernel-based or variational quantum circuits. For instance, variational algorithms often struggle with issues like barren plateaus. Their training, alongside the choice of optimizers and loss functions, greatly influences efficiency. In contrast, with kernel techniques, selecting the right feature map is essential.
Optimization, scalability, and the generalization capabilities of QML models are crucial. It is essential for these algorithms to scale effectively for real-world applications. Moreover, addressing the security and vulnerabilities of QML models is vital to prevent potential adversarial attacks

\textcolor{black}{The papers are grouped based on real-world applications whose groups can be found in Table~\ref{Tab:hep_table} for high energy physics, Table~\ref{Tab:finance_table} for finance, and Table~\ref{Tab:Healthcare_data} for healthcare.} 
\textcolor{black}{Additionally, we include papers using benchmark datasets, such as MNIST and Iris, in Table~\ref{Tab:image}, along with references to papers using quantum datasets in Table~\ref{Tab:quantum_data}, and artificial datasets in Table~~\ref{Tab:artificial_data}.}
%
In this context, quantum data refers to data already embedded in a Hilbert space, represented as quantum states or unitaries. This differs from classical data, which requires quantum system encoding. 
The tables provide a comprehensive overview of the included studies. They detail the reference, the number of qubits used, and the specific problem type addressed. Moreover, we specify the type of hardware employed (e.g., superconducting or ion-trapped) and whether the training was done on a QPU or simulator. It is noteworthy that all tests cited in the papers were performed on a QPU. The tables further delve into the quantum models utilized, including Quantum Generative Adversarial Network (QGAN), Variational Quantum Circuit (VQC), Quantum Tensor Network (QTN), Quantum Principal Component Analysis (qPCA) and Quantum K-means.

The paper is organized as follows: Firstly, we present a summary of the notation used in the paper in Section~\ref{sec:not}. Subsequently, we offer an overview of fundamental concepts in Classical Machine Learning in Section~\ref{sec:cml}, Quantum Computing in Section~\ref{sec:qc}, and Quantum Machine Learning in Section~\ref{sec:qml}. In Section~\ref{sec:applications}, we explore the applications of Quantum Machine Learning techniques, with a specific focus on kernel techniques and variational quantum classifiers, categorized into supervised and unsupervised learning. Section~\ref{sec:limit} delves into the limitations related to hardware and algorithms, concluding with discussions on current bottlenecks and proposing possible solutions for future research in Section~\ref{sec:oq}.

\begin{table}[!ht]
\centering
\caption{The papers listed below deal with QML in the field of High Energy Physics. The table highlights the problem type, number of qubits, quantum hardware type, training approach, and specific QML methodologies employed in each paper.}
\begin{tabular}{lrrrrr}
 \toprule
 Reference & Qubits & Hardware Type & Trained on & Method \\ 
  \midrule
\multicolumn{5}{l}{\textbf{Classification}} \\
 \citelink{muten2021modified}{Muten et al.(2021)}\smallcite{muten2021modified}   & 1 & Superconducting &  Simulator & VQC \\
 \citelink{blance2021quantum}{Blance and Spannowsky (2021)   }\smallcite{blance2021quantum} & 2 & Superconducting & Simulator & VQC \\
 \citelink{terashi2021event}{Terashi et al. (2021)   }\smallcite{terashi2021event}  & 3 & Superconducting &  QPU & VQC\\
 \citelink{wu_vqc}{Wu et al. (2021a)   }\smallcite{wu_vqc}   & 10 & Superconducting & QPU & VQC \\
 \citelink{wu_kernel}{Wu et al. (2021b)   }\smallcite{wu_kernel}  & 15 & Superconducting & QPU & Kernel \\\citelink{araz2022classical}{Araz and Spannowsky (2022)   }\smallcite{araz2022classical} & 6 & Superconducting & QPU& QTN \\
 \citelink{wozniak2023quantum}{Wo{\'z}niak et al. (2023)   }\smallcite{wozniak2023quantum} & 8 & Superconducting & QPU & Kernel \\
\citelink{li2023application}{Li et al. (2023)   }\smallcite{li2023application}  & 5 & Superconducting & QPU & Kernel \\
\citelink{bermot2023quantum}{Bermot et al. (2023)   }\smallcite{bermot2023quantum} & 3 & Superconducting & QPU & QGAN \\
\color{black}\citelink{cugini2023comparing}{Cugini et al.~(2023)   }\smallcite{cugini2023comparing} & 5 & Superconducting & Simulator & VQC\\
\color{black}\citelink{peixoto2023fitting}{Peixoto et al.~(2023)   }\smallcite{peixoto2023fitting} & 5 & Superconducting & Simulator & VQC\\
\citelink{lazar2024new}{Lazar et al.~(2024)   }\smallcite{lazar2024new} & 8 & Superconducting & QPU & Parity\\

\midrule
\multicolumn{5}{l}{\textbf{Data Generation}} \\
\citelink{perez2021determining}{Pérez-Salinas et al. (2021)   }\smallcite{perez2021determining}& 8 & Superconducting  & Simulator &  VQC \\
 \citelink{bravo2022style}{Bravo-Prieto et al. (2022)   }\smallcite{bravo2022style} & 3 & Superconducting/Ion trap & QPU & QGAN\\
\citelink{chang2023running}{Chang et al. (2023)   }\smallcite{chang2023running} & 6 & Superconducting/Ion trap &  QPU & QGAN \\
\citelink{rehm2023precise}{Rehm et al. (2023)   }\smallcite{rehm2023precise}& 8 & Superconducting  & QPU &  VQC \\
\midrule
\multicolumn{5}{l}{\textbf{Clustering}} \\
 \citelink{ngairangbam2022anomaly}{Ngairangbam et al. (2022)   }\smallcite{ngairangbam2022anomaly} & 4 & Superconducting & Simulator& VQC \\
  \bottomrule
\end{tabular}
\label{Tab:hep_table}
\end{table}

\begin{table}[!ht]
\centering
\caption{The papers listed below deal with QML applications in Finance.}
\begin{tabular}{lrrrrr}
 \toprule
 Reference & Qubits & Hardware Type & Trained on & Method \\ 
  \midrule
\multicolumn{5}{l}{\textbf{Data Loading}} \\
 \citelink{zoufal2019quantum}{Zoufal et al. (2019)   }\smallcite{zoufal2019quantum} & 3 & Superconducting & QPU & QGAN \\
\midrule
\multicolumn{5}{l}{\textbf{Classification}} \\
 \citelink{ray2022classical}{Ray et al. (2022)   }\smallcite{ray2022classical} & 3 & Superconducting & QPU & VQC/Kernel \\
 
\citelink{suzuki2023quantum}{Suzuki et al. (2023)   }\smallcite{suzuki2023quantum} & 4 & Ion trap & QPU & Kernel \\
\citelink{thakkar2023improved}{Thakkar et al. (2023)   }\smallcite{thakkar2023improved} & 8 & Superconducting & Simulator & VQC \\
 
\midrule
\multicolumn{5}{l}{\textbf{Dimensionality Reduction}} \\
 \citelink{martin2021toward}{Martin et al. (2021)   }\smallcite{martin2021toward} & 4 & Superconducting & QPU & qPCA \\
\midrule
\multicolumn{5}{l}{\textbf{Feature Selection}} \\
 \citelink{zoufal2023variational}{Zoufal et al. (2023)   }\smallcite{zoufal2023variational} & 20 & Superconducting & QPU & VQC \\

  \bottomrule
\end{tabular}
\label{Tab:finance_table}
\end{table}

\begin{table}[!ht]
\centering
\caption{The papers listed below deal with QML in the field of Healthcare.}
\begin{tabular}{lrrrrr}
 \toprule
 Reference & Qubits & Hardware Type & Trained on & Method \\ 
  \midrule
\multicolumn{5}{l}{\textbf{Classification}} \\
 \citelink{yano2020efficient}{Yano et al. (2020)    }\smallcite{yano2020efficient}  & 2 & Superconducting & QPU & VQC \\
 \citelink{acar2021covid}{Acar et al. (2021)   }\smallcite{acar2021covid} & 4 & Superconducting & QPU & VQC \\
 \citelink{mathur2021medical}{Mathur et al. (2021)   }\smallcite{mathur2021medical} & 9 & Superconducting & QPU & VQC \\
 \citelink{ren2022experimental}{Ren et al. (2022)   }\smallcite{ren2022experimental} & 10 & Superconducting & QPU & VQC \\
\color{black}\citelink{moradi2022clinical}{Moradi et al. (2022)   }\smallcite{moradi2022clinical} & 7 & Superconducting & QPU & Kernel \\
 \citelink{krunic2022quantum}{Krunic et al. (2022)   }\smallcite{krunic2022quantum} & 20 & Superconducting & QPU & Kernel \\\citelink{azevedo2022quantum}{Azevedo et al. (2022)   }\smallcite{azevedo2022quantum} & 4 & Superconducting & Simulator & VQC \\
 \color{black}\citelink{vasques2023application}{Vasques et al. (2023)   }\smallcite{vasques2023application} & 5 & Superconducting & QPU & Kernel \\
 \citelink{mensa2023quantum}{Mensa et al. (2023)   }\smallcite{mensa2023quantum} & 8 & Superconducting & QPU & Kernel \\
 \color{black}
\citelink{innan2024fedqnn}{Innan et al. (2024)   }\smallcite{innan2024fedqnn} & 4 & Superconducting & Simulator & VQC\\
\midrule
\multicolumn{5}{l}{\textbf{Inference}} \\
\citelink{benedetti2021variational}{Benedetti et al. (2021)   }\smallcite{benedetti2021variational} & 5 & Superconducting & QPU & VQC \color{black} \\
  \bottomrule
\end{tabular}
\label{Tab:Healthcare_data}
\end{table}

\begin{table}[!th]
\centering 
\caption[The table lists QML papers using standard datasets such as Iris \cite{anderson1936species, fisher1936use}, MNIST \cite{deng2012mnist}, ;  Fashion-MNIST \cite{xiao2017fashion}, Titanic Survival \cite{titanic} and Wine \cite{cortez2009modeling}.]
{The table lists QML papers using standard datasets such as Iris~\cite{anderson1936species, fisher1936use}, MNIST~\cite{deng2012mnist} , FashionMNIST~\cite{xiao2017fashion},
Titanic Survival~\cite{titanic}, Astronomical~\cite{plasticc} and Wine~\cite{cortez2009modeling}.}
\begin{tabular}{lrrrrr}
 \toprule
 Reference & Qubits & Hardware Type & Trained on & Method & Dataset \\ 
  \midrule
\multicolumn{6}{l}{\textbf{Classification}} \\

 \citelink{li2015experimental}{Li et al. (2015)   }\smallcite{li2015experimental} & 4 & NMR & QPU & Kernel & Handwritten \\
 \citelink{grant2018hierarchical}{Grant et al. (2018)   }\smallcite{grant2018hierarchical} & 4 & Superconducting & Simulator & TN & Iris \\
 \citelink{cappelletti2020polyadic}{Cappelletti et al. (2020)   }\smallcite{cappelletti2020polyadic} & 2 & Superconducting & QPU & VQC & Iris \\
 \citelink{thumwanit2021trainable}{Thumwanit et al. (2021)   }\smallcite{thumwanit2021trainable} & 3 & Superconducting & QPU & VQC & Titanic Survival \\
 \citelink{peters2021machine}{Peters et al. (2021)   }\smallcite{peters2021machine} & 17 & Superconducting & QPU & Kernel & Astronomical \\
 \citelink{abbas2021power}{Abbas et al. (2021)   }\smallcite{abbas2021power} & 4 & Superconducting & QPU & VQC & Iris \\
 \citelink{blank2022compact}{Blank et al. (2022)   }\smallcite{blank2022compact} & 5 & Superconducting & QPU & Kernel & Iris/Wine \\
 \citelink{ren2022experimental}{Ren et al. (2022)   }\smallcite{ren2022experimental} & 10 & Superconducting & QPU & VQC & MNIST/FashionMNIST \\

\citelink{suzuki2023quantum}{Suzuki et al. (2023)   }\smallcite{suzuki2023quantum} & 4 & Ion trap & QPU & Kernel & MNIST/FashionMNIST\\
\color{black}\citelink{kriieeeqce23}{Koyasu et al. (2023)   }\smallcite{kriieeeqce23} & 3 & Superconducting & QPU & VQC & MNIST/FashionMNIST \\
\color{black}\citelink{haug2023quantum}{Haug et al. (2023)   }\smallcite{haug2023quantum} & 8 & Superconducting & QPU & Kernel & MNIST \\
\color{black}\citelink{simoes2023experimental}{Simoes et al. (2023)}\smallcite{simoes2023experimental} & 5 & Superconducting & QPU & Kernel/VQC & Vlds\smallcite{scikit-learn}, Iris \\
\color{black}\citelink{chen2023quantumsea}{Chen et al. (2023)   }\smallcite{chen2023quantumsea} & 4 & Superconducting & QPU & VQC & MNIST/FashionMNIST \\
\color{black}\citelink{melo2023pulse}{Melo et al. (2023)   }\smallcite{melo2023pulse} & 9 & Superconducting & QPU & Kernel & MNIST \\
\color{black}\citelink{anagolum2024elivagar}{Anagolum et al. (2024)   }\smallcite{anagolum2024elivagar} & 10 & Superconducting & QPU  & VQC & MNIST \\
\color{black}
\citelink{shen2024classification}{Shen et al. (2024)   }\smallcite{shen2024classification} & 11 & Superconducting & Simulator & VQC & FashionMNIST\\
\color{black}
\citelink{innan2024fedqnn}{Innan et al. (2024)   }\smallcite{innan2024fedqnn} & 4 & Superconducting & Simulator & VQC & Iris\\

\midrule
\multicolumn{6}{l}{\textbf{Clustering}} \\
 \citelink{khan2019k}{Khan et al. (2019)   }\smallcite{khan2019k} & 4 & Superconducting & QPU & Q-Kmeans & Iris/MNIST \\
 \citelink{johri2021nearest}{Johri et al. (2021)   }\smallcite{johri2021nearest} & 8 & Ion trap & QPU & Nearest centroid & Iris/MNIST \\
\midrule
\multicolumn{6}{l}{\textbf{Simulation}} \\
 \citelink{huang2021experimental}{Huang et al. (2021)   }\smallcite{huang2021experimental} & 5 & Superconducting & QPU & QGAN & Handwritten \\
 \citelink{rudolph2022generation}{Rudolph et al. (2022)   }\smallcite{rudolph2022generation} & 8 & Ion trap & QPU & QGAN & MNIST \\
  \bottomrule
\end{tabular}
\label{Tab:image}
\end{table}

\begin{table}[h!]
\centering
\caption{The QML papers listed below primarily utilize quantum data. This data can consist of intrinsically quantum information or classical information that is transformed into a quantum feature space.}
\begin{tabular}{lrrrrr}
 \toprule
 Reference & Qubits & Hardware Type & Trained on & Method \\ 
  \midrule
\multicolumn{5}{l}{\textbf{Simulation}} \\
\citelink{gibbs2024dynamical}{Gibbs et al. (2022)   }\smallcite{gibbs2024dynamical} & 2 & Superconducting & QPU & VQC \\
\color{black}
\citelink{bartkiewicz2023synergic}{Bartkiewicz et al. (2023)   }\smallcite{bartkiewicz2023synergic} & 3 & Superconducting & QPU & VQC \\

\midrule
\multicolumn{5}{l}{\textbf{Classification}} \\
\citelink{blank2020quantum}{Blank et al. (2020)   }\smallcite{blank2020quantum} & 5 & Superconducting & QPU & Kernel \\
\citelink{herrmann2022realizing}{Herrmann et al. (2022)   }\smallcite{herrmann2022realizing} & 7 & Superconducting & QPU & VQC \\
\citelink{ren2022experimental}{Ren et al. (2022)   }\smallcite{ren2022experimental} & 10 & Superconducting & QPU & VQC \\
\citelink{gong2023quantum}{Gong et al. (2022)   }\smallcite{gong2023quantum} & 61 & Superconducting & QPU & VQC \\
 
\midrule
\multicolumn{5}{l}{\textbf{Data Generation}} \\
\color{black}
\citelink{bartkiewicz2023synergic}{Bartkiewicz et al. (2023)   }\smallcite{bartkiewicz2023synergic} & 3 & Superconducting & QPU & VQC \\
\midrule
\multicolumn{5}{l}{\textbf{Training}} \\ \citelink{pan2023deep}{Pan et al. (2023)   }\smallcite{pan2023deep} & 6 & Superconducting & QPU & VQC \\
\midrule
\multicolumn{5}{l}{\textbf{Clustering}} \\
\citelink{huang2022quantum}{Huang et al. (2022)   }\smallcite{huang2022quantum} & 40 & Superconducting & QPU & Kernel \\ 
\color{black}\citelink{nakayama2023vqe}{Nakayama et al. (2023)   }\smallcite{nakayama2023vqe} & 4 & Superconducting & QPU  & Kernel\\
  \bottomrule
\end{tabular}
\label{Tab:quantum_data}
\end{table}

\begin{table}[!ht]
\centering
\caption{The papers listed below use artificial datasets which are synthetically created to evaluate and benchmark QML algorithms.}
\begin{tabular}{lrrrrr}
 \toprule
 Reference & Qubits & Hardware Type & Trained on & Method \\ 
  \midrule
\multicolumn{5}{l}{\textbf{Classification}} \\
 \citelink{havlivcek2019supervised}{Havl{\'\i}{\v{c}}ek et al. (2019)   }\smallcite{havlivcek2019supervised} & 2 & Superconducting & QPU & Kernel/VQC \\
 \citelink{bartkiewicz2020experimental}{Bartkiewicz et al. (2020)   }\smallcite{bartkiewicz2020experimental} & 2 & Photonic & QPU & Kernel \\
\color{black}\citelink{melo2023pulse}{Melo et al. (2023)   }\smallcite{melo2023pulse} & 5 & Superconducting & QPU & VQC  \\
\color{black}\citelink{simoes2023experimental}{Simoes et al. (2023)   }\smallcite{simoes2023experimental} & 4 & Superconducting & QPU & Kernel/VQC \\
 \citelink{glick2024covariant}{Glick et al. (2022)   }\smallcite{glick2024covariant} & 27 & Superconducting & QPU & Kernel \\\citelink{heese2023explainable}{Heese et al. (2023)   }\smallcite{heese2023explainable} & 3 & Superconducting & Simulator & QGAN \\
 \citelink{gentinetta2023quantum}{Gentinetta et al. (2023)   }\smallcite{gentinetta2023quantum} & 7 & Superconducting & QPU & Kernel \\
\midrule
\multicolumn{5}{l}{\textbf{Data Generation}} \\
 \citelink{huang2021quantum}{Huang et al. (2021)   }\smallcite{huang2021quantum} & 5 & Superconducting & QPU & QGAN \\
\midrule
\multicolumn{5}{l}{\textbf{Clustering}} \\
 \citelink{johri2021nearest}{Johri et al. (2021)   }\smallcite{johri2021nearest} & 8 & Ion trap & QPU & Nearest centroid \\
\midrule
\multicolumn{5}{l}{\textbf{Regression}} \\
 \citelink{kreplin2023reduction}{Kreplin et al. (2023)   }\smallcite{kreplin2023reduction} & 10 & Superconducting & QPU & QNN \\
  \bottomrule
\end{tabular}
\label{Tab:artificial_data}
\end{table}

\section{Notation}
\label{sec:not}
Throughout the paper, the dataset is denoted as $\mathcal{D} = \{(x^1,y^1),\ldots,(x^m,y^m)\}$, where $\mathcal{D}$ represents the dataset for supervised learning containing $m$ samples or observations. For unsupervised learning, the data does not contain labels and is represented as $\mathcal{D} = \{x^1,\ldots,x^m\}$. Each $x^i$ represents the $i$-th input data sample and can be understood as a vector defined as $x^i = [x^{i}_{1}, x^{i}_{2}, \ldots, x^{i}_{d}]$ where $d$ is the number of features in the input data. The corresponding label or output associated with $x^i$ is represented by $y^i$. Moving over to the vector spaces, we represent the $N$-dimensional Hilbert space as $\mathcal{H}^{N}$ for a system with $n$ qubits such that $N = 2^n$. The complex space is represented using $\mathbb{C}$, while the real space is denoted as $\mathcal{R}$. The feature map is denoted as \(\phi\). The quantum gates are represented as \(H\) for the Hadamard gate, \(X\) for the Pauli-X gate, \(Y\) for the Pauli-Y gate, \(Z\) for the Pauli-Z gate, and \(U\) for the unitary operator. The gate parameters are denoted using $\theta$. The depth of the encoder part of the circuit is represented as $N_{\text{depth}}^{\text{in}}$, while $N_{\text{depth}}^{\text{var}}$ is used to represent the depth of the variational part of the circuit.



\begin{table}[!ht]
\centering
\begin{tabular}{ |p{4cm}|p{10cm}|}
\hline
Acronym & Full form  \\
\hline
CC & Classical computing
\\
NN & Neural Network
\\
NN-GD & Neural Network with Gradient Descent
\\
QC & Quantum Computing
\\
VQC & Variational Quantum Circuit
\\
VQC-QGD & Variational Quantum Circuit with Quantum Gradient Descent
\\
VQC-GD & Variational Quantum Circuit with Gradient Descent
\\
QML & Quantum Machine Learning
\\
VQA & Variational Quantum Algorithm
\\
PQC & Parameterized quantum circuit
\\
QKE & Quantum Kernel Estimation
\\
QSVM & Quantum Support Vector Machine
\\
FOE & First Order Expansion
\\
SOE & Second Order Expansion
\\
HEP & High Energy Physics
\\
QAE & Quantum Amplitude Estimation
\\
LHC & Large Hadron Collider
\\
BDT & Boosted Decision Tree
\\
DNN & Deep Neural Network
\\
MSE & Mean Square Error
\\
DRC & Data Re-uploading based classifier
\\
TN & Tensor Networks
\\
QTN & Quantum-inspired Tensor Networks
\\
TPR & True Positive Rate 
\\
FPR & False Positive Rate 
\\
\hline
\end{tabular}
\caption{This table lists the abbreviations used throughout the paper, along with their expansions.}
\end{table}

\color{black}

\section{Classical Machine Learning}
\label{sec:cml}
The field of Artificial intelligence (AI) has become omnipresent with many practical applications such as automation of routine labor, speech recognition, computer vision etc. To avoid depending on hard-coded knowledge, it is essential for these AI systems to acquire knowledge from their surroundings by solving a learning problem \cite{goodfellow2016deep}. Machine learning (ML)\cite{bishop2006pattern, goodfellow2016deep,jordan2015machine} is an evolving branch of Artificial intelligence that is essentially devoted to solving such problems where the goal is to improve some measure of performance when executing a task, through some type of training experience. ML models are trained on sample data, called training data, which enables them to learn properties of the data and make predictions or decisions accordingly. 
\label{sec:headings} In this survey, we look at supervised and unsupervised learning.

\paragraph{Supervised Learning}
Here, the model is provided with labelled data. To measure the performance, the model is evaluated on unseen data called testing data. Two common types of supervised-learning algorithms include classification and regression.
Training involves minimizing the cost function over the input data and adjusting its weights until the model has been fitted appropriately. Examples of common classifiers include linear classifiers, support vector machines (SVM), random forest etc. In regression type problems, the goal is to fit a function over the data (independent variables) to predict the output. Commonly used regression models include linear regression, support vector regression etc. 

\paragraph{Unsupervised Learning}
On the other hand, unsupervised learning involves training the model to analyze and cluster unlabeled datasets with the goal of discovering hidden patterns or structure in the data. In addition to clustering, which involves finding structure in the data by grouping similar points and separating dissimilar points, unsupervised learning also includes dimensionality reduction techniques such as PCA, autoencoders, singular value decomposition etc. Here, the objective is to reduce the dimension of the data without losing too much information.

\begin{figure*}[t]
\centering  
\subfloat[Yearly plot displaying the frequency of works utilizing the search term 'quantum machine learning' in title/abstract \cite{pub.1106289502}.]{%
  \includegraphics[width=9cm]{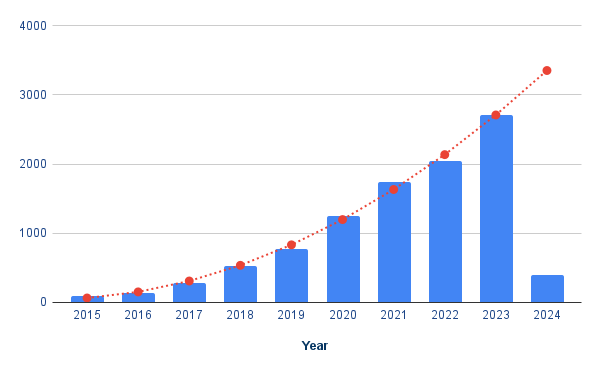}}
  \\
  \vspace{3mm}
  \subfloat[The distribution of search terms related to various real-world applications in the titles and abstracts of around 1000 papers sampled obtained using the arXiv API under the 'quant-ph' category in the last five years. For papers belonging to multiple domains, we include them in each of the respective categories to calculate the final distribution. We see that the collection of Physics, Finance, and Healthcare constitute approximately 34\% of the applications queried using the API.]{%
  \includegraphics[width=10cm]{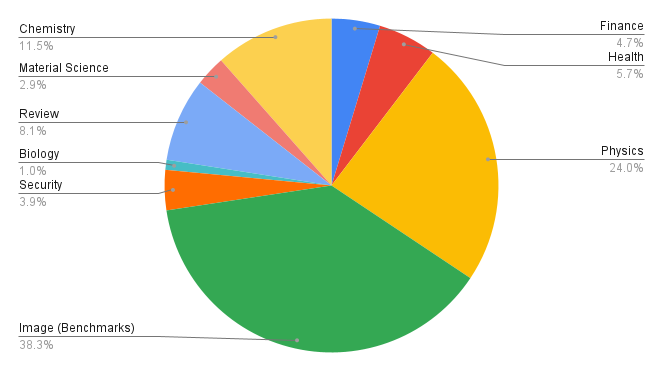}}
\caption{\textbf{(a)} Timeline Trend and \textbf{(b)} Coverage Distribution of Works Related to Quantum Machine Learning on arXiv.}
 \label{fig:Distribution}
\end{figure*}

\section{Quantum Computing}
\label{sec:qc}
\begin{figure}
\vspace{-3cm}
    \centering
\includegraphics{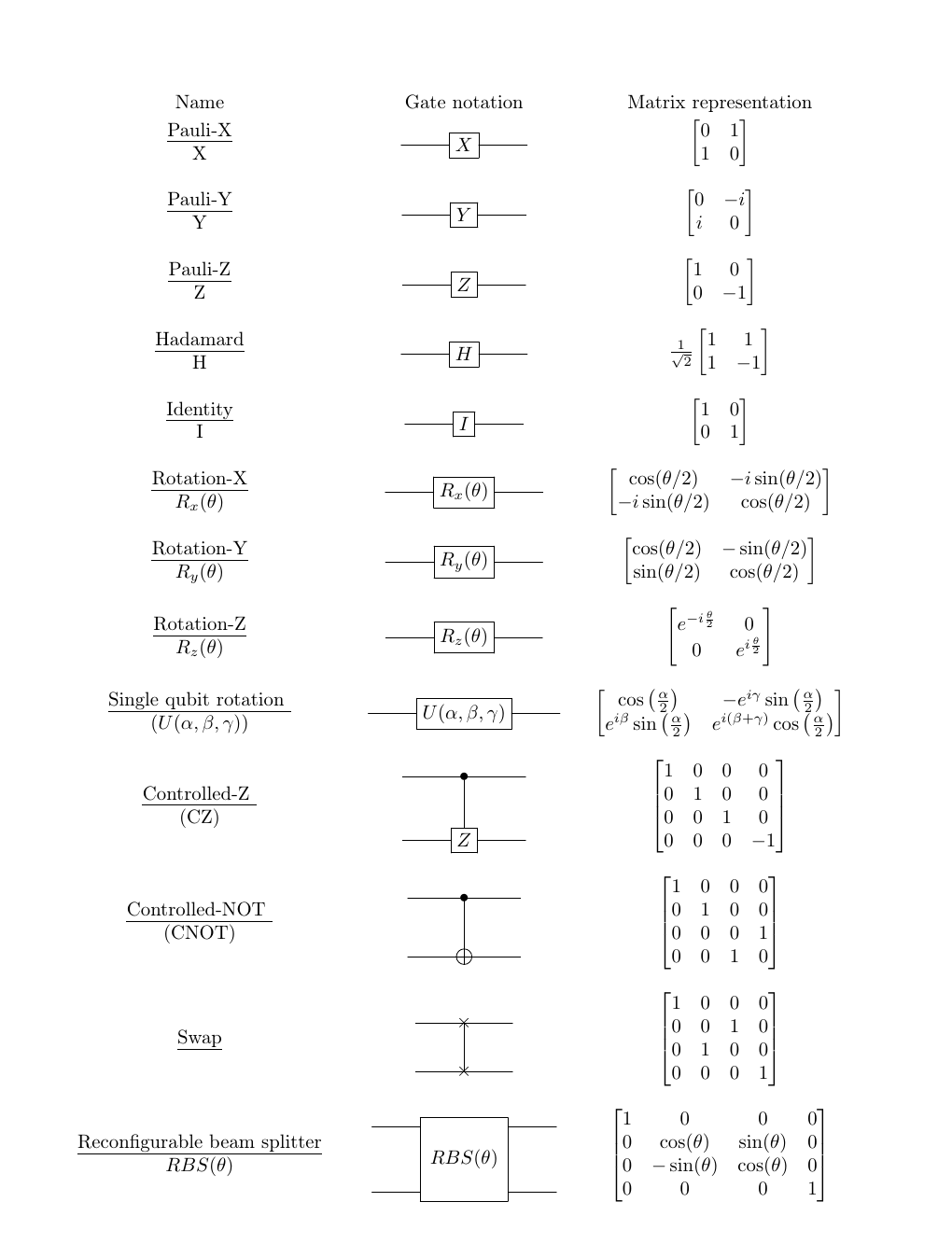}
    \caption{Description of quantum gates in the order of single and multiple qubits}   \label{fig:quantum_gates}
\end{figure}

The phenomenon of quantum superposition and entanglement is what gives quantum computing an edge over classical computing. This can translate to significant speedup or reduced computational resources in terms of time and space. Here, we briefly discuss the basics of quantum computing. The basic unit of quantum computation is the qubit, $$ |\psi\rangle = \alpha |0\rangle + \beta |1\rangle$$ (where $\alpha$ , $\beta$  $\in$ $\mathbb{C}$ and $|0\rangle, |1\rangle$ represent the computational basis in the
two-dimensional Hilbert space $\mathcal{H}$).
The absolute squares of the amplitudes (i.e. $ |\alpha|^2 $ and  $|\beta|^2$)  are the probabilities to measure the qubit in either 0 or 1 state, respectively, such that $ |\alpha|^2 + |\beta|^2 = 1$.
As such, $|\psi\rangle$ as can be rewritten as
    $ |\psi\rangle = \cos \frac{\theta}{2} |0\rangle + e^{i\phi}\sin  \frac{\theta}{2} |1\rangle$
    where $0 \leq \theta \leq \pi$  and $0 \leq \phi \leq 2\pi$ are real numbers. Unitary matrices (quantum gates) can be applied to quantum states to transform into other quantum states to ensure that the condition on the amplitude-based probabilities is maintained even after the transformation. Through single qubit quantum gates we can manipulate the basis state, amplitude or phase of a qubit (for example through the so called $X$ gate,
the $Z$ gate and the $Y$ gate respectively), or put a
qubit with $\beta = 0  (\alpha = 0)$ into an equal superposition. $\alpha = \frac{1}{\sqrt{2}}, \beta = {\displaystyle \pm }\frac{1}{\sqrt{2}}$ (the Hadamard
or $H$-gate). Multi-qubit gates are often based on
controlled operations that execute a single qubit
operation only if another (ancilla or control qubit) is
in a certain state. One of the most important gates
is the two qubit controlled-NOT (CNOT, or CX) gate, which flips the basis state of the target qubit when the control qubit is in
state $|1\rangle$. A set of arbitrary one-qubit rotation gates and two-qubit controlled-NOT (or CNOT) gates is universal which means that any quantum operation can be implemented using a combination of these basic gates. We list typical quantum gates (along with their symbols, and their matrix forms) used in quantum circuits for quantum machine learning in Figure~\ref{fig:quantum_gates}.

\section{Quantum Machine Learning}
\label{sec:qml}
In this study, we focus on two widely used QML algorithms based on variational quantum circuits and quantum kernel methods.
In both of these approaches, we start by encoding $d$-dimensional classical data so that it is embedded as a quantum state vector in the Hilbert space.  By doing so, we can exploit the exponential dimensionality of the Hilbert space that grows with the number of qubits, giving it a stronger representational power over the classical feature space which may help capture strong correlation between variables. Both the models involve data encoding but differ in the way the quantum state is handled. We look at some of the most commonly used techniques for encoding classical data along with the different QML models.

\subsection{Encoding Datasets}
Quantum Machine Learning (QML) involves learning from either classical or quantum data. It is more likely to obtain exponential quantum advantage in machine learning when data comes from quantum-mechanical processes \cite{cerezo2022challenges}. Classical data is encoded in bits (0s and 1s), such as images, texts, medical records, etc. Quantum data, on the other hand, is encoded in quantum bits called qubits, which can represent states beyond 0 and 1. Qubits can contain information from physical processes like quantum sensing or quantum control. While classical data can be efficiently encoded in qubits, the reverse is not true. In QML, quantum data refers to data already in a quantum state, while classical data needs to be encoded into a quantum system. 

A requisite for obtaining quantum advantage in both VQC and QKE \cite{schuldkernel} on classical datasets is that the embedding or encoding the datasets has to be efficiently implementable on quantum circuits to avoid the so-called data-loading problem~\cite{AaronsonFinePrint2015}. The quantum embeddings help represent classical data as quantum states in the Hilbert space thereby allowing us to truly harness the power of quantum systems. Some desirable properties of an encoder are that the number of gates required to implement the encoder must be at most polynomial in the number of qubits, and the intractability by any classical operation to simulate it is preferred. Additionally, it is ideal to have a bijective encoding such that there is a unique quantum state $\rho_{x_i}$ for each sample $x_i$. Finally, the single and two-qubit gates required to implement the encoder should be compatible with the native gate set of the near-term quantum devices so that the compilation of the circuit is hardware efficient \cite{encoding1}. Thus, data  encoding plays an important role as they determine the features that  quantum models represent \cite{schuld2019quantum, havlivcek2019supervised}, the decision boundaries learnt \cite{encoding1}, and measurements that optimally distinguish between data classes \cite{lloyd2020quantum}.

\subsubsection{Basis Encoding}
This is the simplest and one of the most common encodings \cite{basisencoding1, basisencoding2} that maps a binary string classical data $x = x_{1}\ldots x_{n}$ into the computational basis $\ket{x} = \ket{x_{1}\ldots x_{n}}$. It requires $n$ qubits to encode $n$ bits of classical data, and is useful to feed one sample classical bit at a time to a QML model. The power of quantum resource comes when the batches of classical samples are represented as superpositions of basis states~\cite{rebentrost2014quantum}. Quantum bits can be used to create quantum states that are superposition of classical datasets, i.e., \textit{quantum batches}.

In the case of supervised learning, as pointed out in~\cite{basisencoding1}, one can create quantum states $\ket{+1}$ and $\ket{-1}$ each of which is a superposition of the basis encoding of samples with label $l(x)$ as $+1$ and $-1$, respectively, as below (omitting ancilla and working qubits), and use them to train a QML model on superposition states of real world data.  
\begin{eqnarray*}
\ket{+1} &=& \frac{1}{\sqrt{N_{+}}}\sum_{x:l(x)=+1}\ket{x}\\
\ket{-1} &=& \frac{1}{\sqrt{N_{-}}}\sum_{x:l(x)=-1}\ket{x},
\end{eqnarray*}
where $N_{+}$ and $N_{-}$ are, respectively, the number of samples with label $+1$ and $-1$. It is argued in~\cite{basisencoding1} that the above quantum batches can result in training a QML model with smoother loss fluctuation and can be more efficient in the sample complexity for better generalization error than individual samples.

\subsubsection{Amplitude Encoding}

The classical data $x$, which is an $d$-dimensional vector, is encoded into the amplitude of the quantum state\cite{vazquez2021efficient, schuld2017implementing, wiebe2012quantum, zoufal2019quantum}. Namely, for $x = (x_1,\ldots,x_d)$ such that $\sum_i |x_i|^2 = 1$, the corresponding encoding is the quantum state
  $$
  \ket{\psi_x} = \sum_{i=1}^{d} x_i\ket{i},
  $$
 that only requires $\log{d}$ qubits to store $x$. 
  The advantage of this encoding is in the exponential memory saving and, if one can design a QML model that runs in polynomial time in the size of the number of qubits, then there are hopes for exponential quantum advantage. In fact, many QML models promising quantum advantages use this encoding combined with quantum basic linear algebras, such as, HHL~\cite{PhysRevLett.103.150502} and others (see, e.g.,~\cite{grover1996fast,d1998general,brassard2002quantum,gilliam2021grover}). The main drawback is that quantum circuits that generate $\ket{\psi_x}$ can require quantum circuits with exponential number of native gates~\cite{grover2000synthesis, shende2005synthesis, plesch2011quantum,sanders2019black}, and hence the data-loading problem~\cite{AaronsonFinePrint2015}.
   
   To avoid exponential circuit complexity, recent works~\cite{KerenidisDataLoader2020,kerenidis2021classical} propose the use of \textit{unary amplitude encoding} to encode $x$ using an $d$-qubit quantum state (i.e., a qubit per feature) as
   $$
   \ket{\phi_x} = \sum_{i=1}^{d} x_i\ket{e_i},
   $$
   where $\ket{e_i}$ is the $i$-th unary computational basis $\ket{0\ldots{0}1{0}\ldots{0}}$ with "1" only at the $i$-th qubit. It is shown that the depth of the circuit to generate unary encoding is logarithmic in $d$~\cite{KerenidisDataLoader2020}, and linear using cascade of RBS gates~\cite{kerenidis2021classical}. 

\subsubsection{Divide-and-Conquer Approach}
This data loading technique is a modified version of amplitude encoding and is introduced in \cite{araujo2021divide} using controlled swap gates and ancilla qubits. As the name suggests, this method is based on divide-and-conquer approach and derives motivation from \cite{mottonen2004transformation}. The $d$-dimensional input vector is loaded in the probability amplitudes of computational basis state with entangled information in ancillary qubits. The results show exponential time advantage using a quantum circuit with poly-logarithmic depth and $O(d)$ qubits. However, the reduced circuit depth comes at a cost of increasing the circuit width and creating additional entanglement between data register qubits and an ancillary system.

  \subsubsection{Angle Encoding}

While the aforementioned amplitude encodings require at least $O(\log{d})$-depth circuits, one can load $x$ with constant depth quantum circuits by embedding $x_i \in \mathbb{R}$, i.e., the $i$-th element of $x$, as a parameter of Pauli rotational gates $R_X(x_i) \equiv e^{-ix_iX/2}$, or $R_Y(x_i) \equiv e^{-ix_iY/2}$, or $R_Z(x_i) = e^{-ix_iZ/2}$. The data also needs to be normalised or scaled using min-max scaling in a suitable range to be evaluated as gate angles and the choice of this range can influence the performance. For example, in \cite{wu_kernel}, the use of angles in range $[-1, 1]$ was found to be more optimal than  $[-\pi, \pi]$. For example, starting from the all-zero quantum state, one can create the following $n$-qubit quantum state (where $n = d$) representing $x$ by applying $R_Y(x_i)$ to the $i$-th qubit for $i=0\ldots{d-1}$.
$$
\ket{x} \equiv \bigotimes_{i=0}^{d-1} R_Y(x_i) \ket{0}^d = \bigotimes_{i=0}^{d-1} \cos{\left(\frac{x_i}{2}\right)}\ket{0} + \sin{\left(\frac{x_i}{2}\right)}\ket{1}
$$
The above quantum state is a product state that can be represented classically in $O(d)$ computational space and time, but when combined with entanglement layers and their block repetitions,
the angle encoding can be used as a building block to generate sophisticated entangled states that are difficult to compute classically. 
Also worth mentioning are the so-called, \textit{First order encoding} (FOE) and \textit{Second Order Encoding} (SOE) as defined in \cite{havlivcek2019supervised}. In FOE, to encode $x_k \in \mathbb{R}$, the single qubit gates $R_Z(x_k)$ are used. This can be lifted to a higher encoding using SOE where more parameters are used along with entangling gates. For example, to encode $x_l, x_m \in \mathbb{R}$ along with their correlation in the $l$-th and $m$-th qubits, SOE utilizes the gate $e^{i(\pi - x_l)(\pi - x_m)Z_l Z_m}$.

    When the classical data $x$ is a bitstring of length $d$, which is often used to represent discrete features, \cite{yano2020efficient} proposes to utilize the so-caled \textit{Quantum Random Access Codes} (QRAC) to obtain a constant factor saving in the number of qubits. For example, the previous $\ket{e_i}$ is known as one-hot encoding in classical machine learning that requires $d$ qubits. With the QRAC encoding, the bitstring $x = x_0\ldots{x_{d-1}}\in \{0,1\}^d$ can be represented with $\lceil{d/3}\rceil$-qubit quantum state $\rho_x$ as below.
    $$
    \rho_{x} \equiv \ket{\psi_x}\bra{\psi_x} =  \bigotimes_{i=0}^{d/3-1} \frac{1}{2}\left(I + \frac{1}{\sqrt{3}}\left( (-1)^{x_{3i}} X + (-1)^{x_{3i+1}}Y + (-1)^{x_{3i+2}}Z  \right)\right),
    $$
    where for simplicity $d > 0$ is assumed to be divisible by $3$. Notice that the value of $x_{3i+j}$ can be retrieved by measuring the $i$-th qubit of $\rho_x$ in $X, Y$ or $Z$ bases for $j = 0, 1, 2$, respectively. The QRAC encoding can be run with a single-qubit gate for each qubit.

\paragraph {Data Re-uploading}
\label{drc}
Angle encoding applies a Pauli rotation gate whose degree of freedom is one, say for $x_j \in \mathbb{R}$, the $R_Z(x_j)$ at the $j$-th qubit. Meanwhile, it is known that a general single-qubit rotation gate $U(\cdot)$ has three degrees of freedom and is represented by matrix form in Figure \ref{fig:quantum_gates}.

First proposed in \cite{reupload}, the data re-uploading techniques utilizes the above $U(\cdot)$ to encode three elements of $x$ in a qubit. By repeating the application of $U(\cdot)$ each with different three elements of $x$ for $j\in \{0,\ldots,d/3-1\}$, hence the re-uploading, the whole data point $x$ can be encoded in a single qubit. %
We can easily see that the data re-uploading is the angle encoding repeated with different parameters $x_j$'s because the above $U(\cdot)$ can be decomposed into a sequence of Pauli rotation gates as below. 
$$
U\left(x_{3j}, x_{3j+1}, x_{3j+2}\right) = R_Z(x_{3j+1}+\pi)~\sqrt{X}~R_Z(x_{3j}+\pi)~\sqrt{X}~R_Z(x_{3j+2})
$$
The parameters of data re-uploading can be linearly transformed before being used in $U(\cdot)$ or trained to fit the prediction~\cite{reupload}. 
This method has been used in a variety of applications ranging from drug discovery \cite{batra2021quantum, li2021drug}, image classification of MNIST dataset \cite{easom2021depth} and Variational Quantum Eigensolver~\cite{cervera2021meta}. Due to the structure of single-qubit unitary gates, this encoding is particularly suited for data with rotational symmetry.

\subsubsection{Hamiltonian Encoding}
While the encoding quantum state from the angle encoding is obtained by transforming the all-zero quantum state with single-qubit rotational gates which are classically computable, the Hamiltonian encoding evolves the all-zero quantum state according to the Hamiltonian parameterized by $x$ to generate highly entangled states. Namely, let the Hamiltonian be $H(x) = \sum_i f_i(x) h_i$, where $f_i(x)\in \mathbb{R}$ is the weight function and $h_i = \otimes_{j=1}^n\sigma_i^j$ for $\sigma_i^j \in \{I, X, Y, Z\}$. For a fixed $t$, the quantum state $\ket{\psi_t(x)}$ that encodes $x$ is obtained from the time evolution
$$
\ket{\psi_{t}(x)} = e^{-iH(x)t/\hbar}\ket{0}^{\otimes n},
$$
which can be run on gate-based quantum hardware using techniques such as  Trotterization \cite{lloyd1996universal}, variational approaches \cite{barison2021efficient, yao2021adaptive, yuan2019theory} and linear combination of unitaries \cite{childs2012hamiltonian, low2019well}.

Another example of encoding involving the parameters of a tunable Hamiltonian is discussed in \cite{albrecht2023quantum}, where the authors present a quantum feature map for graph data on a neutral atom quantum processor comprising up to 32 qubits. The results demonstrate the ability of the map to effectively capture local and global graph structures while applying this quantum graph kernel to predict toxicity on a real-world dataset of molecules and compare its performance against various classical kernels.

\subsection{Models}
\subsubsection{Quantum Kernel Estimation}

The kernel trick enables one to process higher dimensional data without explicitly computing the feature vector. This method is most commonly used in classification using of the support vector machine (SVM)\cite{svm1, noble2006support}. By means of the kernel, every feature map corresponds to a distance measure in input space by means of the inner product of feature vectors \cite{schuld2019quantum,gentinetta2024complexity}. The key highlight of kernel tricks with quantum states, or quantum kernels, comes from its ability to compute similarities from the encoding of the classical data into the quantum state space through entanglement and interference so as to generate correlations between variables that are classically intractable \cite{huang2021power}. This is expected to give more expressive feature embeddings leading to better performance in pattern recognition and classification tasks compared to the classical counterparts. However, the true advantage does not come from the high dimensional space (which is also possible using classical kernels) but rather from being able to construct complex circuits which are hard to calculate classically. Even so, while the classical kernels can be computed exactly, the quantum kernels are subject to small additive noise in each kernel entry due to finite sampling, while classical kernels can be computed exactly. To tackle this, error-mitigation techniques have been developed \cite{PhysRevX.7.021050,kandala2019error,temme2017error, liu2021rigorous} for cases when the feature map circuit is sufficiently shallow.

The following steps are key components involved in QKE\cite{schuldkernel}:
\begin{itemize}
    \item \textbf{Quantum Feature Map} : A feature map $\phi$ is employed to encode the classical data $x$ to the quantum state space using unitary operations. For any two data points $x^i$, $x^j$ $\in$ $\mathcal{D}$, the encoded data is represented as $\Phi(x^i)$ and $\Phi(x^j)$ respectively.
    \item \textbf{Inner product} : The kernel entry can be obtained as the inner product between two data-encoded feature vectors $\Phi(x^i)$ and $\Phi(x^j)$ i.e.
    \begin{equation} 
\kappa(x^i, x^j) = |\langle{\Phi(x^j)}|\Phi(x^i)\rangle|^2 
\label{eq:kernel}
    \end{equation}
     The kernel entry can be estimated by recording the frequency of the all-zero outcome $0^n$. This procedure is referred to as quantum kernel estimation (QKE). 
    Different methods \cite{cincio2018learning,buhrman2001quantum} can be employed to estimate the fidelity between general quantum states, one of which is the swap test.
\end{itemize}

Quantum Support Vector Machines use the kernel built using QKE with a classical SVM. It was first introduced in \cite{rebentrost2014quantum} while the proof-of-principle was first demonstrated for classifying handwritten characters in \cite{li2015experimental}.

The advantage of using quantum kernels is not so apparent when we have large datasets where the quantum cost scales quadratically with the training dataset size \cite{lloyd2020quantum}. 
Efficient data encoding and generating useful quantum kernels is constrained by the limited number of qubits and heuristic characterization \cite{ reupload, schuldkernel,wu_kernel, havlivcek2019supervised}. Additionally, fewer measurements, and large system noise necessitate error mitigation techniques  requiring significant additional quantum resources  \cite{temme2017error, endo2021hybrid}. In \cite{wang2021towards}, an indefinite kernel learning based method is implemented to demonstrate the advantage of kernel methods for near term quantum devices by suppressing the estimation error. Recently, the work in \cite{haug2023quantum} introduced a novel approach for measuring quantum kernels using randomized measurements showing a linear scaling of features based on circuit depth. The method also incorporates a cost-free error mitigation and offers improved scalability, with the quantum computation time scaling linearly with the dataset size and quadratic scaling for classical post-processing.

{\paragraph{Different types of kernels.}In the previous section, the quantum kernel is computed as the (non-negative) frequency of observing the all-zero bits of running the concatenation of the quantum circuit encoding $x^i$ with the inverse of quantum circuit encoding $x^j$ as in Eq.~(\ref{eq:kernel}). This type of quantum kernels is quite powerful to classify artificial data derived from the discrete-log problems~\cite{liu2021rigorous}, and to classify group-structured data when the initial state $\ket{0^n}$ in Eq.~(\ref{eq:kernel}) is replaced with optimized fiducial quantum states computed from kernel alignment~\cite{glick2024covariant}. At the latter, experimental results on a 27-qubit device, when the data are encoded with single-qubit rotational gates and the fiducial quantum state is matched with connectivity of qubits in the quantum device, are demonstrated.}

There are many other types of quantum kernels available whose elements are not necessarily restricted to be non-negative. For example, the Hadamard-test classifier (HTC), that encodes real-valued vectors with amplitude encoding, computes the weighted sum of inner product between a test data vector with the superposition of training data vectors for binary classification~\cite{schuld2017implementing}. The compact version of HTC is by~\cite{blank2022compact}. While the full quantum space in Eq.~(\ref{eq:kernel}) seems to be powerful, it is pointed in~\cite{huang2021power} that it can fail to learn a simple function. To overcome this, the \textit{projected quantum kernel}, that projects the quantum kernel into classical one and compute the elements of kernels from the functions of reduced density matrices is introduced in~\cite{huang2021power} to obtain better quantum kernels that can also learn the data derived from the discrete-log problems in~\cite{liu2021rigorous}.

\subsubsection{Swap-test Classifier}

The Swap-test classifier as proposed in
\cite{blank2020quantum} is implemented as a distance-based quantum classifier where the kernel is based on the quantum state fidelity raised to a certain power at the cost of using multiple copies of training and test data. The choice of the quantum feature map plays a pivotal role in defining the kernel and the overall efficiency of the classifier. The training and test data are encoded in a specific format following which the classifier is realized by means of the swap-test \cite{buhrman2001quantum}.

\begin{figure}[tb]
\vspace{-1cm}
\centering
\includegraphics{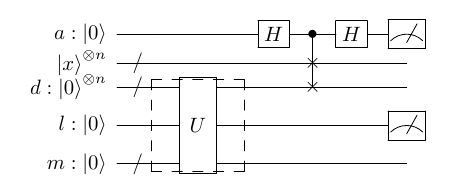}
\caption{Swap test Classifier : The first register is the ancilla qubit $(a)$, the second contains n copies of the test datum $(x)$, the third are the data qubits $(d)$, the fourth is the label qubit $(l)$ and the final register corresponds to the index qubits $(m)$. An operator $U$ creates the input state necessary for the classification protocol. The swap-test and the two-qubit measurement statistics yield the classification outcome.} 
\label{fig:swap_test}
\end{figure}


The swap test measures the similarity between the input quantum state and the reference quantum states for each class using measurements to compute a similarity score that indicates the overlap between the input state and the reference states.

\subsubsection{Variational Quantum Circuits (VQC)}

These algorithms primarily focus on optimizing the parameters of the PQC and known to provide a general framework that is compatible with different classes of problems leading to different structures and grades of complexity. The optimization is performed classically while allowing the circuit to remain shallow making it a versatile tool for near term quantum devices.

This basic structure of VQC involves the following three steps :
\begin{itemize}
    \item \textbf{Quantum Feature Map} : A non-linear feature map $\phi$ is employed to encode the classical data $x$ to the quantum state space. This is done by applying the circuit $U_{\phi(x)}$ to the initial state $|0\rangle^{\otimes n}$ :
    $$ | \Phi({x})\rangle = U_{\phi(x)}|0\rangle^{\otimes n}.$$
    The initial state $\ket{0}^{\otimes{n}}$ can be replaced by any fiducial quantum state as shown in~\cite{glick2024covariant}. The encoding circuit  $U_{\phi(x)}$ can also be applied more than once and/or interleaved with the model circuit described later. 

 \begin{figure}[!thb]
 \hspace{1.5cm}
\centering
\includegraphics{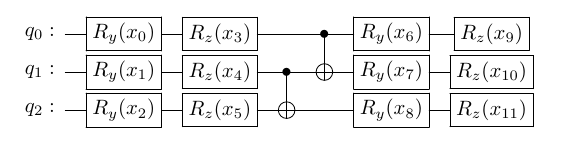}
\caption{The Feature Map $(\Phi(x))$ depicted above utilizes 3 qubits to encode 12 parameters of the data point $x$. It applies Rotation-Y and Rotation-Z operations to the feature values while employing CNOT gates to establish entanglement.}
\label{ref:feature_map_vqc}
 \end{figure}

    \item  \textbf{Model Circuit} : A short-depth parameterised quantum circuit \textbf{$W(\theta)$} is applied on the obtained quantum state with layers that are parameterized by the rotational angles for the gates that needs to be optimized during training. The optimization is performed over a cost function. 
    \item \textbf{Measurement and Preprocessing} : The outcome of the measurement results in a bit string $z \in \{0,1\}^n$ that is mapped to a label. This circuit is re-run multiple times and sampled to estimate the probability of observation $z$ which can be obtained as  
    $$  \langle \Phi(x) | W^{\dagger}(\theta)M_y W(\theta)|\Phi(x)\rangle$$
   which is calculated for each of the different classes $y$ using the measurement operator $M_y$.
   
\end{itemize}

At the aforementioned quantum feature map and the model circuit, the CZ and CNOT (along with Hadamard gate) are commonly used to create entanglement. A common strategy to optimize the sub-circuit for entangling qubits is to entangle adjacent qubits, namely, we first entangle the $2i$-th qubit with the $2i+1$, and then after this, we only entangle the $2i+1$-th qubit with the $2i+2$-th qubit for $i=0,\ldots n$. By doing so, we can parallelize the entanglement operation and reduce the execution dependency \cite{wu_vqc_data}. The circuit for this is shown in Figure \ref{fig:parallel_entanglement}. Based on the depth of the circuit chosen, we can repeat the quantum feature map with entangling sub-circuit, or the model circuit with entangling sub-circuit.

\begin{figure}[!hb]
    \centering
    \includegraphics{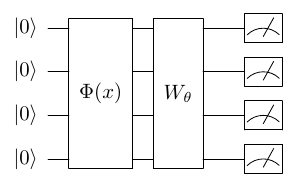}
    \caption{Variational Quantum Classifier : The state preparation $\Phi(x)$ followed by $W(\theta)$ which is the parameterized circuit with parameters $\theta$ followed by measurement in the $Z$-basis.} 
\label{fig:vqc_overall}
\end{figure}
 
\paragraph{Ansatz}

The choice of the ansatz also plays a pivotal role as the parameters $\theta$ of the circuit represented by $W$ are optimized during the training. For example, the experiments in \cite{perez2021determining} showed better performance for the weighted ansatz (Section \ref{drc}) in comparison to the Fourier ansatz (inspired from \cite{schuld2021effect}), which introduces linear and logarithmic dependencies to the same gate without using tunable weights. The authors speculate that using weights allows better representability especially for smaller layers. 
The layered ansatz is another common technique that comprises of layers such that each of the layers is composed of a set of entangling gates preceded by two alternating single-qubit rotation gates. The number of layers is an hyperparameter.

\begin{figure}[!t]
\centering
\includegraphics{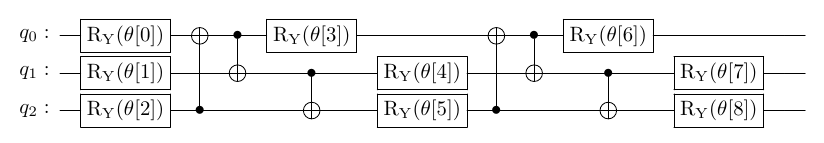}
\caption{The above representation illustrates $W(\theta)$ as a hardware-efficient ansatz for a 3-qubit system, employing a parameterized quantum circuit with 9 parameters.}
\end{figure}

 
\paragraph{Modified Layerwise Learning}
\label{section:layerwiselearning}
 
 As the name suggests, this strategy involves training the circuit layer by layer such that only a small set of parameters are optimized in a single update~\cite{skolik2021layerwise}. Initially, only a small circuit with few start layers is chosen such that all parameters are set to 0. This circuit is optimized by running it for a few epochs. The parameters are now frozen and a new set of layers is added. Now, the new layers' parameters are optimized with the previous layers' frozen parameters until no more improvement is obtained in the cost function or until the desired depth is reached. Then, the circuit depth is fixed and a larger set of the parameters is trained again. This strategy can help avoid the barren plateau due to the small number of layers and also maintains a favorable signal-to-noise ratio. \cite{skolik2021layerwise, muten2021modified}.

\paragraph{Optimizers}
The work in \cite{pellow2021comparison} demonstrated that gradient-free optimizers, Simultaneous Perturbation Stochastic Approximation (SPSA) and Powell’s method, and the gradient-based optimizers, AMSGrad and BFGS performed the best in the noisy simulation, and appeared to be less affected by noise than the rest of the methods. SPSA appeared to be the best performing method while COBYLA, Nelder-Mead and Conjugate-Gradient methods were the most heavily affected by noise even with the slightest noise levels.

In \cite{benedetti2021hardware}, the authors explored hardware-efficient ways to optimize the Ansatz using algorithms based on tensor methods. One of the introduced methods, called coordinate-wise optimization \cite{vidal2018calculus,nakanishi2020sequential,parrish2019jacobi,ostaszewski2021structure}, does not require the use of gradient or Hessian computations. Additionally, the authors of \cite{mc2023classically} presented a non-gradient-based variational algorithm by introducing an unconventional hybrid quantum-classical algorithm that utilizes the quantum part only once after optimizing the circuits entirely on a classical computer. This is in contrast to traditional methods that involve multiple uses of the quantum computer during circuit optimization.

More recently, the work in \cite{wiedmann2023empirical} proposed a novel approach that combines the approximated gradient from Simultaneous Perturbation Stochastic Approximation (SPSA) with classical optimizers. This hybrid approach outperformed standard SPSA and the parameter-shift rule in regression tasks, demonstrating enhanced convergence rate and error reduction, especially when considering noise.


Even the choice of batch size for training affects the convergence rate. In principle, quantum computing allows encoding a batch of training inputs into a quantum state in superposition and feed it into the classifier, which can be used to extract gradients for the updates from the quantum device. However this would extend the time complexity of state preparation routine for general cases, and even worse for more sophisticated feature maps. Single-batch stochastic gradient descent, where only one randomly sampled training input is considered in each iteration, can have favourable convergence properties, especially in cases where there is a lot of data available \cite{bottou2010large}. However in \cite{ngairangbam2022anomaly}, training using single data per update  led to slow convergence with volatile validation loss per epoch which was avoided by increasing the batch size to 64. 

Some of the major limitations associated with classical optimizers are repeated measurements and the complexity of gradient calculation \cite{chen2024pure}. Classical optimizers often require repeatedly measuring the outputs of a quantum circuit and feeding them into the classical computer. This process can lead to slower convergence rates for the optimization algorithm. The complexity of calculating gradients can impact the convergence of the optimization algorithm, particularly as the feature size ($d$) of the input increases. For instance, gradient-based methods like gradient descent have a complexity of $O(d)$ \cite{gao2021quantum}, which can become a scalability bottleneck.
To address these limitations, researchers have proposed quantum gradient methods \cite{jordan2005fast, wiersema2024here, gilyen2019optimizing} in the recent past as potential alternatives. These methods aim to leverage the benefits of quantum computation to overcome the challenges associated with classical optimization. However, their practical implementation still faces challenges related to applicability and complexity.

\paragraph{Parameter shift rule}
To optimize the objective, it is useful to have access to exact gradients of quantum circuits with respect to gate parameters. The parameter update requires computing $\nabla{L}(\theta)$ which in turns requires computing the gradient of the quantum circuit output $f$ due to the chain rule since the loss function is the function of the output of the quantum circuits. The gradient of the quantum circuits output is calculated using the parameter-shift rules~\cite{mitarai2018quantum, schuld2019evaluating} by varying the value of the gate parameters $\theta$ slightly.

For the gates used in angle encoding, the parameter-shift can be applied as
$$\frac{\partial f}{\partial \theta} = \frac{1}{2}[f(\theta + \frac{\pi}{2}) - f(\theta - \frac{\pi}{2})] $$
In other cases, different strategies can be applied as discussed in \cite{schuld2019evaluating}. When the ansatz consists of single-qubit rotation gates $R_x(\theta), R_y(\theta), R_z(\theta)$ as in Figure~\ref{fig:quantum_gates}, the loss function can be optimized with gradient-free optimizers using coordinate descent~\cite{ostaszewski2021structure,nakanishi2020}. While the gradient-based optimizers can be parallelized~\cite{tilly2022variational,QUDIO2021}, the gradient-free coordinate descent methods are sequential but have been shown to converge to local optima faster~\cite{tilly2022variational}, \textcolor{black}{and can be combined with \textit{data parallelism} to run several small quantum circuits on a device with larger number qubits as shown in~\cite{kriieeeqce23}}. Generalization of gradient-free sequential single-qubit gate optimizers are derived in~\cite{watanabe2021,WRSW_FQS_2022}. Nevertheless, within the existing training framework for Quantum Neural Networks (QNNs), it is necessary to compute gradients with respect to the objective function directly on the quantum device. However, this computation faces significant scalability challenges and is susceptible to hardware limitations and sampling noise inherent in the current generation of quantum hardware \cite{abbas2024quantum}.

In a recent study, \cite{kulshrestha2023learning} presented an alternative training algorithm that circumvents the need for gradient information. They introduced a novel meta-optimization algorithm, which involves training a meta-optimizer network to generate optimal parameters for the quantum circuit. These parameters are carefully chosen to minimize the objective function without relying on traditional gradient-based approaches.

\paragraph{Quantum Natural Gradient}
\label{section:quantumgraddescent}
The geometry of the parameter space plays a huge role in the efficient optimization of the VQC parameters\cite{neyshabur2015path}. In \cite{blance2021quantum}, the authors expect that a smaller network structure of the VQC can lead to significant advantage as it allows using a computationally more expensive optimization algorithms resulting in a faster learning rate. This is also advantageous when the training data is limited.

In vanilla gradient descent, the loss function $L(\theta)$ is minimized in the $l2$ vector space by updating the network parameter $\theta^{(t)}$ at time $t$ to $\theta^{(t+1)}$ in the direction of the steepest slope as
$$\theta^{(t+1)} = \theta^{(t)} - \eta\nabla L(\theta)$$

Since each of the model parameters are updated by the same Euclidean distance, there is a possibility of getting stuck in the local minima since the value of $f(\theta)$ varies at different rate with respect to each parameter. This is tackled in natural gradient descent where the parameter space corresponds to Riemannian geometry which is defined by the Fisher Information Matrix \cite{amari1998natural, amari2019fisher} and is invariant under re-parametrisation. The parameters are updated as
$$\theta^{(t+1)} = \theta^{(t)} - \eta F^{-1}\nabla L(\theta)$$
where $F$ is the Fisher Information index. The calculation of $ F^{-1}$ in general is computationally expensive. However, this leads to faster convergence, and can help avoid getting stuck in local minima~\cite{yamamoto2019natural}.

For VQC parameter optimization, it has been shown that using the standard Euclidean geometry is sub-optimal\cite{harrow2021low}. The quantum gradient descent is the quantum version of natural gradient descent which uses the Fubini-Study metric $g$  \cite{cheng2010quantum,brody2001geometric}. This  Fubini-Study metric tensor is unique invariant metric tensor to the space of quantum states and exploits the geometric structure of the VQC's parameter space. The parameters are updated as
$$\theta^{(t+1)} = \theta^{(t)} - \eta g^{+}\nabla L(\theta)$$
where $g^{+}$ is the pseudo-inverse of the Fubini-Study metric $g$.
Faster convergence has been observed for quantum gradient descent compared to the vanilla gradient descent with similar number of trainable parameters~\cite{blance2021quantum}.

\paragraph{Quantum Natural SPSA}

The large computational costs associated with calculating the Quantum Fisher Information(QFI), which scales quadratically in the number of ansatz parameters, limits the advantage of using quantum gradient descent over standard gradients.  To counter this, a new approach is introduced in \cite{gacon2021simultaneous}, Quantum Natural-Simultaneous Perturbation Stochastic Approximation(QN-SPSA), which inherits fast convergence and robustness of quantum natural gradient with respect to the initial parameters, while having the computational cost benefits of SPSA~\cite{spall1998overview}. 

Additionally, it is worth mentioning some recent works, such as the Pure Quantum Gradient Descent Algorithm, which was proposed in a recent study \cite{chen2024pure}. This innovative quantum-based method for gradient calculation claims to provide a theoretical computational complexity of $O(1)$  in contrast to the $O(d)$ complexity of the classical algorithm \cite{gao2021quantum}.

\subsubsection{Quantum Principal Component Analysis (PCA)}
PCA has been used for the optimal low-rank approximation of a matrix through spectral decomposition by setting a threshold on the eigenvalues. By doing so, we only retain the principal components of the spectral decomposition while discarding those with the smaller eigenvalues. However, when the size of the matrix is large, the computational costs increase which is why we look at quantum algorithms.

The implementation of quantum PCA in \cite{lloyd2014quantum} helps construct the eigenvectors and eigenvalues of the unknown density matrix thereby discovering their properties. The authors assume that the matrix can be represented by a quantum state, i.e. it is a non-negative matrix with trace equal to one, which covers a wide range of interesting cases. It uses multiple copies of an unknown density matrix to construct the eigenvectors corresponding to the large eigenvalues of the state (the principal components) in time $O(\log N)$ where $N$ is the dimension of the Hilbert space, resulting in an exponential speed-up over existing algorithms. They provide novel methods of state discrimination and cluster assignment.
 
\subsubsection{Quantum Orthogonal Neural Networks}
Orthogonal neural networks are neural networks with orthogonal trained weight matrices which provide the advantage of avoiding vanishing gradients and improved accuracies~\cite{li2019orthogonal}. The PQC for implementing the orthogonal neural networks was first introduced in \cite{kerenidis2021classical} using unary amplitude encoding and a pyramidal structure using only RBS gates. The orthogonality of the weight matrix is preserved by performing gradient descent on the parameters of the quantum circuit. This works because a quantum circuit with real-valued unitary gates is an orthogonal matrix hence the gradient descent is equivalent to updating the weight matrix. Another feature of the circuit is one-to-one mapping between the parameters of the orthogonal  matrix and the quantum gates of the circuit. The circuit architecture benefits from linear circuit depth and error mitigation due to unary encoding along with nearest neighbor connectivity due to the distribution of the RBS gates. In \cite{mathur2021medical,kerenidis2021classical}, the results show linear scaling of the training run time with respect to the number of parameters.

\subsubsection{Quantum Generative Adversarial Networks}
The primary goal of a classical generative adversarial network (GAN) \cite{goodfellow2020generative} is to generate data by studying a collection of training examples and learning the underlying probability distribution. It typically involves an iterative adversarial training procedure between two neural networks, the discriminator and the generator model. The generator creates fake data with the goal of generating data as close as possible to the real training dataset while the discriminator tries to separate this fake data from the real data. 

The quantum variant of GAN (QGAN) was proposed independently in \cite{QGAN1, QGAN2} where a QNN is used as the discriminator or generator or both. In \cite{bravo2022style}, faster convergence was noted for a classical discrimator in comparison to other architectures. For more details refer to \cite{QGansurvey}. 

\color{black}
The research described in \cite{hu2019quantum} showcased the first proof-of-concept experimental demonstration of Quantum Generative Adversarial Networks (QGAN) on a superconducting processor. The authors successfully trained a quantum-state generator through adversarial learning to replicate quantum data with 98.8\% fidelity. 
However, the scalability of QGANs to noisy intermediate-scale quantum devices was first shown in~\cite{huang2021quantum}   by implementing the QGAN using a programmable superconducting processor.

\color{black}

\paragraph{Quantum Adversarial Learning}
Adversarial machine learning involves assessing vulnerabilities of machine learning in adversarial settings and consequently implementing techniques to make the models more robust to such manipulations. In the quantum setting, \cite{lu2020quantum} shows that a quantum classifier which performs with nearly the state-of-the-art accuracy can be deceived by adding unnoticeable perturbations to the original samples. For more information and discussions on the latest advancements and key challenges in the field of Quantum Adversarial Machine Learning, we refer the reader to \cite{west2023towards}.

\subsubsection{Tensor Networks}
Tensor networks (TN) are a popular method in the field of quantum many-body problems due to their ability to represent many-body localized systems and are already known for their performance in the classical setting for supervised and unsupervised learning tasks. TNs can represent both quantum states and circuits \cite{tnn3, tnn4, tnn5} using VQCs with rules described in \cite{huggins2019towards}. They can also simulate strongly entangled quantum systems \cite{tnn1, tnn2}. Depending on the architecture,  the number of physical qubits scales only logarithmically with, or independently of the input or output data sizes which can be implemented on small, near-term quantum devices using lesser physical qubits. The work in \cite{araz2022classical} shows that classical TNs require exponentially more trainable parameters and higher Hilbert-space mapping to perform on par with the quantum counterparts which makes them vulnerable to a highly flat loss landscape. 
The work in \cite{martin2023barren} explores the trainability of randomly initialized quantum tensor networks with a focus on the different architectures.
The conjecture suggests that classical gradient computation for quantum tensor networks could be more efficient than their quantum counterparts. A review can be found in \cite{liu2023tensor}.

\begin{figure}
\centering
\subfloat[Parallel]{\label{fig:parallel_entanglement}
\includegraphics[width=6.3cm]{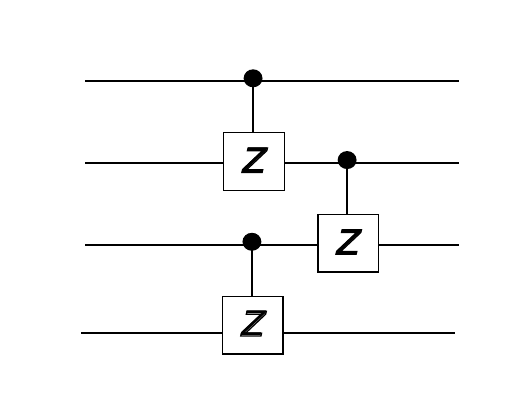}
}\hspace{10mm}
\subfloat[All to all]{\label{fig:all_ent}
\includegraphics[width=7.5cm]{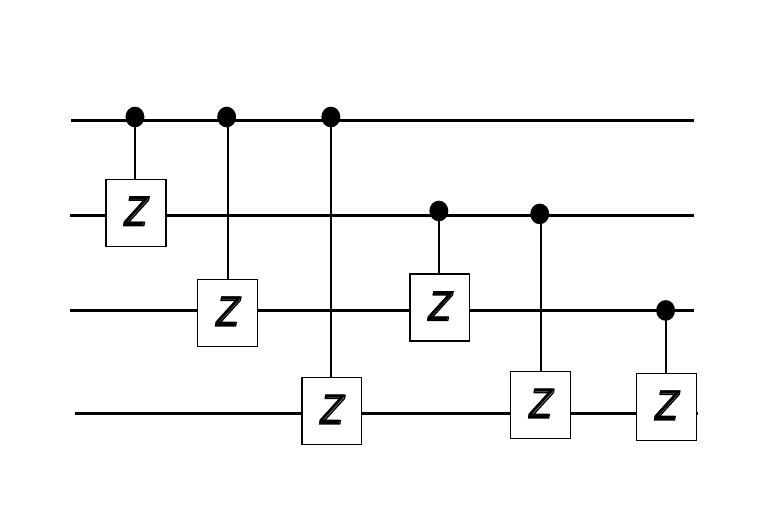}}
\\
\vspace{3mm}
\subfloat[Circular]{\label{fig:circular_ent}
\includegraphics[width=7cm]{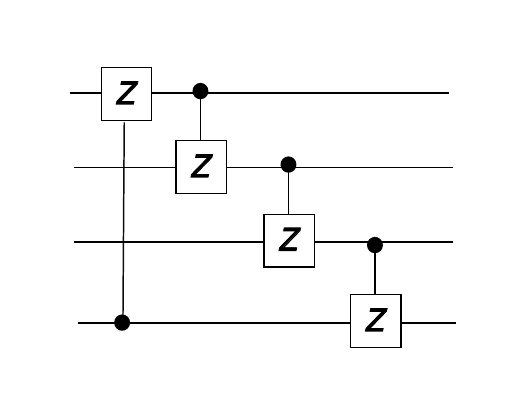}}
\hspace{10mm}
\subfloat[Linear]{\label{fig:linear_ent}
\includegraphics[width=6.8cm]{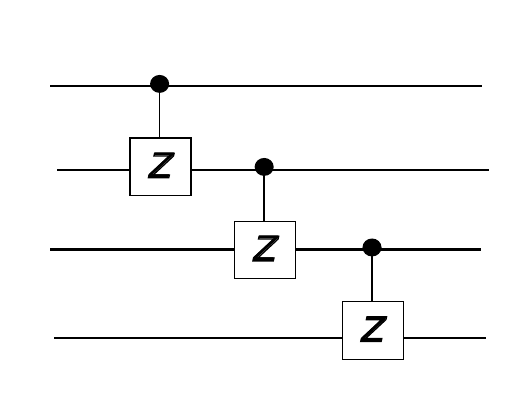}
}
\caption{Different types of entanglement}
\label{fig:ent}
    \end{figure}


\subsubsection{Quantum Autoencoder}

The task of a classical autoencoder is to obtain a low level representation of a given input such that the original data can be recovered. This has applications in dimensionality reduction and generative data model. The quantum version of the classical encoder was first implemented in \cite{qae1} where an ansatz is trained to obtain a compressed version of an ensemble of pure quantum states. Different variants are explored in \cite{autoencoder3, autoencoder4, autoencoder5}. The learning task involves finding unitaries that preserve the quantum information of the input through the smaller intermediate latent space. The PQC initially encodes the input state into an intermediate latent space. Following this, the decoder acts with the goal of being able to reconstruct the input. A cost function is used to estimate the fidelity (distance) between the input and output states.
\color{black}
Recently, quantum autoencoders have found application in quantum error correction \cite{locher2023quantum},  thereby expanding the possibilities for further research and development in the field.
\color{black}

\section{Applications}
\label{sec:applications}
The performance metric is usually measured using AUC-ROC which stands for \textit{Area under the ROC Curve}~\cite{hand2001simple,fawcett2006pattern}. The ROC curve (receiver operating characteristic curve) is a commonly used graph that summarizes the performance of a classifier over all possible probability thresholds. The AUC-ROC provides intuition about the capability of the model to distinguish accurately between true positives and false positives (True Positive Rate (TPR) on the $y$-axis and the False Positive Rate (FPR) on the $x$-axis). The score varies from 0 to 1 where higher score implies better distinction/performance and score 0.5 corresponds to random guessing.

\subsection{High Energy Physics}
Most of the studies focus on obtaining better performance with limited data \cite{terashimachine}. 
Among recent works, VQC has been used widely for HEP based applications with data agnostic techniques for feature encoding like single qubit rotation gate or ZZ-gate \cite{terashimachine}. However, these methods are not suitable for HEP as they end up incurring large overhead on the number of qubits or gates for multi-dimensional data.
Here we review the applications of VQC and QSVM in HEP.

\subsubsection{Classification}
A prominent use case under this category is the Event Classification which involves discriminating signal events from the background events in the context of the Standard Model of particle physics\cite{terashi2021event, muten2021modified}.

In \cite{terashi2021event}, the data is encoded using first-order encoding (FOE) and a variational part based on a simplified version of \cite{havlivcek2019supervised} with depth 1. A combination of three variables is determined using a DNN with AUC-ROC and run on the IBM Quantum device \citep{ibmq} with 3 qubits. The cross-entropy cost function is optimized using COBYLA. Results showed a higher cost function with more fluctuations for the real device compared to the simulator, but both showed consistent AUC (around 0.80) within the standard deviation. Additionally, second-order encoding (SOE) was employed, but no improvement was observed on a real quantum computer. This may be attributed to the 60\% increase in single and two-qubit gates when transitioning from FOE to SOE, resulting in increased hardware noise due to gate errors.

In \cite{muten2021modified}, an improvement over this method is shown using data re-uploading and modified layer-wise learning with only 1 qubit and 5 layers. However, training is performed on the PennyLane simulator, optimizing the MSE cost function with the Adam optimizer. Inference tests on Rigetti's 32-qubit superconducting quantum processor obtained an AUC of 0.830, surpassing \cite{terashi2021event} while using fewer qubits. Training and testing AUC using 2000 samples demonstrate that data re-uploading generalizes well without overfitting or underfitting.

To compare the performance of variational circuit-based and kernel-based methods on the same dataset, we refer to \cite{wu_vqc} and \cite{wu_kernel}, which use VQC and QSVM, respectively. Both works employ PCA~\cite{pca1, pca2} for data preprocessing, matching the number of encoded variables with the available qubits of the IBM Quantum device \cite{ibmq}, followed by angle encoding.

The ansatz in~\cite{wu_vqc} uses parallelized entangling CZ gates with linear qubit connectivity (Figure~\ref{fig:parallel_entanglement}). For training on real hardware, the feature map and variational circuit depth were set to 1. SPSA was used for optimization, and the results were benchmarked against classical SVM~\cite{svm1} and binary decision tree~\cite{xgboost}. The simulator performance was comparable to classical methods (AUC around 0.82). To minimize readout errors, only half the qubits were observed after pairing them with CZ gates. The performance on real quantum hardware was similar (AUCs > 0.80) to the simulator. However, the authors note the long training time on quantum hardware (200 hours) for 500 training iterations on 100 events.

For the QSVM-based approach in~\cite{wu_kernel}, a parallel entangling circuit was constructed using 15 qubits of the IBM Quantum device, similar to~\cite{wu_vqc}, to obtain a short-depth circuit for execution on real quantum hardware. To reduce statistical uncertainties, 8192 measurement shots were performed for each kernel entry. The hardware performance approached that of the noiseless simulator for small training samples of size 100 (average AUC: 0.831).

 The results of testing on IBM Quantum device showed that the simulator-trained model performed well even on the hardware with AUC around 0.78. The authors note that the faster learning rate despite the computationally expensive optimization algorithm is tied to the small structure of the VQC.

Testing the model on an IBM Quantum device showed good performance, with an AUC of approximately 0.78. The authors attribute the faster learning rate, despite the computationally expensive optimization algorithm, to the small structure of the VQC.

\textcolor{black}{In~\cite{huang2022quantum}, the authors note data obtained from quantum-enhanced experiments can achieve quantum advantage for learning tasks of a physical state and a physical process from a perspective of sampling complexity. Quantum-enhanced experiments consist of quantum sensors, quantum memory, and quantum computers.
In quantum-enhanced experiments, quantum information is directly stored in quantum memory while classical experiments require measurements to store classical data in classical memory. 
Quantum-enhanced experiments preserve quantumness of quantum data until performing entanglement measurements on pairs of copies of data in quantum memory. It is expected that research utilizing the power of quantum data will become increasingly active in the future.  
}

\subsubsection{Regression}

\paragraph{Simulation}
The use of different variants of QGAN architectures in HEP can be seen in \cite{chang2023running} and \cite{bravo2022style} for simulation. For example, in \cite{chang2023running}, the proposed QGAN contains a classical discriminator and two parameterized quantum circuits generators for generating images. The performance was measured via relative entropy and individual relative entropy. The model was trained using a simulator while the inference results of the pre-trained model on superconducting chips and ion-trap machines showed low standard deviation and error rates indicating the feasibility of dual-PQC training for superconducting chips. However, the authors note the vulnerability of the training process of falling into  mode collapse \cite{goodfellow2020generative} where the model only reproduces a low variety of samples. They also suggest techniques such as increasing the training set size and adding an additional term to the loss function as possible solutions to ameliorate the problem.
 
 In contrast, the QGAN in \cite{bravo2022style}, named style-QGAN, was implemented using a QNN generator and a classical NN discriminator. The data was encoded using angle encoding and the cross-entropy loss function was optimized for both using Adadelta \cite{adadelta}. While earlier implementations of QGAN provided the prior noise distribution to the generator via the first input gates, the work in \cite{bravo2022style} embeds it on every layer of single qubit and entangling gate in the circuit. The results showed an improvement over the state-of-the-art with shallow circuits on both the 3-qubit superconducting and ion-trapped architectures implying potential hardware independent viability. Additionally, both the quantum hardware are able to capture the correlations even on small sample set.

\subsection{ Healthcare }
A few applications of quantum machine learning in healthcare include Healthcare diagnostics and treatment, cancer detection, prediction of different stages of diabetes and even the security of sensitive information such as healthcare data. 

Given the sensitivity of these applications, the cost of any incorrect predictions may have huge negative consequences and hence requires utmost carefulness. In this regard, binary classification of MRI images using a VQC is performed in \cite{ren2022experimental} to check the vulnerability of quantum learning systems in Healthcare diagnostics. 
 To prepare highly-entangled multi-qubit quantum state, interleaved block-encoding \cite{reupload, blockencoding1, blockencoding2} was used on 10 qubits. The variational parameters are fixed for adversarial perturbations and the results show that the original differ from the adversarial images by a small amount of perturbations. The results shows that the quantum classifier predicts the legitimate states accurately while mispredicting all (half) of the adversarial examples highlighting the vulnerability aspect. Additionally, experiments were performed on quantum data as well with the quantum classifier reaching perfect accuracy in about 30 epochs on both train and test datasets.

In \cite{mathur2021medical}, QNN and quantum orthogonal neural networks are used for Healthcare  image  classification on RetinaMNIST dataset and PneumoniaMNIST~\cite{PNEUMONIA_DS}. The images were pre-processed using PCA and followed by unary amplitude encoding. A series of experiments was performed on the real hardware using 5 and 9 qubits. The results show comparable accuracies for majority of the classification experiments performed on real quantum hardware to that of their classical counterparts. However, the hardware limitations come into picture for more difficult tasks.  Additionally, circuit optimization based on the hardware and translation of the RBS gates into native hardware gates was performed to reduce the overall gate count. The results show better performance for 5-qubit results in contrast to the 9-qubit experiments where the hardware performance seems to diverge from the simulator performance. The authors note the unstable performance of the quantum hardware due to the randomness in the training or inference making it incapable of performing Healthcare image classification on par with the classical models.

To analyse the advantage of using quantum machine learning in terms of sample complexity, the work in \cite{krunic2022quantum} conduct experiments using QSVM on small dataset of size 200–300 training samples uses kernel techniques for prediction on six-month persistance of rheumatoid arthritis. The experiments were conducted on different configurations of features and data sizes to identify cases where quantum kernels could provide advantage. A new metric, Empirical Quantum Advantage (EQA), is proposed to quantitatively estimate the accuracy of the model performance as a function of the number of features and sample size. The estimation of the custom kernel turns out to be the most computationally expensive task. The authors claim to be the first ones to use geometric difference to analyse the relative separation between classical and quantum feature. They note that kernels are noisy and that quantum advantage expressed in terms of the generalization error vanishes with large datasets, fewer measurements, and increased system noise.

\subsection{ Finance }
Some applications of ML operations applicable to finance include regression for asset pricing \cite{bagnara2022asset, nagel2021machine, gu2020empirical},
classification for portfolio optimization \cite{ma2021portfolio,ban2018machine,chen2021mean,perrin2020machine,ta2018prediction,conlon2021machine}, clustering for portfolio risk analysis and stock selection \cite{fu2018machine,rasekhschaffe2019machine,yuan2020integrated,yan2007machine, kumbure2022machine}, generative modeling for market regime
identification, feature extraction for fraud detection\cite{awoyemi2017credit, varmedja2019credit,
perols2011financial,
dornadula2019credit,
thennakoon2019real,
bauder2017medicare,
yee2018credit}, reinforcement
learning for algorithmic trading \cite{pricope2021deep,theate2021application,li2019deep,cumming2015investigation}, and Natural Language Processing
(NLP) for risk assessment \cite{fisher2016natural,
gao2021review}, financial forecasting \cite{di2018multitask,
kamalov2021financial,
sarangi2020literature,
wasserbacher2022machine,
krollner2010financial,
kwong2001financial} and accounting and
auditing \cite{koshiyama2021towards,roszkowska2021fintech,fisher2016natural}. In a similar vein, QML has been used for different applications such as feature selection for fraud detection in \cite{zoufal2023variational} where a PQC was trained on a subset of good features selected based on their performance using a predefined metric. The use of QN-SPSA showed good convergence for training on 20 qubits with potential for deeper circuits. The results on hardware were comparable to state-of-the-art classical methods in certain aspects while in others it showed the potential to find better subsets. The authors of ~\cite{zoufal2023variational} note that model run on IBM Quantum device was able to outperform traditional methods without using error mitigation.
\color{black}
A recent study by \cite{xia2023configured} introduced an innovative method for Quantum Reservoir Computing (QRC) and applied it to the foreign exchange (FX) market. The approach accurately depicted the stochastic evolution of exchange rates compared to classical methods.  In QRC, input signals are transformed into a complex quantum superposition in a high-dimensional space, after which the transformed signals are connected to the desired output through a basic neural networks. The study highlights the learning performance and potential of QRC to be run on near term quantum devices.

\color{black}
Another application of QML was explored in \cite{martin2021toward} to reduce the number of noisy factors for pricing interest-rate financial derivatives using a qPCA. The experiments were performed on 5-qubit IBM Quantum device for $2 \times 2$ and $3 \times 3$ cross-correlation matrices based on historical data for two and three time-maturing forward rates.  However, this method showed difficulty in scaling to larger datasets.

We direct the readers to \cite{herman2023quantum} for a more comprehensive summary of the state of the art of quantum computing for financial applications.

\section{Limitations }
\label{sec:limit}

The current quantum hardware is susceptible to noise resulting in a very low qubit coherence time of the order of a few hundred microseconds.
Common sources of noise include (1) crosstalk due to simultaneous gate execution in quantum algorithm that allow parallel operations (2) quantum decoherence (3) single qubit/rotation and two-qubit gate error rate due to imperfect implementation and (4) shot noise from measurements on quantum
states. Additional limitations due to qubit count and gate fidelity prevent the use of quantum error correction. The use of VQCs provide a framework to enable practical applications of noisy quantum hardware. Here, we briefly look at some of the limitations associated with the current QML approaches.

\paragraph{Hardware limitations}
The common causes of error rates are the State Preparation and Measurement Error Rate (SPAM) and gate errors. The SPAM measures the correctness of the initial calibration settings and the final read out measurement and is indispensible for scaling to hundreds or thousands of qubits. A general strategy to counter the noise in quantum hardware is to increase the number of measurements to help reduce the generalization error \cite{wang2021towards}. However, this may also be counterproductive due to the readout error during measurement. For example, the prediction accuracy dropped on increasing the number of shots from 500 to 1000 in \cite{johri2021nearest}. The authors note that the experiment was already dominated by systematic noise which was prone to changes every time the system was calibrated indicating the variability in the calibration of the system. Other options include using shot-frugal optimizers \cite{kubler2020adaptive,arrasmith2020operator, gu2021adaptive,sweke2020stochastic} which use a stochastic gradient descent based approach while adapting the number of
shots (or measurements) needed at each iteration.
A popular noise mitigation technique is the zero-noise extrapolation to first order for gate-error mitigation described in \cite{kandala2018extending, temme2017error} and can be implemented in software without requiring any prior knowledge of the quantum computer noise parameters. Factors such as qubit life time and coherence time are affected by decoherence. 
Decoherence, characterized by uncontrolled interactions between a quantum system and its environment, poses a significant challenge in quantum computing resulting in the loss of quantum behavior within the quantum processor, nullifying any potential advantages offered by quantum algorithms. The decoherence time limitation significantly restricts the number of operations that can be performed in a quantum algorithm. Additionally, the development of high-fidelity qubits poses another critical hardware challenge. To tackle these issues, an effective approach is to treat qubits as part of an open environment and leverage classical simulation software packages during the design phase.

Superconducting QPUs have a coherence time of around 100 microseconds while certain trapped ions have extended that to 50 seconds. The gate speed along with decoherence need to make sure that the gates are applied before the system decoheres. Superconducting and photonic generally have the fastest gate speeds. The qubit connectivity which is the general layout of the qubits dictates interaction between a given qubit and its neighbours. Due to limited connectivity, SWAP gates can be inserted but can result in additional overhead and subsequent error rates. While some device offer all-to-all connectivity, long-range gates are generally more noisy.
The delay between submitting a circuit to the cloud and receiving a result, without clarity on the calibration timings, can lead to significant statistical errors \cite{blinov2021comparison} as it is unclear on how these errors influence circuit performance between runs on all systems. The lack of information on aspects such as qubit assignment, compiler/transpiler methods, component drift rate, and time since last calibration also affect the analysis as noted in \cite{blinov2021comparison}.

\paragraph{Long running time}
Often, studies require large number of samples and qubits (20 qubits or more) which necessitates large amount of computational power for quantum computer simulations. Long running times have been noted in \cite{araz2022classical, wu2022challenges, wu_vqc, wu_kernel, krunic2022quantum} on current quantum hardware, even when using small data samples, likely due to the initialization, queuing, execution and measurement time in the current quantum hardware.  For example, the study in \cite{wu_vqc} took around 200 hours to run 500 training iterations on 100 events on quantum hardware. This poses a serious limitation for real world applications such as HEP which generally require large training data. In terms of the model performance, using small sample size often leads to significant variance and poor performance. Furthermore, the limited access to QPU resources makes it infeasible to conduct validation on multiple sets \cite{krunic2022quantum}. 

In \cite{wack2021quality}, the authors propose measuring the speed using circuit layer operations per second (CLOPS) by considering the interaction between classical and quantum computing. The CLOPS benchmark consists of 100 parameterized templated circuits and takes into account various factors such as data transfer, run-time compilation, latencies, gate times, measurements, qubit reset time, delays, parameter updates, and result processing. However, CLOPS focuses mainly on the quantum computing aspect and considers classical computation as an auxiliary to quantum computing. Furthermore, factors such as qubit quality and gate operations are not captured in the metric. Experimental results indicate that the execution time of quantum circuits constitutes a small proportion (less than 1\%) of the total execution time \cite{e24101467,wack2021quality}.

Another proposed solution to improve the training time is Quantum Federated Learning (QFL), which uses distributed training across several quantum computer. Federated learning consists of several clients or local nodes learning on their own data and a central node to aggregate the models collected from those local nodes. 
A framework for federated training was presented in \cite{chen2021federated} using hybrid quantum-classical machine learning models. Their simulation results show faster convergence compared to the non-federated training and the same level of trained model accuracies. Other works include \cite{yun2022slimmable} where they introduce slimmable QFL (SlimQFL), a dynamic QFL framework which has shown to achieve higher classification accuracy than the standard QFL. 

In contrast, ensemble learning involves the combination of multiple individual models, referred to as base models or weak learners, to create a more accurate and robust predictive model. These base models can be of the same type or different types, and their predictions are aggregated using methods such as voting, averaging, or weighted averaging. Ensemble learning aims to improve overall performance and accuracy by leveraging the strengths of multiple models.

On the other hand, federated learning is distinct from ensemble learning in that it enables collaborative training across distributed entities without sharing raw data, ensuring privacy and security. While ensemble learning focuses on model aggregation, federated learning emphasizes the distributed nature of training.

Some works that explore ensemble learning in the context of quantum machine learning include \cite{schuld2018quantum, wang2021quantum, izdebski2020improved, macaluso2020quantum, arunachalam2020quantum,west2023boosted}.

\paragraph{Inefficient data loader}
Being able to load classical data as quantum states efficiently is a bottleneck that has often been sidelined in works that discuss speedup using QML algorithms. Given a classical data point, the job of a data loader is to read the data once and output a PQC that prepares an appropriate quantum representation. The encoding part of input data generally consumes a significant portion of the coherence time often leaving little time for actual algorithm to process the data \cite{terashimachine}. Several proposals for more efficient data loading have been made in this regard. For example, the work in \cite{johri2021nearest} tries to tackle this by describing ways to load a classical data point with logarithmic depth quantum circuits while using the same number of qubits as the features dimension. Another technique is described in \cite{mathur2021medical} where a shallow parallel data loader is implemented for $d$-dimensional data points using $d$ qubits, $d$ - 1 RBS gates and circuits of depth only $\log d$. However, the viability of this approach is limited by connectivity requirements beyond those supported by the hardware. 

The idea of Quantum Random Access Memory (QRAM)\cite{rebentrost2014quantum, zhao2019quantum,giovannetti2008quantum} has been proposed for the long-term storage of the state of quantum registers and can be considered to be a specific hardware device that can access classical data in superposition natively, thus having the ability to create quantum states in logarithmic time. Despite challenges in implementation, alternative proposals with similar functionality have emerged. In \cite{arunachalam2015robustness}, a circuit with $O(d)$ qubits and $O(d)$ depth was described to perform the bucket brigade architecture with proven robustness to a certain level of noise.

\paragraph{Barren plateau }
 Flat optimization landscapes, where the gradient variance diminishes exponentially with the number of qubits, are commonly encountered in variational quantum algorithms. Similar to classical machine learning, quantum loss landscapes are susceptible to numerous local minima. Recent studies \cite{larocca2023theory,kiani2020learning} have demonstrated that overparameterization can help alleviate Barren Plateaus by utilizing more parameters than necessary for a given problem. This allows the Quantum Neural Network (QNN) to explore all relevant directions in the state space. However, factors such as ansatz architecture \cite{huang2021near, coyle2021quantum}, cost function \cite{cerezo2021cost}, and parameter initialization contribute to encountering Barren Plateaus \cite{zoufal2023variational}.
For instance, highly expressive ansatz \cite{hayashi2005reexamination} or ansatz with exhaustive entanglement \cite{marrero2021entanglement, sharma2022trainability, holmes2021barren} can result in exponentially flat landscapes as the number of qubits increases \cite{mcclean2018barren}. 
 Addressing these challenges involves employing adaptive initialization methods or informed parameter initialization, problem-dependent ansatz design, circuit pruning, utilizing density matrices and random features for distribution estimation, concurrent optimization of parameters and rotation generators, as well as the incorporation of global optimization techniques like genetic algorithms for enhancing gate or structural optimization. The work in \cite{altares2021automatic} focuses on supervised learning with quantum feature maps optimized using a genetic algorithm which designs feature map circuits with high accuracy, generalization, and minimal size, demonstrated through diverse benchmarks, suggesting potential for hybrid quantum-inspired machine learning strategies.

  Limiting entanglement in the ansatz \cite{patti2021entanglement} can help overcome exhaustive entanglement-induced Barren Plateaus \cite{marrero2021entanglement, sharma2022trainability, holmes2021barren}. The choice of observables to define the loss function also influences the presence of Barren Plateaus. Using global observables that require measuring all $n$ qubits simultaneously \cite{cerezo2021cost} can lead to Barren Plateaus, whereas employing local observables that compare quantum states at the single-qubit level \cite{uvarov2021barren, cerezo2021cost} can avoid this issue. Recent research \cite{garcia2023barren} has shown that local cost functions encounter Barren Plateaus when learning random unitary properties. 
Furthermore, local noise in the hardware \cite{wang2021noise} can affect the optimization process. Techniques such as error mitigation \cite{temme2017error, piveteau2021error} can help reduce the impact of local noise. Different ansatz designs, including Variable ansatz \cite{cerezo2022variational, cincio2018learning}, Hamiltonian Variational Ansatz \cite{hadfield2019quantum, wecker2015progress, park2024hamiltonian}, or Hardware-Efficient ansatz, which aim to reduce gate overhead \cite{google2020hartree, harrigan2021quantum}, can be utilized and optimized using quantum-specific optimizers \cite{kubler2020adaptive, lavrijsen2020classical} for training.

Gradient-based methods are generally preferred for large parameter spaces \cite{mitarai2018quantum}. However, gradient-free methods have also been utilized for optimization, as shown in \cite{peruzzo2014variational}, where Nelder-Mead was employed for QVE optimization. However, scaling results in \cite{mcclean2018barren} indicate that deep versions of randomly initialized hardware-efficient ansatzes suffer from exponentially vanishing gradients. As an alternative, one can opt for barren-plateaus-immune ansatzes \cite{cerezo2021cost, du2022quantum, pesah2021absence, zhang2020toward} instead of hardware-efficient ansatzes. Additionally, using shallow circuits with local cost functions \cite{cerezo2022variational, cerezo2021cost} can help mitigate the presence of Barren Plateaus. In \cite{cerezo2021cost}, an alternating layered ansatz is proposed, which was later proven to have sufficient expressibility \cite{nakaji2021expressibility}. The results in \cite{cerezo2021cost} demonstrate that the barren plateau phenomenon extends to VQAs with randomly initialized shallow alternating layered ansatzes and establish a relationship between locality and trainability of VQCs. They also show that despite using a shallow circuit, defining a cost function using global observables leads to exponentially vanishing gradients. Among other techniques, the initialization strategy using identity blocks described in \cite{grant2019initialization} and layer-wise training can be employed to mitigate Barren Plateaus.

In the study by \cite{heyraud2023efficient}, researchers developed a scalable method to calculate the gradient and its variance by proving that randomly initialized circuits can be exactly mapped to a set of simpler circuits that can be efficiently simulated on a classical computer.

The study in \cite{west2023reflection} highlights geometric quantum machine learning's potential for addressing barren plateaus and overfitting where quantum models are customized to reflect image symmetry. The results show an improvement in the accuracy compared to generic models while using amplitude encoding. The issue of overfitting in VQCs has also been addressed in \cite{west2023boosted}, where the authors implemented boosted ensembles of Quantum Support Vector Machines on HEP datasets. The ensemble classifier was found to double the efficiency of a single QSVM, which the authors claim is highly susceptible to overfitting.

\section{Open questions}
\label{sec:oq}
The key objective in the field of Quantum Machine Learning (QML) is to demonstrate quantum advantage, surpassing classical methods in data science applications either in terms of sample complexity or time complexity. This requires a flexible and exploratory approach to identify the areas where QML can have the greatest impact. Although there are claims of polynomial and exponential speed-ups in QML, empirical evidence establishing a clear advantage over classical algorithms is still limited. Furthermore, providing a robust theoretical foundation for quantum advantage poses significant challenges in the field. It remains unclear whether the observed performance improvements are solely attributed to careful hyperparameter selection, benchmarks, and comparisons, or if there is a fundamental structural advantage \cite{schuld2022quantum}. 
It can be observed that QML as a field is moving towards becoming an empirical science. The theoretical aspect of proving concepts is anticipated to be challenging, and the emphasis is increasingly placed on practical demonstrations. This trend is particularly notable as the number of qubits and circuit depth surpasses $100 \times 100$.
There is a possibility that an efficient classical algorithm exists for a given learning problem that can achieve comparable results to quantum learning algorithms. This is exemplified in \cite{havlivcek2019supervised} where the variational circuits can be replaced by a classical support vector machine if the encoding is classically tractable. 

Furthermore, due to finite sampling noise, none of the heuristic quantum learning algorithms have proven to solve a classically hard learning problem \cite{liu2021rigorous}. These inherent limitations imply that the current benefits of quantum algorithms can only be realized under certain circumstances. Specifically, only a few variational quantum-based algorithms have shown an apparent advantage in a constrained situation \cite{kwak2021quantum}. Recently, \cite{kreplin2023reduction} investigated the impact of finite sampling noise and subsequently introduced a technique called variance regularization based on the expressivity of QNNs to reduce the variance of the output. 

To shed light on the current state of quantum machine learning, several research directions and areas of investigation are identified:

\paragraph{Establishing Standardized Benchmarks}  In order to effectively evaluate the superiority of QML algorithms compared to classical ones, it is crucial to establish standardized benchmarks. Currently, standard classical data benchmarks such as MNIST, Iris etc are used (as shown in Table \ref{Tab:image}). The lack of standardized quantum datasets highlights the need for easily preparable quantum states to serve as benchmarks for evaluating QML models \cite{cerezo2022challenges}. \textcolor{black}{We note that open source software and standard datasets for benchmarking standard QML models for binary classification tasks addressing the claims of QML superiority over its classical counterparts are provided in~\cite{bowles2024better}. Furthermore, proposal to develop practically meaningful quantum datasets using quantum circuits or states where quantum methods are expected to excel compared to classical methods has been addressed in \cite{nakayama2023vqe}.} 

\paragraph{Quantum Data Preparation} While achieving a quantum advantage with classical data is challenging, QML models utilizing quantum data show more promise. Finding the most optimal encoding technique for a given dataset is another crucial challenge that needs to be addressed. These embedding techniques necessitate having features that are classically hard to simulate with practical usefulness. Identifying datasets that can take advantage of quantum computing for computing kernels is an important avenue of research.
Some recent works  \cite{jobst2023efficient} show that when the classical data follows certain patterns, the quantum states can be represented using efficient approximation leading to quantum circuits where the number of gates grows linearly with the number of qubits rather than exponentially making them more feasible. Currently, there is a lack of efficient quantum RAM (qRAM) capable of encoding and reliably storing information as a quantum state. This presents a significant hardware challenge in quantum computing.
 
\paragraph{Error Mitigation and Quantum Error Correction} Error mitigation and error correction are crucial for the long-term viability of fault-tolerant quantum computers. However, the implementation of quantum error correction introduces overhead that can reduce the speedup of quantum computations \cite{cerezo2022challenges}. Therefore, finding efficient quantum error correcting codes and developing methods to generate ground states using QML models are important areas of research.

\color{black}
Recently, studies have investigated the potential of QML, particularly quantum autoencoders (QAEs), for error correction in quantum memory \cite{locher2023quantum}. QAEs offer promise in autonomously correcting errors and extending logical qubit lifetimes, potentially streamlining error correction processes. Future research should explore the fault tolerance of QAEs and their integration into broader error correction frameworks. 
Various error-mitigation techniques, including ensemble learning approaches that combine multiple variational quantum circuits (VQCs), are employed to enhance the precision of classifiers for both classical and quantum datasets. \color{black} One such study by \cite{li2023ensemble} proposes two ensemble-learning error mitigation methods for VQCs: bootstrap aggregating and adaptive boosting. These methods can be applied to classification, kernel learning, regression, and even extended to QSVM. Importantly, their ensemble-learning VQCs are designed to be compatible with near-term quantum devices, distinguishing them from other ensemble-learning proposals that rely on resource-intensive hardware implementations involving multi-qubit controlled unitaries and complex quantum subroutines such as quantum phase estimation \cite{dorner2009optimal,d1998general}, Grover search \cite{biron1999generalized,
cui2010correlations}, and quantum mean estimation \cite{blume2010optimal,carollo2019quantumness,helstrom1969quantum}.

\paragraph{Ansatz Selection and Scalability}  Ansatz selection plays a crucial role in preventing Barren plateaus and achieving efficient scalability. Despite the theoretical work done to demonstrate provable advantage on synthetic datasets \cite{caro2022generalization}, more research is needed to understand the impact of entanglement in the model ansatz. Developing efficient methods to adjust parameter values and train quantum circuits to minimize specific loss functions in VQCs is an active area of research. Parameter initialization strategies for large-scale QNNs need to be explored to improve their scalability.
To comprehend the scalability of QML methods for large problems, analyzing trainability and prediction error is necessary. Access to reliable quantum hardware is also crucial. In QML, training the model involves minimizing a loss function to find the optimal set of parameters. Quantum landscape theory explores the properties of this loss function landscape, focusing on challenges like local minima and barren plateaus \cite{cerezo2022challenges}. \textcolor{black}{Recently, there has been an active line of research, with open source software, on quantum circuit selection tailored to underlying quantum devices~\cite{anagolum2024elivagar,chen2023quantumsea} that partially addresses this question. }

\paragraph{Backpropagation and Scalability}   Backpropagation plays a crucial role in the success of deep neural networks by efficiently computing gradients using the computational graph. This computational advantage allows for the training of deep networks. Recent applications like ChatGPT \cite{chatgpt}, utilize backpropagation during training for the efficient calculation of gradients for batches of input-output pairs which enables scalability in handling large datasets. This technique allows for parallel computation and parameter updates, contributing to the model's ability to handle increased complexity. However, when it comes to parameterized quantum circuits, backpropagation is significantly less efficient compared to classical circuits. This inefficiency directly impacts the trainability of quantum models. The existing gradient methods used in parameterized quantum models lack the scaling properties of backpropagation, raising questions about their computational complexity.

Addressing this issue, a recent study by \cite{abbas2024quantum} highlights the need to explore alternative architectures and optimization methods to improve the scalability of quantum models. The authors suggest that backpropagation may not be the appropriate optimization method for quantum models and propose an alternative while emphasizing on the importance of finding optimization methods that can effectively handle the computational complexity of parameterized quantum circuits to enhance the trainability and scalability of quantum models.

\paragraph{QML Model Security }  The current state of QML lacks privacy-preserving features, raising concerns about the potential exposure of sensitive information in machine learning datasets \cite{watkins2023quantum,kundu2022security,akter2023exploring,tehrani2024stabilized}.
To address this issue, it is crucial to implement privacy-preserving algorithms in QML, such as differential privacy, which minimizes the influence of individual data points on the training process.
However, the application of differential privacy in the context of QML requires further study and exploration to ensure effective privacy protection in machine learning models. Recently, \cite{watkins2023quantum} demonstrated the first proof‐of‐principle of privacy‐preserving QML. 

Additionally, another study \cite{west2023benchmarking} highlights the robustness of VQCs against adversarial attacks, even outperforming classical neural networks in this regard. The authors propose that combining the outcomes of both quantum and classical networks can have significant implications for enhancing security and reliability in applications like autonomous vehicles, cybersecurity, and surveillance robotic systems thereby opening up new possibilities for addressing security by leveraging the power of quantum and classical models together. \color{black}Recent work in \cite{innan2024fedqnn} investigates quantum federated learning in the context of secure data handling and sharing in distributed settings and cooperative learning. \color{black}

\paragraph{Towards Explainable QML models}
Realizing explainable AI (XAI) is a challenging yet crucial research field that provides insights into the decision-making process of machine learning models, addressing aspects such as fairness and security, especially in domains like medical research \cite{saranya2023systematic,islam2022systematic}.  An organic extension to QML also necessitates studying the fundamental aspects of QML \cite{heese2023explainable,steinmuller2022explainable,schuld2022quantum}. Given this context, exploring explainability in QML, called explainable QML (XQML), aims to provide humanly understandable interpretations of QML systems, similar to classical ML. Currently, the field of XQML remains relatively unexplored; however, it holds great potential for yielding fundamental insights, particularly given that QML is still in its early stages. Certain aspects of XQML, such as intuitively explaining quantum feature spaces and understanding the behavior of QML models in relation to QPUs through transformations and operations, go beyond the scope of classical XAI. Addressing these aspects may require the development of entirely new approaches to explainability or interpretability. The exploration of XQML, in conjunction with the prospect of improved hardware, may be considered more promising than solely focusing on identifying quantum advantages \cite{heese2023explainable,schuld2022quantum}.

\paragraph{Hyperparameter Choices and Transparency}   The lack of extensive discussions on hyperparameter choices in current quantum machine learning studies poses challenges to transparency, interpretability, and progress in the field. Many studies that demonstrate promising results on benchmark datasets often fail to provide open-source reference implementations of their competitive algorithms. This lack of accessibility hinders the reproducibility of results and raises concerns about potential positive bias, where only a selected set of experiments showing favorable model performance are reported, while others are disregarded.

Furthermore, reproducibility can be challenging when working with open-source projects like Qiskit, which undergo regular updates and improvements to enhance functionality, address bugs, and introduce new features. These updates can lead to deprecated features and code incompatibility, affecting the ability to reproduce results.

To address this issue, researchers should prioritize providing comprehensive documentation that includes detailed information on hyperparameter selection. This documentation should offer insights into the decision-making processes behind choosing specific hyperparameters and discuss the potential implications of different selections. By sharing this information, researchers can enhance transparency and enable others to replicate and build upon their work effectively. Furthermore, the provision of open-source reference implementations is crucial for fostering collaboration, promoting rigorous evaluation, and advancing the field collectively. Accessible and reproducible code allows researchers to validate and compare different approaches, facilitating the identification of strengths and weaknesses in quantum machine learning algorithms.

Reproducibility in experiments on quantum hardware can be challenging due to noise, limited access, calibration issues, and algorithmic variability. To address these challenges, researchers should thoroughly document the experimental setup, share the source code, use standardized benchmarks, and promote collaboration and open science practices.

\paragraph{Federated Learning and Quantum Boosting}   Thoroughly studying the use of federated learning to distribute computational tasks among limited-capability quantum machines, coupled with investigating the potential of quantum boosting classifiers, can significantly enhance the scalability and utilization of available Near-term noisy devices \cite{chen2021federated,li2021quantum,bhatia2023federated}. These research directions hold promise for leveraging the collective power of distributed quantum resources and improving the overall performance of quantum machine learning systems.

The challenge of scalability in quantum algorithms and its impact on real-world applications is a critical issue that requires further investigation. The recent findings in \cite{Kim2023} demonstrate the successful measurement of accurate expectation values for large circuit volumes using a noisy 127-qubit quantum processor, highlighting the potential of quantum computing in a pre-fault-tolerant era. However, it is important to acknowledge that the error mitigation techniques discussed in \cite{Kim2023} suffer from exponential computational time as the number of qubits increases. Moreover, comparing these techniques to "brute force" classical methods may not be entirely fair, as it fails to acknowledge the significant advancements made by classical methods in simulating quantum dynamics.

To advance the field, it is crucial to establish a shared community consensus on identifying problems that are both interesting for practical applications and genuinely challenging to simulate classically. This requires acknowledging the progress made by classical methods and not solely equating high entanglement with classical simulation difficulty. It is necessary to continue the development and exploration of both quantum and classical approximation methods, as they provide valuable benchmarks for each other's capabilities. By addressing these challenges and fostering collaboration between quantum and classical approaches, we can drive the field forward and unlock the full potential of quantum computing.

\section{Acknowledgement}
We want to emphasize that the list of papers presented in this draft has been compiled to include as many relevant papers as possible. However, we acknowledge that due to the rapid pace of development in this field, our coverage may not be exhaustive. 
\color{black}Thanks to arXiv for use of its open access interoperability that enables us to gather almost all of the papers referred in this paper. \color{black} Special thanks to Tamiya Onodera of IBM Quantum, IBM Research Tokyo, and Shesha Raghunathan of IBM Quantum, IBM Research India for their valuable discussions and comments.\color{black}This work is partially supported by Kakenhi
23K28035, 23K18501, 20H05967 and 21H05052.\color{black}

This paper was prepared for informational purposes by the Global Technology Applied Research center of JPMorgan Chase \& Co. This paper is not a product of the Research Department of JPMorgan Chase \& Co. or its affiliates. Neither JPMorgan Chase \& Co. nor any of its affiliates makes any explicit or implied representation or warranty and none of them accept any liability in connection with this paper, including, without limitation, with respect to the completeness, accuracy, or reliability of the information contained herein and the potential legal, compliance, tax, or accounting effects thereof. This document is not intended as investment research or investment advice, or as a recommendation, offer, or solicitation for the purchase or sale of any security, financial instrument, financial product or service, or to be used in any way for evaluating the merits of participating in any transaction.

\bibliography{Survey_arXiv/main}

\begin{thebibliography}{426}%
\makeatletter
\providecommand \@ifxundefined [1]{%
 \@ifx{#1\undefined}
}%
\providecommand \@ifnum [1]{%
 \ifnum #1\expandafter \@firstoftwo
 \else \expandafter \@secondoftwo
 \fi
}%
\providecommand \@ifx [1]{%
 \ifx #1\expandafter \@firstoftwo
 \else \expandafter \@secondoftwo
 \fi
}%
\providecommand \natexlab [1]{#1}%
\providecommand \enquote  [1]{``#1''}%
\providecommand \bibnamefont  [1]{#1}%
\providecommand \bibfnamefont [1]{#1}%
\providecommand \citenamefont [1]{#1}%
\providecommand \href@noop [0]{\@secondoftwo}%
\providecommand \href [0]{\begingroup \@sanitize@url \@href}%
\providecommand \@href[1]{\@@startlink{#1}\@@href}%
\providecommand \@@href[1]{\endgroup#1\@@endlink}%
\providecommand \@sanitize@url [0]{\catcode `\\12\catcode `\$12\catcode `\&12\catcode `\#12\catcode `\^12\catcode `\_12\catcode `\%12\relax}%
\providecommand \@@startlink[1]{}%
\providecommand \@@endlink[0]{}%
\providecommand \url  [0]{\begingroup\@sanitize@url \@url }%
\providecommand \@url [1]{\endgroup\@href {#1}{\urlprefix }}%
\providecommand \urlprefix  [0]{URL }%
\providecommand \Eprint [0]{\href }%
\providecommand \doibase [0]{https://doi.org/}%
\providecommand \selectlanguage [0]{\@gobble}%
\providecommand \bibinfo  [0]{\@secondoftwo}%
\providecommand \bibfield  [0]{\@secondoftwo}%
\providecommand \translation [1]{[#1]}%
\providecommand \BibitemOpen [0]{}%
\providecommand \bibitemStop [0]{}%
\providecommand \bibitemNoStop [0]{.\EOS\space}%
\providecommand \EOS [0]{\spacefactor3000\relax}%
\providecommand \BibitemShut  [1]{\csname bibitem#1\endcsname}%
\let\auto@bib@innerbib\@empty
\bibitem [{\citenamefont {Liu}\ et~al.(2023)\citenamefont {Liu}, \citenamefont {Li}, \citenamefont {Zhang},\ and\ \citenamefont {Zhang}}]{liu2023tensor}%
  \BibitemOpen
  \bibfield  {author} {\bibinfo {author} {\bibfnamefont {J.}~\bibnamefont {Liu}}, \bibinfo {author} {\bibfnamefont {S.}~\bibnamefont {Li}}, \bibinfo {author} {\bibfnamefont {J.}~\bibnamefont {Zhang}},\ and\ \bibinfo {author} {\bibfnamefont {P.}~\bibnamefont {Zhang}},\ }\bibfield  {title} {{Tensor networks for unsupervised machine learning},\ }\href@noop {} {\bibfield  {journal} {\bibinfo  {journal} {Physical Review E}\ }\textbf {\bibinfo {volume} {107}},\ \bibinfo {pages} {L012103} (\bibinfo {year} {2023})}\BibitemShut {NoStop}%
\bibitem [{\citenamefont {Tehrani}\ et~al.(2024)\citenamefont {Tehrani}, \citenamefont {Sultanow}, \citenamefont {Buchanan}, \citenamefont {Amir}, \citenamefont {Jeschke}, \citenamefont {Houmani}, \citenamefont {Chow},\ and\ \citenamefont {Lemoudden}}]{tehrani2024stabilized}%
  \BibitemOpen
  \bibfield  {author} {\bibinfo {author} {\bibfnamefont {M.~G.}\ \bibnamefont {Tehrani}}, \bibinfo {author} {\bibfnamefont {E.}~\bibnamefont {Sultanow}}, \bibinfo {author} {\bibfnamefont {W.~J.}\ \bibnamefont {Buchanan}}, \bibinfo {author} {\bibfnamefont {M.}~\bibnamefont {Amir}}, \bibinfo {author} {\bibfnamefont {A.}~\bibnamefont {Jeschke}}, \bibinfo {author} {\bibfnamefont {M.}~\bibnamefont {Houmani}}, \bibinfo {author} {\bibfnamefont {R.}~\bibnamefont {Chow}},\ and\ \bibinfo {author} {\bibfnamefont {M.}~\bibnamefont {Lemoudden}},\ }\bibfield  {title} {{Stabilized quantum-enhanced SIEM architecture and speed-up through Hoeffding tree algorithms enable quantum cybersecurity analytics in botnet detection},\ }\href@noop {} {\bibfield  {journal} {\bibinfo  {journal} {Scientific Reports}\ }\textbf {\bibinfo {volume} {14}},\ \bibinfo {pages} {1732} (\bibinfo {year} {2024})}\BibitemShut {NoStop}%
\bibitem [{\citenamefont {Hu}\ et~al.(2019)\citenamefont {Hu}, \citenamefont {Wu}, \citenamefont {Cai}, \citenamefont {Ma}, \citenamefont {Mu}, \citenamefont {Xu}, \citenamefont {Wang}, \citenamefont {Song}, \citenamefont {Deng}, \citenamefont {Zou} et~al.}]{hu2019quantum}%
  \BibitemOpen
  \bibfield  {author} {\bibinfo {author} {\bibfnamefont {L.}~\bibnamefont {Hu}}, \bibinfo {author} {\bibfnamefont {S.-H.}\ \bibnamefont {Wu}}, \bibinfo {author} {\bibfnamefont {W.}~\bibnamefont {Cai}}, \bibinfo {author} {\bibfnamefont {Y.}~\bibnamefont {Ma}}, \bibinfo {author} {\bibfnamefont {X.}~\bibnamefont {Mu}}, \bibinfo {author} {\bibfnamefont {Y.}~\bibnamefont {Xu}}, \bibinfo {author} {\bibfnamefont {H.}~\bibnamefont {Wang}}, \bibinfo {author} {\bibfnamefont {Y.}~\bibnamefont {Song}}, \bibinfo {author} {\bibfnamefont {D.-L.}\ \bibnamefont {Deng}}, \bibinfo {author} {\bibfnamefont {C.-L.}\ \bibnamefont {Zou}}, et~al.,\ }\bibfield  {title} {{Quantum generative adversarial learning in a superconducting quantum circuit},\ }\href@noop {} {\bibfield  {journal} {\bibinfo  {journal} {Science advances}\ }\textbf {\bibinfo {volume} {5}},\ \bibinfo {pages} {eaav2761} (\bibinfo {year} {2019})}\BibitemShut {NoStop}%
\bibitem [{\citenamefont {Das~Sarma}\ et~al.(2019)\citenamefont {Das~Sarma}, \citenamefont {Deng},\ and\ \citenamefont {Duan}}]{das2019machine}%
  \BibitemOpen
  \bibfield  {author} {\bibinfo {author} {\bibfnamefont {S.}~\bibnamefont {Das~Sarma}}, \bibinfo {author} {\bibfnamefont {D.-L.}\ \bibnamefont {Deng}},\ and\ \bibinfo {author} {\bibfnamefont {L.-M.}\ \bibnamefont {Duan}},\ }\bibfield  {title} {{Machine learning meets quantum physics},\ }\href@noop {} {\bibfield  {journal} {\bibinfo  {journal} {Physics Today}\ }\textbf {\bibinfo {volume} {72}},\ \bibinfo {pages} {48} (\bibinfo {year} {2019})}\BibitemShut {NoStop}%
\bibitem [{\citenamefont {Gily{\'e}n}\ et~al.(2019)\citenamefont {Gily{\'e}n}, \citenamefont {Arunachalam},\ and\ \citenamefont {Wiebe}}]{gilyen2019optimizing}%
  \BibitemOpen
  \bibfield  {author} {\bibinfo {author} {\bibfnamefont {A.}~\bibnamefont {Gily{\'e}n}}, \bibinfo {author} {\bibfnamefont {S.}~\bibnamefont {Arunachalam}},\ and\ \bibinfo {author} {\bibfnamefont {N.}~\bibnamefont {Wiebe}},\ }in\ \href@noop {} {\bibinfo {booktitle} {Proceedings of the Thirtieth Annual ACM-SIAM Symposium on Discrete Algorithms}}\ (\bibinfo {organization} {SIAM},\ \bibinfo {year} {2019})\ pp.\ \bibinfo {pages} {1425--1444}\BibitemShut {NoStop}%
\bibitem [{\citenamefont {{Hull}}(1994)}]{uspsdataset}%
  \BibitemOpen
  \bibfield  {author} {\bibinfo {author} {\bibfnamefont {J.~J.}\ \bibnamefont {{Hull}}},\ }\bibfield  {title} {{A database for handwritten text recognition research},\ }\href {https://doi.org/10.1109/34.291440} {\bibfield  {journal} {\bibinfo  {journal} {IEEE Transactions on Pattern Analysis and Machine Intelligence}\ }\textbf {\bibinfo {volume} {16}},\ \bibinfo {pages} {550} (\bibinfo {year} {1994})}\BibitemShut {NoStop}%
\bibitem [{\citenamefont {Deng}(2012)}]{deng2012mnist}%
  \BibitemOpen
  \bibfield  {author} {\bibinfo {author} {\bibfnamefont {L.}~\bibnamefont {Deng}},\ }\bibfield  {title} {{The mnist database of handwritten digit images for machine learning research},\ }\href@noop {} {\bibfield  {journal} {\bibinfo  {journal} {IEEE Signal Processing Magazine}\ }\textbf {\bibinfo {volume} {29}},\ \bibinfo {pages} {141} (\bibinfo {year} {2012})}\BibitemShut {NoStop}%
\bibitem [{\citenamefont {Fisher}(1936)}]{fisher1936use}%
  \BibitemOpen
  \bibfield  {author} {\bibinfo {author} {\bibfnamefont {R.~A.}\ \bibnamefont {Fisher}},\ }\bibfield  {title} {{The use of multiple measurements in taxonomic problems},\ }\href@noop {} {\bibfield  {journal} {\bibinfo  {journal} {Annals of eugenics}\ }\textbf {\bibinfo {volume} {7}},\ \bibinfo {pages} {179} (\bibinfo {year} {1936})}\BibitemShut {NoStop}%
\bibitem [{\citenamefont {Anderson}(1936)}]{anderson1936species}%
  \BibitemOpen
  \bibfield  {author} {\bibinfo {author} {\bibfnamefont {E.}~\bibnamefont {Anderson}},\ }\bibfield  {title} {{{The species problem in Iris}},\ }\href@noop {} {\bibfield  {journal} {\bibinfo  {journal} {Annals of the Missouri Botanical Garden}\ }\textbf {\bibinfo {volume} {23}},\ \bibinfo {pages} {457} (\bibinfo {year} {1936})}\BibitemShut {NoStop}%
\bibitem [{\citenamefont {Cortez}\ et~al.(2009)\citenamefont {Cortez}, \citenamefont {Cerdeira}, \citenamefont {Almeida}, \citenamefont {Matos},\ and\ \citenamefont {Reis}}]{cortez2009modeling}%
  \BibitemOpen
  \bibfield  {author} {\bibinfo {author} {\bibfnamefont {P.}~\bibnamefont {Cortez}}, \bibinfo {author} {\bibfnamefont {A.}~\bibnamefont {Cerdeira}}, \bibinfo {author} {\bibfnamefont {F.}~\bibnamefont {Almeida}}, \bibinfo {author} {\bibfnamefont {T.}~\bibnamefont {Matos}},\ and\ \bibinfo {author} {\bibfnamefont {J.}~\bibnamefont {Reis}},\ }\bibfield  {title} {{Modeling wine preferences by data mining from physicochemical properties},\ }\href@noop {} {\bibfield  {journal} {\bibinfo  {journal} {Decision support systems}\ }\textbf {\bibinfo {volume} {47}},\ \bibinfo {pages} {547} (\bibinfo {year} {2009})}\BibitemShut {NoStop}%
\bibitem [{\citenamefont {Xiao}\ et~al.(2017)\citenamefont {Xiao}, \citenamefont {Rasul},\ and\ \citenamefont {Vollgraf}}]{xiao2017fashion}%
  \BibitemOpen
  \bibfield  {author} {\bibinfo {author} {\bibfnamefont {H.}~\bibnamefont {Xiao}}, \bibinfo {author} {\bibfnamefont {K.}~\bibnamefont {Rasul}},\ and\ \bibinfo {author} {\bibfnamefont {R.}~\bibnamefont {Vollgraf}},\ }\bibfield  {title} {{Fashion-mnist: a novel image dataset for benchmarking machine learning algorithms},\ }\href@noop {} {\bibfield  {journal} {\bibinfo  {journal} {arXiv preprint arXiv:1708.07747}\ } (\bibinfo {year} {2017})}\BibitemShut {NoStop}%
\bibitem [{\citenamefont {Mart{\'\i}n}\ et~al.(2023)\citenamefont {Mart{\'\i}n}, \citenamefont {Plekhanov},\ and\ \citenamefont {Lubasch}}]{martin2023barren}%
  \BibitemOpen
  \bibfield  {author} {\bibinfo {author} {\bibfnamefont {E.~C.}\ \bibnamefont {Mart{\'\i}n}}, \bibinfo {author} {\bibfnamefont {K.}~\bibnamefont {Plekhanov}},\ and\ \bibinfo {author} {\bibfnamefont {M.}~\bibnamefont {Lubasch}},\ }\bibfield  {title} {{Barren plateaus in quantum tensor network optimization},\ }\href@noop {} {\bibfield  {journal} {\bibinfo  {journal} {Quantum}\ }\textbf {\bibinfo {volume} {7}},\ \bibinfo {pages} {974} (\bibinfo {year} {2023})}\BibitemShut {NoStop}%
\bibitem [{\citenamefont {West}\ et~al.(2023{\natexlab{a}})\citenamefont {West}, \citenamefont {Erfani}, \citenamefont {Leckie}, \citenamefont {Sevior}, \citenamefont {Hollenberg},\ and\ \citenamefont {Usman}}]{west2023benchmarking}%
  \BibitemOpen
  \bibfield  {author} {\bibinfo {author} {\bibfnamefont {M.~T.}\ \bibnamefont {West}}, \bibinfo {author} {\bibfnamefont {S.~M.}\ \bibnamefont {Erfani}}, \bibinfo {author} {\bibfnamefont {C.}~\bibnamefont {Leckie}}, \bibinfo {author} {\bibfnamefont {M.}~\bibnamefont {Sevior}}, \bibinfo {author} {\bibfnamefont {L.~C.}\ \bibnamefont {Hollenberg}},\ and\ \bibinfo {author} {\bibfnamefont {M.}~\bibnamefont {Usman}},\ }\bibfield  {title} {{Benchmarking adversarially robust quantum machine learning at scale},\ }\href@noop {} {\bibfield  {journal} {\bibinfo  {journal} {Physical Review Research}\ }\textbf {\bibinfo {volume} {5}},\ \bibinfo {pages} {023186} (\bibinfo {year} {2023}{\natexlab{a}})}\BibitemShut {NoStop}%
\bibitem [{\citenamefont {Li}\ et~al.(2021{\natexlab{a}})\citenamefont {Li}, \citenamefont {Lu},\ and\ \citenamefont {Deng}}]{li2021quantum}%
  \BibitemOpen
  \bibfield  {author} {\bibinfo {author} {\bibfnamefont {W.}~\bibnamefont {Li}}, \bibinfo {author} {\bibfnamefont {S.}~\bibnamefont {Lu}},\ and\ \bibinfo {author} {\bibfnamefont {D.-L.}\ \bibnamefont {Deng}},\ }\bibfield  {title} {{Quantum federated learning through blind quantum computing},\ }\href@noop {} {\bibfield  {journal} {\bibinfo  {journal} {Science China Physics, Mechanics \& Astronomy}\ }\textbf {\bibinfo {volume} {64}},\ \bibinfo {pages} {100312} (\bibinfo {year} {2021}{\natexlab{a}})}\BibitemShut {NoStop}%
\bibitem [{\citenamefont {Jobst}\ et~al.(2023)\citenamefont {Jobst}, \citenamefont {Shen}, \citenamefont {Riofr{\'\i}o}, \citenamefont {Shishenina},\ and\ \citenamefont {Pollmann}}]{jobst2023efficient}%
  \BibitemOpen
  \bibfield  {author} {\bibinfo {author} {\bibfnamefont {B.}~\bibnamefont {Jobst}}, \bibinfo {author} {\bibfnamefont {K.}~\bibnamefont {Shen}}, \bibinfo {author} {\bibfnamefont {C.~A.}\ \bibnamefont {Riofr{\'\i}o}}, \bibinfo {author} {\bibfnamefont {E.}~\bibnamefont {Shishenina}},\ and\ \bibinfo {author} {\bibfnamefont {F.}~\bibnamefont {Pollmann}},\ }\bibfield  {title} {{{Efficient MPS representations and quantum circuits from the Fourier modes of classical image data}},\ }\href@noop {} {\bibfield  {journal} {\bibinfo  {journal} {arXiv preprint arXiv:2311.07666}\ } (\bibinfo {year} {2023})}\BibitemShut {NoStop}%
\bibitem [{\citenamefont {Du}\ et~al.(2022)\citenamefont {Du}, \citenamefont {Huang}, \citenamefont {You}, \citenamefont {Hsieh},\ and\ \citenamefont {Tao}}]{du2022quantum}%
  \BibitemOpen
  \bibfield  {author} {\bibinfo {author} {\bibfnamefont {Y.}~\bibnamefont {Du}}, \bibinfo {author} {\bibfnamefont {T.}~\bibnamefont {Huang}}, \bibinfo {author} {\bibfnamefont {S.}~\bibnamefont {You}}, \bibinfo {author} {\bibfnamefont {M.-H.}\ \bibnamefont {Hsieh}},\ and\ \bibinfo {author} {\bibfnamefont {D.}~\bibnamefont {Tao}},\ }\bibfield  {title} {{Quantum circuit architecture search for variational quantum algorithms},\ }\href@noop {} {\bibfield  {journal} {\bibinfo  {journal} {npj Quantum Information}\ }\textbf {\bibinfo {volume} {8}},\ \bibinfo {pages} {62} (\bibinfo {year} {2022})}\BibitemShut {NoStop}%
\bibitem [{\citenamefont {West}\ et~al.(2023{\natexlab{b}})\citenamefont {West}, \citenamefont {Sevior},\ and\ \citenamefont {Usman}}]{west2023reflection}%
  \BibitemOpen
  \bibfield  {author} {\bibinfo {author} {\bibfnamefont {M.~T.}\ \bibnamefont {West}}, \bibinfo {author} {\bibfnamefont {M.}~\bibnamefont {Sevior}},\ and\ \bibinfo {author} {\bibfnamefont {M.}~\bibnamefont {Usman}},\ }\bibfield  {title} {{Reflection equivariant quantum neural networks for enhanced image classification},\ }\href@noop {} {\bibfield  {journal} {\bibinfo  {journal} {Machine Learning: Science and Technology}\ }\textbf {\bibinfo {volume} {4}},\ \bibinfo {pages} {035027} (\bibinfo {year} {2023}{\natexlab{b}})}\BibitemShut {NoStop}%
\bibitem [{\citenamefont {Altares-L{\'o}pez}\ et~al.(2021)\citenamefont {Altares-L{\'o}pez}, \citenamefont {Ribeiro},\ and\ \citenamefont {Garc{\'\i}a-Ripoll}}]{altares2021automatic}%
  \BibitemOpen
  \bibfield  {author} {\bibinfo {author} {\bibfnamefont {S.}~\bibnamefont {Altares-L{\'o}pez}}, \bibinfo {author} {\bibfnamefont {A.}~\bibnamefont {Ribeiro}},\ and\ \bibinfo {author} {\bibfnamefont {J.~J.}\ \bibnamefont {Garc{\'\i}a-Ripoll}},\ }\bibfield  {title} {{Automatic design of quantum feature maps},\ }\href@noop {} {\bibfield  {journal} {\bibinfo  {journal} {Quantum Science and Technology}\ }\textbf {\bibinfo {volume} {6}},\ \bibinfo {pages} {045015} (\bibinfo {year} {2021})}\BibitemShut {NoStop}%
\bibitem [{\citenamefont {Benedetti}\ et~al.(2021{\natexlab{a}})\citenamefont {Benedetti}, \citenamefont {Fiorentini},\ and\ \citenamefont {Lubasch}}]{benedetti2021hardware}%
  \BibitemOpen
  \bibfield  {author} {\bibinfo {author} {\bibfnamefont {M.}~\bibnamefont {Benedetti}}, \bibinfo {author} {\bibfnamefont {M.}~\bibnamefont {Fiorentini}},\ and\ \bibinfo {author} {\bibfnamefont {M.}~\bibnamefont {Lubasch}},\ }\bibfield  {title} {{Hardware-efficient variational quantum algorithms for time evolution},\ }\href@noop {} {\bibfield  {journal} {\bibinfo  {journal} {Physical Review Research}\ }\textbf {\bibinfo {volume} {3}},\ \bibinfo {pages} {033083} (\bibinfo {year} {2021}{\natexlab{a}})}\BibitemShut {NoStop}%
\bibitem [{\citenamefont {Huang}\ et~al.(2021{\natexlab{a}})\citenamefont {Huang}, \citenamefont {Bharti},\ and\ \citenamefont {Rebentrost}}]{huang2021near}%
  \BibitemOpen
  \bibfield  {author} {\bibinfo {author} {\bibfnamefont {H.-Y.}\ \bibnamefont {Huang}}, \bibinfo {author} {\bibfnamefont {K.}~\bibnamefont {Bharti}},\ and\ \bibinfo {author} {\bibfnamefont {P.}~\bibnamefont {Rebentrost}},\ }\bibfield  {title} {{Near-term quantum algorithms for linear systems of equations with regression loss functions},\ }\href@noop {} {\bibfield  {journal} {\bibinfo  {journal} {New Journal of Physics}\ }\textbf {\bibinfo {volume} {23}},\ \bibinfo {pages} {113021} (\bibinfo {year} {2021}{\natexlab{a}})}\BibitemShut {NoStop}%
\bibitem [{\citenamefont {Wiersema}\ et~al.(2024)\citenamefont {Wiersema}, \citenamefont {Lewis}, \citenamefont {Wierichs}, \citenamefont {Carrasquilla},\ and\ \citenamefont {Killoran}}]{wiersema2024here}%
  \BibitemOpen
  \bibfield  {author} {\bibinfo {author} {\bibfnamefont {R.}~\bibnamefont {Wiersema}}, \bibinfo {author} {\bibfnamefont {D.}~\bibnamefont {Lewis}}, \bibinfo {author} {\bibfnamefont {D.}~\bibnamefont {Wierichs}}, \bibinfo {author} {\bibfnamefont {J.}~\bibnamefont {Carrasquilla}},\ and\ \bibinfo {author} {\bibfnamefont {N.}~\bibnamefont {Killoran}},\ }\bibfield  {title} {{{Here comes the SU (N): multivariate quantum gates and gradients}},\ }\href@noop {} {\bibfield  {journal} {\bibinfo  {journal} {Quantum}\ }\textbf {\bibinfo {volume} {8}},\ \bibinfo {pages} {1275} (\bibinfo {year} {2024})}\BibitemShut {NoStop}%
\bibitem [{\citenamefont {Wiedmann}\ et~al.(2023)\citenamefont {Wiedmann}, \citenamefont {H{\"o}lle}, \citenamefont {Periyasamy}, \citenamefont {Meyer}, \citenamefont {Ufrecht}, \citenamefont {Scherer}, \citenamefont {Plinge},\ and\ \citenamefont {Mutschler}}]{wiedmann2023empirical}%
  \BibitemOpen
  \bibfield  {author} {\bibinfo {author} {\bibfnamefont {M.}~\bibnamefont {Wiedmann}}, \bibinfo {author} {\bibfnamefont {M.}~\bibnamefont {H{\"o}lle}}, \bibinfo {author} {\bibfnamefont {M.}~\bibnamefont {Periyasamy}}, \bibinfo {author} {\bibfnamefont {N.}~\bibnamefont {Meyer}}, \bibinfo {author} {\bibfnamefont {C.}~\bibnamefont {Ufrecht}}, \bibinfo {author} {\bibfnamefont {D.~D.}\ \bibnamefont {Scherer}}, \bibinfo {author} {\bibfnamefont {A.}~\bibnamefont {Plinge}},\ and\ \bibinfo {author} {\bibfnamefont {C.}~\bibnamefont {Mutschler}},\ }in\ \href@noop {} {\bibinfo {booktitle} {2023 IEEE International Conference on Quantum Computing and Engineering (QCE)}},\ Vol.~\bibinfo {volume} {1}\ (\bibinfo {organization} {IEEE},\ \bibinfo {year} {2023})\ pp.\ \bibinfo {pages} {450--456}\BibitemShut {NoStop}%
\bibitem [{\citenamefont {West}\ et~al.(2023{\natexlab{c}})\citenamefont {West}, \citenamefont {Sevior},\ and\ \citenamefont {Usman}}]{west2023boosted}%
  \BibitemOpen
  \bibfield  {author} {\bibinfo {author} {\bibfnamefont {M.~T.}\ \bibnamefont {West}}, \bibinfo {author} {\bibfnamefont {M.}~\bibnamefont {Sevior}},\ and\ \bibinfo {author} {\bibfnamefont {M.}~\bibnamefont {Usman}},\ }\bibfield  {title} {{{Boosted Ensembles of Qubit and Continuous Variable Quantum Support Vector Machines for B Meson Flavor Tagging}},\ }\href@noop {} {\bibfield  {journal} {\bibinfo  {journal} {Advanced Quantum Technologies}\ }\textbf {\bibinfo {volume} {6}},\ \bibinfo {pages} {2300130} (\bibinfo {year} {2023}{\natexlab{c}})}\BibitemShut {NoStop}%
\bibitem [{\citenamefont {West}\ et~al.(2023{\natexlab{d}})\citenamefont {West}, \citenamefont {Tsang}, \citenamefont {Low}, \citenamefont {Hill}, \citenamefont {Leckie}, \citenamefont {Hollenberg}, \citenamefont {Erfani},\ and\ \citenamefont {Usman}}]{west2023towards}%
  \BibitemOpen
  \bibfield  {author} {\bibinfo {author} {\bibfnamefont {M.~T.}\ \bibnamefont {West}}, \bibinfo {author} {\bibfnamefont {S.-L.}\ \bibnamefont {Tsang}}, \bibinfo {author} {\bibfnamefont {J.~S.}\ \bibnamefont {Low}}, \bibinfo {author} {\bibfnamefont {C.~D.}\ \bibnamefont {Hill}}, \bibinfo {author} {\bibfnamefont {C.}~\bibnamefont {Leckie}}, \bibinfo {author} {\bibfnamefont {L.~C.}\ \bibnamefont {Hollenberg}}, \bibinfo {author} {\bibfnamefont {S.~M.}\ \bibnamefont {Erfani}},\ and\ \bibinfo {author} {\bibfnamefont {M.}~\bibnamefont {Usman}},\ }\bibfield  {title} {{Towards quantum enhanced adversarial robustness in machine learning},\ }\href@noop {} {\bibfield  {journal} {\bibinfo  {journal} {Nature Machine Intelligence}\ ,\ \bibinfo {pages} {1}} (\bibinfo {year} {2023}{\natexlab{d}})}\BibitemShut {NoStop}%
\bibitem [{\citenamefont {Tilly}\ et~al.(2022)\citenamefont {Tilly}, \citenamefont {Chen}, \citenamefont {Cao}, \citenamefont {Picozzi}, \citenamefont {Setia}, \citenamefont {Li}, \citenamefont {Grant}, \citenamefont {Wossnig}, \citenamefont {Rungger}, \citenamefont {Booth} et~al.}]{tilly2022variational}%
  \BibitemOpen
  \bibfield  {author} {\bibinfo {author} {\bibfnamefont {J.}~\bibnamefont {Tilly}}, \bibinfo {author} {\bibfnamefont {H.}~\bibnamefont {Chen}}, \bibinfo {author} {\bibfnamefont {S.}~\bibnamefont {Cao}}, \bibinfo {author} {\bibfnamefont {D.}~\bibnamefont {Picozzi}}, \bibinfo {author} {\bibfnamefont {K.}~\bibnamefont {Setia}}, \bibinfo {author} {\bibfnamefont {Y.}~\bibnamefont {Li}}, \bibinfo {author} {\bibfnamefont {E.}~\bibnamefont {Grant}}, \bibinfo {author} {\bibfnamefont {L.}~\bibnamefont {Wossnig}}, \bibinfo {author} {\bibfnamefont {I.}~\bibnamefont {Rungger}}, \bibinfo {author} {\bibfnamefont {G.~H.}\ \bibnamefont {Booth}}, et~al.,\ }\bibfield  {title} {{The variational quantum eigensolver: a review of methods and best practices},\ }\href@noop {} {\bibfield  {journal} {\bibinfo  {journal} {Physics Reports}\ }\textbf {\bibinfo {volume} {986}},\ \bibinfo {pages} {1} (\bibinfo {year} {2022})}\BibitemShut {NoStop}%
\bibitem [{\citenamefont {Abbas}\ et~al.(2024)\citenamefont {Abbas}, \citenamefont {King}, \citenamefont {Huang}, \citenamefont {Huggins}, \citenamefont {Movassagh}, \citenamefont {Gilboa},\ and\ \citenamefont {McClean}}]{abbas2024quantum}%
  \BibitemOpen
  \bibfield  {author} {\bibinfo {author} {\bibfnamefont {A.}~\bibnamefont {Abbas}}, \bibinfo {author} {\bibfnamefont {R.}~\bibnamefont {King}}, \bibinfo {author} {\bibfnamefont {H.-Y.}\ \bibnamefont {Huang}}, \bibinfo {author} {\bibfnamefont {W.~J.}\ \bibnamefont {Huggins}}, \bibinfo {author} {\bibfnamefont {R.}~\bibnamefont {Movassagh}}, \bibinfo {author} {\bibfnamefont {D.}~\bibnamefont {Gilboa}},\ and\ \bibinfo {author} {\bibfnamefont {J.}~\bibnamefont {McClean}},\ }\bibfield  {title} {{On quantum backpropagation, information reuse, and cheating measurement collapse},\ }\href@noop {} {\bibfield  {journal} {\bibinfo  {journal} {Advances in Neural Information Processing Systems}\ }\textbf {\bibinfo {volume} {36}} (\bibinfo {year} {2024})}\BibitemShut {NoStop}%
\bibitem [{\citenamefont {Watkins}\ et~al.(2023)\citenamefont {Watkins}, \citenamefont {Chen},\ and\ \citenamefont {Yoo}}]{watkins2023quantum}%
  \BibitemOpen
  \bibfield  {author} {\bibinfo {author} {\bibfnamefont {W.~M.}\ \bibnamefont {Watkins}}, \bibinfo {author} {\bibfnamefont {S.~Y.-C.}\ \bibnamefont {Chen}},\ and\ \bibinfo {author} {\bibfnamefont {S.}~\bibnamefont {Yoo}},\ }\bibfield  {title} {{Quantum machine learning with differential privacy},\ }\href@noop {} {\bibfield  {journal} {\bibinfo  {journal} {Scientific Reports}\ }\textbf {\bibinfo {volume} {13}},\ \bibinfo {pages} {2453} (\bibinfo {year} {2023})}\BibitemShut {NoStop}%
\bibitem [{\citenamefont {Garcia}\ et~al.(2023)\citenamefont {Garcia}, \citenamefont {Zhao}, \citenamefont {Bu},\ and\ \citenamefont {Jaffe}}]{garcia2023barren}%
  \BibitemOpen
  \bibfield  {author} {\bibinfo {author} {\bibfnamefont {R.~J.}\ \bibnamefont {Garcia}}, \bibinfo {author} {\bibfnamefont {C.}~\bibnamefont {Zhao}}, \bibinfo {author} {\bibfnamefont {K.}~\bibnamefont {Bu}},\ and\ \bibinfo {author} {\bibfnamefont {A.}~\bibnamefont {Jaffe}},\ }\bibfield  {title} {{Barren plateaus from learning scramblers with local cost functions},\ }\href@noop {} {\bibfield  {journal} {\bibinfo  {journal} {Journal of High Energy Physics}\ }\textbf {\bibinfo {volume} {2023}},\ \bibinfo {pages} {1} (\bibinfo {year} {2023})}\BibitemShut {NoStop}%
\bibitem [{\citenamefont {Gao}\ et~al.(2021{\natexlab{a}})\citenamefont {Gao}, \citenamefont {Li}, \citenamefont {Wei}, \citenamefont {Gao},\ and\ \citenamefont {Long}}]{gao2021quantum}%
  \BibitemOpen
  \bibfield  {author} {\bibinfo {author} {\bibfnamefont {P.}~\bibnamefont {Gao}}, \bibinfo {author} {\bibfnamefont {K.}~\bibnamefont {Li}}, \bibinfo {author} {\bibfnamefont {S.}~\bibnamefont {Wei}}, \bibinfo {author} {\bibfnamefont {J.}~\bibnamefont {Gao}},\ and\ \bibinfo {author} {\bibfnamefont {G.}~\bibnamefont {Long}},\ }\bibfield  {title} {{Quantum gradient algorithm for general polynomials},\ }\href@noop {} {\bibfield  {journal} {\bibinfo  {journal} {Physical Review A}\ }\textbf {\bibinfo {volume} {103}},\ \bibinfo {pages} {042403} (\bibinfo {year} {2021}{\natexlab{a}})}\BibitemShut {NoStop}%
\bibitem [{\citenamefont {Buhrman}\ et~al.(2001)\citenamefont {Buhrman}, \citenamefont {Cleve}, \citenamefont {Watrous},\ and\ \citenamefont {De~Wolf}}]{buhrman2001quantum}%
  \BibitemOpen
  \bibfield  {author} {\bibinfo {author} {\bibfnamefont {H.}~\bibnamefont {Buhrman}}, \bibinfo {author} {\bibfnamefont {R.}~\bibnamefont {Cleve}}, \bibinfo {author} {\bibfnamefont {J.}~\bibnamefont {Watrous}},\ and\ \bibinfo {author} {\bibfnamefont {R.}~\bibnamefont {De~Wolf}},\ }\bibfield  {title} {{Quantum fingerprinting},\ }\href@noop {} {\bibfield  {journal} {\bibinfo  {journal} {Physical Review Letters}\ }\textbf {\bibinfo {volume} {87}},\ \bibinfo {pages} {167902} (\bibinfo {year} {2001})}\BibitemShut {NoStop}%
\bibitem [{\citenamefont {Jordan}(2005)}]{jordan2005fast}%
  \BibitemOpen
  \bibfield  {author} {\bibinfo {author} {\bibfnamefont {S.~P.}\ \bibnamefont {Jordan}},\ }\bibfield  {title} {{Fast quantum algorithm for numerical gradient estimation},\ }\href@noop {} {\bibfield  {journal} {\bibinfo  {journal} {Physical review letters}\ }\textbf {\bibinfo {volume} {95}},\ \bibinfo {pages} {050501} (\bibinfo {year} {2005})}\BibitemShut {NoStop}%
\bibitem [{\citenamefont {Kwong}(2001)}]{kwong2001financial}%
  \BibitemOpen
  \bibfield  {author} {\bibinfo {author} {\bibfnamefont {K.}~\bibnamefont {Kwong}},\ }\bibfield  {title} {{Financial forecasting using neural network or machine learning techniques},\ }\href@noop {} {\bibfield  {journal} {\bibinfo  {journal} {University of Queensland}\ }\textbf {\bibinfo {volume} {13}},\ \bibinfo {pages} {221} (\bibinfo {year} {2001})}\BibitemShut {NoStop}%
\bibitem [{\citenamefont {Krollner}\ et~al.(2010)\citenamefont {Krollner}, \citenamefont {Vanstone},\ and\ \citenamefont {Finnie}}]{krollner2010financial}%
  \BibitemOpen
  \bibfield  {author} {\bibinfo {author} {\bibfnamefont {B.}~\bibnamefont {Krollner}}, \bibinfo {author} {\bibfnamefont {B.}~\bibnamefont {Vanstone}},\ and\ \bibinfo {author} {\bibfnamefont {G.}~\bibnamefont {Finnie}},\ }in\ \href@noop {} {\bibinfo {booktitle} {European Symposium on Artificial Neural Networks: Computational Intelligence and Machine Learning}}\ (\bibinfo {year} {2010})\ pp.\ \bibinfo {pages} {25--30}\BibitemShut {NoStop}%
\bibitem [{\citenamefont {Wasserbacher}\ and\ \citenamefont {Spindler}(2022)}]{wasserbacher2022machine}%
  \BibitemOpen
  \bibfield  {author} {\bibinfo {author} {\bibfnamefont {H.}~\bibnamefont {Wasserbacher}}\ and\ \bibinfo {author} {\bibfnamefont {M.}~\bibnamefont {Spindler}},\ }\bibfield  {title} {{Machine learning for financial forecasting, planning and analysis: recent developments and pitfalls},\ }\href@noop {} {\bibfield  {journal} {\bibinfo  {journal} {Digital Finance}\ }\textbf {\bibinfo {volume} {4}},\ \bibinfo {pages} {63} (\bibinfo {year} {2022})}\BibitemShut {NoStop}%
\bibitem [{\citenamefont {Chawla}\ and\ \citenamefont {Sarangi}(2020)}]{sarangi2020literature}%
  \BibitemOpen
  \bibfield  {author} {\bibinfo {author} {\bibfnamefont {M.}~\bibnamefont {Chawla}}\ and\ \bibinfo {author} {\bibfnamefont {P.}~\bibnamefont {Sarangi}},\ }\bibfield  {title} {{A Literature Review on Machine Learning Applications in Financial Forecasting},\ }\href {https://doi.org/10.15415/jtmge.2020.111004} {\bibfield  {journal} {\bibinfo  {journal} {Journal of Technology Management for Growing Economies}\ }\textbf {\bibinfo {volume} {11}},\ \bibinfo {pages} {23} (\bibinfo {year} {2020})}\BibitemShut {NoStop}%
\bibitem [{\citenamefont {Kamalov}\ et~al.(2021)\citenamefont {Kamalov}, \citenamefont {Gurrib},\ and\ \citenamefont {Rajab}}]{kamalov2021financial}%
  \BibitemOpen
  \bibfield  {author} {\bibinfo {author} {\bibfnamefont {F.}~\bibnamefont {Kamalov}}, \bibinfo {author} {\bibfnamefont {I.}~\bibnamefont {Gurrib}},\ and\ \bibinfo {author} {\bibfnamefont {K.}~\bibnamefont {Rajab}},\ }\bibfield  {title} {{Financial forecasting with machine learning: price vs return},\ }\href@noop {} {\bibfield  {journal} {\bibinfo  {journal} {Kamalov, F., Gurrib, I. \& Rajab, K.(2021). Financial Forecasting with Machine Learning: Price Vs Return. Journal of Computer Science}\ }\textbf {\bibinfo {volume} {17}},\ \bibinfo {pages} {251} (\bibinfo {year} {2021})}\BibitemShut {NoStop}%
\bibitem [{\citenamefont {Di~Persio}\ and\ \citenamefont {Honchar}(2018)}]{di2018multitask}%
  \BibitemOpen
  \bibfield  {author} {\bibinfo {author} {\bibfnamefont {L.}~\bibnamefont {Di~Persio}}\ and\ \bibinfo {author} {\bibfnamefont {O.}~\bibnamefont {Honchar}},\ }\bibfield  {title} {{Multitask machine learning for financial forecasting},\ }\href@noop {} {\bibfield  {journal} {\bibinfo  {journal} {International Journal of Circuits, Systems and Signal Processing}\ }\textbf {\bibinfo {volume} {12}},\ \bibinfo {pages} {444} (\bibinfo {year} {2018})}\BibitemShut {NoStop}%
\bibitem [{\citenamefont {Gao}\ et~al.(2021{\natexlab{b}})\citenamefont {Gao}, \citenamefont {Zhang}, \citenamefont {Shi}, \citenamefont {Xu}, \citenamefont {Zhang},\ and\ \citenamefont {Zhu}}]{gao2021review}%
  \BibitemOpen
  \bibfield  {author} {\bibinfo {author} {\bibfnamefont {R.}~\bibnamefont {Gao}}, \bibinfo {author} {\bibfnamefont {Z.}~\bibnamefont {Zhang}}, \bibinfo {author} {\bibfnamefont {Z.}~\bibnamefont {Shi}}, \bibinfo {author} {\bibfnamefont {D.}~\bibnamefont {Xu}}, \bibinfo {author} {\bibfnamefont {W.}~\bibnamefont {Zhang}},\ and\ \bibinfo {author} {\bibfnamefont {D.}~\bibnamefont {Zhu}},\ }in\ \href@noop {} {\bibinfo {booktitle} {International Symposium on Artificial Intelligence and Robotics 2021}},\ Vol.\ \bibinfo {volume} {11884}\ (\bibinfo {organization} {SPIE},\ \bibinfo {year} {2021})\ pp.\ \bibinfo {pages} {262--277}\BibitemShut {NoStop}%
\bibitem [{\citenamefont {Fisher}\ et~al.(2016)\citenamefont {Fisher}, \citenamefont {Garnsey},\ and\ \citenamefont {Hughes}}]{fisher2016natural}%
  \BibitemOpen
  \bibfield  {author} {\bibinfo {author} {\bibfnamefont {I.~E.}\ \bibnamefont {Fisher}}, \bibinfo {author} {\bibfnamefont {M.~R.}\ \bibnamefont {Garnsey}},\ and\ \bibinfo {author} {\bibfnamefont {M.~E.}\ \bibnamefont {Hughes}},\ }\bibfield  {title} {{Natural language processing in accounting, auditing and finance: A synthesis of the literature with a roadmap for future research},\ }\href@noop {} {\bibfield  {journal} {\bibinfo  {journal} {Intelligent Systems in Accounting, Finance and Management}\ }\textbf {\bibinfo {volume} {23}},\ \bibinfo {pages} {157} (\bibinfo {year} {2016})}\BibitemShut {NoStop}%
\bibitem [{\citenamefont {Yee}\ et~al.(2018)\citenamefont {Yee}, \citenamefont {Sagadevan},\ and\ \citenamefont {Malim}}]{yee2018credit}%
  \BibitemOpen
  \bibfield  {author} {\bibinfo {author} {\bibfnamefont {O.~S.}\ \bibnamefont {Yee}}, \bibinfo {author} {\bibfnamefont {S.}~\bibnamefont {Sagadevan}},\ and\ \bibinfo {author} {\bibfnamefont {N.~H. A.~H.}\ \bibnamefont {Malim}},\ }\bibfield  {title} {{Credit card fraud detection using machine learning as data mining technique},\ }\href@noop {} {\bibfield  {journal} {\bibinfo  {journal} {Journal of Telecommunication, Electronic and Computer Engineering (JTEC)}\ }\textbf {\bibinfo {volume} {10}},\ \bibinfo {pages} {23} (\bibinfo {year} {2018})}\BibitemShut {NoStop}%
\bibitem [{\citenamefont {Bauder}\ and\ \citenamefont {Khoshgoftaar}(2017)}]{bauder2017medicare}%
  \BibitemOpen
  \bibfield  {author} {\bibinfo {author} {\bibfnamefont {R.~A.}\ \bibnamefont {Bauder}}\ and\ \bibinfo {author} {\bibfnamefont {T.~M.}\ \bibnamefont {Khoshgoftaar}},\ }in\ \href@noop {} {\bibinfo {booktitle} {2017 16th IEEE international conference on machine learning and applications (ICMLA)}}\ (\bibinfo {organization} {IEEE},\ \bibinfo {year} {2017})\ pp.\ \bibinfo {pages} {858--865}\BibitemShut {NoStop}%
\bibitem [{\citenamefont {Thennakoon}\ et~al.(2019)\citenamefont {Thennakoon}, \citenamefont {Bhagyani}, \citenamefont {Premadasa}, \citenamefont {Mihiranga},\ and\ \citenamefont {Kuruwitaarachchi}}]{thennakoon2019real}%
  \BibitemOpen
  \bibfield  {author} {\bibinfo {author} {\bibfnamefont {A.}~\bibnamefont {Thennakoon}}, \bibinfo {author} {\bibfnamefont {C.}~\bibnamefont {Bhagyani}}, \bibinfo {author} {\bibfnamefont {S.}~\bibnamefont {Premadasa}}, \bibinfo {author} {\bibfnamefont {S.}~\bibnamefont {Mihiranga}},\ and\ \bibinfo {author} {\bibfnamefont {N.}~\bibnamefont {Kuruwitaarachchi}},\ }in\ \href@noop {} {\bibinfo {booktitle} {2019 9th International Conference on Cloud Computing, Data Science \& Engineering (Confluence)}}\ (\bibinfo {organization} {IEEE},\ \bibinfo {year} {2019})\ pp.\ \bibinfo {pages} {488--493}\BibitemShut {NoStop}%
\bibitem [{\citenamefont {Dornadula}\ and\ \citenamefont {Geetha}(2019)}]{dornadula2019credit}%
  \BibitemOpen
  \bibfield  {author} {\bibinfo {author} {\bibfnamefont {V.~N.}\ \bibnamefont {Dornadula}}\ and\ \bibinfo {author} {\bibfnamefont {S.}~\bibnamefont {Geetha}},\ }\bibfield  {title} {{Credit card fraud detection using machine learning algorithms},\ }\href@noop {} {\bibfield  {journal} {\bibinfo  {journal} {Procedia computer science}\ }\textbf {\bibinfo {volume} {165}},\ \bibinfo {pages} {631} (\bibinfo {year} {2019})}\BibitemShut {NoStop}%
\bibitem [{\citenamefont {Perols}(2011)}]{perols2011financial}%
  \BibitemOpen
  \bibfield  {author} {\bibinfo {author} {\bibfnamefont {J.}~\bibnamefont {Perols}},\ }\bibfield  {title} {{Financial statement fraud detection: An analysis of statistical and machine learning algorithms},\ }\href@noop {} {\bibfield  {journal} {\bibinfo  {journal} {Auditing: A Journal of Practice \& Theory}\ }\textbf {\bibinfo {volume} {30}},\ \bibinfo {pages} {19} (\bibinfo {year} {2011})}\BibitemShut {NoStop}%
\bibitem [{\citenamefont {Varmedja}\ et~al.(2019)\citenamefont {Varmedja}, \citenamefont {Karanovic}, \citenamefont {Sladojevic}, \citenamefont {Arsenovic},\ and\ \citenamefont {Anderla}}]{varmedja2019credit}%
  \BibitemOpen
  \bibfield  {author} {\bibinfo {author} {\bibfnamefont {D.}~\bibnamefont {Varmedja}}, \bibinfo {author} {\bibfnamefont {M.}~\bibnamefont {Karanovic}}, \bibinfo {author} {\bibfnamefont {S.}~\bibnamefont {Sladojevic}}, \bibinfo {author} {\bibfnamefont {M.}~\bibnamefont {Arsenovic}},\ and\ \bibinfo {author} {\bibfnamefont {A.}~\bibnamefont {Anderla}},\ }in\ \href@noop {} {\bibinfo {booktitle} {2019 18th International Symposium INFOTEH-JAHORINA (INFOTEH)}}\ (\bibinfo {organization} {IEEE},\ \bibinfo {year} {2019})\ pp.\ \bibinfo {pages} {1--5}\BibitemShut {NoStop}%
\bibitem [{\citenamefont {Awoyemi}\ et~al.(2017)\citenamefont {Awoyemi}, \citenamefont {Adetunmbi},\ and\ \citenamefont {Oluwadare}}]{awoyemi2017credit}%
  \BibitemOpen
  \bibfield  {author} {\bibinfo {author} {\bibfnamefont {J.~O.}\ \bibnamefont {Awoyemi}}, \bibinfo {author} {\bibfnamefont {A.~O.}\ \bibnamefont {Adetunmbi}},\ and\ \bibinfo {author} {\bibfnamefont {S.~A.}\ \bibnamefont {Oluwadare}},\ }in\ \href@noop {} {\bibinfo {booktitle} {2017 international conference on computing networking and informatics (ICCNI)}}\ (\bibinfo {organization} {IEEE},\ \bibinfo {year} {2017})\ pp.\ \bibinfo {pages} {1--9}\BibitemShut {NoStop}%
\bibitem [{\citenamefont {Kumbure}\ et~al.(2022)\citenamefont {Kumbure}, \citenamefont {Lohrmann}, \citenamefont {Luukka},\ and\ \citenamefont {Porras}}]{kumbure2022machine}%
  \BibitemOpen
  \bibfield  {author} {\bibinfo {author} {\bibfnamefont {M.~M.}\ \bibnamefont {Kumbure}}, \bibinfo {author} {\bibfnamefont {C.}~\bibnamefont {Lohrmann}}, \bibinfo {author} {\bibfnamefont {P.}~\bibnamefont {Luukka}},\ and\ \bibinfo {author} {\bibfnamefont {J.}~\bibnamefont {Porras}},\ }\bibfield  {title} {{Machine learning techniques and data for stock market forecasting: A literature review},\ }\href@noop {} {\bibfield  {journal} {\bibinfo  {journal} {Expert Systems with Applications}\ ,\ \bibinfo {pages} {116659}} (\bibinfo {year} {2022})}\BibitemShut {NoStop}%
\bibitem [{\citenamefont {Yan}\ and\ \citenamefont {Ling}(2007)}]{yan2007machine}%
  \BibitemOpen
  \bibfield  {author} {\bibinfo {author} {\bibfnamefont {R.~J.}\ \bibnamefont {Yan}}\ and\ \bibinfo {author} {\bibfnamefont {C.~X.}\ \bibnamefont {Ling}},\ }in\ \href@noop {} {\bibinfo {booktitle} {Proceedings of the 13th ACM SIGKDD international conference on knowledge discovery and data mining}}\ (\bibinfo {year} {2007})\ pp.\ \bibinfo {pages} {1038--1042}\BibitemShut {NoStop}%
\bibitem [{\citenamefont {Yuan}\ et~al.(2020)\citenamefont {Yuan}, \citenamefont {Yuan}, \citenamefont {Jiang},\ and\ \citenamefont {Ain}}]{yuan2020integrated}%
  \BibitemOpen
  \bibfield  {author} {\bibinfo {author} {\bibfnamefont {X.}~\bibnamefont {Yuan}}, \bibinfo {author} {\bibfnamefont {J.}~\bibnamefont {Yuan}}, \bibinfo {author} {\bibfnamefont {T.}~\bibnamefont {Jiang}},\ and\ \bibinfo {author} {\bibfnamefont {Q.~U.}\ \bibnamefont {Ain}},\ }\bibfield  {title} {{Integrated long-term stock selection models based on feature selection and machine learning algorithms for China stock market},\ }\href@noop {} {\bibfield  {journal} {\bibinfo  {journal} {IEEE Access}\ }\textbf {\bibinfo {volume} {8}},\ \bibinfo {pages} {22672} (\bibinfo {year} {2020})}\BibitemShut {NoStop}%
\bibitem [{\citenamefont {Rasekhschaffe}\ and\ \citenamefont {Jones}(2019)}]{rasekhschaffe2019machine}%
  \BibitemOpen
  \bibfield  {author} {\bibinfo {author} {\bibfnamefont {K.~C.}\ \bibnamefont {Rasekhschaffe}}\ and\ \bibinfo {author} {\bibfnamefont {R.~C.}\ \bibnamefont {Jones}},\ }\bibfield  {title} {{Machine learning for stock selection},\ }\href@noop {} {\bibfield  {journal} {\bibinfo  {journal} {Financial Analysts Journal}\ }\textbf {\bibinfo {volume} {75}},\ \bibinfo {pages} {70} (\bibinfo {year} {2019})}\BibitemShut {NoStop}%
\bibitem [{\citenamefont {Fu}\ et~al.(2018)\citenamefont {Fu}, \citenamefont {Du}, \citenamefont {Guo}, \citenamefont {Liu}, \citenamefont {Dong},\ and\ \citenamefont {Duan}}]{fu2018machine}%
  \BibitemOpen
  \bibfield  {author} {\bibinfo {author} {\bibfnamefont {X.}~\bibnamefont {Fu}}, \bibinfo {author} {\bibfnamefont {J.}~\bibnamefont {Du}}, \bibinfo {author} {\bibfnamefont {Y.}~\bibnamefont {Guo}}, \bibinfo {author} {\bibfnamefont {M.}~\bibnamefont {Liu}}, \bibinfo {author} {\bibfnamefont {T.}~\bibnamefont {Dong}},\ and\ \bibinfo {author} {\bibfnamefont {X.}~\bibnamefont {Duan}},\ }\bibfield  {title} {{A machine learning framework for stock selection},\ }\href@noop {} {\bibfield  {journal} {\bibinfo  {journal} {arXiv preprint arXiv:1806.01743}\ } (\bibinfo {year} {2018})}\BibitemShut {NoStop}%
\bibitem [{\citenamefont {Conlon}\ et~al.(2021)\citenamefont {Conlon}, \citenamefont {Cotter},\ and\ \citenamefont {Kynigakis}}]{conlon2021machine}%
  \BibitemOpen
  \bibfield  {author} {\bibinfo {author} {\bibfnamefont {T.}~\bibnamefont {Conlon}}, \bibinfo {author} {\bibfnamefont {J.}~\bibnamefont {Cotter}},\ and\ \bibinfo {author} {\bibfnamefont {I.}~\bibnamefont {Kynigakis}},\ }\bibfield  {title} {{Machine learning and factor-based portfolio optimization},\ }\href@noop {} {\bibfield  {journal} {\bibinfo  {journal} {arXiv preprint arXiv:2107.13866}\ } (\bibinfo {year} {2021})}\BibitemShut {NoStop}%
\bibitem [{\citenamefont {Ta}\ et~al.(2018)\citenamefont {Ta}, \citenamefont {Liu},\ and\ \citenamefont {Addis}}]{ta2018prediction}%
  \BibitemOpen
  \bibfield  {author} {\bibinfo {author} {\bibfnamefont {V.-D.}\ \bibnamefont {Ta}}, \bibinfo {author} {\bibfnamefont {C.-M.}\ \bibnamefont {Liu}},\ and\ \bibinfo {author} {\bibfnamefont {D.}~\bibnamefont {Addis}},\ }in\ \href@noop {} {\bibinfo {booktitle} {Proceedings of the 9th International Symposium on Information and Communication Technology}}\ (\bibinfo {year} {2018})\ pp.\ \bibinfo {pages} {98--105}\BibitemShut {NoStop}%
\bibitem [{\citenamefont {Perrin}\ and\ \citenamefont {Roncalli}(2020)}]{perrin2020machine}%
  \BibitemOpen
  \bibfield  {author} {\bibinfo {author} {\bibfnamefont {S.}~\bibnamefont {Perrin}}\ and\ \bibinfo {author} {\bibfnamefont {T.}~\bibnamefont {Roncalli}},\ }\bibfield  {title} {{Machine learning optimization algorithms \& portfolio allocation},\ }\href@noop {} {\bibfield  {journal} {\bibinfo  {journal} {Machine Learning for Asset Management: New Developments and Financial Applications}\ ,\ \bibinfo {pages} {261}} (\bibinfo {year} {2020})}\BibitemShut {NoStop}%
\bibitem [{\citenamefont {Chen}\ et~al.(2021)\citenamefont {Chen}, \citenamefont {Zhang}, \citenamefont {Mehlawat},\ and\ \citenamefont {Jia}}]{chen2021mean}%
  \BibitemOpen
  \bibfield  {author} {\bibinfo {author} {\bibfnamefont {W.}~\bibnamefont {Chen}}, \bibinfo {author} {\bibfnamefont {H.}~\bibnamefont {Zhang}}, \bibinfo {author} {\bibfnamefont {M.~K.}\ \bibnamefont {Mehlawat}},\ and\ \bibinfo {author} {\bibfnamefont {L.}~\bibnamefont {Jia}},\ }\bibfield  {title} {{Mean--variance portfolio optimization using machine learning-based stock price prediction},\ }\href@noop {} {\bibfield  {journal} {\bibinfo  {journal} {Applied Soft Computing}\ }\textbf {\bibinfo {volume} {100}},\ \bibinfo {pages} {106943} (\bibinfo {year} {2021})}\BibitemShut {NoStop}%
\bibitem [{\citenamefont {Ban}\ et~al.(2018)\citenamefont {Ban}, \citenamefont {El~Karoui},\ and\ \citenamefont {Lim}}]{ban2018machine}%
  \BibitemOpen
  \bibfield  {author} {\bibinfo {author} {\bibfnamefont {G.-Y.}\ \bibnamefont {Ban}}, \bibinfo {author} {\bibfnamefont {N.}~\bibnamefont {El~Karoui}},\ and\ \bibinfo {author} {\bibfnamefont {A.~E.}\ \bibnamefont {Lim}},\ }\bibfield  {title} {{Machine learning and portfolio optimization},\ }\href@noop {} {\bibfield  {journal} {\bibinfo  {journal} {Management Science}\ }\textbf {\bibinfo {volume} {64}},\ \bibinfo {pages} {1136} (\bibinfo {year} {2018})}\BibitemShut {NoStop}%
\bibitem [{\citenamefont {Ma}\ et~al.(2021)\citenamefont {Ma}, \citenamefont {Han},\ and\ \citenamefont {Wang}}]{ma2021portfolio}%
  \BibitemOpen
  \bibfield  {author} {\bibinfo {author} {\bibfnamefont {Y.}~\bibnamefont {Ma}}, \bibinfo {author} {\bibfnamefont {R.}~\bibnamefont {Han}},\ and\ \bibinfo {author} {\bibfnamefont {W.}~\bibnamefont {Wang}},\ }\bibfield  {title} {{Portfolio optimization with return prediction using deep learning and machine learning},\ }\href@noop {} {\bibfield  {journal} {\bibinfo  {journal} {Expert Systems with Applications}\ }\textbf {\bibinfo {volume} {165}},\ \bibinfo {pages} {113973} (\bibinfo {year} {2021})}\BibitemShut {NoStop}%
\bibitem [{\citenamefont {Gu}\ et~al.(2020)\citenamefont {Gu}, \citenamefont {Kelly},\ and\ \citenamefont {Xiu}}]{gu2020empirical}%
  \BibitemOpen
  \bibfield  {author} {\bibinfo {author} {\bibfnamefont {S.}~\bibnamefont {Gu}}, \bibinfo {author} {\bibfnamefont {B.}~\bibnamefont {Kelly}},\ and\ \bibinfo {author} {\bibfnamefont {D.}~\bibnamefont {Xiu}},\ }\bibfield  {title} {{Empirical asset pricing via machine learning},\ }\href@noop {} {\bibfield  {journal} {\bibinfo  {journal} {The Review of Financial Studies}\ }\textbf {\bibinfo {volume} {33}},\ \bibinfo {pages} {2223} (\bibinfo {year} {2020})}\BibitemShut {NoStop}%
\bibitem [{\citenamefont {Nagel}(2021)}]{nagel2021machine}%
  \BibitemOpen
  \bibfield  {author} {\bibinfo {author} {\bibfnamefont {S.}~\bibnamefont {Nagel}},\ }\href@noop {} {\bibinfo {title} {Machine learning in asset pricing}},\ Vol.~\bibinfo {volume} {1}\ (\bibinfo  {publisher} {Princeton University Press},\ \bibinfo {year} {2021})\BibitemShut {NoStop}%
\bibitem [{\citenamefont {Bagnara}(2022)}]{bagnara2022asset}%
  \BibitemOpen
  \bibfield  {author} {\bibinfo {author} {\bibfnamefont {M.}~\bibnamefont {Bagnara}},\ }\bibfield  {title} {{Asset Pricing and Machine Learning: A critical review},\ }\href@noop {} {\bibfield  {journal} {\bibinfo  {journal} {Journal of Economic Surveys}\ } (\bibinfo {year} {2022})}\BibitemShut {NoStop}%
\bibitem [{\citenamefont {Gibbs}\ et~al.(2024)\citenamefont {Gibbs}, \citenamefont {Holmes}, \citenamefont {Caro}, \citenamefont {Ezzell}, \citenamefont {Huang}, \citenamefont {Cincio}, \citenamefont {Sornborger},\ and\ \citenamefont {Coles}}]{gibbs2024dynamical}%
  \BibitemOpen
  \bibfield  {author} {\bibinfo {author} {\bibfnamefont {J.}~\bibnamefont {Gibbs}}, \bibinfo {author} {\bibfnamefont {Z.}~\bibnamefont {Holmes}}, \bibinfo {author} {\bibfnamefont {M.~C.}\ \bibnamefont {Caro}}, \bibinfo {author} {\bibfnamefont {N.}~\bibnamefont {Ezzell}}, \bibinfo {author} {\bibfnamefont {H.-Y.}\ \bibnamefont {Huang}}, \bibinfo {author} {\bibfnamefont {L.}~\bibnamefont {Cincio}}, \bibinfo {author} {\bibfnamefont {A.~T.}\ \bibnamefont {Sornborger}},\ and\ \bibinfo {author} {\bibfnamefont {P.~J.}\ \bibnamefont {Coles}},\ }\bibfield  {title} {{Dynamical simulation via quantum machine learning with provable generalization},\ }\href@noop {} {\bibfield  {journal} {\bibinfo  {journal} {Physical Review Research}\ }\textbf {\bibinfo {volume} {6}},\ \bibinfo {pages} {013241} (\bibinfo {year} {2024})}\BibitemShut {NoStop}%
\bibitem [{\citenamefont {Mensa}\ et~al.(2023)\citenamefont {Mensa}, \citenamefont {Sahin}, \citenamefont {Tacchino}, \citenamefont {Barkoutsos},\ and\ \citenamefont {Tavernelli}}]{mensa2023quantum}%
  \BibitemOpen
  \bibfield  {author} {\bibinfo {author} {\bibfnamefont {S.}~\bibnamefont {Mensa}}, \bibinfo {author} {\bibfnamefont {E.}~\bibnamefont {Sahin}}, \bibinfo {author} {\bibfnamefont {F.}~\bibnamefont {Tacchino}}, \bibinfo {author} {\bibfnamefont {P.~K.}\ \bibnamefont {Barkoutsos}},\ and\ \bibinfo {author} {\bibfnamefont {I.}~\bibnamefont {Tavernelli}},\ }\bibfield  {title} {{Quantum machine learning framework for virtual screening in drug discovery: a prospective quantum advantage},\ }\href@noop {} {\bibfield  {journal} {\bibinfo  {journal} {Machine Learning: Science and Technology}\ }\textbf {\bibinfo {volume} {4}},\ \bibinfo {pages} {015023} (\bibinfo {year} {2023})}\BibitemShut {NoStop}%
\bibitem [{\citenamefont {Kasieczka}\ et~al.(2019{\natexlab{a}})\citenamefont {Kasieczka}, \citenamefont {Plehn}, \citenamefont {Butter}, \citenamefont {Cranmer}, \citenamefont {Debnath}, \citenamefont {Dillon}, \citenamefont {Fairbairn}, \citenamefont {Faroughy}, \citenamefont {Fedorko}, \citenamefont {Gay} et~al.}]{araz_ds1}%
  \BibitemOpen
  \bibfield  {author} {\bibinfo {author} {\bibfnamefont {G.}~\bibnamefont {Kasieczka}}, \bibinfo {author} {\bibfnamefont {T.}~\bibnamefont {Plehn}}, \bibinfo {author} {\bibfnamefont {A.}~\bibnamefont {Butter}}, \bibinfo {author} {\bibfnamefont {K.}~\bibnamefont {Cranmer}}, \bibinfo {author} {\bibfnamefont {D.}~\bibnamefont {Debnath}}, \bibinfo {author} {\bibfnamefont {B.~M.}\ \bibnamefont {Dillon}}, \bibinfo {author} {\bibfnamefont {M.}~\bibnamefont {Fairbairn}}, \bibinfo {author} {\bibfnamefont {D.~A.}\ \bibnamefont {Faroughy}}, \bibinfo {author} {\bibfnamefont {W.}~\bibnamefont {Fedorko}}, \bibinfo {author} {\bibfnamefont {C.}~\bibnamefont {Gay}}, et~al.,\ }\bibfield  {title} {{The machine learning landscape of top taggers},\ }\href@noop {} {\bibfield  {journal} {\bibinfo  {journal} {SciPost Physics}\ }\textbf {\bibinfo {volume} {7}},\ \bibinfo {pages} {014} (\bibinfo {year} {2019}{\natexlab{a}})}\BibitemShut {NoStop}%
\bibitem [{\citenamefont {Amari}\ et~al.(2019)\citenamefont {Amari}, \citenamefont {Karakida},\ and\ \citenamefont {Oizumi}}]{amari2019fisher}%
  \BibitemOpen
  \bibfield  {author} {\bibinfo {author} {\bibfnamefont {S.-i.}\ \bibnamefont {Amari}}, \bibinfo {author} {\bibfnamefont {R.}~\bibnamefont {Karakida}},\ and\ \bibinfo {author} {\bibfnamefont {M.}~\bibnamefont {Oizumi}},\ }in\ \href@noop {} {\bibinfo {booktitle} {The 22nd International Conference on Artificial Intelligence and Statistics}}\ (\bibinfo {organization} {PMLR},\ \bibinfo {year} {2019})\ pp.\ \bibinfo {pages} {694--702}\BibitemShut {NoStop}%
\bibitem [{\citenamefont {Harrow}\ et~al.(2009)\citenamefont {Harrow}, \citenamefont {Hassidim},\ and\ \citenamefont {Lloyd}}]{PhysRevLett.103.150502}%
  \BibitemOpen
  \bibfield  {author} {\bibinfo {author} {\bibfnamefont {A.~W.}\ \bibnamefont {Harrow}}, \bibinfo {author} {\bibfnamefont {A.}~\bibnamefont {Hassidim}},\ and\ \bibinfo {author} {\bibfnamefont {S.}~\bibnamefont {Lloyd}},\ }\bibfield  {title} {{Quantum Algorithm for Linear Systems of Equations},\ }\href {https://doi.org/10.1103/PhysRevLett.103.150502} {\bibfield  {journal} {\bibinfo  {journal} {Phys. Rev. Lett.}\ }\textbf {\bibinfo {volume} {103}},\ \bibinfo {pages} {150502} (\bibinfo {year} {2009})}\BibitemShut {NoStop}%
\bibitem [{\citenamefont {Harrow}\ and\ \citenamefont {Napp}(2021)}]{harrow2021low}%
  \BibitemOpen
  \bibfield  {author} {\bibinfo {author} {\bibfnamefont {A.~W.}\ \bibnamefont {Harrow}}\ and\ \bibinfo {author} {\bibfnamefont {J.~C.}\ \bibnamefont {Napp}},\ }\bibfield  {title} {{Low-depth gradient measurements can improve convergence in variational hybrid quantum-classical algorithms},\ }\href@noop {} {\bibfield  {journal} {\bibinfo  {journal} {Physical Review Letters}\ }\textbf {\bibinfo {volume} {126}},\ \bibinfo {pages} {140502} (\bibinfo {year} {2021})}\BibitemShut {NoStop}%
\bibitem [{\citenamefont {Yamamoto}(2019)}]{yamamoto2019natural}%
  \BibitemOpen
  \bibfield  {author} {\bibinfo {author} {\bibfnamefont {N.}~\bibnamefont {Yamamoto}},\ }\bibfield  {title} {{On the natural gradient for variational quantum eigensolver},\ }\href@noop {} {\bibfield  {journal} {\bibinfo  {journal} {arXiv preprint arXiv:1909.05074}\ } (\bibinfo {year} {2019})}\BibitemShut {NoStop}%
\bibitem [{\citenamefont {Huggins}\ et~al.(2019)\citenamefont {Huggins}, \citenamefont {Patil}, \citenamefont {Mitchell}, \citenamefont {Whaley},\ and\ \citenamefont {Stoudenmire}}]{huggins2019towards}%
  \BibitemOpen
  \bibfield  {author} {\bibinfo {author} {\bibfnamefont {W.}~\bibnamefont {Huggins}}, \bibinfo {author} {\bibfnamefont {P.}~\bibnamefont {Patil}}, \bibinfo {author} {\bibfnamefont {B.}~\bibnamefont {Mitchell}}, \bibinfo {author} {\bibfnamefont {K.~B.}\ \bibnamefont {Whaley}},\ and\ \bibinfo {author} {\bibfnamefont {E.~M.}\ \bibnamefont {Stoudenmire}},\ }\bibfield  {title} {{Towards quantum machine learning with tensor networks},\ }\href@noop {} {\bibfield  {journal} {\bibinfo  {journal} {Quantum Science and technology}\ }\textbf {\bibinfo {volume} {4}},\ \bibinfo {pages} {024001} (\bibinfo {year} {2019})}\BibitemShut {NoStop}%
\bibitem [{\citenamefont {Brody}\ and\ \citenamefont {Hughston}(2001)}]{brody2001geometric}%
  \BibitemOpen
  \bibfield  {author} {\bibinfo {author} {\bibfnamefont {D.~C.}\ \bibnamefont {Brody}}\ and\ \bibinfo {author} {\bibfnamefont {L.~P.}\ \bibnamefont {Hughston}},\ }\bibfield  {title} {{Geometric quantum mechanics},\ }\href@noop {} {\bibfield  {journal} {\bibinfo  {journal} {Journal of geometry and physics}\ }\textbf {\bibinfo {volume} {38}},\ \bibinfo {pages} {19} (\bibinfo {year} {2001})}\BibitemShut {NoStop}%
\bibitem [{\citenamefont {Cheng}(2010)}]{cheng2010quantum}%
  \BibitemOpen
  \bibfield  {author} {\bibinfo {author} {\bibfnamefont {R.}~\bibnamefont {Cheng}},\ }\bibfield  {title} {{Quantum geometric tensor (fubini-study metric) in simple quantum system: A pedagogical introduction},\ }\href@noop {} {\bibfield  {journal} {\bibinfo  {journal} {arXiv preprint arXiv:1012.1337}\ } (\bibinfo {year} {2010})}\BibitemShut {NoStop}%
\bibitem [{\citenamefont {Amari}(1998)}]{amari1998natural}%
  \BibitemOpen
  \bibfield  {author} {\bibinfo {author} {\bibfnamefont {S.-I.}\ \bibnamefont {Amari}},\ }\bibfield  {title} {{Natural gradient works efficiently in learning},\ }\href@noop {} {\bibfield  {journal} {\bibinfo  {journal} {Neural computation}\ }\textbf {\bibinfo {volume} {10}},\ \bibinfo {pages} {251} (\bibinfo {year} {1998})}\BibitemShut {NoStop}%
\bibitem [{\citenamefont {Neyshabur}\ et~al.(2015)\citenamefont {Neyshabur}, \citenamefont {Salakhutdinov},\ and\ \citenamefont {Srebro}}]{neyshabur2015path}%
  \BibitemOpen
  \bibfield  {author} {\bibinfo {author} {\bibfnamefont {B.}~\bibnamefont {Neyshabur}}, \bibinfo {author} {\bibfnamefont {R.~R.}\ \bibnamefont {Salakhutdinov}},\ and\ \bibinfo {author} {\bibfnamefont {N.}~\bibnamefont {Srebro}},\ }\bibfield  {title} {{Path-sgd: Path-normalized optimization in deep neural networks},\ }\href@noop {} {\bibfield  {journal} {\bibinfo  {journal} {Advances in neural information processing systems}\ }\textbf {\bibinfo {volume} {28}} (\bibinfo {year} {2015})}\BibitemShut {NoStop}%
\bibitem [{\citenamefont {Caro}\ et~al.(2022)\citenamefont {Caro}, \citenamefont {Huang}, \citenamefont {Cerezo}, \citenamefont {Sharma}, \citenamefont {Sornborger}, \citenamefont {Cincio},\ and\ \citenamefont {Coles}}]{caro2022generalization}%
  \BibitemOpen
  \bibfield  {author} {\bibinfo {author} {\bibfnamefont {M.~C.}\ \bibnamefont {Caro}}, \bibinfo {author} {\bibfnamefont {H.-Y.}\ \bibnamefont {Huang}}, \bibinfo {author} {\bibfnamefont {M.}~\bibnamefont {Cerezo}}, \bibinfo {author} {\bibfnamefont {K.}~\bibnamefont {Sharma}}, \bibinfo {author} {\bibfnamefont {A.}~\bibnamefont {Sornborger}}, \bibinfo {author} {\bibfnamefont {L.}~\bibnamefont {Cincio}},\ and\ \bibinfo {author} {\bibfnamefont {P.~J.}\ \bibnamefont {Coles}},\ }\bibfield  {title} {{Generalization in quantum machine learning from few training data},\ }\href@noop {} {\bibfield  {journal} {\bibinfo  {journal} {Nature communications}\ }\textbf {\bibinfo {volume} {13}},\ \bibinfo {pages} {1} (\bibinfo {year} {2022})}\BibitemShut {NoStop}%
\bibitem [{\citenamefont {Piveteau}\ et~al.(2021)\citenamefont {Piveteau}, \citenamefont {Sutter}, \citenamefont {Bravyi}, \citenamefont {Gambetta},\ and\ \citenamefont {Temme}}]{piveteau2021error}%
  \BibitemOpen
  \bibfield  {author} {\bibinfo {author} {\bibfnamefont {C.}~\bibnamefont {Piveteau}}, \bibinfo {author} {\bibfnamefont {D.}~\bibnamefont {Sutter}}, \bibinfo {author} {\bibfnamefont {S.}~\bibnamefont {Bravyi}}, \bibinfo {author} {\bibfnamefont {J.~M.}\ \bibnamefont {Gambetta}},\ and\ \bibinfo {author} {\bibfnamefont {K.}~\bibnamefont {Temme}},\ }\bibfield  {title} {{Error mitigation for universal gates on encoded qubits},\ }\href@noop {} {\bibfield  {journal} {\bibinfo  {journal} {Physical Review Letters}\ }\textbf {\bibinfo {volume} {127}},\ \bibinfo {pages} {200505} (\bibinfo {year} {2021})}\BibitemShut {NoStop}%
\bibitem [{\citenamefont {Wang}\ et~al.(2021{\natexlab{a}})\citenamefont {Wang}, \citenamefont {Fontana}, \citenamefont {Cerezo}, \citenamefont {Sharma}, \citenamefont {Sone}, \citenamefont {Cincio},\ and\ \citenamefont {Coles}}]{wang2021noise}%
  \BibitemOpen
  \bibfield  {author} {\bibinfo {author} {\bibfnamefont {S.}~\bibnamefont {Wang}}, \bibinfo {author} {\bibfnamefont {E.}~\bibnamefont {Fontana}}, \bibinfo {author} {\bibfnamefont {M.}~\bibnamefont {Cerezo}}, \bibinfo {author} {\bibfnamefont {K.}~\bibnamefont {Sharma}}, \bibinfo {author} {\bibfnamefont {A.}~\bibnamefont {Sone}}, \bibinfo {author} {\bibfnamefont {L.}~\bibnamefont {Cincio}},\ and\ \bibinfo {author} {\bibfnamefont {P.~J.}\ \bibnamefont {Coles}},\ }\bibfield  {title} {{Noise-induced barren plateaus in variational quantum algorithms},\ }\href@noop {} {\bibfield  {journal} {\bibinfo  {journal} {Nature communications}\ }\textbf {\bibinfo {volume} {12}},\ \bibinfo {pages} {1} (\bibinfo {year} {2021}{\natexlab{a}})}\BibitemShut {NoStop}%
\bibitem [{\citenamefont {Patti}\ et~al.(2021)\citenamefont {Patti}, \citenamefont {Najafi}, \citenamefont {Gao},\ and\ \citenamefont {Yelin}}]{patti2021entanglement}%
  \BibitemOpen
  \bibfield  {author} {\bibinfo {author} {\bibfnamefont {T.~L.}\ \bibnamefont {Patti}}, \bibinfo {author} {\bibfnamefont {K.}~\bibnamefont {Najafi}}, \bibinfo {author} {\bibfnamefont {X.}~\bibnamefont {Gao}},\ and\ \bibinfo {author} {\bibfnamefont {S.~F.}\ \bibnamefont {Yelin}},\ }\bibfield  {title} {{Entanglement devised barren plateau mitigation},\ }\href@noop {} {\bibfield  {journal} {\bibinfo  {journal} {Physical Review Research}\ }\textbf {\bibinfo {volume} {3}},\ \bibinfo {pages} {033090} (\bibinfo {year} {2021})}\BibitemShut {NoStop}%
\bibitem [{\citenamefont {Holmes}\ et~al.(2021)\citenamefont {Holmes}, \citenamefont {Arrasmith}, \citenamefont {Yan}, \citenamefont {Coles}, \citenamefont {Albrecht},\ and\ \citenamefont {Sornborger}}]{holmes2021barren}%
  \BibitemOpen
  \bibfield  {author} {\bibinfo {author} {\bibfnamefont {Z.}~\bibnamefont {Holmes}}, \bibinfo {author} {\bibfnamefont {A.}~\bibnamefont {Arrasmith}}, \bibinfo {author} {\bibfnamefont {B.}~\bibnamefont {Yan}}, \bibinfo {author} {\bibfnamefont {P.~J.}\ \bibnamefont {Coles}}, \bibinfo {author} {\bibfnamefont {A.}~\bibnamefont {Albrecht}},\ and\ \bibinfo {author} {\bibfnamefont {A.~T.}\ \bibnamefont {Sornborger}},\ }\bibfield  {title} {{Barren plateaus preclude learning scramblers},\ }\href@noop {} {\bibfield  {journal} {\bibinfo  {journal} {Physical Review Letters}\ }\textbf {\bibinfo {volume} {126}},\ \bibinfo {pages} {190501} (\bibinfo {year} {2021})}\BibitemShut {NoStop}%
\bibitem [{\citenamefont {Sharma}\ et~al.(2022)\citenamefont {Sharma}, \citenamefont {Cerezo}, \citenamefont {Cincio},\ and\ \citenamefont {Coles}}]{sharma2022trainability}%
  \BibitemOpen
  \bibfield  {author} {\bibinfo {author} {\bibfnamefont {K.}~\bibnamefont {Sharma}}, \bibinfo {author} {\bibfnamefont {M.}~\bibnamefont {Cerezo}}, \bibinfo {author} {\bibfnamefont {L.}~\bibnamefont {Cincio}},\ and\ \bibinfo {author} {\bibfnamefont {P.~J.}\ \bibnamefont {Coles}},\ }\bibfield  {title} {{Trainability of dissipative perceptron-based quantum neural networks},\ }\href@noop {} {\bibfield  {journal} {\bibinfo  {journal} {Physical Review Letters}\ }\textbf {\bibinfo {volume} {128}},\ \bibinfo {pages} {180505} (\bibinfo {year} {2022})}\BibitemShut {NoStop}%
\bibitem [{\citenamefont {Kandala}\ et~al.(2018)\citenamefont {Kandala}, \citenamefont {Temme}, \citenamefont {Corcoles}, \citenamefont {Mezzacapo}, \citenamefont {Chow},\ and\ \citenamefont {Gambetta}}]{kandala2018extending}%
  \BibitemOpen
  \bibfield  {author} {\bibinfo {author} {\bibfnamefont {A.}~\bibnamefont {Kandala}}, \bibinfo {author} {\bibfnamefont {K.}~\bibnamefont {Temme}}, \bibinfo {author} {\bibfnamefont {A.~D.}\ \bibnamefont {Corcoles}}, \bibinfo {author} {\bibfnamefont {A.}~\bibnamefont {Mezzacapo}}, \bibinfo {author} {\bibfnamefont {J.~M.}\ \bibnamefont {Chow}},\ and\ \bibinfo {author} {\bibfnamefont {J.~M.}\ \bibnamefont {Gambetta}},\ }\bibfield  {title} {{Extending the computational reach of a noisy superconducting quantum processor},\ }\href@noop {} {\bibfield  {journal} {\bibinfo  {journal} {arXiv preprint arXiv:1805.04492}\ } (\bibinfo {year} {2018})}\BibitemShut {NoStop}%
\bibitem [{\citenamefont {Wang}\ et~al.(2021{\natexlab{b}})\citenamefont {Wang}, \citenamefont {Du}, \citenamefont {Luo},\ and\ \citenamefont {Tao}}]{wang2021towards}%
  \BibitemOpen
  \bibfield  {author} {\bibinfo {author} {\bibfnamefont {X.}~\bibnamefont {Wang}}, \bibinfo {author} {\bibfnamefont {Y.}~\bibnamefont {Du}}, \bibinfo {author} {\bibfnamefont {Y.}~\bibnamefont {Luo}},\ and\ \bibinfo {author} {\bibfnamefont {D.}~\bibnamefont {Tao}},\ }\bibfield  {title} {{Towards understanding the power of quantum kernels in the NISQ era},\ }\href@noop {} {\bibfield  {journal} {\bibinfo  {journal} {Quantum}\ }\textbf {\bibinfo {volume} {5}},\ \bibinfo {pages} {531} (\bibinfo {year} {2021}{\natexlab{b}})}\BibitemShut {NoStop}%
\bibitem [{\citenamefont {Blank}\ et~al.(2020)\citenamefont {Blank}, \citenamefont {Park}, \citenamefont {Rhee},\ and\ \citenamefont {Petruccione}}]{blank2020quantum}%
  \BibitemOpen
  \bibfield  {author} {\bibinfo {author} {\bibfnamefont {C.}~\bibnamefont {Blank}}, \bibinfo {author} {\bibfnamefont {D.~K.}\ \bibnamefont {Park}}, \bibinfo {author} {\bibfnamefont {J.-K.~K.}\ \bibnamefont {Rhee}},\ and\ \bibinfo {author} {\bibfnamefont {F.}~\bibnamefont {Petruccione}},\ }\bibfield  {title} {{Quantum classifier with tailored quantum kernel},\ }\href@noop {} {\bibfield  {journal} {\bibinfo  {journal} {npj Quantum Information}\ }\textbf {\bibinfo {volume} {6}},\ \bibinfo {pages} {1} (\bibinfo {year} {2020})}\BibitemShut {NoStop}%
\bibitem [{\citenamefont {Huang}\ et~al.(2021{\natexlab{b}})\citenamefont {Huang}, \citenamefont {Broughton}, \citenamefont {Mohseni}, \citenamefont {Babbush}, \citenamefont {Boixo}, \citenamefont {Neven},\ and\ \citenamefont {McClean}}]{huang2021power}%
  \BibitemOpen
  \bibfield  {author} {\bibinfo {author} {\bibfnamefont {H.-Y.}\ \bibnamefont {Huang}}, \bibinfo {author} {\bibfnamefont {M.}~\bibnamefont {Broughton}}, \bibinfo {author} {\bibfnamefont {M.}~\bibnamefont {Mohseni}}, \bibinfo {author} {\bibfnamefont {R.}~\bibnamefont {Babbush}}, \bibinfo {author} {\bibfnamefont {S.}~\bibnamefont {Boixo}}, \bibinfo {author} {\bibfnamefont {H.}~\bibnamefont {Neven}},\ and\ \bibinfo {author} {\bibfnamefont {J.~R.}\ \bibnamefont {McClean}},\ }\bibfield  {title} {{Power of data in quantum machine learning},\ }\href@noop {} {\bibfield  {journal} {\bibinfo  {journal} {Nature communications}\ }\textbf {\bibinfo {volume} {12}},\ \bibinfo {pages} {1} (\bibinfo {year} {2021}{\natexlab{b}})}\BibitemShut {NoStop}%
\bibitem [{\citenamefont {Rist{\`e}}\ et~al.(2017)\citenamefont {Rist{\`e}}, \citenamefont {Da~Silva}, \citenamefont {Ryan}, \citenamefont {Cross}, \citenamefont {C{\'o}rcoles}, \citenamefont {Smolin}, \citenamefont {Gambetta}, \citenamefont {Chow},\ and\ \citenamefont {Johnson}}]{riste2017demonstration}%
  \BibitemOpen
  \bibfield  {author} {\bibinfo {author} {\bibfnamefont {D.}~\bibnamefont {Rist{\`e}}}, \bibinfo {author} {\bibfnamefont {M.~P.}\ \bibnamefont {Da~Silva}}, \bibinfo {author} {\bibfnamefont {C.~A.}\ \bibnamefont {Ryan}}, \bibinfo {author} {\bibfnamefont {A.~W.}\ \bibnamefont {Cross}}, \bibinfo {author} {\bibfnamefont {A.~D.}\ \bibnamefont {C{\'o}rcoles}}, \bibinfo {author} {\bibfnamefont {J.~A.}\ \bibnamefont {Smolin}}, \bibinfo {author} {\bibfnamefont {J.~M.}\ \bibnamefont {Gambetta}}, \bibinfo {author} {\bibfnamefont {J.~M.}\ \bibnamefont {Chow}},\ and\ \bibinfo {author} {\bibfnamefont {B.~R.}\ \bibnamefont {Johnson}},\ }\bibfield  {title} {{Demonstration of quantum advantage in machine learning},\ }\href@noop {} {\bibfield  {journal} {\bibinfo  {journal} {npj Quantum Information}\ }\textbf {\bibinfo {volume} {3}},\ \bibinfo {pages} {1} (\bibinfo {year} {2017})}\BibitemShut {NoStop}%
\bibitem [{\citenamefont {LaRose}\ and\ \citenamefont {Coyle}(2020)}]{encoding1}%
  \BibitemOpen
  \bibfield  {author} {\bibinfo {author} {\bibfnamefont {R.}~\bibnamefont {LaRose}}\ and\ \bibinfo {author} {\bibfnamefont {B.}~\bibnamefont {Coyle}},\ }\bibfield  {title} {{Robust data encodings for quantum classifiers},\ }\href@noop {} {\bibfield  {journal} {\bibinfo  {journal} {Physical Review A}\ }\textbf {\bibinfo {volume} {102}},\ \bibinfo {pages} {032420} (\bibinfo {year} {2020})}\BibitemShut {NoStop}%
\bibitem [{\citenamefont {Cai}\ et~al.(2015)\citenamefont {Cai}, \citenamefont {Wu}, \citenamefont {Su}, \citenamefont {Chen}, \citenamefont {Wang}, \citenamefont {Li}, \citenamefont {Liu}, \citenamefont {Lu},\ and\ \citenamefont {Pan}}]{cai2015entanglement}%
  \BibitemOpen
  \bibfield  {author} {\bibinfo {author} {\bibfnamefont {X.-D.}\ \bibnamefont {Cai}}, \bibinfo {author} {\bibfnamefont {D.}~\bibnamefont {Wu}}, \bibinfo {author} {\bibfnamefont {Z.-E.}\ \bibnamefont {Su}}, \bibinfo {author} {\bibfnamefont {M.-C.}\ \bibnamefont {Chen}}, \bibinfo {author} {\bibfnamefont {X.-L.}\ \bibnamefont {Wang}}, \bibinfo {author} {\bibfnamefont {L.}~\bibnamefont {Li}}, \bibinfo {author} {\bibfnamefont {N.-L.}\ \bibnamefont {Liu}}, \bibinfo {author} {\bibfnamefont {C.-Y.}\ \bibnamefont {Lu}},\ and\ \bibinfo {author} {\bibfnamefont {J.-W.}\ \bibnamefont {Pan}},\ }\bibfield  {title} {{Entanglement-based machine learning on a quantum computer},\ }\href@noop {} {\bibfield  {journal} {\bibinfo  {journal} {Physical review letters}\ }\textbf {\bibinfo {volume} {114}},\ \bibinfo {pages} {110504} (\bibinfo {year} {2015})}\BibitemShut {NoStop}%
\bibitem [{\citenamefont {Periyasamy}\ et~al.(2022)\citenamefont {Periyasamy}, \citenamefont {Meyer}, \citenamefont {Ufrecht}, \citenamefont {Scherer}, \citenamefont {Plinge},\ and\ \citenamefont {Mutschler}}]{periyasamy2022incremental}%
  \BibitemOpen
  \bibfield  {author} {\bibinfo {author} {\bibfnamefont {M.}~\bibnamefont {Periyasamy}}, \bibinfo {author} {\bibfnamefont {N.}~\bibnamefont {Meyer}}, \bibinfo {author} {\bibfnamefont {C.}~\bibnamefont {Ufrecht}}, \bibinfo {author} {\bibfnamefont {D.~D.}\ \bibnamefont {Scherer}}, \bibinfo {author} {\bibfnamefont {A.}~\bibnamefont {Plinge}},\ and\ \bibinfo {author} {\bibfnamefont {C.}~\bibnamefont {Mutschler}},\ }in\ \href@noop {} {\bibinfo {booktitle} {2022 IEEE International Conference on Quantum Computing and Engineering (QCE)}}\ (\bibinfo {organization} {IEEE},\ \bibinfo {year} {2022})\ pp.\ \bibinfo {pages} {31--37}\BibitemShut {NoStop}%
\bibitem [{\citenamefont {Arute}\ et~al.(2019)\citenamefont {Arute}, \citenamefont {Arya}, \citenamefont {Babbush}, \citenamefont {Bacon}, \citenamefont {Bardin}, \citenamefont {Barends}, \citenamefont {Biswas}, \citenamefont {Boixo}, \citenamefont {Brandao}, \citenamefont {Buell} et~al.}]{quantumsupremacy}%
  \BibitemOpen
  \bibfield  {author} {\bibinfo {author} {\bibfnamefont {F.}~\bibnamefont {Arute}}, \bibinfo {author} {\bibfnamefont {K.}~\bibnamefont {Arya}}, \bibinfo {author} {\bibfnamefont {R.}~\bibnamefont {Babbush}}, \bibinfo {author} {\bibfnamefont {D.}~\bibnamefont {Bacon}}, \bibinfo {author} {\bibfnamefont {J.~C.}\ \bibnamefont {Bardin}}, \bibinfo {author} {\bibfnamefont {R.}~\bibnamefont {Barends}}, \bibinfo {author} {\bibfnamefont {R.}~\bibnamefont {Biswas}}, \bibinfo {author} {\bibfnamefont {S.}~\bibnamefont {Boixo}}, \bibinfo {author} {\bibfnamefont {F.~G.}\ \bibnamefont {Brandao}}, \bibinfo {author} {\bibfnamefont {D.~A.}\ \bibnamefont {Buell}}, et~al.,\ }\bibfield  {title} {{Quantum supremacy using a programmable superconducting processor},\ }\href@noop {} {\bibfield  {journal} {\bibinfo  {journal} {Nature}\ }\textbf {\bibinfo {volume} {574}},\ \bibinfo {pages} {505} (\bibinfo {year} {2019})}\BibitemShut {NoStop}%
\bibitem [{\citenamefont {Preskill}(2023)}]{preskill2023quantum}%
  \BibitemOpen
  \bibfield  {author} {\bibinfo {author} {\bibfnamefont {J.}~\bibnamefont {Preskill}},\ }in\ \href@noop {} {\bibinfo {booktitle} {Feynman Lectures on Computation}}\ (\bibinfo  {publisher} {CRC Press},\ \bibinfo {year} {2023})\ pp.\ \bibinfo {pages} {193--244}\BibitemShut {NoStop}%
\bibitem [{\citenamefont {Preskill}(2018)}]{preskill2018quantum}%
  \BibitemOpen
  \bibfield  {author} {\bibinfo {author} {\bibfnamefont {J.}~\bibnamefont {Preskill}},\ }\bibfield  {title} {{Quantum computing in the {NISQ} era and beyond},\ }\href@noop {} {\bibfield  {journal} {\bibinfo  {journal} {Quantum}\ }\textbf {\bibinfo {volume} {2}},\ \bibinfo {pages} {79} (\bibinfo {year} {2018})}\BibitemShut {NoStop}%
\bibitem [{\citenamefont {Pearson}(1901)}]{pca1}%
  \BibitemOpen
  \bibfield  {author} {\bibinfo {author} {\bibfnamefont {K.}~\bibnamefont {Pearson}},\ }\bibfield  {title} {{LIII. On lines and planes of closest fit to systems of points in space},\ }\href@noop {} {\bibfield  {journal} {\bibinfo  {journal} {The London, Edinburgh, and Dublin philosophical magazine and journal of science}\ }\textbf {\bibinfo {volume} {2}},\ \bibinfo {pages} {559} (\bibinfo {year} {1901})}\BibitemShut {NoStop}%
\bibitem [{\citenamefont {Pedregosa}\ et~al.(2011)\citenamefont {Pedregosa}, \citenamefont {Varoquaux}, \citenamefont {Gramfort}, \citenamefont {Michel}, \citenamefont {Thirion}, \citenamefont {Grisel}, \citenamefont {Blondel}, \citenamefont {Prettenhofer}, \citenamefont {Weiss}, \citenamefont {Dubourg} et~al.}]{min_max_scikit}%
  \BibitemOpen
  \bibfield  {author} {\bibinfo {author} {\bibfnamefont {F.}~\bibnamefont {Pedregosa}}, \bibinfo {author} {\bibfnamefont {G.}~\bibnamefont {Varoquaux}}, \bibinfo {author} {\bibfnamefont {A.}~\bibnamefont {Gramfort}}, \bibinfo {author} {\bibfnamefont {V.}~\bibnamefont {Michel}}, \bibinfo {author} {\bibfnamefont {B.}~\bibnamefont {Thirion}}, \bibinfo {author} {\bibfnamefont {O.}~\bibnamefont {Grisel}}, \bibinfo {author} {\bibfnamefont {M.}~\bibnamefont {Blondel}}, \bibinfo {author} {\bibfnamefont {P.}~\bibnamefont {Prettenhofer}}, \bibinfo {author} {\bibfnamefont {R.}~\bibnamefont {Weiss}}, \bibinfo {author} {\bibfnamefont {V.}~\bibnamefont {Dubourg}}, et~al.,\ }\bibfield  {title} {{Scikit-learn: Machine learning in Python},\ }\href@noop {} {\bibfield  {journal} {\bibinfo  {journal} {the Journal of machine Learning research}\ }\textbf {\bibinfo {volume} {12}},\ \bibinfo {pages} {2825} (\bibinfo {year} {2011})}\BibitemShut {NoStop}%
\bibitem [{\citenamefont {Alwall}\ et~al.(2014)\citenamefont {Alwall}, \citenamefont {Frederix}, \citenamefont {Frixione}, \citenamefont {Hirschi}, \citenamefont {Maltoni}, \citenamefont {Mattelaer}, \citenamefont {Shao}, \citenamefont {Stelzer}, \citenamefont {Torrielli},\ and\ \citenamefont {Zaro}}]{wu_vqc_data}%
  \BibitemOpen
  \bibfield  {author} {\bibinfo {author} {\bibfnamefont {J.}~\bibnamefont {Alwall}}, \bibinfo {author} {\bibfnamefont {R.}~\bibnamefont {Frederix}}, \bibinfo {author} {\bibfnamefont {S.}~\bibnamefont {Frixione}}, \bibinfo {author} {\bibfnamefont {V.}~\bibnamefont {Hirschi}}, \bibinfo {author} {\bibfnamefont {F.}~\bibnamefont {Maltoni}}, \bibinfo {author} {\bibfnamefont {O.}~\bibnamefont {Mattelaer}}, \bibinfo {author} {\bibfnamefont {H.-S.}\ \bibnamefont {Shao}}, \bibinfo {author} {\bibfnamefont {T.}~\bibnamefont {Stelzer}}, \bibinfo {author} {\bibfnamefont {P.}~\bibnamefont {Torrielli}},\ and\ \bibinfo {author} {\bibfnamefont {M.}~\bibnamefont {Zaro}},\ }\bibfield  {title} {{The automated computation of tree-level and next-to-leading order differential cross sections, and their matching to parton shower simulations},\ }\href@noop {} {\bibfield  {journal} {\bibinfo  {journal} {Journal of High Energy Physics}\ }\textbf {\bibinfo {volume} {2014}},\ \bibinfo {pages} {1} (\bibinfo {year} {2014})}\BibitemShut
  {NoStop}%
\bibitem [{\citenamefont {P{\'e}rez-Salinas}\ et~al.(2021)\citenamefont {P{\'e}rez-Salinas}, \citenamefont {Cruz-Martinez}, \citenamefont {Alhajri},\ and\ \citenamefont {Carrazza}}]{perez2021determining}%
  \BibitemOpen
  \bibfield  {author} {\bibinfo {author} {\bibfnamefont {A.}~\bibnamefont {P{\'e}rez-Salinas}}, \bibinfo {author} {\bibfnamefont {J.}~\bibnamefont {Cruz-Martinez}}, \bibinfo {author} {\bibfnamefont {A.~A.}\ \bibnamefont {Alhajri}},\ and\ \bibinfo {author} {\bibfnamefont {S.}~\bibnamefont {Carrazza}},\ }\bibfield  {title} {{Determining the proton content with a quantum computer},\ }\href@noop {} {\bibfield  {journal} {\bibinfo  {journal} {Physical Review D}\ }\textbf {\bibinfo {volume} {103}},\ \bibinfo {pages} {034027} (\bibinfo {year} {2021})}\BibitemShut {NoStop}%
\bibitem [{\citenamefont {Aleksandrowicz}\ et~al.(2019)\citenamefont {Aleksandrowicz}, \citenamefont {Alexander}, \citenamefont {Barkoutsos}, \citenamefont {Bello}, \citenamefont {Ben-Haim}, \citenamefont {Bucher}, \citenamefont {Cabrera-Hernández}, \citenamefont {Carballo-Franquis}, \citenamefont {Chen}, \citenamefont {Chen}, \citenamefont {Chow}, \citenamefont {Córcoles-Gonzales}, \citenamefont {Cross}, \citenamefont {Cross}, \citenamefont {Cruz-Benito}, \citenamefont {Culver}, \citenamefont {González}, \citenamefont {Torre}, \citenamefont {Ding}, \citenamefont {Dumitrescu}, \citenamefont {Duran}, \citenamefont {Eendebak}, \citenamefont {Everitt}, \citenamefont {Sertage}, \citenamefont {Frisch}, \citenamefont {Fuhrer}, \citenamefont {Gambetta}, \citenamefont {Gago}, \citenamefont {Gomez-Mosquera}, \citenamefont {Greenberg}, \citenamefont {Hamamura}, \citenamefont {Havlicek}, \citenamefont {Hellmers}, \citenamefont {Łukasz Herok}, \citenamefont {Horii}, \citenamefont {Hu}, \citenamefont {Imamichi},
  \citenamefont {Itoko}, \citenamefont {Javadi-Abhari}, \citenamefont {Kanazawa}, \citenamefont {Karazeev}, \citenamefont {Krsulich}, \citenamefont {Liu}, \citenamefont {Luh}, \citenamefont {Maeng}, \citenamefont {Marques}, \citenamefont {Martín-Fernández}, \citenamefont {McClure}, \citenamefont {McKay}, \citenamefont {Meesala}, \citenamefont {Mezzacapo}, \citenamefont {Moll}, \citenamefont {Rodríguez}, \citenamefont {Nannicini}, \citenamefont {Nation}, \citenamefont {Ollitrault}, \citenamefont {O'Riordan}, \citenamefont {Paik}, \citenamefont {Pérez}, \citenamefont {Phan}, \citenamefont {Pistoia}, \citenamefont {Prutyanov}, \citenamefont {Reuter}, \citenamefont {Rice}, \citenamefont {Davila}, \citenamefont {Rudy}, \citenamefont {Ryu}, \citenamefont {Sathaye}, \citenamefont {Schnabel}, \citenamefont {Schoute}, \citenamefont {Setia}, \citenamefont {Shi}, \citenamefont {Silva}, \citenamefont {Siraichi}, \citenamefont {Sivarajah}, \citenamefont {Smolin}, \citenamefont {Soeken}, \citenamefont {Takahashi},
  \citenamefont {Tavernelli}, \citenamefont {Taylor}, \citenamefont {Taylour}, \citenamefont {Trabing}, \citenamefont {Treinish}, \citenamefont {Turner}, \citenamefont {Vogt-Lee}, \citenamefont {Vuillot}, \citenamefont {Wildstrom}, \citenamefont {Wilson}, \citenamefont {Winston}, \citenamefont {Wood}, \citenamefont {Wood}, \citenamefont {Wörner}, \citenamefont {Akhalwaya},\ and\ \citenamefont {Zoufal}}]{qiskit2}%
  \BibitemOpen
  \bibfield  {author} {\bibinfo {author} {\bibfnamefont {G.}~\bibnamefont {Aleksandrowicz}}, \bibinfo {author} {\bibfnamefont {T.}~\bibnamefont {Alexander}}, \bibinfo {author} {\bibfnamefont {P.}~\bibnamefont {Barkoutsos}}, \bibinfo {author} {\bibfnamefont {L.}~\bibnamefont {Bello}}, \bibinfo {author} {\bibfnamefont {Y.}~\bibnamefont {Ben-Haim}}, \bibinfo {author} {\bibfnamefont {D.}~\bibnamefont {Bucher}}, \bibinfo {author} {\bibfnamefont {F.~J.}\ \bibnamefont {Cabrera-Hernández}}, \bibinfo {author} {\bibfnamefont {J.}~\bibnamefont {Carballo-Franquis}}, \bibinfo {author} {\bibfnamefont {A.}~\bibnamefont {Chen}}, \bibinfo {author} {\bibfnamefont {C.-F.}\ \bibnamefont {Chen}}, \bibinfo {author} {\bibfnamefont {J.~M.}\ \bibnamefont {Chow}}, \bibinfo {author} {\bibfnamefont {A.~D.}\ \bibnamefont {Córcoles-Gonzales}}, \bibinfo {author} {\bibfnamefont {A.~J.}\ \bibnamefont {Cross}}, \bibinfo {author} {\bibfnamefont {A.}~\bibnamefont {Cross}}, \bibinfo {author} {\bibfnamefont {J.}~\bibnamefont {Cruz-Benito}},
  \bibinfo {author} {\bibfnamefont {C.}~\bibnamefont {Culver}}, \bibinfo {author} {\bibfnamefont {S.~D. L.~P.}\ \bibnamefont {González}}, \bibinfo {author} {\bibfnamefont {E.~D.~L.}\ \bibnamefont {Torre}}, \bibinfo {author} {\bibfnamefont {D.}~\bibnamefont {Ding}}, \bibinfo {author} {\bibfnamefont {E.}~\bibnamefont {Dumitrescu}}, \bibinfo {author} {\bibfnamefont {I.}~\bibnamefont {Duran}}, \bibinfo {author} {\bibfnamefont {P.}~\bibnamefont {Eendebak}}, \bibinfo {author} {\bibfnamefont {M.}~\bibnamefont {Everitt}}, \bibinfo {author} {\bibfnamefont {I.~F.}\ \bibnamefont {Sertage}}, \bibinfo {author} {\bibfnamefont {A.}~\bibnamefont {Frisch}}, \bibinfo {author} {\bibfnamefont {A.}~\bibnamefont {Fuhrer}}, \bibinfo {author} {\bibfnamefont {J.}~\bibnamefont {Gambetta}}, \bibinfo {author} {\bibfnamefont {B.~G.}\ \bibnamefont {Gago}}, \bibinfo {author} {\bibfnamefont {J.}~\bibnamefont {Gomez-Mosquera}}, \bibinfo {author} {\bibfnamefont {D.}~\bibnamefont {Greenberg}}, \bibinfo {author} {\bibfnamefont
  {I.}~\bibnamefont {Hamamura}}, \bibinfo {author} {\bibfnamefont {V.}~\bibnamefont {Havlicek}}, \bibinfo {author} {\bibfnamefont {J.}~\bibnamefont {Hellmers}}, \bibinfo {author} {\bibnamefont {Łukasz Herok}}, \bibinfo {author} {\bibfnamefont {H.}~\bibnamefont {Horii}}, \bibinfo {author} {\bibfnamefont {S.}~\bibnamefont {Hu}}, \bibinfo {author} {\bibfnamefont {T.}~\bibnamefont {Imamichi}}, \bibinfo {author} {\bibfnamefont {T.}~\bibnamefont {Itoko}}, \bibinfo {author} {\bibfnamefont {A.}~\bibnamefont {Javadi-Abhari}}, \bibinfo {author} {\bibfnamefont {N.}~\bibnamefont {Kanazawa}}, \bibinfo {author} {\bibfnamefont {A.}~\bibnamefont {Karazeev}}, \bibinfo {author} {\bibfnamefont {K.}~\bibnamefont {Krsulich}}, \bibinfo {author} {\bibfnamefont {P.}~\bibnamefont {Liu}}, \bibinfo {author} {\bibfnamefont {Y.}~\bibnamefont {Luh}}, \bibinfo {author} {\bibfnamefont {Y.}~\bibnamefont {Maeng}}, \bibinfo {author} {\bibfnamefont {M.}~\bibnamefont {Marques}}, \bibinfo {author} {\bibfnamefont {F.~J.}\ \bibnamefont
  {Martín-Fernández}}, \bibinfo {author} {\bibfnamefont {D.~T.}\ \bibnamefont {McClure}}, \bibinfo {author} {\bibfnamefont {D.}~\bibnamefont {McKay}}, \bibinfo {author} {\bibfnamefont {S.}~\bibnamefont {Meesala}}, \bibinfo {author} {\bibfnamefont {A.}~\bibnamefont {Mezzacapo}}, \bibinfo {author} {\bibfnamefont {N.}~\bibnamefont {Moll}}, \bibinfo {author} {\bibfnamefont {D.~M.}\ \bibnamefont {Rodríguez}}, \bibinfo {author} {\bibfnamefont {G.}~\bibnamefont {Nannicini}}, \bibinfo {author} {\bibfnamefont {P.}~\bibnamefont {Nation}}, \bibinfo {author} {\bibfnamefont {P.}~\bibnamefont {Ollitrault}}, \bibinfo {author} {\bibfnamefont {L.~J.}\ \bibnamefont {O'Riordan}}, \bibinfo {author} {\bibfnamefont {H.}~\bibnamefont {Paik}}, \bibinfo {author} {\bibfnamefont {J.}~\bibnamefont {Pérez}}, \bibinfo {author} {\bibfnamefont {A.}~\bibnamefont {Phan}}, \bibinfo {author} {\bibfnamefont {M.}~\bibnamefont {Pistoia}}, \bibinfo {author} {\bibfnamefont {V.}~\bibnamefont {Prutyanov}}, \bibinfo {author} {\bibfnamefont
  {M.}~\bibnamefont {Reuter}}, \bibinfo {author} {\bibfnamefont {J.}~\bibnamefont {Rice}}, \bibinfo {author} {\bibfnamefont {A.~R.}\ \bibnamefont {Davila}}, \bibinfo {author} {\bibfnamefont {R.~H.~P.}\ \bibnamefont {Rudy}}, \bibinfo {author} {\bibfnamefont {M.}~\bibnamefont {Ryu}}, \bibinfo {author} {\bibfnamefont {N.}~\bibnamefont {Sathaye}}, \bibinfo {author} {\bibfnamefont {C.}~\bibnamefont {Schnabel}}, \bibinfo {author} {\bibfnamefont {E.}~\bibnamefont {Schoute}}, \bibinfo {author} {\bibfnamefont {K.}~\bibnamefont {Setia}}, \bibinfo {author} {\bibfnamefont {Y.}~\bibnamefont {Shi}}, \bibinfo {author} {\bibfnamefont {A.}~\bibnamefont {Silva}}, \bibinfo {author} {\bibfnamefont {Y.}~\bibnamefont {Siraichi}}, \bibinfo {author} {\bibfnamefont {S.}~\bibnamefont {Sivarajah}}, \bibinfo {author} {\bibfnamefont {J.~A.}\ \bibnamefont {Smolin}}, \bibinfo {author} {\bibfnamefont {M.}~\bibnamefont {Soeken}}, \bibinfo {author} {\bibfnamefont {H.}~\bibnamefont {Takahashi}}, \bibinfo {author} {\bibfnamefont
  {I.}~\bibnamefont {Tavernelli}}, \bibinfo {author} {\bibfnamefont {C.}~\bibnamefont {Taylor}}, \bibinfo {author} {\bibfnamefont {P.}~\bibnamefont {Taylour}}, \bibinfo {author} {\bibfnamefont {K.}~\bibnamefont {Trabing}}, \bibinfo {author} {\bibfnamefont {M.}~\bibnamefont {Treinish}}, \bibinfo {author} {\bibfnamefont {W.}~\bibnamefont {Turner}}, \bibinfo {author} {\bibfnamefont {D.}~\bibnamefont {Vogt-Lee}}, \bibinfo {author} {\bibfnamefont {C.}~\bibnamefont {Vuillot}}, \bibinfo {author} {\bibfnamefont {J.~A.}\ \bibnamefont {Wildstrom}}, \bibinfo {author} {\bibfnamefont {J.}~\bibnamefont {Wilson}}, \bibinfo {author} {\bibfnamefont {E.}~\bibnamefont {Winston}}, \bibinfo {author} {\bibfnamefont {C.}~\bibnamefont {Wood}}, \bibinfo {author} {\bibfnamefont {S.}~\bibnamefont {Wood}}, \bibinfo {author} {\bibfnamefont {S.}~\bibnamefont {Wörner}}, \bibinfo {author} {\bibfnamefont {I.~Y.}\ \bibnamefont {Akhalwaya}},\ and\ \bibinfo {author} {\bibfnamefont {C.}~\bibnamefont {Zoufal}},\ }\href
  {https://doi.org/10.5281/zenodo.2562111} {{{Qiskit: An Open-source Framework for Quantum Computing}}} (\bibinfo {year} {2019})\BibitemShut {NoStop}%
\bibitem [{\citenamefont {Bhatia}\ et~al.(2023)\citenamefont {Bhatia}, \citenamefont {Kais},\ and\ \citenamefont {Alam}}]{bhatia2023federated}%
  \BibitemOpen
  \bibfield  {author} {\bibinfo {author} {\bibfnamefont {A.~S.}\ \bibnamefont {Bhatia}}, \bibinfo {author} {\bibfnamefont {S.}~\bibnamefont {Kais}},\ and\ \bibinfo {author} {\bibfnamefont {M.~A.}\ \bibnamefont {Alam}},\ }\bibfield  {title} {{Federated quanvolutional neural network: a new paradigm for collaborative quantum learning},\ }\href@noop {} {\bibfield  {journal} {\bibinfo  {journal} {Quantum Science and Technology}\ }\textbf {\bibinfo {volume} {8}},\ \bibinfo {pages} {045032} (\bibinfo {year} {2023})}\BibitemShut {NoStop}%
\bibitem [{\citenamefont {Kasieczka}\ et~al.(2019{\natexlab{b}})\citenamefont {Kasieczka}, \citenamefont {Plehn}, \citenamefont {Thompson},\ and\ \citenamefont {Russel}}]{araz_ds2}%
  \BibitemOpen
  \bibfield  {author} {\bibinfo {author} {\bibfnamefont {G.}~\bibnamefont {Kasieczka}}, \bibinfo {author} {\bibfnamefont {T.}~\bibnamefont {Plehn}}, \bibinfo {author} {\bibfnamefont {J.}~\bibnamefont {Thompson}},\ and\ \bibinfo {author} {\bibfnamefont {M.}~\bibnamefont {Russel}},\ }\href {https://doi.org/10.5281/zenodo.2603256} {{Top Quark Tagging Reference Dataset}} (\bibinfo {year} {2019}{\natexlab{b}})\BibitemShut {NoStop}%
\bibitem [{\citenamefont {McKay}\ et~al.(2018)\citenamefont {McKay}, \citenamefont {Alexander}, \citenamefont {Bello}, \citenamefont {Biercuk}, \citenamefont {Bishop}, \citenamefont {Chen}, \citenamefont {Chow}, \citenamefont {C{\'o}rcoles}, \citenamefont {Egger}, \citenamefont {Filipp} et~al.}]{qiskit1}%
  \BibitemOpen
  \bibfield  {author} {\bibinfo {author} {\bibfnamefont {D.~C.}\ \bibnamefont {McKay}}, \bibinfo {author} {\bibfnamefont {T.}~\bibnamefont {Alexander}}, \bibinfo {author} {\bibfnamefont {L.}~\bibnamefont {Bello}}, \bibinfo {author} {\bibfnamefont {M.~J.}\ \bibnamefont {Biercuk}}, \bibinfo {author} {\bibfnamefont {L.}~\bibnamefont {Bishop}}, \bibinfo {author} {\bibfnamefont {J.}~\bibnamefont {Chen}}, \bibinfo {author} {\bibfnamefont {J.~M.}\ \bibnamefont {Chow}}, \bibinfo {author} {\bibfnamefont {A.~D.}\ \bibnamefont {C{\'o}rcoles}}, \bibinfo {author} {\bibfnamefont {D.}~\bibnamefont {Egger}}, \bibinfo {author} {\bibfnamefont {S.}~\bibnamefont {Filipp}}, et~al.,\ }\bibfield  {title} {{Qiskit backend specifications for openqasm and openpulse experiments},\ }\href@noop {} {\bibfield  {journal} {\bibinfo  {journal} {arXiv preprint arXiv:1809.03452}\ } (\bibinfo {year} {2018})}\BibitemShut {NoStop}%
\bibitem [{\citenamefont {Clark}\ et~al.(2018)\citenamefont {Clark}, \citenamefont {Farrington}, \citenamefont {Faucci~Giannelli}, \citenamefont {Gao}, \citenamefont {Hasib}, \citenamefont {Martin},\ and\ \citenamefont {Mijovic}}]{clark2018search}%
  \BibitemOpen
  \bibfield  {author} {\bibinfo {author} {\bibfnamefont {P.~J.}\ \bibnamefont {Clark}}, \bibinfo {author} {\bibfnamefont {S.}~\bibnamefont {Farrington}}, \bibinfo {author} {\bibfnamefont {M.}~\bibnamefont {Faucci~Giannelli}}, \bibinfo {author} {\bibfnamefont {Y.}~\bibnamefont {Gao}}, \bibinfo {author} {\bibfnamefont {A.}~\bibnamefont {Hasib}}, \bibinfo {author} {\bibfnamefont {V.~J.}\ \bibnamefont {Martin}},\ and\ \bibinfo {author} {\bibfnamefont {L.}~\bibnamefont {Mijovic}},\ }\bibfield  {title} {{Search for heavy particles decaying into top-quark pairs using lepton-plus-jets events in proton--proton collisions at $\sqrt{s} = 13\,\text{TeV}$ with the {ATLAS} detector},\ }\href@noop {} {\bibfield  {journal} {\bibinfo  {journal} {European Physical Journal C: Particles and Fields}\ }\textbf {\bibinfo {volume} {78}},\ \bibinfo {pages} {565} (\bibinfo {year} {2018})}\BibitemShut {NoStop}%
\bibitem [{\citenamefont {Lazar}\ et~al.(2024)\citenamefont {Lazar}, \citenamefont {Olavarrieta}, \citenamefont {Gatti}, \citenamefont {Arg{\"u}elles},\ and\ \citenamefont {Sanz}}]{lazar2024new}%
  \BibitemOpen
  \bibfield  {author} {\bibinfo {author} {\bibfnamefont {J.}~\bibnamefont {Lazar}}, \bibinfo {author} {\bibfnamefont {S.~G.}\ \bibnamefont {Olavarrieta}}, \bibinfo {author} {\bibfnamefont {G.}~\bibnamefont {Gatti}}, \bibinfo {author} {\bibfnamefont {C.~A.}\ \bibnamefont {Arg{\"u}elles}},\ and\ \bibinfo {author} {\bibfnamefont {M.}~\bibnamefont {Sanz}},\ }\bibfield  {title} {{New Pathways in Neutrino Physics via Quantum-Encoded Data Analysis},\ }\href@noop {} {\bibfield  {journal} {\bibinfo  {journal} {arXiv preprint arXiv:2402.19306}\ } (\bibinfo {year} {2024})}\BibitemShut {NoStop}%
\bibitem [{\citenamefont {Chen}\ et~al.(2023)\citenamefont {Chen}, \citenamefont {Zhang}, \citenamefont {Wang}, \citenamefont {Gu}, \citenamefont {Li}, \citenamefont {Pan}, \citenamefont {Chong}, \citenamefont {Han},\ and\ \citenamefont {Wang}}]{chen2023quantumsea}%
  \BibitemOpen
  \bibfield  {author} {\bibinfo {author} {\bibfnamefont {T.}~\bibnamefont {Chen}}, \bibinfo {author} {\bibfnamefont {Z.}~\bibnamefont {Zhang}}, \bibinfo {author} {\bibfnamefont {H.}~\bibnamefont {Wang}}, \bibinfo {author} {\bibfnamefont {J.}~\bibnamefont {Gu}}, \bibinfo {author} {\bibfnamefont {Z.}~\bibnamefont {Li}}, \bibinfo {author} {\bibfnamefont {D.~Z.}\ \bibnamefont {Pan}}, \bibinfo {author} {\bibfnamefont {F.~T.}\ \bibnamefont {Chong}}, \bibinfo {author} {\bibfnamefont {S.}~\bibnamefont {Han}},\ and\ \bibinfo {author} {\bibfnamefont {Z.}~\bibnamefont {Wang}},\ }in\ \href@noop {} {\bibinfo {booktitle} {2023 IEEE International Conference on Quantum Computing and Engineering (QCE)}},\ Vol.~\bibinfo {volume} {1}\ (\bibinfo {organization} {IEEE},\ \bibinfo {year} {2023})\ pp.\ \bibinfo {pages} {51--62}\BibitemShut {NoStop}%
\bibitem [{\citenamefont {Vakili}\ et~al.(2024)\citenamefont {Vakili}, \citenamefont {Gorgulla}, \citenamefont {Nigam}, \citenamefont {Bezrukov}, \citenamefont {Varoli}, \citenamefont {Aliper}, \citenamefont {Polykovsky}, \citenamefont {Das}, \citenamefont {Snider}, \citenamefont {Lyakisheva} et~al.}]{vakili2024quantum}%
  \BibitemOpen
  \bibfield  {author} {\bibinfo {author} {\bibfnamefont {M.~G.}\ \bibnamefont {Vakili}}, \bibinfo {author} {\bibfnamefont {C.}~\bibnamefont {Gorgulla}}, \bibinfo {author} {\bibfnamefont {A.}~\bibnamefont {Nigam}}, \bibinfo {author} {\bibfnamefont {D.}~\bibnamefont {Bezrukov}}, \bibinfo {author} {\bibfnamefont {D.}~\bibnamefont {Varoli}}, \bibinfo {author} {\bibfnamefont {A.}~\bibnamefont {Aliper}}, \bibinfo {author} {\bibfnamefont {D.}~\bibnamefont {Polykovsky}}, \bibinfo {author} {\bibfnamefont {K.~M.~P.}\ \bibnamefont {Das}}, \bibinfo {author} {\bibfnamefont {J.}~\bibnamefont {Snider}}, \bibinfo {author} {\bibfnamefont {A.}~\bibnamefont {Lyakisheva}}, et~al.,\ }\bibfield  {title} {{Quantum Computing-Enhanced Algorithm Unveils Novel Inhibitors for KRAS},\ }\href@noop {} {\bibfield  {journal} {\bibinfo  {journal} {arXiv preprint arXiv:2402.08210}\ } (\bibinfo {year} {2024})}\BibitemShut {NoStop}%
\bibitem [{\citenamefont {Goodfellow}\ et~al.(2020)\citenamefont {Goodfellow}, \citenamefont {Pouget-Abadie}, \citenamefont {Mirza}, \citenamefont {Xu}, \citenamefont {Warde-Farley}, \citenamefont {Ozair}, \citenamefont {Courville},\ and\ \citenamefont {Bengio}}]{goodfellow2020generative}%
  \BibitemOpen
  \bibfield  {author} {\bibinfo {author} {\bibfnamefont {I.}~\bibnamefont {Goodfellow}}, \bibinfo {author} {\bibfnamefont {J.}~\bibnamefont {Pouget-Abadie}}, \bibinfo {author} {\bibfnamefont {M.}~\bibnamefont {Mirza}}, \bibinfo {author} {\bibfnamefont {B.}~\bibnamefont {Xu}}, \bibinfo {author} {\bibfnamefont {D.}~\bibnamefont {Warde-Farley}}, \bibinfo {author} {\bibfnamefont {S.}~\bibnamefont {Ozair}}, \bibinfo {author} {\bibfnamefont {A.}~\bibnamefont {Courville}},\ and\ \bibinfo {author} {\bibfnamefont {Y.}~\bibnamefont {Bengio}},\ }\bibfield  {title} {{Generative adversarial networks},\ }\href@noop {} {\bibfield  {journal} {\bibinfo  {journal} {Communications of the ACM}\ }\textbf {\bibinfo {volume} {63}},\ \bibinfo {pages} {139} (\bibinfo {year} {2020})}\BibitemShut {NoStop}%
\bibitem [{\citenamefont {Temme}\ et~al.(2017)\citenamefont {Temme}, \citenamefont {Bravyi},\ and\ \citenamefont {Gambetta}}]{temme2017error}%
  \BibitemOpen
  \bibfield  {author} {\bibinfo {author} {\bibfnamefont {K.}~\bibnamefont {Temme}}, \bibinfo {author} {\bibfnamefont {S.}~\bibnamefont {Bravyi}},\ and\ \bibinfo {author} {\bibfnamefont {J.~M.}\ \bibnamefont {Gambetta}},\ }\bibfield  {title} {{Error mitigation for short-depth quantum circuits},\ }\href@noop {} {\bibfield  {journal} {\bibinfo  {journal} {Physical review letters}\ }\textbf {\bibinfo {volume} {119}},\ \bibinfo {pages} {180509} (\bibinfo {year} {2017})}\BibitemShut {NoStop}%
\bibitem [{\citenamefont {Kandala}\ et~al.(2019)\citenamefont {Kandala}, \citenamefont {Temme}, \citenamefont {C{\'o}rcoles}, \citenamefont {Mezzacapo}, \citenamefont {Chow},\ and\ \citenamefont {Gambetta}}]{kandala2019error}%
  \BibitemOpen
  \bibfield  {author} {\bibinfo {author} {\bibfnamefont {A.}~\bibnamefont {Kandala}}, \bibinfo {author} {\bibfnamefont {K.}~\bibnamefont {Temme}}, \bibinfo {author} {\bibfnamefont {A.~D.}\ \bibnamefont {C{\'o}rcoles}}, \bibinfo {author} {\bibfnamefont {A.}~\bibnamefont {Mezzacapo}}, \bibinfo {author} {\bibfnamefont {J.~M.}\ \bibnamefont {Chow}},\ and\ \bibinfo {author} {\bibfnamefont {J.~M.}\ \bibnamefont {Gambetta}},\ }\bibfield  {title} {{Error mitigation extends the computational reach of a noisy quantum processor},\ }\href@noop {} {\bibfield  {journal} {\bibinfo  {journal} {Nature}\ }\textbf {\bibinfo {volume} {567}},\ \bibinfo {pages} {491} (\bibinfo {year} {2019})}\BibitemShut {NoStop}%
\bibitem [{\citenamefont {Chatrchyan}\ et~al.(2012)\citenamefont {Chatrchyan}, \citenamefont {Khachatryan}, \citenamefont {Sirunyan}, \citenamefont {Tumasyan}, \citenamefont {Adam}, \citenamefont {Bergauer}, \citenamefont {Dragicevic}, \citenamefont {Eroe}, \citenamefont {Fabjan}, \citenamefont {Friedl} et~al.}]{dipole1}%
  \BibitemOpen
  \bibfield  {author} {\bibinfo {author} {\bibfnamefont {S.}~\bibnamefont {Chatrchyan}}, \bibinfo {author} {\bibfnamefont {V.}~\bibnamefont {Khachatryan}}, \bibinfo {author} {\bibfnamefont {A.}~\bibnamefont {Sirunyan}}, \bibinfo {author} {\bibfnamefont {A.}~\bibnamefont {Tumasyan}}, \bibinfo {author} {\bibfnamefont {W.}~\bibnamefont {Adam}}, \bibinfo {author} {\bibfnamefont {T.}~\bibnamefont {Bergauer}}, \bibinfo {author} {\bibfnamefont {M.}~\bibnamefont {Dragicevic}}, \bibinfo {author} {\bibfnamefont {J.}~\bibnamefont {Eroe}}, \bibinfo {author} {\bibfnamefont {C.}~\bibnamefont {Fabjan}}, \bibinfo {author} {\bibfnamefont {M.}~\bibnamefont {Friedl}}, et~al.,\ }\bibfield  {title} {{Search for anomalous $ t\overline t $ production in the highly-boosted all-hadronic final state},\ }\href@noop {} {\bibfield  {journal} {\bibinfo  {journal} {Journal of High Energy Physics}\ }\textbf {\bibinfo {volume} {2012}},\ \bibinfo {pages} {1} (\bibinfo {year} {2012})}\BibitemShut {NoStop}%
\bibitem [{\citenamefont {Bottou}(2010)}]{bottou2010large}%
  \BibitemOpen
  \bibfield  {author} {\bibinfo {author} {\bibfnamefont {L.}~\bibnamefont {Bottou}},\ }in\ \href@noop {} {\bibinfo {booktitle} {Proceedings of COMPSTAT'2010}}\ (\bibinfo  {publisher} {Springer},\ \bibinfo {year} {2010})\ pp.\ \bibinfo {pages} {177--186}\BibitemShut {NoStop}%
\bibitem [{\citenamefont {Tumer}\ and\ \citenamefont {Ghosh}(1996)}]{tumer1996error}%
  \BibitemOpen
  \bibfield  {author} {\bibinfo {author} {\bibfnamefont {K.}~\bibnamefont {Tumer}}\ and\ \bibinfo {author} {\bibfnamefont {J.}~\bibnamefont {Ghosh}},\ }\bibfield  {title} {{Error correlation and error reduction in ensemble classifiers},\ }\href@noop {} {\bibfield  {journal} {\bibinfo  {journal} {Connection science}\ }\textbf {\bibinfo {volume} {8}},\ \bibinfo {pages} {385} (\bibinfo {year} {1996})}\BibitemShut {NoStop}%
\bibitem [{\citenamefont {Domingos}(2000)}]{domingos2000bayesian}%
  \BibitemOpen
  \bibfield  {author} {\bibinfo {author} {\bibfnamefont {P.}~\bibnamefont {Domingos}},\ }in\ \href@noop {} {\bibinfo {booktitle} {ICML}},\ Vol.\ \bibinfo {volume} {747}\ (\bibinfo {organization} {Citeseer},\ \bibinfo {year} {2000})\ pp.\ \bibinfo {pages} {223--230}\BibitemShut {NoStop}%
\bibitem [{\citenamefont {Schuld}\ and\ \citenamefont {Killoran}(2019)}]{schuld2019quantum}%
  \BibitemOpen
  \bibfield  {author} {\bibinfo {author} {\bibfnamefont {M.}~\bibnamefont {Schuld}}\ and\ \bibinfo {author} {\bibfnamefont {N.}~\bibnamefont {Killoran}},\ }\bibfield  {title} {{Quantum machine learning in feature Hilbert spaces},\ }\href@noop {} {\bibfield  {journal} {\bibinfo  {journal} {Physical review letters}\ }\textbf {\bibinfo {volume} {122}},\ \bibinfo {pages} {040504} (\bibinfo {year} {2019})}\BibitemShut {NoStop}%
\bibitem [{\citenamefont {Schuld}\ and\ \citenamefont {Petruccione}(2018{\natexlab{a}})}]{schuld2018quantum}%
  \BibitemOpen
  \bibfield  {author} {\bibinfo {author} {\bibfnamefont {M.}~\bibnamefont {Schuld}}\ and\ \bibinfo {author} {\bibfnamefont {F.}~\bibnamefont {Petruccione}},\ }\bibfield  {title} {{Quantum ensembles of quantum classifiers},\ }\href@noop {} {\bibfield  {journal} {\bibinfo  {journal} {Scientific reports}\ }\textbf {\bibinfo {volume} {8}},\ \bibinfo {pages} {1} (\bibinfo {year} {2018}{\natexlab{a}})}\BibitemShut {NoStop}%
\bibitem [{\citenamefont {Abbas}\ et~al.(2020)\citenamefont {Abbas}, \citenamefont {Schuld},\ and\ \citenamefont {Petruccione}}]{abbas2020quantum}%
  \BibitemOpen
  \bibfield  {author} {\bibinfo {author} {\bibfnamefont {A.}~\bibnamefont {Abbas}}, \bibinfo {author} {\bibfnamefont {M.}~\bibnamefont {Schuld}},\ and\ \bibinfo {author} {\bibfnamefont {F.}~\bibnamefont {Petruccione}},\ }\bibfield  {title} {{On quantum ensembles of quantum classifiers},\ }\href@noop {} {\bibfield  {journal} {\bibinfo  {journal} {Quantum Machine Intelligence}\ }\textbf {\bibinfo {volume} {2}},\ \bibinfo {pages} {1} (\bibinfo {year} {2020})}\BibitemShut {NoStop}%
\bibitem [{\citenamefont {Araujo}\ et~al.(2021)\citenamefont {Araujo}, \citenamefont {Park}, \citenamefont {Petruccione},\ and\ \citenamefont {da~Silva}}]{araujo2021divide}%
  \BibitemOpen
  \bibfield  {author} {\bibinfo {author} {\bibfnamefont {I.~F.}\ \bibnamefont {Araujo}}, \bibinfo {author} {\bibfnamefont {D.~K.}\ \bibnamefont {Park}}, \bibinfo {author} {\bibfnamefont {F.}~\bibnamefont {Petruccione}},\ and\ \bibinfo {author} {\bibfnamefont {A.~J.}\ \bibnamefont {da~Silva}},\ }\bibfield  {title} {{A divide-and-conquer algorithm for quantum state preparation},\ }\href@noop {} {\bibfield  {journal} {\bibinfo  {journal} {Scientific reports}\ }\textbf {\bibinfo {volume} {11}},\ \bibinfo {pages} {1} (\bibinfo {year} {2021})}\BibitemShut {NoStop}%
\bibitem [{\citenamefont {Pellow-Jarman}\ et~al.(2021)\citenamefont {Pellow-Jarman}, \citenamefont {Sinayskiy}, \citenamefont {Pillay},\ and\ \citenamefont {Petruccione}}]{pellow2021comparison}%
  \BibitemOpen
  \bibfield  {author} {\bibinfo {author} {\bibfnamefont {A.}~\bibnamefont {Pellow-Jarman}}, \bibinfo {author} {\bibfnamefont {I.}~\bibnamefont {Sinayskiy}}, \bibinfo {author} {\bibfnamefont {A.}~\bibnamefont {Pillay}},\ and\ \bibinfo {author} {\bibfnamefont {F.}~\bibnamefont {Petruccione}},\ }\bibfield  {title} {{A comparison of various classical optimizers for a variational quantum linear solver},\ }\href@noop {} {\bibfield  {journal} {\bibinfo  {journal} {Quantum Information Processing}\ }\textbf {\bibinfo {volume} {20}},\ \bibinfo {pages} {1} (\bibinfo {year} {2021})}\BibitemShut {NoStop}%
\bibitem [{\citenamefont {Blank}\ et~al.(2022)\citenamefont {Blank}, \citenamefont {da~Silva}, \citenamefont {de~Albuquerque}, \citenamefont {Petruccione},\ and\ \citenamefont {Park}}]{blank2022compact}%
  \BibitemOpen
  \bibfield  {author} {\bibinfo {author} {\bibfnamefont {C.}~\bibnamefont {Blank}}, \bibinfo {author} {\bibfnamefont {A.~J.}\ \bibnamefont {da~Silva}}, \bibinfo {author} {\bibfnamefont {L.~P.}\ \bibnamefont {de~Albuquerque}}, \bibinfo {author} {\bibfnamefont {F.}~\bibnamefont {Petruccione}},\ and\ \bibinfo {author} {\bibfnamefont {D.~K.}\ \bibnamefont {Park}},\ }\bibfield  {title} {{Compact quantum kernel-based binary classifier},\ }\href@noop {} {\bibfield  {journal} {\bibinfo  {journal} {Quantum Science and Technology}\ }\textbf {\bibinfo {volume} {7}},\ \bibinfo {pages} {045007} (\bibinfo {year} {2022})}\BibitemShut {NoStop}%
\bibitem [{\citenamefont {Li}\ and\ \citenamefont {Benjamin}(2017)}]{PhysRevX.7.021050}%
  \BibitemOpen
  \bibfield  {author} {\bibinfo {author} {\bibfnamefont {Y.}~\bibnamefont {Li}}\ and\ \bibinfo {author} {\bibfnamefont {S.~C.}\ \bibnamefont {Benjamin}},\ }\bibfield  {title} {{Efficient Variational Quantum Simulator Incorporating Active Error Minimization},\ }\href {https://doi.org/10.1103/PhysRevX.7.021050} {\bibfield  {journal} {\bibinfo  {journal} {Phys. Rev. X}\ }\textbf {\bibinfo {volume} {7}},\ \bibinfo {pages} {021050} (\bibinfo {year} {2017})}\BibitemShut {NoStop}%
\bibitem [{\citenamefont {Giurgica-Tiron}\ et~al.(2020)\citenamefont {Giurgica-Tiron}, \citenamefont {Hindy}, \citenamefont {LaRose}, \citenamefont {Mari},\ and\ \citenamefont {Zeng}}]{giurgica2020digital}%
  \BibitemOpen
  \bibfield  {author} {\bibinfo {author} {\bibfnamefont {T.}~\bibnamefont {Giurgica-Tiron}}, \bibinfo {author} {\bibfnamefont {Y.}~\bibnamefont {Hindy}}, \bibinfo {author} {\bibfnamefont {R.}~\bibnamefont {LaRose}}, \bibinfo {author} {\bibfnamefont {A.}~\bibnamefont {Mari}},\ and\ \bibinfo {author} {\bibfnamefont {W.~J.}\ \bibnamefont {Zeng}},\ }in\ \href@noop {} {\bibinfo {booktitle} {2020 IEEE International Conference on Quantum Computing and Engineering (QCE)}}\ (\bibinfo {organization} {IEEE},\ \bibinfo {year} {2020})\ pp.\ \bibinfo {pages} {306--316}\BibitemShut {NoStop}%
\bibitem [{\citenamefont {Peruzzo}\ et~al.(2014)\citenamefont {Peruzzo}, \citenamefont {McClean}, \citenamefont {Shadbolt}, \citenamefont {Yung}, \citenamefont {Zhou}, \citenamefont {Love}, \citenamefont {Aspuru-Guzik},\ and\ \citenamefont {O’brien}}]{peruzzo2014variational}%
  \BibitemOpen
  \bibfield  {author} {\bibinfo {author} {\bibfnamefont {A.}~\bibnamefont {Peruzzo}}, \bibinfo {author} {\bibfnamefont {J.}~\bibnamefont {McClean}}, \bibinfo {author} {\bibfnamefont {P.}~\bibnamefont {Shadbolt}}, \bibinfo {author} {\bibfnamefont {M.-H.}\ \bibnamefont {Yung}}, \bibinfo {author} {\bibfnamefont {X.-Q.}\ \bibnamefont {Zhou}}, \bibinfo {author} {\bibfnamefont {P.~J.}\ \bibnamefont {Love}}, \bibinfo {author} {\bibfnamefont {A.}~\bibnamefont {Aspuru-Guzik}},\ and\ \bibinfo {author} {\bibfnamefont {J.~L.}\ \bibnamefont {O’brien}},\ }\bibfield  {title} {{A variational eigenvalue solver on a photonic quantum processor},\ }\href@noop {} {\bibfield  {journal} {\bibinfo  {journal} {Nature communications}\ }\textbf {\bibinfo {volume} {5}},\ \bibinfo {pages} {1} (\bibinfo {year} {2014})}\BibitemShut {NoStop}%
\bibitem [{\citenamefont {Mitarai}\ et~al.(2018)\citenamefont {Mitarai}, \citenamefont {Negoro}, \citenamefont {Kitagawa},\ and\ \citenamefont {Fujii}}]{mitarai2018quantum}%
  \BibitemOpen
  \bibfield  {author} {\bibinfo {author} {\bibfnamefont {K.}~\bibnamefont {Mitarai}}, \bibinfo {author} {\bibfnamefont {M.}~\bibnamefont {Negoro}}, \bibinfo {author} {\bibfnamefont {M.}~\bibnamefont {Kitagawa}},\ and\ \bibinfo {author} {\bibfnamefont {K.}~\bibnamefont {Fujii}},\ }\bibfield  {title} {{Quantum circuit learning},\ }\href@noop {} {\bibfield  {journal} {\bibinfo  {journal} {Physical Review A}\ }\textbf {\bibinfo {volume} {98}},\ \bibinfo {pages} {032309} (\bibinfo {year} {2018})}\BibitemShut {NoStop}%
\bibitem [{\citenamefont {Liu}\ et~al.(2021)\citenamefont {Liu}, \citenamefont {Arunachalam},\ and\ \citenamefont {Temme}}]{liu2021rigorous}%
  \BibitemOpen
  \bibfield  {author} {\bibinfo {author} {\bibfnamefont {Y.}~\bibnamefont {Liu}}, \bibinfo {author} {\bibfnamefont {S.}~\bibnamefont {Arunachalam}},\ and\ \bibinfo {author} {\bibfnamefont {K.}~\bibnamefont {Temme}},\ }\bibfield  {title} {{A rigorous and robust quantum speed-up in supervised machine learning},\ }\href@noop {} {\bibfield  {journal} {\bibinfo  {journal} {Nature Physics}\ }\textbf {\bibinfo {volume} {17}},\ \bibinfo {pages} {1013} (\bibinfo {year} {2021})}\BibitemShut {NoStop}%
\bibitem [{\citenamefont {Schuld}(2021)}]{schuldkernel}%
  \BibitemOpen
  \bibfield  {author} {\bibinfo {author} {\bibfnamefont {M.}~\bibnamefont {Schuld}},\ }\bibfield  {title} {{Supervised quantum machine learning models are kernel methods},\ }\href@noop {} {\bibfield  {journal} {\bibinfo  {journal} {arXiv preprint arXiv:2101.11020}\ } (\bibinfo {year} {2021})}\BibitemShut {NoStop}%
\bibitem [{\citenamefont {Huang}\ et~al.(2021{\natexlab{c}})\citenamefont {Huang}, \citenamefont {Du}, \citenamefont {Gong}, \citenamefont {Zhao}, \citenamefont {Wu}, \citenamefont {Wang}, \citenamefont {Li}, \citenamefont {Liang}, \citenamefont {Lin}, \citenamefont {Xu} et~al.}]{huang2021experimental}%
  \BibitemOpen
  \bibfield  {author} {\bibinfo {author} {\bibfnamefont {H.-L.}\ \bibnamefont {Huang}}, \bibinfo {author} {\bibfnamefont {Y.}~\bibnamefont {Du}}, \bibinfo {author} {\bibfnamefont {M.}~\bibnamefont {Gong}}, \bibinfo {author} {\bibfnamefont {Y.}~\bibnamefont {Zhao}}, \bibinfo {author} {\bibfnamefont {Y.}~\bibnamefont {Wu}}, \bibinfo {author} {\bibfnamefont {C.}~\bibnamefont {Wang}}, \bibinfo {author} {\bibfnamefont {S.}~\bibnamefont {Li}}, \bibinfo {author} {\bibfnamefont {F.}~\bibnamefont {Liang}}, \bibinfo {author} {\bibfnamefont {J.}~\bibnamefont {Lin}}, \bibinfo {author} {\bibfnamefont {Y.}~\bibnamefont {Xu}}, et~al.,\ }\bibfield  {title} {{Experimental quantum generative adversarial networks for image generation},\ }\href@noop {} {\bibfield  {journal} {\bibinfo  {journal} {Physical Review Applied}\ }\textbf {\bibinfo {volume} {16}},\ \bibinfo {pages} {024051} (\bibinfo {year} {2021}{\natexlab{c}})}\BibitemShut {NoStop}%
\bibitem [{\citenamefont {Li}\ et~al.(2020)\citenamefont {Li}, \citenamefont {Zhang},\ and\ \citenamefont {Xia}}]{QGansurvey}%
  \BibitemOpen
  \bibfield  {author} {\bibinfo {author} {\bibfnamefont {T.}~\bibnamefont {Li}}, \bibinfo {author} {\bibfnamefont {S.}~\bibnamefont {Zhang}},\ and\ \bibinfo {author} {\bibfnamefont {J.}~\bibnamefont {Xia}},\ }\bibfield  {title} {{Quantum Generative Adversarial Network: A Survey},\ }\href@noop {} {\bibfield  {journal} {\bibinfo  {journal} {CMC-COMPUTERS MATERIALS \& CONTINUA}\ }\textbf {\bibinfo {volume} {64}},\ \bibinfo {pages} {401} (\bibinfo {year} {2020})}\BibitemShut {NoStop}%
\bibitem [{\citenamefont {Gacon}\ et~al.(2021)\citenamefont {Gacon}, \citenamefont {Zoufal}, \citenamefont {Carleo},\ and\ \citenamefont {Woerner}}]{gacon2021simultaneous}%
  \BibitemOpen
  \bibfield  {author} {\bibinfo {author} {\bibfnamefont {J.}~\bibnamefont {Gacon}}, \bibinfo {author} {\bibfnamefont {C.}~\bibnamefont {Zoufal}}, \bibinfo {author} {\bibfnamefont {G.}~\bibnamefont {Carleo}},\ and\ \bibinfo {author} {\bibfnamefont {S.}~\bibnamefont {Woerner}},\ }\bibfield  {title} {{Simultaneous perturbation stochastic approximation of the quantum fisher information},\ }\href@noop {} {\bibfield  {journal} {\bibinfo  {journal} {Quantum}\ }\textbf {\bibinfo {volume} {5}},\ \bibinfo {pages} {567} (\bibinfo {year} {2021})}\BibitemShut {NoStop}%
\bibitem [{\citenamefont {Kim}\ et~al.(2023)\citenamefont {Kim}, \citenamefont {Eddins}, \citenamefont {Anand}, \citenamefont {Wei}, \citenamefont {van~den Berg}, \citenamefont {Rosenblatt}, \citenamefont {Nayfeh}, \citenamefont {Wu}, \citenamefont {Zaletel}, \citenamefont {Temme},\ and\ \citenamefont {Kandala}}]{Kim2023}%
  \BibitemOpen
  \bibfield  {author} {\bibinfo {author} {\bibfnamefont {Y.}~\bibnamefont {Kim}}, \bibinfo {author} {\bibfnamefont {A.}~\bibnamefont {Eddins}}, \bibinfo {author} {\bibfnamefont {S.}~\bibnamefont {Anand}}, \bibinfo {author} {\bibfnamefont {K.~X.}\ \bibnamefont {Wei}}, \bibinfo {author} {\bibfnamefont {E.}~\bibnamefont {van~den Berg}}, \bibinfo {author} {\bibfnamefont {S.}~\bibnamefont {Rosenblatt}}, \bibinfo {author} {\bibfnamefont {H.}~\bibnamefont {Nayfeh}}, \bibinfo {author} {\bibfnamefont {Y.}~\bibnamefont {Wu}}, \bibinfo {author} {\bibfnamefont {M.}~\bibnamefont {Zaletel}}, \bibinfo {author} {\bibfnamefont {K.}~\bibnamefont {Temme}},\ and\ \bibinfo {author} {\bibfnamefont {A.}~\bibnamefont {Kandala}},\ }\bibfield  {title} {{Evidence for the utility of quantum computing before fault tolerance},\ }\href {https://doi.org/10.1038/s41586-023-06096-3} {\bibfield  {journal} {\bibinfo  {journal} {Nature}\ }\textbf {\bibinfo {volume} {618}},\ \bibinfo {pages} {500} (\bibinfo {year} {2023})}\BibitemShut {NoStop}%
\bibitem [{\citenamefont {Suzuki}\ et~al.(2023)\citenamefont {Suzuki}, \citenamefont {Hasebe},\ and\ \citenamefont {Miyazaki}}]{suzuki2023quantum}%
  \BibitemOpen
  \bibfield  {author} {\bibinfo {author} {\bibfnamefont {T.}~\bibnamefont {Suzuki}}, \bibinfo {author} {\bibfnamefont {T.}~\bibnamefont {Hasebe}},\ and\ \bibinfo {author} {\bibfnamefont {T.}~\bibnamefont {Miyazaki}},\ }\bibfield  {title} {{Quantum support vector machines for classification and regression on a trapped-ion quantum computer},\ }\href@noop {} {\bibfield  {journal} {\bibinfo  {journal} {arXiv preprint arXiv:2307.02091}\ } (\bibinfo {year} {2023})}\BibitemShut {NoStop}%
\bibitem [{\citenamefont {Herman}\ et~al.(2023)\citenamefont {Herman}, \citenamefont {Googin}, \citenamefont {Liu}, \citenamefont {Sun}, \citenamefont {Galda}, \citenamefont {Safro}, \citenamefont {Pistoia},\ and\ \citenamefont {Alexeev}}]{herman2023quantum}%
  \BibitemOpen
  \bibfield  {author} {\bibinfo {author} {\bibfnamefont {D.}~\bibnamefont {Herman}}, \bibinfo {author} {\bibfnamefont {C.}~\bibnamefont {Googin}}, \bibinfo {author} {\bibfnamefont {X.}~\bibnamefont {Liu}}, \bibinfo {author} {\bibfnamefont {Y.}~\bibnamefont {Sun}}, \bibinfo {author} {\bibfnamefont {A.}~\bibnamefont {Galda}}, \bibinfo {author} {\bibfnamefont {I.}~\bibnamefont {Safro}}, \bibinfo {author} {\bibfnamefont {M.}~\bibnamefont {Pistoia}},\ and\ \bibinfo {author} {\bibfnamefont {Y.}~\bibnamefont {Alexeev}},\ }\bibfield  {title} {{Quantum computing for finance},\ }\href@noop {} {\bibfield  {journal} {\bibinfo  {journal} {Nature Reviews Physics}\ }\textbf {\bibinfo {volume} {5}},\ \bibinfo {pages} {450} (\bibinfo {year} {2023})}\BibitemShut {NoStop}%
\bibitem [{\citenamefont {Albrecht}\ et~al.(2023)\citenamefont {Albrecht}, \citenamefont {Dalyac}, \citenamefont {Leclerc}, \citenamefont {Ortiz-Guti{\'e}rrez}, \citenamefont {Thabet}, \citenamefont {D'Arcangelo}, \citenamefont {Cline}, \citenamefont {Elfving}, \citenamefont {Lassabli{\`e}re}, \citenamefont {Silv{\'e}rio} et~al.}]{albrecht2023quantum}%
  \BibitemOpen
  \bibfield  {author} {\bibinfo {author} {\bibfnamefont {B.}~\bibnamefont {Albrecht}}, \bibinfo {author} {\bibfnamefont {C.}~\bibnamefont {Dalyac}}, \bibinfo {author} {\bibfnamefont {L.}~\bibnamefont {Leclerc}}, \bibinfo {author} {\bibfnamefont {L.}~\bibnamefont {Ortiz-Guti{\'e}rrez}}, \bibinfo {author} {\bibfnamefont {S.}~\bibnamefont {Thabet}}, \bibinfo {author} {\bibfnamefont {M.}~\bibnamefont {D'Arcangelo}}, \bibinfo {author} {\bibfnamefont {J.~R.}\ \bibnamefont {Cline}}, \bibinfo {author} {\bibfnamefont {V.~E.}\ \bibnamefont {Elfving}}, \bibinfo {author} {\bibfnamefont {L.}~\bibnamefont {Lassabli{\`e}re}}, \bibinfo {author} {\bibfnamefont {H.}~\bibnamefont {Silv{\'e}rio}}, et~al.,\ }\bibfield  {title} {{Quantum feature maps for graph machine learning on a neutral atom quantum processor},\ }\href@noop {} {\bibfield  {journal} {\bibinfo  {journal} {Physical Review A}\ }\textbf {\bibinfo {volume} {107}},\ \bibinfo {pages} {042615} (\bibinfo {year} {2023})}\BibitemShut {NoStop}%
\bibitem [{\citenamefont {Glick}\ et~al.(2024)\citenamefont {Glick}, \citenamefont {Gujarati}, \citenamefont {Corcoles}, \citenamefont {Kim}, \citenamefont {Kandala}, \citenamefont {Gambetta},\ and\ \citenamefont {Temme}}]{glick2024covariant}%
  \BibitemOpen
  \bibfield  {author} {\bibinfo {author} {\bibfnamefont {J.~R.}\ \bibnamefont {Glick}}, \bibinfo {author} {\bibfnamefont {T.~P.}\ \bibnamefont {Gujarati}}, \bibinfo {author} {\bibfnamefont {A.~D.}\ \bibnamefont {Corcoles}}, \bibinfo {author} {\bibfnamefont {Y.}~\bibnamefont {Kim}}, \bibinfo {author} {\bibfnamefont {A.}~\bibnamefont {Kandala}}, \bibinfo {author} {\bibfnamefont {J.~M.}\ \bibnamefont {Gambetta}},\ and\ \bibinfo {author} {\bibfnamefont {K.}~\bibnamefont {Temme}},\ }\bibfield  {title} {{Covariant quantum kernels for data with group structure},\ }\href@noop {} {\bibfield  {journal} {\bibinfo  {journal} {Nature Physics}\ ,\ \bibinfo {pages} {1}} (\bibinfo {year} {2024})}\BibitemShut {NoStop}%
\bibitem [{\citenamefont {Pan}\ et~al.(2023)\citenamefont {Pan}, \citenamefont {Lu}, \citenamefont {Wang}, \citenamefont {Hua}, \citenamefont {Xu}, \citenamefont {Li}, \citenamefont {Cai}, \citenamefont {Li}, \citenamefont {Wang}, \citenamefont {Song} et~al.}]{pan2023deep}%
  \BibitemOpen
  \bibfield  {author} {\bibinfo {author} {\bibfnamefont {X.}~\bibnamefont {Pan}}, \bibinfo {author} {\bibfnamefont {Z.}~\bibnamefont {Lu}}, \bibinfo {author} {\bibfnamefont {W.}~\bibnamefont {Wang}}, \bibinfo {author} {\bibfnamefont {Z.}~\bibnamefont {Hua}}, \bibinfo {author} {\bibfnamefont {Y.}~\bibnamefont {Xu}}, \bibinfo {author} {\bibfnamefont {W.}~\bibnamefont {Li}}, \bibinfo {author} {\bibfnamefont {W.}~\bibnamefont {Cai}}, \bibinfo {author} {\bibfnamefont {X.}~\bibnamefont {Li}}, \bibinfo {author} {\bibfnamefont {H.}~\bibnamefont {Wang}}, \bibinfo {author} {\bibfnamefont {Y.-P.}\ \bibnamefont {Song}}, et~al.,\ }\bibfield  {title} {{Deep quantum neural networks on a superconducting processor},\ }\href@noop {} {\bibfield  {journal} {\bibinfo  {journal} {Nature Communications}\ }\textbf {\bibinfo {volume} {14}},\ \bibinfo {pages} {4006} (\bibinfo {year} {2023})}\BibitemShut {NoStop}%
\bibitem [{\citenamefont {Gong}\ et~al.(2023)\citenamefont {Gong}, \citenamefont {Huang}, \citenamefont {Wang}, \citenamefont {Guo}, \citenamefont {Li}, \citenamefont {Wu}, \citenamefont {Zhu}, \citenamefont {Zhao}, \citenamefont {Guo}, \citenamefont {Qian} et~al.}]{gong2023quantum}%
  \BibitemOpen
  \bibfield  {author} {\bibinfo {author} {\bibfnamefont {M.}~\bibnamefont {Gong}}, \bibinfo {author} {\bibfnamefont {H.-L.}\ \bibnamefont {Huang}}, \bibinfo {author} {\bibfnamefont {S.}~\bibnamefont {Wang}}, \bibinfo {author} {\bibfnamefont {C.}~\bibnamefont {Guo}}, \bibinfo {author} {\bibfnamefont {S.}~\bibnamefont {Li}}, \bibinfo {author} {\bibfnamefont {Y.}~\bibnamefont {Wu}}, \bibinfo {author} {\bibfnamefont {Q.}~\bibnamefont {Zhu}}, \bibinfo {author} {\bibfnamefont {Y.}~\bibnamefont {Zhao}}, \bibinfo {author} {\bibfnamefont {S.}~\bibnamefont {Guo}}, \bibinfo {author} {\bibfnamefont {H.}~\bibnamefont {Qian}}, et~al.,\ }\bibfield  {title} {{Quantum neuronal sensing of quantum many-body states on a 61-qubit programmable superconducting processor},\ }\href@noop {} {\bibfield  {journal} {\bibinfo  {journal} {Science Bulletin}\ }\textbf {\bibinfo {volume} {68}},\ \bibinfo {pages} {906} (\bibinfo {year} {2023})}\BibitemShut {NoStop}%
\bibitem [{\citenamefont {Benedetti}\ et~al.(2021{\natexlab{b}})\citenamefont {Benedetti}, \citenamefont {Coyle}, \citenamefont {Fiorentini}, \citenamefont {Lubasch},\ and\ \citenamefont {Rosenkranz}}]{benedetti2021variational}%
  \BibitemOpen
  \bibfield  {author} {\bibinfo {author} {\bibfnamefont {M.}~\bibnamefont {Benedetti}}, \bibinfo {author} {\bibfnamefont {B.}~\bibnamefont {Coyle}}, \bibinfo {author} {\bibfnamefont {M.}~\bibnamefont {Fiorentini}}, \bibinfo {author} {\bibfnamefont {M.}~\bibnamefont {Lubasch}},\ and\ \bibinfo {author} {\bibfnamefont {M.}~\bibnamefont {Rosenkranz}},\ }\bibfield  {title} {{Variational inference with a quantum computer},\ }\href@noop {} {\bibfield  {journal} {\bibinfo  {journal} {Physical Review Applied}\ }\textbf {\bibinfo {volume} {16}},\ \bibinfo {pages} {044057} (\bibinfo {year} {2021}{\natexlab{b}})}\BibitemShut {NoStop}%
\bibitem [{\citenamefont {Rehm}\ et~al.(2023)\citenamefont {Rehm}, \citenamefont {Vallecorsa}, \citenamefont {Borras}, \citenamefont {Kr{\"u}cker}, \citenamefont {Grossi},\ and\ \citenamefont {Varo}}]{rehm2023precise}%
  \BibitemOpen
  \bibfield  {author} {\bibinfo {author} {\bibfnamefont {F.}~\bibnamefont {Rehm}}, \bibinfo {author} {\bibfnamefont {S.}~\bibnamefont {Vallecorsa}}, \bibinfo {author} {\bibfnamefont {K.}~\bibnamefont {Borras}}, \bibinfo {author} {\bibfnamefont {D.}~\bibnamefont {Kr{\"u}cker}}, \bibinfo {author} {\bibfnamefont {M.}~\bibnamefont {Grossi}},\ and\ \bibinfo {author} {\bibfnamefont {V.}~\bibnamefont {Varo}},\ }\bibfield  {title} {{Precise image generation on current noisy quantum computing devices},\ }\href@noop {} {\bibfield  {journal} {\bibinfo  {journal} {Quantum Science and Technology}\ }\textbf {\bibinfo {volume} {9}},\ \bibinfo {pages} {015009} (\bibinfo {year} {2023})}\BibitemShut {NoStop}%
\bibitem [{\citenamefont {Li}\ et~al.(2023{\natexlab{a}})\citenamefont {Li}, \citenamefont {Yao}, \citenamefont {Huang}, \citenamefont {Zou}, \citenamefont {Lin},\ and\ \citenamefont {Li}}]{li2023application}%
  \BibitemOpen
  \bibfield  {author} {\bibinfo {author} {\bibfnamefont {T.}~\bibnamefont {Li}}, \bibinfo {author} {\bibfnamefont {Z.}~\bibnamefont {Yao}}, \bibinfo {author} {\bibfnamefont {X.}~\bibnamefont {Huang}}, \bibinfo {author} {\bibfnamefont {J.}~\bibnamefont {Zou}}, \bibinfo {author} {\bibfnamefont {T.}~\bibnamefont {Lin}},\ and\ \bibinfo {author} {\bibfnamefont {W.}~\bibnamefont {Li}},\ }in\ \href@noop {} {\bibinfo {booktitle} {Journal of Physics: Conference Series}},\ Vol.\ \bibinfo {volume} {2438}\ (\bibinfo {organization} {IOP Publishing},\ \bibinfo {year} {2023})\ p.\ \bibinfo {pages} {012071}\BibitemShut {NoStop}%
\bibitem [{\citenamefont {Thakkar}\ et~al.(2023)\citenamefont {Thakkar}, \citenamefont {Kazdaghli}, \citenamefont {Mathur}, \citenamefont {Kerenidis}, \citenamefont {Ferreira-Martins},\ and\ \citenamefont {Brito}}]{thakkar2023improved}%
  \BibitemOpen
  \bibfield  {author} {\bibinfo {author} {\bibfnamefont {S.}~\bibnamefont {Thakkar}}, \bibinfo {author} {\bibfnamefont {S.}~\bibnamefont {Kazdaghli}}, \bibinfo {author} {\bibfnamefont {N.}~\bibnamefont {Mathur}}, \bibinfo {author} {\bibfnamefont {I.}~\bibnamefont {Kerenidis}}, \bibinfo {author} {\bibfnamefont {A.~J.}\ \bibnamefont {Ferreira-Martins}},\ and\ \bibinfo {author} {\bibfnamefont {S.}~\bibnamefont {Brito}},\ }\bibfield  {title} {{{Improved Financial Forecasting via Quantum Machine Learning}},\ }\href@noop {} {\bibfield  {journal} {\bibinfo  {journal} {arXiv preprint arXiv:2306.12965}\ } (\bibinfo {year} {2023})}\BibitemShut {NoStop}%
\bibitem [{\citenamefont {Zoufal}\ et~al.(2023)\citenamefont {Zoufal}, \citenamefont {Mishmash}, \citenamefont {Sharma}, \citenamefont {Kumar}, \citenamefont {Sheshadri}, \citenamefont {Deshmukh}, \citenamefont {Ibrahim}, \citenamefont {Gacon},\ and\ \citenamefont {Woerner}}]{zoufal2023variational}%
  \BibitemOpen
  \bibfield  {author} {\bibinfo {author} {\bibfnamefont {C.}~\bibnamefont {Zoufal}}, \bibinfo {author} {\bibfnamefont {R.~V.}\ \bibnamefont {Mishmash}}, \bibinfo {author} {\bibfnamefont {N.}~\bibnamefont {Sharma}}, \bibinfo {author} {\bibfnamefont {N.}~\bibnamefont {Kumar}}, \bibinfo {author} {\bibfnamefont {A.}~\bibnamefont {Sheshadri}}, \bibinfo {author} {\bibfnamefont {A.}~\bibnamefont {Deshmukh}}, \bibinfo {author} {\bibfnamefont {N.}~\bibnamefont {Ibrahim}}, \bibinfo {author} {\bibfnamefont {J.}~\bibnamefont {Gacon}},\ and\ \bibinfo {author} {\bibfnamefont {S.}~\bibnamefont {Woerner}},\ }\bibfield  {title} {{Variational quantum algorithm for unconstrained black box binary optimization: Application to feature selection},\ }\href@noop {} {\bibfield  {journal} {\bibinfo  {journal} {Quantum}\ }\textbf {\bibinfo {volume} {7}},\ \bibinfo {pages} {909} (\bibinfo {year} {2023})}\BibitemShut {NoStop}%
\bibitem [{\citenamefont {Bermot}\ et~al.(2023)\citenamefont {Bermot}, \citenamefont {Zoufal}, \citenamefont {Grossi}, \citenamefont {Schuhmacher}, \citenamefont {Tacchino}, \citenamefont {Vallecorsa},\ and\ \citenamefont {Tavernelli}}]{bermot2023quantum}%
  \BibitemOpen
  \bibfield  {author} {\bibinfo {author} {\bibfnamefont {E.}~\bibnamefont {Bermot}}, \bibinfo {author} {\bibfnamefont {C.}~\bibnamefont {Zoufal}}, \bibinfo {author} {\bibfnamefont {M.}~\bibnamefont {Grossi}}, \bibinfo {author} {\bibfnamefont {J.}~\bibnamefont {Schuhmacher}}, \bibinfo {author} {\bibfnamefont {F.}~\bibnamefont {Tacchino}}, \bibinfo {author} {\bibfnamefont {S.}~\bibnamefont {Vallecorsa}},\ and\ \bibinfo {author} {\bibfnamefont {I.}~\bibnamefont {Tavernelli}},\ }in\ \href@noop {} {\bibinfo {booktitle} {2023 IEEE International Conference on Quantum Computing and Engineering (QCE)}},\ Vol.~\bibinfo {volume} {1}\ (\bibinfo {organization} {IEEE},\ \bibinfo {year} {2023})\ pp.\ \bibinfo {pages} {331--341}\BibitemShut {NoStop}%
\bibitem [{\citenamefont {Herrmann}\ et~al.(2022)\citenamefont {Herrmann}, \citenamefont {Llima}, \citenamefont {Remm}, \citenamefont {Zapletal}, \citenamefont {McMahon}, \citenamefont {Scarato}, \citenamefont {Swiadek}, \citenamefont {Andersen}, \citenamefont {Hellings}, \citenamefont {Krinner} et~al.}]{herrmann2022realizing}%
  \BibitemOpen
  \bibfield  {author} {\bibinfo {author} {\bibfnamefont {J.}~\bibnamefont {Herrmann}}, \bibinfo {author} {\bibfnamefont {S.~M.}\ \bibnamefont {Llima}}, \bibinfo {author} {\bibfnamefont {A.}~\bibnamefont {Remm}}, \bibinfo {author} {\bibfnamefont {P.}~\bibnamefont {Zapletal}}, \bibinfo {author} {\bibfnamefont {N.~A.}\ \bibnamefont {McMahon}}, \bibinfo {author} {\bibfnamefont {C.}~\bibnamefont {Scarato}}, \bibinfo {author} {\bibfnamefont {F.}~\bibnamefont {Swiadek}}, \bibinfo {author} {\bibfnamefont {C.~K.}\ \bibnamefont {Andersen}}, \bibinfo {author} {\bibfnamefont {C.}~\bibnamefont {Hellings}}, \bibinfo {author} {\bibfnamefont {S.}~\bibnamefont {Krinner}}, et~al.,\ }\bibfield  {title} {{Realizing quantum convolutional neural networks on a superconducting quantum processor to recognize quantum phases},\ }\href@noop {} {\bibfield  {journal} {\bibinfo  {journal} {Nature communications}\ }\textbf {\bibinfo {volume} {13}},\ \bibinfo {pages} {1} (\bibinfo {year} {2022})}\BibitemShut {NoStop}%
\bibitem [{\citenamefont {Bishop}\ and\ \citenamefont {Nasrabadi}(2006)}]{bishop2006pattern}%
  \BibitemOpen
  \bibfield  {author} {\bibinfo {author} {\bibfnamefont {C.~M.}\ \bibnamefont {Bishop}}\ and\ \bibinfo {author} {\bibfnamefont {N.~M.}\ \bibnamefont {Nasrabadi}},\ }\href@noop {} {\bibinfo {title} {Pattern recognition and machine learning}}\ (\bibinfo  {publisher} {Springer},\ \bibinfo {year} {2006})\BibitemShut {NoStop}%
\bibitem [{\citenamefont {Goodfellow}\ et~al.(2016)\citenamefont {Goodfellow}, \citenamefont {Bengio},\ and\ \citenamefont {Courville}}]{goodfellow2016deep}%
  \BibitemOpen
  \bibfield  {author} {\bibinfo {author} {\bibfnamefont {I.}~\bibnamefont {Goodfellow}}, \bibinfo {author} {\bibfnamefont {Y.}~\bibnamefont {Bengio}},\ and\ \bibinfo {author} {\bibfnamefont {A.}~\bibnamefont {Courville}},\ }\href@noop {} {\bibinfo {title} {Deep learning}}\ (\bibinfo  {publisher} {MIT press},\ \bibinfo {year} {2016})\BibitemShut {NoStop}%
\bibitem [{\citenamefont {Lloyd}\ and\ \citenamefont {Weedbrook}(2018)}]{QGAN2}%
  \BibitemOpen
  \bibfield  {author} {\bibinfo {author} {\bibfnamefont {S.}~\bibnamefont {Lloyd}}\ and\ \bibinfo {author} {\bibfnamefont {C.}~\bibnamefont {Weedbrook}},\ }\bibfield  {title} {{Quantum generative adversarial learning},\ }\href@noop {} {\bibfield  {journal} {\bibinfo  {journal} {Physical review letters}\ }\textbf {\bibinfo {volume} {121}},\ \bibinfo {pages} {040502} (\bibinfo {year} {2018})}\BibitemShut {NoStop}%
\bibitem [{\citenamefont {Skolik}\ et~al.(2021)\citenamefont {Skolik}, \citenamefont {McClean}, \citenamefont {Mohseni}, \citenamefont {van~der Smagt},\ and\ \citenamefont {Leib}}]{skolik2021layerwise}%
  \BibitemOpen
  \bibfield  {author} {\bibinfo {author} {\bibfnamefont {A.}~\bibnamefont {Skolik}}, \bibinfo {author} {\bibfnamefont {J.~R.}\ \bibnamefont {McClean}}, \bibinfo {author} {\bibfnamefont {M.}~\bibnamefont {Mohseni}}, \bibinfo {author} {\bibfnamefont {P.}~\bibnamefont {van~der Smagt}},\ and\ \bibinfo {author} {\bibfnamefont {M.}~\bibnamefont {Leib}},\ }\bibfield  {title} {{Layerwise learning for quantum neural networks},\ }\href@noop {} {\bibfield  {journal} {\bibinfo  {journal} {Quantum Machine Intelligence}\ }\textbf {\bibinfo {volume} {3}},\ \bibinfo {pages} {1} (\bibinfo {year} {2021})}\BibitemShut {NoStop}%
\bibitem [{\citenamefont {Baldi}\ et~al.(2014)\citenamefont {Baldi}, \citenamefont {Sadowski},\ and\ \citenamefont {Whiteson}}]{baldi2014searching}%
  \BibitemOpen
  \bibfield  {author} {\bibinfo {author} {\bibfnamefont {P.}~\bibnamefont {Baldi}}, \bibinfo {author} {\bibfnamefont {P.}~\bibnamefont {Sadowski}},\ and\ \bibinfo {author} {\bibfnamefont {D.}~\bibnamefont {Whiteson}},\ }\bibfield  {title} {{Searching for exotic particles in high-energy physics with deep learning},\ }\href@noop {} {\bibfield  {journal} {\bibinfo  {journal} {Nature communications}\ }\textbf {\bibinfo {volume} {5}},\ \bibinfo {pages} {1} (\bibinfo {year} {2014})}\BibitemShut {NoStop}%
\bibitem [{\citenamefont {Dallaire-Demers}\ and\ \citenamefont {Killoran}(2018)}]{QGAN1}%
  \BibitemOpen
  \bibfield  {author} {\bibinfo {author} {\bibfnamefont {P.-L.}\ \bibnamefont {Dallaire-Demers}}\ and\ \bibinfo {author} {\bibfnamefont {N.}~\bibnamefont {Killoran}},\ }\bibfield  {title} {{Quantum generative adversarial networks},\ }\href@noop {} {\bibfield  {journal} {\bibinfo  {journal} {Physical Review A}\ }\textbf {\bibinfo {volume} {98}},\ \bibinfo {pages} {012324} (\bibinfo {year} {2018})}\BibitemShut {NoStop}%
\bibitem [{\citenamefont {Zeiler}(2012)}]{adadelta}%
  \BibitemOpen
  \bibfield  {author} {\bibinfo {author} {\bibfnamefont {M.~D.}\ \bibnamefont {Zeiler}},\ }\bibfield  {title} {{Adadelta: an adaptive learning rate method},\ }\href@noop {} {\bibfield  {journal} {\bibinfo  {journal} {arXiv preprint arXiv:1212.5701}\ } (\bibinfo {year} {2012})}\BibitemShut {NoStop}%
\bibitem [{\citenamefont {Bravo-Prieto}\ et~al.(2022)\citenamefont {Bravo-Prieto}, \citenamefont {Baglio}, \citenamefont {C{\`e}}, \citenamefont {Francis}, \citenamefont {Grabowska},\ and\ \citenamefont {Carrazza}}]{bravo2022style}%
  \BibitemOpen
  \bibfield  {author} {\bibinfo {author} {\bibfnamefont {C.}~\bibnamefont {Bravo-Prieto}}, \bibinfo {author} {\bibfnamefont {J.}~\bibnamefont {Baglio}}, \bibinfo {author} {\bibfnamefont {M.}~\bibnamefont {C{\`e}}}, \bibinfo {author} {\bibfnamefont {A.}~\bibnamefont {Francis}}, \bibinfo {author} {\bibfnamefont {D.~M.}\ \bibnamefont {Grabowska}},\ and\ \bibinfo {author} {\bibfnamefont {S.}~\bibnamefont {Carrazza}},\ }\bibfield  {title} {{{Style-based quantum generative adversarial networks for Monte Carlo events}},\ }\href@noop {} {\bibfield  {journal} {\bibinfo  {journal} {Quantum}\ }\textbf {\bibinfo {volume} {6}},\ \bibinfo {pages} {777} (\bibinfo {year} {2022})}\BibitemShut {NoStop}%
\bibitem [{\citenamefont {Cirstoiu}\ et~al.(2020)\citenamefont {Cirstoiu}, \citenamefont {Holmes}, \citenamefont {Iosue}, \citenamefont {Cincio}, \citenamefont {Coles},\ and\ \citenamefont {Sornborger}}]{cirstoiu2020variational}%
  \BibitemOpen
  \bibfield  {author} {\bibinfo {author} {\bibfnamefont {C.}~\bibnamefont {Cirstoiu}}, \bibinfo {author} {\bibfnamefont {Z.}~\bibnamefont {Holmes}}, \bibinfo {author} {\bibfnamefont {J.}~\bibnamefont {Iosue}}, \bibinfo {author} {\bibfnamefont {L.}~\bibnamefont {Cincio}}, \bibinfo {author} {\bibfnamefont {P.~J.}\ \bibnamefont {Coles}},\ and\ \bibinfo {author} {\bibfnamefont {A.}~\bibnamefont {Sornborger}},\ }\bibfield  {title} {{Variational fast forwarding for quantum simulation beyond the coherence time},\ }\href@noop {} {\bibfield  {journal} {\bibinfo  {journal} {npj Quantum Information}\ }\textbf {\bibinfo {volume} {6}},\ \bibinfo {pages} {1} (\bibinfo {year} {2020})}\BibitemShut {NoStop}%
\bibitem [{\citenamefont {Wu}\ et~al.(2020)\citenamefont {Wu}, \citenamefont {Cao}, \citenamefont {Xie},\ and\ \citenamefont {Liu}}]{wu2020end}%
  \BibitemOpen
  \bibfield  {author} {\bibinfo {author} {\bibfnamefont {R.-B.}\ \bibnamefont {Wu}}, \bibinfo {author} {\bibfnamefont {X.}~\bibnamefont {Cao}}, \bibinfo {author} {\bibfnamefont {P.}~\bibnamefont {Xie}},\ and\ \bibinfo {author} {\bibfnamefont {Y.-x.}\ \bibnamefont {Liu}},\ }\bibfield  {title} {{End-to-end quantum machine learning implemented with controlled quantum dynamics},\ }\href@noop {} {\bibfield  {journal} {\bibinfo  {journal} {Physical Review Applied}\ }\textbf {\bibinfo {volume} {14}},\ \bibinfo {pages} {064020} (\bibinfo {year} {2020})}\BibitemShut {NoStop}%
\bibitem [{\citenamefont {Wittek}(2014)}]{wittek2014quantum}%
  \BibitemOpen
  \bibfield  {author} {\bibinfo {author} {\bibfnamefont {P.}~\bibnamefont {Wittek}},\ }\href@noop {} {\bibinfo {title} {Quantum machine learning: what quantum computing means to data mining}}\ (\bibinfo  {publisher} {Academic Press},\ \bibinfo {year} {2014})\BibitemShut {NoStop}%
\bibitem [{\citenamefont {Powell}(1994)}]{powell1994direct}%
  \BibitemOpen
  \bibfield  {author} {\bibinfo {author} {\bibfnamefont {M.~J.}\ \bibnamefont {Powell}},\ }in\ \href@noop {} {\bibinfo {booktitle} {Advances in optimization and numerical analysis}}\ (\bibinfo  {publisher} {Springer},\ \bibinfo {year} {1994})\ pp.\ \bibinfo {pages} {51--67}\BibitemShut {NoStop}%
\bibitem [{\citenamefont {Gu}\ et~al.(2021)\citenamefont {Gu}, \citenamefont {Lowe}, \citenamefont {Dub}, \citenamefont {Coles},\ and\ \citenamefont {Arrasmith}}]{gu2021adaptive}%
  \BibitemOpen
  \bibfield  {author} {\bibinfo {author} {\bibfnamefont {A.}~\bibnamefont {Gu}}, \bibinfo {author} {\bibfnamefont {A.}~\bibnamefont {Lowe}}, \bibinfo {author} {\bibfnamefont {P.~A.}\ \bibnamefont {Dub}}, \bibinfo {author} {\bibfnamefont {P.~J.}\ \bibnamefont {Coles}},\ and\ \bibinfo {author} {\bibfnamefont {A.}~\bibnamefont {Arrasmith}},\ }\bibfield  {title} {{Adaptive shot allocation for fast convergence in variational quantum algorithms},\ }\href@noop {} {\bibfield  {journal} {\bibinfo  {journal} {arXiv preprint arXiv:2108.10434}\ } (\bibinfo {year} {2021})}\BibitemShut {NoStop}%
\bibitem [{\citenamefont {K{\"u}bler}\ et~al.(2020)\citenamefont {K{\"u}bler}, \citenamefont {Arrasmith}, \citenamefont {Cincio},\ and\ \citenamefont {Coles}}]{kubler2020adaptive}%
  \BibitemOpen
  \bibfield  {author} {\bibinfo {author} {\bibfnamefont {J.~M.}\ \bibnamefont {K{\"u}bler}}, \bibinfo {author} {\bibfnamefont {A.}~\bibnamefont {Arrasmith}}, \bibinfo {author} {\bibfnamefont {L.}~\bibnamefont {Cincio}},\ and\ \bibinfo {author} {\bibfnamefont {P.~J.}\ \bibnamefont {Coles}},\ }\bibfield  {title} {{An adaptive optimizer for measurement-frugal variational algorithms},\ }\href@noop {} {\bibfield  {journal} {\bibinfo  {journal} {Quantum}\ }\textbf {\bibinfo {volume} {4}},\ \bibinfo {pages} {263} (\bibinfo {year} {2020})}\BibitemShut {NoStop}%
\bibitem [{QuantumHardware()}]{qcmodalities}%
  \BibitemOpen
  QuantumHardware,\ \href@noop {} {{{Quantum hardware}}},\ \bibinfo {howpublished} {\url{https://quantumtech.blog/2022/10/20/quantum-computing-modalities-a-qubit-primer-revisited/}} (\bibinfo {year} {2022})\BibitemShut {NoStop}%
\bibitem [{\citenamefont {Powell}(2007)}]{powell2007view}%
  \BibitemOpen
  \bibfield  {author} {\bibinfo {author} {\bibfnamefont {M.~J.}\ \bibnamefont {Powell}},\ }\bibfield  {title} {{A view of algorithms for optimization without derivatives},\ }\href@noop {} {\bibfield  {journal} {\bibinfo  {journal} {Mathematics Today-Bulletin of the Institute of Mathematics and its Applications}\ }\textbf {\bibinfo {volume} {43}},\ \bibinfo {pages} {170} (\bibinfo {year} {2007})}\BibitemShut {NoStop}%
\bibitem [{\citenamefont {Powell}(1998)}]{powell1998direct}%
  \BibitemOpen
  \bibfield  {author} {\bibinfo {author} {\bibfnamefont {M.~J.}\ \bibnamefont {Powell}},\ }\bibfield  {title} {{Direct search algorithms for optimization calculations},\ }\href@noop {} {\bibfield  {journal} {\bibinfo  {journal} {Acta numerica}\ }\textbf {\bibinfo {volume} {7}},\ \bibinfo {pages} {287} (\bibinfo {year} {1998})}\BibitemShut {NoStop}%
\bibitem [{\citenamefont {Terashi}\ et~al.(2021{\natexlab{a}})\citenamefont {Terashi}, \citenamefont {Ganguly}, \citenamefont {Iiyama}, \citenamefont {Inada}, \citenamefont {Jang}, \citenamefont {Mizuhara}, \citenamefont {Nagano}, \citenamefont {Okubo}, \citenamefont {Sawada}, \citenamefont {Tanaka} et~al.}]{terashimachine}%
  \BibitemOpen
  \bibfield  {author} {\bibinfo {author} {\bibfnamefont {K.}~\bibnamefont {Terashi}}, \bibinfo {author} {\bibfnamefont {S.}~\bibnamefont {Ganguly}}, \bibinfo {author} {\bibfnamefont {Y.}~\bibnamefont {Iiyama}}, \bibinfo {author} {\bibfnamefont {T.}~\bibnamefont {Inada}}, \bibinfo {author} {\bibfnamefont {W.}~\bibnamefont {Jang}}, \bibinfo {author} {\bibfnamefont {S.}~\bibnamefont {Mizuhara}}, \bibinfo {author} {\bibfnamefont {L.}~\bibnamefont {Nagano}}, \bibinfo {author} {\bibfnamefont {R.}~\bibnamefont {Okubo}}, \bibinfo {author} {\bibfnamefont {R.}~\bibnamefont {Sawada}}, \bibinfo {author} {\bibfnamefont {J.}~\bibnamefont {Tanaka}}, et~al.,\ }\href@noop {} {{Machine {L}earning and {A}rchitectural {P}erspectives for {Q}uantum {HEP} {A}pplications}} (\bibinfo {year} {2021}{\natexlab{a}})\BibitemShut {NoStop}%
\bibitem [{\citenamefont {Alam}\ et~al.(2022)\citenamefont {Alam}, \citenamefont {Belomestnykh}, \citenamefont {Bornman}, \citenamefont {Cancelo}, \citenamefont {Chao}, \citenamefont {Checchin}, \citenamefont {Dinh}, \citenamefont {Grassellino}, \citenamefont {Gustafson}, \citenamefont {Harnik} et~al.}]{alam2022quantum}%
  \BibitemOpen
  \bibfield  {author} {\bibinfo {author} {\bibfnamefont {M.~S.}\ \bibnamefont {Alam}}, \bibinfo {author} {\bibfnamefont {S.}~\bibnamefont {Belomestnykh}}, \bibinfo {author} {\bibfnamefont {N.}~\bibnamefont {Bornman}}, \bibinfo {author} {\bibfnamefont {G.}~\bibnamefont {Cancelo}}, \bibinfo {author} {\bibfnamefont {Y.-C.}\ \bibnamefont {Chao}}, \bibinfo {author} {\bibfnamefont {M.}~\bibnamefont {Checchin}}, \bibinfo {author} {\bibfnamefont {V.~S.}\ \bibnamefont {Dinh}}, \bibinfo {author} {\bibfnamefont {A.}~\bibnamefont {Grassellino}}, \bibinfo {author} {\bibfnamefont {E.~J.}\ \bibnamefont {Gustafson}}, \bibinfo {author} {\bibfnamefont {R.}~\bibnamefont {Harnik}}, et~al.,\ }\bibfield  {title} {{{Quantum computing hardware for HEP algorithms and sensing}},\ }\href@noop {} {\bibfield  {journal} {\bibinfo  {journal} {arXiv preprint arXiv:2204.08605}\ } (\bibinfo {year} {2022})}\BibitemShut {NoStop}%
\bibitem [{\citenamefont {Biamonte}\ et~al.(2017)\citenamefont {Biamonte}, \citenamefont {Wittek}, \citenamefont {Pancotti}, \citenamefont {Rebentrost}, \citenamefont {Wiebe},\ and\ \citenamefont {Lloyd}}]{biamonte2017quantum}%
  \BibitemOpen
  \bibfield  {author} {\bibinfo {author} {\bibfnamefont {J.}~\bibnamefont {Biamonte}}, \bibinfo {author} {\bibfnamefont {P.}~\bibnamefont {Wittek}}, \bibinfo {author} {\bibfnamefont {N.}~\bibnamefont {Pancotti}}, \bibinfo {author} {\bibfnamefont {P.}~\bibnamefont {Rebentrost}}, \bibinfo {author} {\bibfnamefont {N.}~\bibnamefont {Wiebe}},\ and\ \bibinfo {author} {\bibfnamefont {S.}~\bibnamefont {Lloyd}},\ }\bibfield  {title} {{Quantum machine learning},\ }\href@noop {} {\bibfield  {journal} {\bibinfo  {journal} {Nature}\ }\textbf {\bibinfo {volume} {549}},\ \bibinfo {pages} {195} (\bibinfo {year} {2017})}\BibitemShut {NoStop}%
\bibitem [{\citenamefont {Arunachalam}\ et~al.(2015)\citenamefont {Arunachalam}, \citenamefont {Gheorghiu}, \citenamefont {Jochym-O’Connor}, \citenamefont {Mosca},\ and\ \citenamefont {Srinivasan}}]{arunachalam2015robustness}%
  \BibitemOpen
  \bibfield  {author} {\bibinfo {author} {\bibfnamefont {S.}~\bibnamefont {Arunachalam}}, \bibinfo {author} {\bibfnamefont {V.}~\bibnamefont {Gheorghiu}}, \bibinfo {author} {\bibfnamefont {T.}~\bibnamefont {Jochym-O’Connor}}, \bibinfo {author} {\bibfnamefont {M.}~\bibnamefont {Mosca}},\ and\ \bibinfo {author} {\bibfnamefont {P.~V.}\ \bibnamefont {Srinivasan}},\ }\bibfield  {title} {{{On the robustness of bucket brigade quantum RAM}},\ }\href@noop {} {\bibfield  {journal} {\bibinfo  {journal} {New Journal of Physics}\ }\textbf {\bibinfo {volume} {17}},\ \bibinfo {pages} {123010} (\bibinfo {year} {2015})}\BibitemShut {NoStop}%
\bibitem [{\citenamefont {Giovannetti}\ et~al.(2008{\natexlab{a}})\citenamefont {Giovannetti}, \citenamefont {Lloyd},\ and\ \citenamefont {Maccone}}]{giovannetti2008architectures}%
  \BibitemOpen
  \bibfield  {author} {\bibinfo {author} {\bibfnamefont {V.}~\bibnamefont {Giovannetti}}, \bibinfo {author} {\bibfnamefont {S.}~\bibnamefont {Lloyd}},\ and\ \bibinfo {author} {\bibfnamefont {L.}~\bibnamefont {Maccone}},\ }\bibfield  {title} {{Architectures for a quantum random access memory},\ }\href@noop {} {\bibfield  {journal} {\bibinfo  {journal} {Physical Review A}\ }\textbf {\bibinfo {volume} {78}},\ \bibinfo {pages} {052310} (\bibinfo {year} {2008}{\natexlab{a}})}\BibitemShut {NoStop}%
\bibitem [{\citenamefont {Rudolph}\ et~al.(2022)\citenamefont {Rudolph}, \citenamefont {Toussaint}, \citenamefont {Katabarwa}, \citenamefont {Johri}, \citenamefont {Peropadre},\ and\ \citenamefont {Perdomo-Ortiz}}]{rudolph2022generation}%
  \BibitemOpen
  \bibfield  {author} {\bibinfo {author} {\bibfnamefont {M.~S.}\ \bibnamefont {Rudolph}}, \bibinfo {author} {\bibfnamefont {N.~B.}\ \bibnamefont {Toussaint}}, \bibinfo {author} {\bibfnamefont {A.}~\bibnamefont {Katabarwa}}, \bibinfo {author} {\bibfnamefont {S.}~\bibnamefont {Johri}}, \bibinfo {author} {\bibfnamefont {B.}~\bibnamefont {Peropadre}},\ and\ \bibinfo {author} {\bibfnamefont {A.}~\bibnamefont {Perdomo-Ortiz}},\ }\bibfield  {title} {{Generation of high-resolution handwritten digits with an ion-trap quantum computer},\ }\href@noop {} {\bibfield  {journal} {\bibinfo  {journal} {Physical Review X}\ }\textbf {\bibinfo {volume} {12}},\ \bibinfo {pages} {031010} (\bibinfo {year} {2022})}\BibitemShut {NoStop}%
\bibitem [{\citenamefont {Coyle}\ et~al.(2021)\citenamefont {Coyle}, \citenamefont {Henderson}, \citenamefont {Le}, \citenamefont {Kumar}, \citenamefont {Paini},\ and\ \citenamefont {Kashefi}}]{coyle2021quantum}%
  \BibitemOpen
  \bibfield  {author} {\bibinfo {author} {\bibfnamefont {B.}~\bibnamefont {Coyle}}, \bibinfo {author} {\bibfnamefont {M.}~\bibnamefont {Henderson}}, \bibinfo {author} {\bibfnamefont {J.~C.~J.}\ \bibnamefont {Le}}, \bibinfo {author} {\bibfnamefont {N.}~\bibnamefont {Kumar}}, \bibinfo {author} {\bibfnamefont {M.}~\bibnamefont {Paini}},\ and\ \bibinfo {author} {\bibfnamefont {E.}~\bibnamefont {Kashefi}},\ }\bibfield  {title} {{Quantum versus classical generative modelling in finance},\ }\href@noop {} {\bibfield  {journal} {\bibinfo  {journal} {Quantum Science and Technology}\ }\textbf {\bibinfo {volume} {6}},\ \bibinfo {pages} {024013} (\bibinfo {year} {2021})}\BibitemShut {NoStop}%
\bibitem [{\citenamefont {Azevedo}\ et~al.(2022)\citenamefont {Azevedo}, \citenamefont {Silva},\ and\ \citenamefont {Dutra}}]{azevedo2022quantum}%
  \BibitemOpen
  \bibfield  {author} {\bibinfo {author} {\bibfnamefont {V.}~\bibnamefont {Azevedo}}, \bibinfo {author} {\bibfnamefont {C.}~\bibnamefont {Silva}},\ and\ \bibinfo {author} {\bibfnamefont {I.}~\bibnamefont {Dutra}},\ }\bibfield  {title} {{Quantum transfer learning for breast cancer detection},\ }\href@noop {} {\bibfield  {journal} {\bibinfo  {journal} {Quantum Machine Intelligence}\ }\textbf {\bibinfo {volume} {4}},\ \bibinfo {pages} {5} (\bibinfo {year} {2022})}\BibitemShut {NoStop}%
\bibitem [{\citenamefont {Fedorov}\ et~al.(2022)\citenamefont {Fedorov}, \citenamefont {Gisin}, \citenamefont {Beloussov},\ and\ \citenamefont {Lvovsky}}]{fedorov2022quantum}%
  \BibitemOpen
  \bibfield  {author} {\bibinfo {author} {\bibfnamefont {A.}~\bibnamefont {Fedorov}}, \bibinfo {author} {\bibfnamefont {N.}~\bibnamefont {Gisin}}, \bibinfo {author} {\bibfnamefont {S.}~\bibnamefont {Beloussov}},\ and\ \bibinfo {author} {\bibfnamefont {A.}~\bibnamefont {Lvovsky}},\ }\bibfield  {title} {{Quantum computing at the quantum advantage threshold: a down-to-business review},\ }\href@noop {} {\bibfield  {journal} {\bibinfo  {journal} {arXiv preprint arXiv:2203.17181}\ } (\bibinfo {year} {2022})}\BibitemShut {NoStop}%
\bibitem [{\citenamefont {Sharma}(2021)}]{sharma2021quantum1}%
  \BibitemOpen
  \bibfield  {author} {\bibinfo {author} {\bibfnamefont {K.~K.}\ \bibnamefont {Sharma}},\ }\bibfield  {title} {{Quantum machine learning and its supremacy in high energy physics},\ }\href@noop {} {\bibfield  {journal} {\bibinfo  {journal} {Modern Physics Letters A}\ }\textbf {\bibinfo {volume} {36}},\ \bibinfo {pages} {2030024} (\bibinfo {year} {2021})}\BibitemShut {NoStop}%
\bibitem [{\citenamefont {Guan}\ et~al.(2021)\citenamefont {Guan}, \citenamefont {Perdue}, \citenamefont {Pesah}, \citenamefont {Schuld}, \citenamefont {Terashi}, \citenamefont {Vallecorsa},\ and\ \citenamefont {Vlimant}}]{guan2021quantum}%
  \BibitemOpen
  \bibfield  {author} {\bibinfo {author} {\bibfnamefont {W.}~\bibnamefont {Guan}}, \bibinfo {author} {\bibfnamefont {G.}~\bibnamefont {Perdue}}, \bibinfo {author} {\bibfnamefont {A.}~\bibnamefont {Pesah}}, \bibinfo {author} {\bibfnamefont {M.}~\bibnamefont {Schuld}}, \bibinfo {author} {\bibfnamefont {K.}~\bibnamefont {Terashi}}, \bibinfo {author} {\bibfnamefont {S.}~\bibnamefont {Vallecorsa}},\ and\ \bibinfo {author} {\bibfnamefont {J.-R.}\ \bibnamefont {Vlimant}},\ }\bibfield  {title} {{Quantum machine learning in high energy physics},\ }\href@noop {} {\bibfield  {journal} {\bibinfo  {journal} {Machine Learning: Science and Technology}\ }\textbf {\bibinfo {volume} {2}},\ \bibinfo {pages} {011003} (\bibinfo {year} {2021})}\BibitemShut {NoStop}%
\bibitem [{\citenamefont {Humble}\ et~al.(2022)\citenamefont {Humble}, \citenamefont {Delgado}, \citenamefont {Pooser}, \citenamefont {Seck}, \citenamefont {Bennink}, \citenamefont {Leyton-Ortega}, \citenamefont {Wang}, \citenamefont {Dumitrescu}, \citenamefont {Morris}, \citenamefont {Hamilton} et~al.}]{humble2022snowmass}%
  \BibitemOpen
  \bibfield  {author} {\bibinfo {author} {\bibfnamefont {T.~S.}\ \bibnamefont {Humble}}, \bibinfo {author} {\bibfnamefont {A.}~\bibnamefont {Delgado}}, \bibinfo {author} {\bibfnamefont {R.}~\bibnamefont {Pooser}}, \bibinfo {author} {\bibfnamefont {C.}~\bibnamefont {Seck}}, \bibinfo {author} {\bibfnamefont {R.}~\bibnamefont {Bennink}}, \bibinfo {author} {\bibfnamefont {V.}~\bibnamefont {Leyton-Ortega}}, \bibinfo {author} {\bibfnamefont {C.-C.~J.}\ \bibnamefont {Wang}}, \bibinfo {author} {\bibfnamefont {E.}~\bibnamefont {Dumitrescu}}, \bibinfo {author} {\bibfnamefont {T.}~\bibnamefont {Morris}}, \bibinfo {author} {\bibfnamefont {K.}~\bibnamefont {Hamilton}}, et~al.,\ }\bibfield  {title} {{Snowmass white paper: Quantum computing systems and software for high-energy physics research},\ }\href@noop {} {\bibfield  {journal} {\bibinfo  {journal} {arXiv preprint arXiv:2203.07091}\ } (\bibinfo {year} {2022})}\BibitemShut {NoStop}%
\bibitem [{\citenamefont {Tomesh}\ et~al.(2021)\citenamefont {Tomesh}, \citenamefont {Gokhale}, \citenamefont {Anschuetz},\ and\ \citenamefont {Chong}}]{tomesh2021coreset}%
  \BibitemOpen
  \bibfield  {author} {\bibinfo {author} {\bibfnamefont {T.}~\bibnamefont {Tomesh}}, \bibinfo {author} {\bibfnamefont {P.}~\bibnamefont {Gokhale}}, \bibinfo {author} {\bibfnamefont {E.~R.}\ \bibnamefont {Anschuetz}},\ and\ \bibinfo {author} {\bibfnamefont {F.~T.}\ \bibnamefont {Chong}},\ }\bibfield  {title} {{Coreset clustering on small quantum computers},\ }\href@noop {} {\bibfield  {journal} {\bibinfo  {journal} {Electronics}\ }\textbf {\bibinfo {volume} {10}},\ \bibinfo {pages} {1690} (\bibinfo {year} {2021})}\BibitemShut {NoStop}%
\bibitem [{\citenamefont {Egger}\ et~al.(2020)\citenamefont {Egger}, \citenamefont {Gambella}, \citenamefont {Marecek}, \citenamefont {McFaddin}, \citenamefont {Mevissen}, \citenamefont {Raymond}, \citenamefont {Simonetto}, \citenamefont {Woerner},\ and\ \citenamefont {Yndurain}}]{egger2020quantum}%
  \BibitemOpen
  \bibfield  {author} {\bibinfo {author} {\bibfnamefont {D.~J.}\ \bibnamefont {Egger}}, \bibinfo {author} {\bibfnamefont {C.}~\bibnamefont {Gambella}}, \bibinfo {author} {\bibfnamefont {J.}~\bibnamefont {Marecek}}, \bibinfo {author} {\bibfnamefont {S.}~\bibnamefont {McFaddin}}, \bibinfo {author} {\bibfnamefont {M.}~\bibnamefont {Mevissen}}, \bibinfo {author} {\bibfnamefont {R.}~\bibnamefont {Raymond}}, \bibinfo {author} {\bibfnamefont {A.}~\bibnamefont {Simonetto}}, \bibinfo {author} {\bibfnamefont {S.}~\bibnamefont {Woerner}},\ and\ \bibinfo {author} {\bibfnamefont {E.}~\bibnamefont {Yndurain}},\ }\bibfield  {title} {{Quantum computing for finance: State-of-the-art and future prospects},\ }\href@noop {} {\bibfield  {journal} {\bibinfo  {journal} {IEEE Transactions on Quantum Engineering}\ }\textbf {\bibinfo {volume} {1}},\ \bibinfo {pages} {1} (\bibinfo {year} {2020})}\BibitemShut {NoStop}%
\bibitem [{\citenamefont {Herman}\ et~al.(2022)\citenamefont {Herman}, \citenamefont {Googin}, \citenamefont {Liu}, \citenamefont {Galda}, \citenamefont {Safro}, \citenamefont {Sun}, \citenamefont {Pistoia},\ and\ \citenamefont {Alexeev}}]{herman2022survey}%
  \BibitemOpen
  \bibfield  {author} {\bibinfo {author} {\bibfnamefont {D.}~\bibnamefont {Herman}}, \bibinfo {author} {\bibfnamefont {C.}~\bibnamefont {Googin}}, \bibinfo {author} {\bibfnamefont {X.}~\bibnamefont {Liu}}, \bibinfo {author} {\bibfnamefont {A.}~\bibnamefont {Galda}}, \bibinfo {author} {\bibfnamefont {I.}~\bibnamefont {Safro}}, \bibinfo {author} {\bibfnamefont {Y.}~\bibnamefont {Sun}}, \bibinfo {author} {\bibfnamefont {M.}~\bibnamefont {Pistoia}},\ and\ \bibinfo {author} {\bibfnamefont {Y.}~\bibnamefont {Alexeev}},\ }\href@noop {} {{A Survey of Quantum Computing for Finance}} (\bibinfo {year} {2022}),\ \Eprint {https://arxiv.org/abs/2201.02773} {arXiv:2201.02773 [quant-ph]} \BibitemShut {NoStop}%
\bibitem [{\citenamefont {Cumming}\ et~al.(2015)\citenamefont {Cumming}, \citenamefont {Alrajeh},\ and\ \citenamefont {Dickens}}]{cumming2015investigation}%
  \BibitemOpen
  \bibfield  {author} {\bibinfo {author} {\bibfnamefont {J.}~\bibnamefont {Cumming}}, \bibinfo {author} {\bibfnamefont {D.~D.}\ \bibnamefont {Alrajeh}},\ and\ \bibinfo {author} {\bibfnamefont {L.}~\bibnamefont {Dickens}},\ }\bibfield  {title} {{An investigation into the use of reinforcement learning techniques within the algorithmic trading domain},\ }\href@noop {} {\bibfield  {journal} {\bibinfo  {journal} {Imperial College London: London, UK}\ }\textbf {\bibinfo {volume} {58}} (\bibinfo {year} {2015})}\BibitemShut {NoStop}%
\bibitem [{\citenamefont {Li}\ et~al.(2019{\natexlab{a}})\citenamefont {Li}, \citenamefont {Zheng},\ and\ \citenamefont {Zheng}}]{li2019deep}%
  \BibitemOpen
  \bibfield  {author} {\bibinfo {author} {\bibfnamefont {Y.}~\bibnamefont {Li}}, \bibinfo {author} {\bibfnamefont {W.}~\bibnamefont {Zheng}},\ and\ \bibinfo {author} {\bibfnamefont {Z.}~\bibnamefont {Zheng}},\ }\bibfield  {title} {{Deep robust reinforcement learning for practical algorithmic trading},\ }\href@noop {} {\bibfield  {journal} {\bibinfo  {journal} {IEEE Access}\ }\textbf {\bibinfo {volume} {7}},\ \bibinfo {pages} {108014} (\bibinfo {year} {2019}{\natexlab{a}})}\BibitemShut {NoStop}%
\bibitem [{\citenamefont {Th{\'e}ate}\ and\ \citenamefont {Ernst}(2021)}]{theate2021application}%
  \BibitemOpen
  \bibfield  {author} {\bibinfo {author} {\bibfnamefont {T.}~\bibnamefont {Th{\'e}ate}}\ and\ \bibinfo {author} {\bibfnamefont {D.}~\bibnamefont {Ernst}},\ }\bibfield  {title} {{An application of deep reinforcement learning to algorithmic trading},\ }\href@noop {} {\bibfield  {journal} {\bibinfo  {journal} {Expert Systems with Applications}\ }\textbf {\bibinfo {volume} {173}},\ \bibinfo {pages} {114632} (\bibinfo {year} {2021})}\BibitemShut {NoStop}%
\bibitem [{\citenamefont {Cao}\ et~al.(2022)\citenamefont {Cao}, \citenamefont {Zhang}, \citenamefont {Wu}, \citenamefont {Grassl},\ and\ \citenamefont {Zeng}}]{cao2022quantum}%
  \BibitemOpen
  \bibfield  {author} {\bibinfo {author} {\bibfnamefont {C.}~\bibnamefont {Cao}}, \bibinfo {author} {\bibfnamefont {C.}~\bibnamefont {Zhang}}, \bibinfo {author} {\bibfnamefont {Z.}~\bibnamefont {Wu}}, \bibinfo {author} {\bibfnamefont {M.}~\bibnamefont {Grassl}},\ and\ \bibinfo {author} {\bibfnamefont {B.}~\bibnamefont {Zeng}},\ }\bibfield  {title} {{Quantum variational learning for quantum error-correcting codes},\ }\href@noop {} {\bibfield  {journal} {\bibinfo  {journal} {Quantum}\ }\textbf {\bibinfo {volume} {6}},\ \bibinfo {pages} {828} (\bibinfo {year} {2022})}\BibitemShut {NoStop}%
\bibitem [{\citenamefont {Liu}\ and\ \citenamefont {Poulin}(2019)}]{liu2019neural}%
  \BibitemOpen
  \bibfield  {author} {\bibinfo {author} {\bibfnamefont {Y.-H.}\ \bibnamefont {Liu}}\ and\ \bibinfo {author} {\bibfnamefont {D.}~\bibnamefont {Poulin}},\ }\bibfield  {title} {{Neural belief-propagation decoders for quantum error-correcting codes},\ }\href@noop {} {\bibfield  {journal} {\bibinfo  {journal} {Physical review letters}\ }\textbf {\bibinfo {volume} {122}},\ \bibinfo {pages} {200501} (\bibinfo {year} {2019})}\BibitemShut {NoStop}%
\bibitem [{\citenamefont {Sweke}\ et~al.(2020{\natexlab{a}})\citenamefont {Sweke}, \citenamefont {Kesselring}, \citenamefont {van Nieuwenburg},\ and\ \citenamefont {Eisert}}]{sweke2020reinforcement}%
  \BibitemOpen
  \bibfield  {author} {\bibinfo {author} {\bibfnamefont {R.}~\bibnamefont {Sweke}}, \bibinfo {author} {\bibfnamefont {M.~S.}\ \bibnamefont {Kesselring}}, \bibinfo {author} {\bibfnamefont {E.~P.}\ \bibnamefont {van Nieuwenburg}},\ and\ \bibinfo {author} {\bibfnamefont {J.}~\bibnamefont {Eisert}},\ }\bibfield  {title} {{Reinforcement learning decoders for fault-tolerant quantum computation},\ }\href@noop {} {\bibfield  {journal} {\bibinfo  {journal} {Machine Learning: Science and Technology}\ }\textbf {\bibinfo {volume} {2}},\ \bibinfo {pages} {025005} (\bibinfo {year} {2020}{\natexlab{a}})}\BibitemShut {NoStop}%
\bibitem [{\citenamefont {Zhou}\ et~al.(2022)\citenamefont {Zhou}, \citenamefont {Tian}, \citenamefont {Song}, \citenamefont {Qiu}, \citenamefont {Wang}, \citenamefont {Zhou}, \citenamefont {Chen}, \citenamefont {Xu},\ and\ \citenamefont {Lu}}]{zhou2022preserving}%
  \BibitemOpen
  \bibfield  {author} {\bibinfo {author} {\bibfnamefont {F.}~\bibnamefont {Zhou}}, \bibinfo {author} {\bibfnamefont {Y.}~\bibnamefont {Tian}}, \bibinfo {author} {\bibfnamefont {Y.}~\bibnamefont {Song}}, \bibinfo {author} {\bibfnamefont {C.}~\bibnamefont {Qiu}}, \bibinfo {author} {\bibfnamefont {X.}~\bibnamefont {Wang}}, \bibinfo {author} {\bibfnamefont {M.}~\bibnamefont {Zhou}}, \bibinfo {author} {\bibfnamefont {B.}~\bibnamefont {Chen}}, \bibinfo {author} {\bibfnamefont {N.}~\bibnamefont {Xu}},\ and\ \bibinfo {author} {\bibfnamefont {D.}~\bibnamefont {Lu}},\ }\bibfield  {title} {{Preserving entanglement in a solid-spin system using quantum autoencoders},\ }\href@noop {} {\bibfield  {journal} {\bibinfo  {journal} {Applied Physics Letters}\ }\textbf {\bibinfo {volume} {121}} (\bibinfo {year} {2022})}\BibitemShut {NoStop}%
\bibitem [{\citenamefont {Bondarenko}\ and\ \citenamefont {Feldmann}(2020)}]{bondarenko2020quantum}%
  \BibitemOpen
  \bibfield  {author} {\bibinfo {author} {\bibfnamefont {D.}~\bibnamefont {Bondarenko}}\ and\ \bibinfo {author} {\bibfnamefont {P.}~\bibnamefont {Feldmann}},\ }\bibfield  {title} {{Quantum autoencoders to denoise quantum data},\ }\href@noop {} {\bibfield  {journal} {\bibinfo  {journal} {Physical review letters}\ }\textbf {\bibinfo {volume} {124}},\ \bibinfo {pages} {130502} (\bibinfo {year} {2020})}\BibitemShut {NoStop}%
\bibitem [{\citenamefont {Huang}\ et~al.(2020)\citenamefont {Huang}, \citenamefont {Ma}, \citenamefont {Yin}, \citenamefont {Tang}, \citenamefont {Dong}, \citenamefont {Chen}, \citenamefont {Xiang}, \citenamefont {Li},\ and\ \citenamefont {Guo}}]{huang2020realization}%
  \BibitemOpen
  \bibfield  {author} {\bibinfo {author} {\bibfnamefont {C.-J.}\ \bibnamefont {Huang}}, \bibinfo {author} {\bibfnamefont {H.}~\bibnamefont {Ma}}, \bibinfo {author} {\bibfnamefont {Q.}~\bibnamefont {Yin}}, \bibinfo {author} {\bibfnamefont {J.-F.}\ \bibnamefont {Tang}}, \bibinfo {author} {\bibfnamefont {D.}~\bibnamefont {Dong}}, \bibinfo {author} {\bibfnamefont {C.}~\bibnamefont {Chen}}, \bibinfo {author} {\bibfnamefont {G.-Y.}\ \bibnamefont {Xiang}}, \bibinfo {author} {\bibfnamefont {C.-F.}\ \bibnamefont {Li}},\ and\ \bibinfo {author} {\bibfnamefont {G.-C.}\ \bibnamefont {Guo}},\ }\bibfield  {title} {{Realization of a quantum autoencoder for lossless compression of quantum data},\ }\href@noop {} {\bibfield  {journal} {\bibinfo  {journal} {Physical Review A}\ }\textbf {\bibinfo {volume} {102}},\ \bibinfo {pages} {032412} (\bibinfo {year} {2020})}\BibitemShut {NoStop}%
\bibitem [{\citenamefont {Locher}\ et~al.(2023)\citenamefont {Locher}, \citenamefont {Cardarelli},\ and\ \citenamefont {M{\"u}ller}}]{locher2023quantum}%
  \BibitemOpen
  \bibfield  {author} {\bibinfo {author} {\bibfnamefont {D.~F.}\ \bibnamefont {Locher}}, \bibinfo {author} {\bibfnamefont {L.}~\bibnamefont {Cardarelli}},\ and\ \bibinfo {author} {\bibfnamefont {M.}~\bibnamefont {M{\"u}ller}},\ }\bibfield  {title} {{Quantum error correction with quantum autoencoders},\ }\href@noop {} {\bibfield  {journal} {\bibinfo  {journal} {Quantum}\ }\textbf {\bibinfo {volume} {7}},\ \bibinfo {pages} {942} (\bibinfo {year} {2023})}\BibitemShut {NoStop}%
\bibitem [{\citenamefont {Meyer}\ et~al.(2022)\citenamefont {Meyer}, \citenamefont {Ufrecht}, \citenamefont {Periyasamy}, \citenamefont {Scherer}, \citenamefont {Plinge},\ and\ \citenamefont {Mutschler}}]{meyer2022survey}%
  \BibitemOpen
  \bibfield  {author} {\bibinfo {author} {\bibfnamefont {N.}~\bibnamefont {Meyer}}, \bibinfo {author} {\bibfnamefont {C.}~\bibnamefont {Ufrecht}}, \bibinfo {author} {\bibfnamefont {M.}~\bibnamefont {Periyasamy}}, \bibinfo {author} {\bibfnamefont {D.~D.}\ \bibnamefont {Scherer}}, \bibinfo {author} {\bibfnamefont {A.}~\bibnamefont {Plinge}},\ and\ \bibinfo {author} {\bibfnamefont {C.}~\bibnamefont {Mutschler}},\ }\bibfield  {title} {{A survey on quantum reinforcement learning},\ }\href@noop {} {\bibfield  {journal} {\bibinfo  {journal} {arXiv preprint arXiv:2211.03464}\ } (\bibinfo {year} {2022})}\BibitemShut {NoStop}%
\bibitem [{\citenamefont {Dunjko}\ et~al.(2017)\citenamefont {Dunjko}, \citenamefont {Taylor},\ and\ \citenamefont {Briegel}}]{dunjko2017advances}%
  \BibitemOpen
  \bibfield  {author} {\bibinfo {author} {\bibfnamefont {V.}~\bibnamefont {Dunjko}}, \bibinfo {author} {\bibfnamefont {J.~M.}\ \bibnamefont {Taylor}},\ and\ \bibinfo {author} {\bibfnamefont {H.~J.}\ \bibnamefont {Briegel}},\ }in\ \href@noop {} {\bibinfo {booktitle} {2017 IEEE International Conference on Systems, Man, and Cybernetics (SMC)}}\ (\bibinfo {organization} {IEEE},\ \bibinfo {year} {2017})\ pp.\ \bibinfo {pages} {282--287}\BibitemShut {NoStop}%
\bibitem [{\citenamefont {Dong}\ et~al.(2008)\citenamefont {Dong}, \citenamefont {Chen}, \citenamefont {Li},\ and\ \citenamefont {Tarn}}]{dong2008quantum}%
  \BibitemOpen
  \bibfield  {author} {\bibinfo {author} {\bibfnamefont {D.}~\bibnamefont {Dong}}, \bibinfo {author} {\bibfnamefont {C.}~\bibnamefont {Chen}}, \bibinfo {author} {\bibfnamefont {H.}~\bibnamefont {Li}},\ and\ \bibinfo {author} {\bibfnamefont {T.-J.}\ \bibnamefont {Tarn}},\ }\bibfield  {title} {{Quantum reinforcement learning},\ }\href@noop {} {\bibfield  {journal} {\bibinfo  {journal} {IEEE Transactions on Systems, Man, and Cybernetics, Part B (Cybernetics)}\ }\textbf {\bibinfo {volume} {38}},\ \bibinfo {pages} {1207} (\bibinfo {year} {2008})}\BibitemShut {NoStop}%
\bibitem [{\citenamefont {Pricope}(2021)}]{pricope2021deep}%
  \BibitemOpen
  \bibfield  {author} {\bibinfo {author} {\bibfnamefont {T.-V.}\ \bibnamefont {Pricope}},\ }\bibfield  {title} {{Deep reinforcement learning in quantitative algorithmic trading: A review},\ }\href@noop {} {\bibfield  {journal} {\bibinfo  {journal} {arXiv preprint arXiv:2106.00123}\ } (\bibinfo {year} {2021})}\BibitemShut {NoStop}%
\bibitem [{\citenamefont {Gilliam}\ et~al.(2021)\citenamefont {Gilliam}, \citenamefont {Woerner},\ and\ \citenamefont {Gonciulea}}]{gilliam2021grover}%
  \BibitemOpen
  \bibfield  {author} {\bibinfo {author} {\bibfnamefont {A.}~\bibnamefont {Gilliam}}, \bibinfo {author} {\bibfnamefont {S.}~\bibnamefont {Woerner}},\ and\ \bibinfo {author} {\bibfnamefont {C.}~\bibnamefont {Gonciulea}},\ }\bibfield  {title} {{Grover adaptive search for constrained polynomial binary optimization},\ }\href@noop {} {\bibfield  {journal} {\bibinfo  {journal} {Quantum}\ }\textbf {\bibinfo {volume} {5}},\ \bibinfo {pages} {428} (\bibinfo {year} {2021})}\BibitemShut {NoStop}%
\bibitem [{\citenamefont {Brassard}\ et~al.(2002)\citenamefont {Brassard}, \citenamefont {Hoyer}, \citenamefont {Mosca},\ and\ \citenamefont {Tapp}}]{brassard2002quantum}%
  \BibitemOpen
  \bibfield  {author} {\bibinfo {author} {\bibfnamefont {G.}~\bibnamefont {Brassard}}, \bibinfo {author} {\bibfnamefont {P.}~\bibnamefont {Hoyer}}, \bibinfo {author} {\bibfnamefont {M.}~\bibnamefont {Mosca}},\ and\ \bibinfo {author} {\bibfnamefont {A.}~\bibnamefont {Tapp}},\ }\bibfield  {title} {{Quantum amplitude amplification and estimation},\ }\href@noop {} {\bibfield  {journal} {\bibinfo  {journal} {Contemporary Mathematics}\ }\textbf {\bibinfo {volume} {305}},\ \bibinfo {pages} {53} (\bibinfo {year} {2002})}\BibitemShut {NoStop}%
\bibitem [{\citenamefont {Orús}\ et~al.(2019)\citenamefont {Orús}, \citenamefont {Mugel},\ and\ \citenamefont {Lizaso}}]{ORUS2019100028}%
  \BibitemOpen
  \bibfield  {author} {\bibinfo {author} {\bibfnamefont {R.}~\bibnamefont {Orús}}, \bibinfo {author} {\bibfnamefont {S.}~\bibnamefont {Mugel}},\ and\ \bibinfo {author} {\bibfnamefont {E.}~\bibnamefont {Lizaso}},\ }\bibfield  {title} {{Quantum computing for finance: Overview and prospects},\ }\href {https://doi.org/https://doi.org/10.1016/j.revip.2019.100028} {\bibfield  {journal} {\bibinfo  {journal} {Reviews in Physics}\ }\textbf {\bibinfo {volume} {4}},\ \bibinfo {pages} {100028} (\bibinfo {year} {2019})}\BibitemShut {NoStop}%
\bibitem [{\citenamefont {Wilkens}\ and\ \citenamefont {Moorhouse}(2023)}]{wilkens2023quantum}%
  \BibitemOpen
  \bibfield  {author} {\bibinfo {author} {\bibfnamefont {S.}~\bibnamefont {Wilkens}}\ and\ \bibinfo {author} {\bibfnamefont {J.}~\bibnamefont {Moorhouse}},\ }\bibfield  {title} {{Quantum computing for financial risk measurement},\ }\href@noop {} {\bibfield  {journal} {\bibinfo  {journal} {Quantum Information Processing}\ }\textbf {\bibinfo {volume} {22}},\ \bibinfo {pages} {51} (\bibinfo {year} {2023})}\BibitemShut {NoStop}%
\bibitem [{\citenamefont {Fl{\"o}ther}(2023)}]{flother2023state}%
  \BibitemOpen
  \bibfield  {author} {\bibinfo {author} {\bibfnamefont {F.~F.}\ \bibnamefont {Fl{\"o}ther}},\ }\bibfield  {title} {{The state of quantum computing applications in health and medicine},\ }\href@noop {} {\bibfield  {journal} {\bibinfo  {journal} {arXiv preprint arXiv:2301.09106}\ } (\bibinfo {year} {2023})}\BibitemShut {NoStop}%
\bibitem [{\citenamefont {Wei}\ et~al.(2023)\citenamefont {Wei}, \citenamefont {Liu}, \citenamefont {Xu}, \citenamefont {Shi}, \citenamefont {Shan}, \citenamefont {Zhao},\ and\ \citenamefont {Gao}}]{wei2023quantum}%
  \BibitemOpen
  \bibfield  {author} {\bibinfo {author} {\bibfnamefont {L.}~\bibnamefont {Wei}}, \bibinfo {author} {\bibfnamefont {H.}~\bibnamefont {Liu}}, \bibinfo {author} {\bibfnamefont {J.}~\bibnamefont {Xu}}, \bibinfo {author} {\bibfnamefont {L.}~\bibnamefont {Shi}}, \bibinfo {author} {\bibfnamefont {Z.}~\bibnamefont {Shan}}, \bibinfo {author} {\bibfnamefont {B.}~\bibnamefont {Zhao}},\ and\ \bibinfo {author} {\bibfnamefont {Y.}~\bibnamefont {Gao}},\ }\bibfield  {title} {{Quantum machine learning in medical image analysis: A Survey},\ }\href@noop {} {\bibfield  {journal} {\bibinfo  {journal} {Neurocomputing}\ } (\bibinfo {year} {2023})}\BibitemShut {NoStop}%
\bibitem [{\citenamefont {Ur~Rasool}\ et~al.(2023)\citenamefont {Ur~Rasool}, \citenamefont {Ahmad}, \citenamefont {Rafique}, \citenamefont {Qayyum}, \citenamefont {Qadir},\ and\ \citenamefont {Anwar}}]{fi15030094}%
  \BibitemOpen
  \bibfield  {author} {\bibinfo {author} {\bibfnamefont {R.}~\bibnamefont {Ur~Rasool}}, \bibinfo {author} {\bibfnamefont {H.~F.}\ \bibnamefont {Ahmad}}, \bibinfo {author} {\bibfnamefont {W.}~\bibnamefont {Rafique}}, \bibinfo {author} {\bibfnamefont {A.}~\bibnamefont {Qayyum}}, \bibinfo {author} {\bibfnamefont {J.}~\bibnamefont {Qadir}},\ and\ \bibinfo {author} {\bibfnamefont {Z.}~\bibnamefont {Anwar}},\ }\bibfield  {title} {{Quantum Computing for Healthcare: A Review},\ }\bibfield  {journal} {\bibinfo  {journal} {Future Internet}\ }\textbf {\bibinfo {volume} {15}},\ \href {https://doi.org/10.3390/fi15030094} {10.3390/fi15030094} (\bibinfo {year} {2023})\BibitemShut {NoStop}%
\bibitem [{\citenamefont {Schuld}\ et~al.(2019)\citenamefont {Schuld}, \citenamefont {Bergholm}, \citenamefont {Gogolin}, \citenamefont {Izaac},\ and\ \citenamefont {Killoran}}]{schuld2019evaluating}%
  \BibitemOpen
  \bibfield  {author} {\bibinfo {author} {\bibfnamefont {M.}~\bibnamefont {Schuld}}, \bibinfo {author} {\bibfnamefont {V.}~\bibnamefont {Bergholm}}, \bibinfo {author} {\bibfnamefont {C.}~\bibnamefont {Gogolin}}, \bibinfo {author} {\bibfnamefont {J.}~\bibnamefont {Izaac}},\ and\ \bibinfo {author} {\bibfnamefont {N.}~\bibnamefont {Killoran}},\ }\bibfield  {title} {{Evaluating analytic gradients on quantum hardware},\ }\href@noop {} {\bibfield  {journal} {\bibinfo  {journal} {Physical Review A}\ }\textbf {\bibinfo {volume} {99}},\ \bibinfo {pages} {032331} (\bibinfo {year} {2019})}\BibitemShut {NoStop}%
\bibitem [{\citenamefont {Jadhav}\ et~al.(2023)\citenamefont {Jadhav}, \citenamefont {Rasool},\ and\ \citenamefont {Gyanchandani}}]{jadhav2023quantum}%
  \BibitemOpen
  \bibfield  {author} {\bibinfo {author} {\bibfnamefont {A.}~\bibnamefont {Jadhav}}, \bibinfo {author} {\bibfnamefont {A.}~\bibnamefont {Rasool}},\ and\ \bibinfo {author} {\bibfnamefont {M.}~\bibnamefont {Gyanchandani}},\ }\bibfield  {title} {{Quantum Machine Learning: Scope for real-world problems},\ }\href@noop {} {\bibfield  {journal} {\bibinfo  {journal} {Procedia Computer Science}\ }\textbf {\bibinfo {volume} {218}},\ \bibinfo {pages} {2612} (\bibinfo {year} {2023})}\BibitemShut {NoStop}%
\bibitem [{\citenamefont {Singh}\ and\ \citenamefont {Bhangu}(2023)}]{singh2023contemporary}%
  \BibitemOpen
  \bibfield  {author} {\bibinfo {author} {\bibfnamefont {J.}~\bibnamefont {Singh}}\ and\ \bibinfo {author} {\bibfnamefont {K.~S.}\ \bibnamefont {Bhangu}},\ }\bibfield  {title} {{Contemporary Quantum Computing Use Cases: Taxonomy, Review and Challenges},\ }\href@noop {} {\bibfield  {journal} {\bibinfo  {journal} {Archives of Computational Methods in Engineering}\ }\textbf {\bibinfo {volume} {30}},\ \bibinfo {pages} {615} (\bibinfo {year} {2023})}\BibitemShut {NoStop}%
\bibitem [{\citenamefont {Bova}\ et~al.(2021)\citenamefont {Bova}, \citenamefont {Goldfarb},\ and\ \citenamefont {Melko}}]{Bova2021}%
  \BibitemOpen
  \bibfield  {author} {\bibinfo {author} {\bibfnamefont {F.}~\bibnamefont {Bova}}, \bibinfo {author} {\bibfnamefont {A.}~\bibnamefont {Goldfarb}},\ and\ \bibinfo {author} {\bibfnamefont {R.~G.}\ \bibnamefont {Melko}},\ }\bibfield  {title} {{Commercial applications of quantum computing},\ }\href {https://doi.org/10.1140/epjqt/s40507-021-00091-1} {\bibfield  {journal} {\bibinfo  {journal} {EPJ Quantum Technology}\ }\textbf {\bibinfo {volume} {8}},\ \bibinfo {pages} {2} (\bibinfo {year} {2021})}\BibitemShut {NoStop}%
\bibitem [{\citenamefont {Hassija}\ et~al.(2020)\citenamefont {Hassija}, \citenamefont {Chamola}, \citenamefont {Saxena}, \citenamefont {Chanana}, \citenamefont {Parashari}, \citenamefont {Mumtaz},\ and\ \citenamefont {Guizani}}]{hassija2020present}%
  \BibitemOpen
  \bibfield  {author} {\bibinfo {author} {\bibfnamefont {V.}~\bibnamefont {Hassija}}, \bibinfo {author} {\bibfnamefont {V.}~\bibnamefont {Chamola}}, \bibinfo {author} {\bibfnamefont {V.}~\bibnamefont {Saxena}}, \bibinfo {author} {\bibfnamefont {V.}~\bibnamefont {Chanana}}, \bibinfo {author} {\bibfnamefont {P.}~\bibnamefont {Parashari}}, \bibinfo {author} {\bibfnamefont {S.}~\bibnamefont {Mumtaz}},\ and\ \bibinfo {author} {\bibfnamefont {M.}~\bibnamefont {Guizani}},\ }\bibfield  {title} {{Present landscape of quantum computing},\ }\href@noop {} {\bibfield  {journal} {\bibinfo  {journal} {IET Quantum Communication}\ }\textbf {\bibinfo {volume} {1}},\ \bibinfo {pages} {42} (\bibinfo {year} {2020})}\BibitemShut {NoStop}%
\bibitem [{\citenamefont {Bayerstadler}\ et~al.(2021)\citenamefont {Bayerstadler}, \citenamefont {Becquin}, \citenamefont {Binder}, \citenamefont {Botter}, \citenamefont {Ehm}, \citenamefont {Ehmer}, \citenamefont {Erdmann}, \citenamefont {Gaus}, \citenamefont {Harbach}, \citenamefont {Hess}, \citenamefont {Klepsch}, \citenamefont {Leib}, \citenamefont {Luber}, \citenamefont {Luckow}, \citenamefont {Mansky}, \citenamefont {Mauerer}, \citenamefont {Neukart}, \citenamefont {Niedermeier}, \citenamefont {Palackal}, \citenamefont {Pfeiffer}, \citenamefont {Polenz}, \citenamefont {Sepulveda}, \citenamefont {Sievers}, \citenamefont {Standen}, \citenamefont {Streif}, \citenamefont {Strohm}, \citenamefont {Utschig-Utschig}, \citenamefont {Volz}, \citenamefont {Weiss}, \citenamefont {Winter}, \citenamefont {Technology},\ and\ \citenamefont {QUTAC}}]{Bayerstadler2021}%
  \BibitemOpen
  \bibfield  {author} {\bibinfo {author} {\bibfnamefont {A.}~\bibnamefont {Bayerstadler}}, \bibinfo {author} {\bibfnamefont {G.}~\bibnamefont {Becquin}}, \bibinfo {author} {\bibfnamefont {J.}~\bibnamefont {Binder}}, \bibinfo {author} {\bibfnamefont {T.}~\bibnamefont {Botter}}, \bibinfo {author} {\bibfnamefont {H.}~\bibnamefont {Ehm}}, \bibinfo {author} {\bibfnamefont {T.}~\bibnamefont {Ehmer}}, \bibinfo {author} {\bibfnamefont {M.}~\bibnamefont {Erdmann}}, \bibinfo {author} {\bibfnamefont {N.}~\bibnamefont {Gaus}}, \bibinfo {author} {\bibfnamefont {P.}~\bibnamefont {Harbach}}, \bibinfo {author} {\bibfnamefont {M.}~\bibnamefont {Hess}}, \bibinfo {author} {\bibfnamefont {J.}~\bibnamefont {Klepsch}}, \bibinfo {author} {\bibfnamefont {M.}~\bibnamefont {Leib}}, \bibinfo {author} {\bibfnamefont {S.}~\bibnamefont {Luber}}, \bibinfo {author} {\bibfnamefont {A.}~\bibnamefont {Luckow}}, \bibinfo {author} {\bibfnamefont {M.}~\bibnamefont {Mansky}}, \bibinfo {author} {\bibfnamefont {W.}~\bibnamefont {Mauerer}}, \bibinfo
  {author} {\bibfnamefont {F.}~\bibnamefont {Neukart}}, \bibinfo {author} {\bibfnamefont {C.}~\bibnamefont {Niedermeier}}, \bibinfo {author} {\bibfnamefont {L.}~\bibnamefont {Palackal}}, \bibinfo {author} {\bibfnamefont {R.}~\bibnamefont {Pfeiffer}}, \bibinfo {author} {\bibfnamefont {C.}~\bibnamefont {Polenz}}, \bibinfo {author} {\bibfnamefont {J.}~\bibnamefont {Sepulveda}}, \bibinfo {author} {\bibfnamefont {T.}~\bibnamefont {Sievers}}, \bibinfo {author} {\bibfnamefont {B.}~\bibnamefont {Standen}}, \bibinfo {author} {\bibfnamefont {M.}~\bibnamefont {Streif}}, \bibinfo {author} {\bibfnamefont {T.}~\bibnamefont {Strohm}}, \bibinfo {author} {\bibfnamefont {C.}~\bibnamefont {Utschig-Utschig}}, \bibinfo {author} {\bibfnamefont {D.}~\bibnamefont {Volz}}, \bibinfo {author} {\bibfnamefont {H.}~\bibnamefont {Weiss}}, \bibinfo {author} {\bibfnamefont {F.}~\bibnamefont {Winter}}, \bibinfo {author} {\bibfnamefont {Q.}~\bibnamefont {Technology}},\ and\ \bibinfo {author} {\bibfnamefont {A.~C.}\ \bibnamefont {QUTAC}},\
  }\bibfield  {title} {{Industry quantum computing applications},\ }\href {https://doi.org/10.1140/epjqt/s40507-021-00114-x} {\bibfield  {journal} {\bibinfo  {journal} {EPJ Quantum Technology}\ }\textbf {\bibinfo {volume} {8}},\ \bibinfo {pages} {25} (\bibinfo {year} {2021})}\BibitemShut {NoStop}%
\bibitem [{\citenamefont {Preskill}(1998)}]{Preskill1998}%
  \BibitemOpen
  \bibfield  {author} {\bibinfo {author} {\bibfnamefont {J.}~\bibnamefont {Preskill}},\ }\bibfield  {title} {{Quantum computing: pro and con},\ }\href {https://doi.org/10.1098/rspa.1998.0171} {\bibfield  {journal} {\bibinfo  {journal} {Proceedings of the Royal Society of London. Series A: Mathematical, Physical and Engineering Sciences}\ }\textbf {\bibinfo {volume} {454}},\ \bibinfo {pages} {469} (\bibinfo {year} {1998})}\BibitemShut {NoStop}%
\bibitem [{\citenamefont {Gupta}\ et~al.(2023)\citenamefont {Gupta}, \citenamefont {Modgil}, \citenamefont {Bhatt}, \citenamefont {{Chiappetta Jabbour}},\ and\ \citenamefont {Kamble}}]{GUPTA2023102544}%
  \BibitemOpen
  \bibfield  {author} {\bibinfo {author} {\bibfnamefont {S.}~\bibnamefont {Gupta}}, \bibinfo {author} {\bibfnamefont {S.}~\bibnamefont {Modgil}}, \bibinfo {author} {\bibfnamefont {P.~C.}\ \bibnamefont {Bhatt}}, \bibinfo {author} {\bibfnamefont {C.~J.}\ \bibnamefont {{Chiappetta Jabbour}}},\ and\ \bibinfo {author} {\bibfnamefont {S.}~\bibnamefont {Kamble}},\ }\bibfield  {title} {{Quantum computing led innovation for achieving a more sustainable Covid-19 healthcare industry},\ }\href {https://doi.org/https://doi.org/10.1016/j.technovation.2022.102544} {\bibfield  {journal} {\bibinfo  {journal} {Technovation}\ }\textbf {\bibinfo {volume} {120}},\ \bibinfo {pages} {102544} (\bibinfo {year} {2023})}\BibitemShut {NoStop}%
\bibitem [{\citenamefont {Coccia}(2022)}]{doi:10.1080/09537325.2022.2110056}%
  \BibitemOpen
  \bibfield  {author} {\bibinfo {author} {\bibfnamefont {M.}~\bibnamefont {Coccia}},\ }\bibfield  {title} {{Technological trajectories in quantum computing to design a quantum ecosystem for industrial change},\ }\href {https://doi.org/10.1080/09537325.2022.2110056} {\bibfield  {journal} {\bibinfo  {journal} {Technology Analysis \& Strategic Management}\ }\textbf {\bibinfo {volume} {0}},\ \bibinfo {pages} {1} (\bibinfo {year} {2022})},\ \Eprint {https://arxiv.org/abs/https://doi.org/10.1080/09537325.2022.2110056} {https://doi.org/10.1080/09537325.2022.2110056} \BibitemShut {NoStop}%
\bibitem [{\citenamefont {Wurtz}\ et~al.(2022)\citenamefont {Wurtz}, \citenamefont {Lopes}, \citenamefont {Gemelke}, \citenamefont {Keesling},\ and\ \citenamefont {Wang}}]{wurtz2022industry}%
  \BibitemOpen
  \bibfield  {author} {\bibinfo {author} {\bibfnamefont {J.}~\bibnamefont {Wurtz}}, \bibinfo {author} {\bibfnamefont {P.}~\bibnamefont {Lopes}}, \bibinfo {author} {\bibfnamefont {N.}~\bibnamefont {Gemelke}}, \bibinfo {author} {\bibfnamefont {A.}~\bibnamefont {Keesling}},\ and\ \bibinfo {author} {\bibfnamefont {S.}~\bibnamefont {Wang}},\ }\bibfield  {title} {{Industry applications of neutral-atom quantum computing solving independent set problems},\ }\href@noop {} {\bibfield  {journal} {\bibinfo  {journal} {arXiv preprint arXiv:2205.08500}\ } (\bibinfo {year} {2022})}\BibitemShut {NoStop}%
\bibitem [{\citenamefont {Ray}\ et~al.(2022)\citenamefont {Ray}, \citenamefont {Guddanti}, \citenamefont {Ajith},\ and\ \citenamefont {Vinayagamurthy}}]{ray2022classical}%
  \BibitemOpen
  \bibfield  {author} {\bibinfo {author} {\bibfnamefont {A.}~\bibnamefont {Ray}}, \bibinfo {author} {\bibfnamefont {S.~S.}\ \bibnamefont {Guddanti}}, \bibinfo {author} {\bibfnamefont {V.}~\bibnamefont {Ajith}},\ and\ \bibinfo {author} {\bibfnamefont {D.}~\bibnamefont {Vinayagamurthy}},\ }\bibfield  {title} {{{Classical ensemble of Quantum-classical ML algorithms for Phishing detection in Ethereum transaction networks}},\ }\href@noop {} {\bibfield  {journal} {\bibinfo  {journal} {arXiv preprint arXiv:2211.00004}\ } (\bibinfo {year} {2022})}\BibitemShut {NoStop}%
\bibitem [{\citenamefont {Prajapati}\ et~al.(2023)\citenamefont {Prajapati}, \citenamefont {Paliwal}, \citenamefont {Prajapati}, \citenamefont {Saikia},\ and\ \citenamefont {Pandey}}]{prajapati2023quantum}%
  \BibitemOpen
  \bibfield  {author} {\bibinfo {author} {\bibfnamefont {J.~B.}\ \bibnamefont {Prajapati}}, \bibinfo {author} {\bibfnamefont {H.}~\bibnamefont {Paliwal}}, \bibinfo {author} {\bibfnamefont {B.~G.}\ \bibnamefont {Prajapati}}, \bibinfo {author} {\bibfnamefont {S.}~\bibnamefont {Saikia}},\ and\ \bibinfo {author} {\bibfnamefont {R.}~\bibnamefont {Pandey}},\ }in\ \href@noop {} {\bibinfo {booktitle} {Quantum Computing: A Shift from Bits to Qubits}}\ (\bibinfo  {publisher} {Springer},\ \bibinfo {year} {2023})\ pp.\ \bibinfo {pages} {351--382}\BibitemShut {NoStop}%
\bibitem [{\citenamefont {Khan}\ et~al.(2019)\citenamefont {Khan}, \citenamefont {Awan},\ and\ \citenamefont {Vall-Llosera}}]{khan2019k}%
  \BibitemOpen
  \bibfield  {author} {\bibinfo {author} {\bibfnamefont {S.~U.}\ \bibnamefont {Khan}}, \bibinfo {author} {\bibfnamefont {A.~J.}\ \bibnamefont {Awan}},\ and\ \bibinfo {author} {\bibfnamefont {G.}~\bibnamefont {Vall-Llosera}},\ }\bibfield  {title} {{K-means clustering on noisy intermediate scale quantum computers},\ }\href@noop {} {\bibfield  {journal} {\bibinfo  {journal} {arXiv preprint arXiv:1909.12183}\ } (\bibinfo {year} {2019})}\BibitemShut {NoStop}%
\bibitem [{\citenamefont {Cerezo}\ et~al.(2022{\natexlab{a}})\citenamefont {Cerezo}, \citenamefont {Verdon}, \citenamefont {Huang}, \citenamefont {Cincio},\ and\ \citenamefont {Coles}}]{cerezo2022challenges}%
  \BibitemOpen
  \bibfield  {author} {\bibinfo {author} {\bibfnamefont {M.}~\bibnamefont {Cerezo}}, \bibinfo {author} {\bibfnamefont {G.}~\bibnamefont {Verdon}}, \bibinfo {author} {\bibfnamefont {H.-Y.}\ \bibnamefont {Huang}}, \bibinfo {author} {\bibfnamefont {L.}~\bibnamefont {Cincio}},\ and\ \bibinfo {author} {\bibfnamefont {P.~J.}\ \bibnamefont {Coles}},\ }\bibfield  {title} {{Challenges and opportunities in quantum machine learning},\ }\href@noop {} {\bibfield  {journal} {\bibinfo  {journal} {Nature Computational Science}\ }\textbf {\bibinfo {volume} {2}},\ \bibinfo {pages} {567} (\bibinfo {year} {2022}{\natexlab{a}})}\BibitemShut {NoStop}%
\bibitem [{\citenamefont {Martin}\ et~al.(2021)\citenamefont {Martin}, \citenamefont {Candelas}, \citenamefont {Rodr{\'\i}guez-Rozas}, \citenamefont {Mart{\'\i}n-Guerrero}, \citenamefont {Chen}, \citenamefont {Lamata}, \citenamefont {Or{\'u}s}, \citenamefont {Solano},\ and\ \citenamefont {Sanz}}]{martin2021toward}%
  \BibitemOpen
  \bibfield  {author} {\bibinfo {author} {\bibfnamefont {A.}~\bibnamefont {Martin}}, \bibinfo {author} {\bibfnamefont {B.}~\bibnamefont {Candelas}}, \bibinfo {author} {\bibfnamefont {{\'A}.}~\bibnamefont {Rodr{\'\i}guez-Rozas}}, \bibinfo {author} {\bibfnamefont {J.~D.}\ \bibnamefont {Mart{\'\i}n-Guerrero}}, \bibinfo {author} {\bibfnamefont {X.}~\bibnamefont {Chen}}, \bibinfo {author} {\bibfnamefont {L.}~\bibnamefont {Lamata}}, \bibinfo {author} {\bibfnamefont {R.}~\bibnamefont {Or{\'u}s}}, \bibinfo {author} {\bibfnamefont {E.}~\bibnamefont {Solano}},\ and\ \bibinfo {author} {\bibfnamefont {M.}~\bibnamefont {Sanz}},\ }\bibfield  {title} {{Toward pricing financial derivatives with an ibm quantum computer},\ }\href@noop {} {\bibfield  {journal} {\bibinfo  {journal} {Physical Review Research}\ }\textbf {\bibinfo {volume} {3}},\ \bibinfo {pages} {013167} (\bibinfo {year} {2021})}\BibitemShut {NoStop}%
\bibitem [{\citenamefont {McClean}\ et~al.(2018)\citenamefont {McClean}, \citenamefont {Boixo}, \citenamefont {Smelyanskiy}, \citenamefont {Babbush},\ and\ \citenamefont {Neven}}]{mcclean2018barren}%
  \BibitemOpen
  \bibfield  {author} {\bibinfo {author} {\bibfnamefont {J.~R.}\ \bibnamefont {McClean}}, \bibinfo {author} {\bibfnamefont {S.}~\bibnamefont {Boixo}}, \bibinfo {author} {\bibfnamefont {V.~N.}\ \bibnamefont {Smelyanskiy}}, \bibinfo {author} {\bibfnamefont {R.}~\bibnamefont {Babbush}},\ and\ \bibinfo {author} {\bibfnamefont {H.}~\bibnamefont {Neven}},\ }\bibfield  {title} {{Barren plateaus in quantum neural network training landscapes},\ }\href@noop {} {\bibfield  {journal} {\bibinfo  {journal} {Nature communications}\ }\textbf {\bibinfo {volume} {9}},\ \bibinfo {pages} {1} (\bibinfo {year} {2018})}\BibitemShut {NoStop}%
\bibitem [{\citenamefont {Grant}\ et~al.(2019)\citenamefont {Grant}, \citenamefont {Wossnig}, \citenamefont {Ostaszewski},\ and\ \citenamefont {Benedetti}}]{grant2019initialization}%
  \BibitemOpen
  \bibfield  {author} {\bibinfo {author} {\bibfnamefont {E.}~\bibnamefont {Grant}}, \bibinfo {author} {\bibfnamefont {L.}~\bibnamefont {Wossnig}}, \bibinfo {author} {\bibfnamefont {M.}~\bibnamefont {Ostaszewski}},\ and\ \bibinfo {author} {\bibfnamefont {M.}~\bibnamefont {Benedetti}},\ }\bibfield  {title} {{An initialization strategy for addressing barren plateaus in parametrized quantum circuits},\ }\href@noop {} {\bibfield  {journal} {\bibinfo  {journal} {Quantum}\ }\textbf {\bibinfo {volume} {3}},\ \bibinfo {pages} {214} (\bibinfo {year} {2019})}\BibitemShut {NoStop}%
\bibitem [{\citenamefont {Mottonen}\ et~al.(2004)\citenamefont {Mottonen}, \citenamefont {Vartiainen}, \citenamefont {Bergholm},\ and\ \citenamefont {Salomaa}}]{mottonen2004transformation}%
  \BibitemOpen
  \bibfield  {author} {\bibinfo {author} {\bibfnamefont {M.}~\bibnamefont {Mottonen}}, \bibinfo {author} {\bibfnamefont {J.~J.}\ \bibnamefont {Vartiainen}}, \bibinfo {author} {\bibfnamefont {V.}~\bibnamefont {Bergholm}},\ and\ \bibinfo {author} {\bibfnamefont {M.~M.}\ \bibnamefont {Salomaa}},\ }\bibfield  {title} {{Transformation of quantum states using uniformly controlled rotations},\ }\href@noop {} {\bibfield  {journal} {\bibinfo  {journal} {arXiv preprint quant-ph/0407010}\ } (\bibinfo {year} {2004})}\BibitemShut {NoStop}%
\bibitem [{\citenamefont {Chen}\ and\ \citenamefont {Yoo}(2021)}]{chen2021federated}%
  \BibitemOpen
  \bibfield  {author} {\bibinfo {author} {\bibfnamefont {S.~Y.-C.}\ \bibnamefont {Chen}}\ and\ \bibinfo {author} {\bibfnamefont {S.}~\bibnamefont {Yoo}},\ }\bibfield  {title} {{Federated quantum machine learning},\ }\href@noop {} {\bibfield  {journal} {\bibinfo  {journal} {Entropy}\ }\textbf {\bibinfo {volume} {23}},\ \bibinfo {pages} {460} (\bibinfo {year} {2021})}\BibitemShut {NoStop}%
\bibitem [{\citenamefont {Yun}\ et~al.(2022)\citenamefont {Yun}, \citenamefont {Kim}, \citenamefont {Jung}, \citenamefont {Park}, \citenamefont {Bennis},\ and\ \citenamefont {Kim}}]{yun2022slimmable}%
  \BibitemOpen
  \bibfield  {author} {\bibinfo {author} {\bibfnamefont {W.~J.}\ \bibnamefont {Yun}}, \bibinfo {author} {\bibfnamefont {J.~P.}\ \bibnamefont {Kim}}, \bibinfo {author} {\bibfnamefont {S.}~\bibnamefont {Jung}}, \bibinfo {author} {\bibfnamefont {J.}~\bibnamefont {Park}}, \bibinfo {author} {\bibfnamefont {M.}~\bibnamefont {Bennis}},\ and\ \bibinfo {author} {\bibfnamefont {J.}~\bibnamefont {Kim}},\ }\bibfield  {title} {{Slimmable quantum federated learning},\ }\href@noop {} {\bibfield  {journal} {\bibinfo  {journal} {arXiv preprint arXiv:2207.10221}\ } (\bibinfo {year} {2022})}\BibitemShut {NoStop}%
\bibitem [{\citenamefont {Xia}\ and\ \citenamefont {Li}(2021)}]{xia2021quantumfed}%
  \BibitemOpen
  \bibfield  {author} {\bibinfo {author} {\bibfnamefont {Q.}~\bibnamefont {Xia}}\ and\ \bibinfo {author} {\bibfnamefont {Q.}~\bibnamefont {Li}},\ }in\ \href@noop {} {\bibinfo {booktitle} {2021 IEEE Global Communications Conference (GLOBECOM)}}\ (\bibinfo {organization} {IEEE},\ \bibinfo {year} {2021})\ pp.\ \bibinfo {pages} {1--6}\BibitemShut {NoStop}%
\bibitem [{\citenamefont {Lavrijsen}\ et~al.(2020)\citenamefont {Lavrijsen}, \citenamefont {Tudor}, \citenamefont {M{\"u}ller}, \citenamefont {Iancu},\ and\ \citenamefont {De~Jong}}]{lavrijsen2020classical}%
  \BibitemOpen
  \bibfield  {author} {\bibinfo {author} {\bibfnamefont {W.}~\bibnamefont {Lavrijsen}}, \bibinfo {author} {\bibfnamefont {A.}~\bibnamefont {Tudor}}, \bibinfo {author} {\bibfnamefont {J.}~\bibnamefont {M{\"u}ller}}, \bibinfo {author} {\bibfnamefont {C.}~\bibnamefont {Iancu}},\ and\ \bibinfo {author} {\bibfnamefont {W.}~\bibnamefont {De~Jong}},\ }in\ \href@noop {} {\bibinfo {booktitle} {2020 IEEE international conference on quantum computing and engineering (QCE)}}\ (\bibinfo {organization} {IEEE},\ \bibinfo {year} {2020})\ pp.\ \bibinfo {pages} {267--277}\BibitemShut {NoStop}%
\bibitem [{\citenamefont {Cincio}\ et~al.(2018)\citenamefont {Cincio}, \citenamefont {Suba{\c{s}}{\i}}, \citenamefont {Sornborger},\ and\ \citenamefont {Coles}}]{cincio2018learning}%
  \BibitemOpen
  \bibfield  {author} {\bibinfo {author} {\bibfnamefont {L.}~\bibnamefont {Cincio}}, \bibinfo {author} {\bibfnamefont {Y.}~\bibnamefont {Suba{\c{s}}{\i}}}, \bibinfo {author} {\bibfnamefont {A.~T.}\ \bibnamefont {Sornborger}},\ and\ \bibinfo {author} {\bibfnamefont {P.~J.}\ \bibnamefont {Coles}},\ }\bibfield  {title} {{Learning the quantum algorithm for state overlap},\ }\href@noop {} {\bibfield  {journal} {\bibinfo  {journal} {New Journal of Physics}\ }\textbf {\bibinfo {volume} {20}},\ \bibinfo {pages} {113022} (\bibinfo {year} {2018})}\BibitemShut {NoStop}%
\bibitem [{\citenamefont {Cerezo}\ et~al.(2022{\natexlab{b}})\citenamefont {Cerezo}, \citenamefont {Sharma}, \citenamefont {Arrasmith},\ and\ \citenamefont {Coles}}]{cerezo2022variational}%
  \BibitemOpen
  \bibfield  {author} {\bibinfo {author} {\bibfnamefont {M.}~\bibnamefont {Cerezo}}, \bibinfo {author} {\bibfnamefont {K.}~\bibnamefont {Sharma}}, \bibinfo {author} {\bibfnamefont {A.}~\bibnamefont {Arrasmith}},\ and\ \bibinfo {author} {\bibfnamefont {P.~J.}\ \bibnamefont {Coles}},\ }\bibfield  {title} {{Variational quantum state eigensolver},\ }\href@noop {} {\bibfield  {journal} {\bibinfo  {journal} {npj Quantum Information}\ }\textbf {\bibinfo {volume} {8}},\ \bibinfo {pages} {1} (\bibinfo {year} {2022}{\natexlab{b}})}\BibitemShut {NoStop}%
\bibitem [{\citenamefont {Chang}\ et~al.(2023)\citenamefont {Chang}, \citenamefont {Agnew}, \citenamefont {Combarro}, \citenamefont {Grossi}, \citenamefont {Herbert},\ and\ \citenamefont {Vallecorsa}}]{chang2023running}%
  \BibitemOpen
  \bibfield  {author} {\bibinfo {author} {\bibfnamefont {S.~Y.}\ \bibnamefont {Chang}}, \bibinfo {author} {\bibfnamefont {E.}~\bibnamefont {Agnew}}, \bibinfo {author} {\bibfnamefont {E.}~\bibnamefont {Combarro}}, \bibinfo {author} {\bibfnamefont {M.}~\bibnamefont {Grossi}}, \bibinfo {author} {\bibfnamefont {S.}~\bibnamefont {Herbert}},\ and\ \bibinfo {author} {\bibfnamefont {S.}~\bibnamefont {Vallecorsa}},\ }in\ \href@noop {} {\bibinfo {booktitle} {Journal of Physics: Conference Series}},\ Vol.\ \bibinfo {volume} {2438}\ (\bibinfo {organization} {IOP Publishing},\ \bibinfo {year} {2023})\ p.\ \bibinfo {pages} {012062}\BibitemShut {NoStop}%
\bibitem [{\citenamefont {Cross}\ et~al.(2019)\citenamefont {Cross}, \citenamefont {Bishop}, \citenamefont {Sheldon}, \citenamefont {Nation},\ and\ \citenamefont {Gambetta}}]{volume1}%
  \BibitemOpen
  \bibfield  {author} {\bibinfo {author} {\bibfnamefont {A.~W.}\ \bibnamefont {Cross}}, \bibinfo {author} {\bibfnamefont {L.~S.}\ \bibnamefont {Bishop}}, \bibinfo {author} {\bibfnamefont {S.}~\bibnamefont {Sheldon}}, \bibinfo {author} {\bibfnamefont {P.~D.}\ \bibnamefont {Nation}},\ and\ \bibinfo {author} {\bibfnamefont {J.~M.}\ \bibnamefont {Gambetta}},\ }\bibfield  {title} {{Validating quantum computers using randomized model circuits},\ }\href@noop {} {\bibfield  {journal} {\bibinfo  {journal} {Physical Review A}\ }\textbf {\bibinfo {volume} {100}},\ \bibinfo {pages} {032328} (\bibinfo {year} {2019})}\BibitemShut {NoStop}%
\bibitem [{\citenamefont {Wan}\ et~al.(2017)\citenamefont {Wan}, \citenamefont {Dahlsten}, \citenamefont {Kristj{\'a}nsson}, \citenamefont {Gardner},\ and\ \citenamefont {Kim}}]{autoencoder4}%
  \BibitemOpen
  \bibfield  {author} {\bibinfo {author} {\bibfnamefont {K.~H.}\ \bibnamefont {Wan}}, \bibinfo {author} {\bibfnamefont {O.}~\bibnamefont {Dahlsten}}, \bibinfo {author} {\bibfnamefont {H.}~\bibnamefont {Kristj{\'a}nsson}}, \bibinfo {author} {\bibfnamefont {R.}~\bibnamefont {Gardner}},\ and\ \bibinfo {author} {\bibfnamefont {M.}~\bibnamefont {Kim}},\ }\bibfield  {title} {{Quantum generalisation of feedforward neural networks},\ }\href@noop {} {\bibfield  {journal} {\bibinfo  {journal} {npj Quantum information}\ }\textbf {\bibinfo {volume} {3}},\ \bibinfo {pages} {1} (\bibinfo {year} {2017})}\BibitemShut {NoStop}%
\bibitem [{\citenamefont {Wo{\'z}niak}\ et~al.(2023)\citenamefont {Wo{\'z}niak}, \citenamefont {Belis}, \citenamefont {Puljak}, \citenamefont {Barkoutsos}, \citenamefont {Dissertori}, \citenamefont {Grossi}, \citenamefont {Pierini}, \citenamefont {Reiter}, \citenamefont {Tavernelli},\ and\ \citenamefont {Vallecorsa}}]{wozniak2023quantum}%
  \BibitemOpen
  \bibfield  {author} {\bibinfo {author} {\bibfnamefont {K.~A.}\ \bibnamefont {Wo{\'z}niak}}, \bibinfo {author} {\bibfnamefont {V.}~\bibnamefont {Belis}}, \bibinfo {author} {\bibfnamefont {E.}~\bibnamefont {Puljak}}, \bibinfo {author} {\bibfnamefont {P.}~\bibnamefont {Barkoutsos}}, \bibinfo {author} {\bibfnamefont {G.}~\bibnamefont {Dissertori}}, \bibinfo {author} {\bibfnamefont {M.}~\bibnamefont {Grossi}}, \bibinfo {author} {\bibfnamefont {M.}~\bibnamefont {Pierini}}, \bibinfo {author} {\bibfnamefont {F.}~\bibnamefont {Reiter}}, \bibinfo {author} {\bibfnamefont {I.}~\bibnamefont {Tavernelli}},\ and\ \bibinfo {author} {\bibfnamefont {S.}~\bibnamefont {Vallecorsa}},\ }\bibfield  {title} {{Quantum anomaly detection in the latent space of proton collision events at the {LHC}},\ }\href@noop {} {\bibfield  {journal} {\bibinfo  {journal} {arXiv preprint arXiv:2301.10780}\ } (\bibinfo {year} {2023})}\BibitemShut {NoStop}%
\bibitem [{\citenamefont {Park}\ and\ \citenamefont {Killoran}(2024)}]{park2024hamiltonian}%
  \BibitemOpen
  \bibfield  {author} {\bibinfo {author} {\bibfnamefont {C.-Y.}\ \bibnamefont {Park}}\ and\ \bibinfo {author} {\bibfnamefont {N.}~\bibnamefont {Killoran}},\ }\bibfield  {title} {{Hamiltonian variational ansatz without barren plateaus},\ }\href@noop {} {\bibfield  {journal} {\bibinfo  {journal} {Quantum}\ }\textbf {\bibinfo {volume} {8}},\ \bibinfo {pages} {1239} (\bibinfo {year} {2024})}\BibitemShut {NoStop}%
\bibitem [{\citenamefont {Li}\ et~al.(2023{\natexlab{b}})\citenamefont {Li}, \citenamefont {Huang}, \citenamefont {Hou}, \citenamefont {Li}, \citenamefont {Wang},\ and\ \citenamefont {Bayat}}]{li2023ensemble}%
  \BibitemOpen
  \bibfield  {author} {\bibinfo {author} {\bibfnamefont {Q.}~\bibnamefont {Li}}, \bibinfo {author} {\bibfnamefont {Y.}~\bibnamefont {Huang}}, \bibinfo {author} {\bibfnamefont {X.}~\bibnamefont {Hou}}, \bibinfo {author} {\bibfnamefont {Y.}~\bibnamefont {Li}}, \bibinfo {author} {\bibfnamefont {X.}~\bibnamefont {Wang}},\ and\ \bibinfo {author} {\bibfnamefont {A.}~\bibnamefont {Bayat}},\ }\bibfield  {title} {{Ensemble-learning variational shallow-circuit quantum classifiers},\ }\href@noop {} {\bibfield  {journal} {\bibinfo  {journal} {arXiv preprint arXiv:2301.12707}\ } (\bibinfo {year} {2023}{\natexlab{b}})}\BibitemShut {NoStop}%
\bibitem [{\citenamefont {Kulshrestha}\ et~al.(2023)\citenamefont {Kulshrestha}, \citenamefont {Liu}, \citenamefont {Ushijima-Mwesigwa},\ and\ \citenamefont {Safro}}]{kulshrestha2023learning}%
  \BibitemOpen
  \bibfield  {author} {\bibinfo {author} {\bibfnamefont {A.}~\bibnamefont {Kulshrestha}}, \bibinfo {author} {\bibfnamefont {X.}~\bibnamefont {Liu}}, \bibinfo {author} {\bibfnamefont {H.}~\bibnamefont {Ushijima-Mwesigwa}},\ and\ \bibinfo {author} {\bibfnamefont {I.}~\bibnamefont {Safro}},\ }\bibfield  {title} {{Learning To Optimize Quantum Neural Network Without Gradients},\ }\href@noop {} {\bibfield  {journal} {\bibinfo  {journal} {arXiv preprint arXiv:2304.07442}\ } (\bibinfo {year} {2023})}\BibitemShut {NoStop}%
\bibitem [{\citenamefont {Haug}\ et~al.(2023)\citenamefont {Haug}, \citenamefont {Self},\ and\ \citenamefont {Kim}}]{haug2023quantum}%
  \BibitemOpen
  \bibfield  {author} {\bibinfo {author} {\bibfnamefont {T.}~\bibnamefont {Haug}}, \bibinfo {author} {\bibfnamefont {C.~N.}\ \bibnamefont {Self}},\ and\ \bibinfo {author} {\bibfnamefont {M.}~\bibnamefont {Kim}},\ }\bibfield  {title} {{Quantum machine learning of large datasets using randomized measurements},\ }\href@noop {} {\bibfield  {journal} {\bibinfo  {journal} {Machine Learning: Science and Technology}\ }\textbf {\bibinfo {volume} {4}},\ \bibinfo {pages} {015005} (\bibinfo {year} {2023})}\BibitemShut {NoStop}%
\bibitem [{\citenamefont {Gentinetta}\ et~al.(2024)\citenamefont {Gentinetta}, \citenamefont {Thomsen}, \citenamefont {Sutter},\ and\ \citenamefont {Woerner}}]{gentinetta2024complexity}%
  \BibitemOpen
  \bibfield  {author} {\bibinfo {author} {\bibfnamefont {G.}~\bibnamefont {Gentinetta}}, \bibinfo {author} {\bibfnamefont {A.}~\bibnamefont {Thomsen}}, \bibinfo {author} {\bibfnamefont {D.}~\bibnamefont {Sutter}},\ and\ \bibinfo {author} {\bibfnamefont {S.}~\bibnamefont {Woerner}},\ }\bibfield  {title} {{The complexity of quantum support vector machines},\ }\href@noop {} {\bibfield  {journal} {\bibinfo  {journal} {Quantum}\ }\textbf {\bibinfo {volume} {8}},\ \bibinfo {pages} {1225} (\bibinfo {year} {2024})}\BibitemShut {NoStop}%
\bibitem [{\citenamefont {Heyraud}\ et~al.(2023)\citenamefont {Heyraud}, \citenamefont {Li}, \citenamefont {Donatella}, \citenamefont {Le~Boit{\'e}},\ and\ \citenamefont {Ciuti}}]{heyraud2023efficient}%
  \BibitemOpen
  \bibfield  {author} {\bibinfo {author} {\bibfnamefont {V.}~\bibnamefont {Heyraud}}, \bibinfo {author} {\bibfnamefont {Z.}~\bibnamefont {Li}}, \bibinfo {author} {\bibfnamefont {K.}~\bibnamefont {Donatella}}, \bibinfo {author} {\bibfnamefont {A.}~\bibnamefont {Le~Boit{\'e}}},\ and\ \bibinfo {author} {\bibfnamefont {C.}~\bibnamefont {Ciuti}},\ }\bibfield  {title} {{Efficient estimation of trainability for variational quantum circuits},\ }\href@noop {} {\bibfield  {journal} {\bibinfo  {journal} {PRX Quantum}\ }\textbf {\bibinfo {volume} {4}},\ \bibinfo {pages} {040335} (\bibinfo {year} {2023})}\BibitemShut {NoStop}%
\bibitem [{\citenamefont {Akter}\ et~al.(2023)\citenamefont {Akter}, \citenamefont {Shahriar}, \citenamefont {Iqbal}, \citenamefont {Hossain}, \citenamefont {Karim}, \citenamefont {Clincy},\ and\ \citenamefont {Voicu}}]{akter2023exploring}%
  \BibitemOpen
  \bibfield  {author} {\bibinfo {author} {\bibfnamefont {M.}~\bibnamefont {Akter}}, \bibinfo {author} {\bibfnamefont {H.}~\bibnamefont {Shahriar}}, \bibinfo {author} {\bibfnamefont {A.}~\bibnamefont {Iqbal}}, \bibinfo {author} {\bibfnamefont {M.}~\bibnamefont {Hossain}}, \bibinfo {author} {\bibfnamefont {M.}~\bibnamefont {Karim}}, \bibinfo {author} {\bibfnamefont {V.}~\bibnamefont {Clincy}},\ and\ \bibinfo {author} {\bibfnamefont {R.}~\bibnamefont {Voicu}},\ }in\ \href@noop {} {\bibinfo {booktitle} {IEEE CARL K. CHANG SYMPOSIUM ON SOFTWARE SERVICES ENGINEERING}}\ (\bibinfo {year} {2023})\BibitemShut {NoStop}%
\bibitem [{\citenamefont {Kundu}\ and\ \citenamefont {Ghosh}(2022)}]{kundu2022security}%
  \BibitemOpen
  \bibfield  {author} {\bibinfo {author} {\bibfnamefont {S.}~\bibnamefont {Kundu}}\ and\ \bibinfo {author} {\bibfnamefont {S.}~\bibnamefont {Ghosh}},\ }in\ \href@noop {} {\bibinfo {booktitle} {Proceedings of the Great Lakes Symposium on VLSI 2022}}\ (\bibinfo {year} {2022})\ pp.\ \bibinfo {pages} {463--468}\BibitemShut {NoStop}%
\bibitem [{\citenamefont {Von~Lilienfeld}(2018)}]{von2018quantum}%
  \BibitemOpen
  \bibfield  {author} {\bibinfo {author} {\bibfnamefont {O.~A.}\ \bibnamefont {Von~Lilienfeld}},\ }\bibfield  {title} {{Quantum machine learning in chemical compound space},\ }\href@noop {} {\bibfield  {journal} {\bibinfo  {journal} {Angewandte Chemie International Edition}\ }\textbf {\bibinfo {volume} {57}},\ \bibinfo {pages} {4164} (\bibinfo {year} {2018})}\BibitemShut {NoStop}%
\bibitem [{\citenamefont {von Lilienfeld}\ et~al.(2020)\citenamefont {von Lilienfeld}, \citenamefont {M{\"u}ller},\ and\ \citenamefont {Tkatchenko}}]{von2020exploring}%
  \BibitemOpen
  \bibfield  {author} {\bibinfo {author} {\bibfnamefont {O.~A.}\ \bibnamefont {von Lilienfeld}}, \bibinfo {author} {\bibfnamefont {K.-R.}\ \bibnamefont {M{\"u}ller}},\ and\ \bibinfo {author} {\bibfnamefont {A.}~\bibnamefont {Tkatchenko}},\ }\bibfield  {title} {{Exploring chemical compound space with quantum-based machine learning},\ }\href@noop {} {\bibfield  {journal} {\bibinfo  {journal} {Nature Reviews Chemistry}\ }\textbf {\bibinfo {volume} {4}},\ \bibinfo {pages} {347} (\bibinfo {year} {2020})}\BibitemShut {NoStop}%
\bibitem [{\citenamefont {Vandersypen}\ et~al.(2001)\citenamefont {Vandersypen}, \citenamefont {Steffen}, \citenamefont {Breyta}, \citenamefont {Yannoni}, \citenamefont {Sherwood},\ and\ \citenamefont {Chuang}}]{vandersypen2001experimental}%
  \BibitemOpen
  \bibfield  {author} {\bibinfo {author} {\bibfnamefont {L.~M.}\ \bibnamefont {Vandersypen}}, \bibinfo {author} {\bibfnamefont {M.}~\bibnamefont {Steffen}}, \bibinfo {author} {\bibfnamefont {G.}~\bibnamefont {Breyta}}, \bibinfo {author} {\bibfnamefont {C.~S.}\ \bibnamefont {Yannoni}}, \bibinfo {author} {\bibfnamefont {M.~H.}\ \bibnamefont {Sherwood}},\ and\ \bibinfo {author} {\bibfnamefont {I.~L.}\ \bibnamefont {Chuang}},\ }\bibfield  {title} {{{Experimental} realization of {Shor's} quantum factoring algorithm using nuclear magnetic resonance},\ }\href@noop {} {\bibfield  {journal} {\bibinfo  {journal} {Nature}\ }\textbf {\bibinfo {volume} {414}},\ \bibinfo {pages} {883} (\bibinfo {year} {2001})}\BibitemShut {NoStop}%
\bibitem [{\citenamefont {Heese}\ et~al.(2023)\citenamefont {Heese}, \citenamefont {Gerlach}, \citenamefont {M{\"u}cke}, \citenamefont {M{\"u}ller}, \citenamefont {Jakobs},\ and\ \citenamefont {Piatkowski}}]{heese2023explainable}%
  \BibitemOpen
  \bibfield  {author} {\bibinfo {author} {\bibfnamefont {R.}~\bibnamefont {Heese}}, \bibinfo {author} {\bibfnamefont {T.}~\bibnamefont {Gerlach}}, \bibinfo {author} {\bibfnamefont {S.}~\bibnamefont {M{\"u}cke}}, \bibinfo {author} {\bibfnamefont {S.}~\bibnamefont {M{\"u}ller}}, \bibinfo {author} {\bibfnamefont {M.}~\bibnamefont {Jakobs}},\ and\ \bibinfo {author} {\bibfnamefont {N.}~\bibnamefont {Piatkowski}},\ }\bibfield  {title} {{Explainable Quantum Machine Learning},\ }\href@noop {} {\bibfield  {journal} {\bibinfo  {journal} {arXiv preprint arXiv:2301.09138}\ } (\bibinfo {year} {2023})}\BibitemShut {NoStop}%
\bibitem [{\citenamefont {Hook}\ et~al.(2018)\citenamefont {Hook}, \citenamefont {Porter},\ and\ \citenamefont {Herzog}}]{pub.1106289502}%
  \BibitemOpen
  \bibfield  {author} {\bibinfo {author} {\bibfnamefont {D.~W.}\ \bibnamefont {Hook}}, \bibinfo {author} {\bibfnamefont {S.~J.}\ \bibnamefont {Porter}},\ and\ \bibinfo {author} {\bibfnamefont {C.}~\bibnamefont {Herzog}},\ }\bibfield  {title} {{Dimensions: Building Context for Search and Evaluation},\ }\href {https://doi.org/10.3389/frma.2018.00023} {\bibfield  {journal} {\bibinfo  {journal} {Frontiers in Research Metrics and Analytics}\ }\textbf {\bibinfo {volume} {3}},\ \bibinfo {pages} {23} (\bibinfo {year} {2018})},\ \bibinfo {note} {https://www.frontiersin.org/articles/10.3389/frma.2018.00023/pdf}\BibitemShut {NoStop}%
\bibitem [{\citenamefont {Gentinetta}\ et~al.(2023)\citenamefont {Gentinetta}, \citenamefont {Sutter}, \citenamefont {Zoufal}, \citenamefont {Fuller},\ and\ \citenamefont {Woerner}}]{gentinetta2023quantum}%
  \BibitemOpen
  \bibfield  {author} {\bibinfo {author} {\bibfnamefont {G.}~\bibnamefont {Gentinetta}}, \bibinfo {author} {\bibfnamefont {D.}~\bibnamefont {Sutter}}, \bibinfo {author} {\bibfnamefont {C.}~\bibnamefont {Zoufal}}, \bibinfo {author} {\bibfnamefont {B.}~\bibnamefont {Fuller}},\ and\ \bibinfo {author} {\bibfnamefont {S.}~\bibnamefont {Woerner}},\ }in\ \href@noop {} {\bibinfo {booktitle} {2023 IEEE International Conference on Quantum Computing and Engineering (QCE)}},\ Vol.~\bibinfo {volume} {1}\ (\bibinfo {organization} {IEEE},\ \bibinfo {year} {2023})\ pp.\ \bibinfo {pages} {256--262}\BibitemShut {NoStop}%
\bibitem [{\citenamefont {Garc{\'\i}a-Mart{\'\i}n}\ et~al.(2023)\citenamefont {Garc{\'\i}a-Mart{\'\i}n}, \citenamefont {Larocca},\ and\ \citenamefont {Cerezo}}]{garcia2023effects}%
  \BibitemOpen
  \bibfield  {author} {\bibinfo {author} {\bibfnamefont {D.}~\bibnamefont {Garc{\'\i}a-Mart{\'\i}n}}, \bibinfo {author} {\bibfnamefont {M.}~\bibnamefont {Larocca}},\ and\ \bibinfo {author} {\bibfnamefont {M.}~\bibnamefont {Cerezo}},\ }\bibfield  {title} {{Effects of noise on the overparametrization of quantum neural networks},\ }\href@noop {} {\bibfield  {journal} {\bibinfo  {journal} {arXiv preprint arXiv:2302.05059}\ } (\bibinfo {year} {2023})}\BibitemShut {NoStop}%
\bibitem [{\citenamefont {Dri}\ et~al.(2023)\citenamefont {Dri}, \citenamefont {Aita}, \citenamefont {Giusto}, \citenamefont {Ricossa}, \citenamefont {Corbelletto}, \citenamefont {Montrucchio},\ and\ \citenamefont {Ugoccioni}}]{e25040593}%
  \BibitemOpen
  \bibfield  {author} {\bibinfo {author} {\bibfnamefont {E.}~\bibnamefont {Dri}}, \bibinfo {author} {\bibfnamefont {A.}~\bibnamefont {Aita}}, \bibinfo {author} {\bibfnamefont {E.}~\bibnamefont {Giusto}}, \bibinfo {author} {\bibfnamefont {D.}~\bibnamefont {Ricossa}}, \bibinfo {author} {\bibfnamefont {D.}~\bibnamefont {Corbelletto}}, \bibinfo {author} {\bibfnamefont {B.}~\bibnamefont {Montrucchio}},\ and\ \bibinfo {author} {\bibfnamefont {R.}~\bibnamefont {Ugoccioni}},\ }\bibfield  {title} {{A More General Quantum Credit Risk Analysis Framework},\ }\bibfield  {journal} {\bibinfo  {journal} {Entropy}\ }\textbf {\bibinfo {volume} {25}},\ \href {https://doi.org/10.3390/e25040593} {10.3390/e25040593} (\bibinfo {year} {2023})\BibitemShut {NoStop}%
\bibitem [{\citenamefont {Ciliberto}\ et~al.(2018)\citenamefont {Ciliberto}, \citenamefont {Herbster}, \citenamefont {Ialongo}, \citenamefont {Pontil}, \citenamefont {Rocchetto}, \citenamefont {Severini},\ and\ \citenamefont {Wossnig}}]{ciliberto2018quantum}%
  \BibitemOpen
  \bibfield  {author} {\bibinfo {author} {\bibfnamefont {C.}~\bibnamefont {Ciliberto}}, \bibinfo {author} {\bibfnamefont {M.}~\bibnamefont {Herbster}}, \bibinfo {author} {\bibfnamefont {A.~D.}\ \bibnamefont {Ialongo}}, \bibinfo {author} {\bibfnamefont {M.}~\bibnamefont {Pontil}}, \bibinfo {author} {\bibfnamefont {A.}~\bibnamefont {Rocchetto}}, \bibinfo {author} {\bibfnamefont {S.}~\bibnamefont {Severini}},\ and\ \bibinfo {author} {\bibfnamefont {L.}~\bibnamefont {Wossnig}},\ }\bibfield  {title} {{Quantum machine learning: a classical perspective},\ }\href@noop {} {\bibfield  {journal} {\bibinfo  {journal} {Proceedings of the Royal Society A: Mathematical, Physical and Engineering Sciences}\ }\textbf {\bibinfo {volume} {474}},\ \bibinfo {pages} {20170551} (\bibinfo {year} {2018})}\BibitemShut {NoStop}%
\bibitem [{\citenamefont {Verdon}\ et~al.(2018)\citenamefont {Verdon}, \citenamefont {Pye},\ and\ \citenamefont {Broughton}}]{autoencoder5}%
  \BibitemOpen
  \bibfield  {author} {\bibinfo {author} {\bibfnamefont {G.}~\bibnamefont {Verdon}}, \bibinfo {author} {\bibfnamefont {J.}~\bibnamefont {Pye}},\ and\ \bibinfo {author} {\bibfnamefont {M.}~\bibnamefont {Broughton}},\ }\bibfield  {title} {{A universal training algorithm for quantum deep learning},\ }\href@noop {} {\bibfield  {journal} {\bibinfo  {journal} {arXiv preprint arXiv:1806.09729}\ } (\bibinfo {year} {2018})}\BibitemShut {NoStop}%
\bibitem [{\citenamefont {Romero}\ et~al.(2017)\citenamefont {Romero}, \citenamefont {Olson},\ and\ \citenamefont {Aspuru-Guzik}}]{qae1}%
  \BibitemOpen
  \bibfield  {author} {\bibinfo {author} {\bibfnamefont {J.}~\bibnamefont {Romero}}, \bibinfo {author} {\bibfnamefont {J.~P.}\ \bibnamefont {Olson}},\ and\ \bibinfo {author} {\bibfnamefont {A.}~\bibnamefont {Aspuru-Guzik}},\ }\bibfield  {title} {{Quantum autoencoders for efficient compression of quantum data},\ }\href@noop {} {\bibfield  {journal} {\bibinfo  {journal} {Quantum Science and Technology}\ }\textbf {\bibinfo {volume} {2}},\ \bibinfo {pages} {045001} (\bibinfo {year} {2017})}\BibitemShut {NoStop}%
\bibitem [{\citenamefont {Stokes}\ et~al.(2020)\citenamefont {Stokes}, \citenamefont {Izaac}, \citenamefont {Killoran},\ and\ \citenamefont {Carleo}}]{qgd}%
  \BibitemOpen
  \bibfield  {author} {\bibinfo {author} {\bibfnamefont {J.}~\bibnamefont {Stokes}}, \bibinfo {author} {\bibfnamefont {J.}~\bibnamefont {Izaac}}, \bibinfo {author} {\bibfnamefont {N.}~\bibnamefont {Killoran}},\ and\ \bibinfo {author} {\bibfnamefont {G.}~\bibnamefont {Carleo}},\ }\bibfield  {title} {{Quantum natural gradient},\ }\href@noop {} {\bibfield  {journal} {\bibinfo  {journal} {Quantum}\ }\textbf {\bibinfo {volume} {4}},\ \bibinfo {pages} {269} (\bibinfo {year} {2020})}\BibitemShut {NoStop}%
\bibitem [{\citenamefont {Low}\ et~al.(2019)\citenamefont {Low}, \citenamefont {Kliuchnikov},\ and\ \citenamefont {Wiebe}}]{low2019well}%
  \BibitemOpen
  \bibfield  {author} {\bibinfo {author} {\bibfnamefont {G.~H.}\ \bibnamefont {Low}}, \bibinfo {author} {\bibfnamefont {V.}~\bibnamefont {Kliuchnikov}},\ and\ \bibinfo {author} {\bibfnamefont {N.}~\bibnamefont {Wiebe}},\ }\bibfield  {title} {{Well-conditioned multiproduct Hamiltonian simulation},\ }\href@noop {} {\bibfield  {journal} {\bibinfo  {journal} {arXiv preprint arXiv:1907.11679}\ } (\bibinfo {year} {2019})}\BibitemShut {NoStop}%
\bibitem [{\citenamefont {Childs}\ and\ \citenamefont {Wiebe}(2012)}]{childs2012hamiltonian}%
  \BibitemOpen
  \bibfield  {author} {\bibinfo {author} {\bibfnamefont {A.~M.}\ \bibnamefont {Childs}}\ and\ \bibinfo {author} {\bibfnamefont {N.}~\bibnamefont {Wiebe}},\ }\bibfield  {title} {{Hamiltonian simulation using linear combinations of unitary operations},\ }\href@noop {} {\bibfield  {journal} {\bibinfo  {journal} {arXiv preprint arXiv:1202.5822}\ } (\bibinfo {year} {2012})}\BibitemShut {NoStop}%
\bibitem [{\citenamefont {Yuan}\ et~al.(2019)\citenamefont {Yuan}, \citenamefont {Endo}, \citenamefont {Zhao}, \citenamefont {Li},\ and\ \citenamefont {Benjamin}}]{yuan2019theory}%
  \BibitemOpen
  \bibfield  {author} {\bibinfo {author} {\bibfnamefont {X.}~\bibnamefont {Yuan}}, \bibinfo {author} {\bibfnamefont {S.}~\bibnamefont {Endo}}, \bibinfo {author} {\bibfnamefont {Q.}~\bibnamefont {Zhao}}, \bibinfo {author} {\bibfnamefont {Y.}~\bibnamefont {Li}},\ and\ \bibinfo {author} {\bibfnamefont {S.~C.}\ \bibnamefont {Benjamin}},\ }\bibfield  {title} {{Theory of variational quantum simulation},\ }\href@noop {} {\bibfield  {journal} {\bibinfo  {journal} {Quantum}\ }\textbf {\bibinfo {volume} {3}},\ \bibinfo {pages} {191} (\bibinfo {year} {2019})}\BibitemShut {NoStop}%
\bibitem [{\citenamefont {Yao}\ et~al.(2021)\citenamefont {Yao}, \citenamefont {Gomes}, \citenamefont {Zhang}, \citenamefont {Wang}, \citenamefont {Ho}, \citenamefont {Iadecola},\ and\ \citenamefont {Orth}}]{yao2021adaptive}%
  \BibitemOpen
  \bibfield  {author} {\bibinfo {author} {\bibfnamefont {Y.-X.}\ \bibnamefont {Yao}}, \bibinfo {author} {\bibfnamefont {N.}~\bibnamefont {Gomes}}, \bibinfo {author} {\bibfnamefont {F.}~\bibnamefont {Zhang}}, \bibinfo {author} {\bibfnamefont {C.-Z.}\ \bibnamefont {Wang}}, \bibinfo {author} {\bibfnamefont {K.-M.}\ \bibnamefont {Ho}}, \bibinfo {author} {\bibfnamefont {T.}~\bibnamefont {Iadecola}},\ and\ \bibinfo {author} {\bibfnamefont {P.~P.}\ \bibnamefont {Orth}},\ }\bibfield  {title} {{Adaptive variational quantum dynamics simulations},\ }\href@noop {} {\bibfield  {journal} {\bibinfo  {journal} {PRX Quantum}\ }\textbf {\bibinfo {volume} {2}},\ \bibinfo {pages} {030307} (\bibinfo {year} {2021})}\BibitemShut {NoStop}%
\bibitem [{\citenamefont {Barison}\ et~al.(2021)\citenamefont {Barison}, \citenamefont {Vicentini},\ and\ \citenamefont {Carleo}}]{barison2021efficient}%
  \BibitemOpen
  \bibfield  {author} {\bibinfo {author} {\bibfnamefont {S.}~\bibnamefont {Barison}}, \bibinfo {author} {\bibfnamefont {F.}~\bibnamefont {Vicentini}},\ and\ \bibinfo {author} {\bibfnamefont {G.}~\bibnamefont {Carleo}},\ }\bibfield  {title} {{An efficient quantum algorithm for the time evolution of parameterized circuits},\ }\href@noop {} {\bibfield  {journal} {\bibinfo  {journal} {Quantum}\ }\textbf {\bibinfo {volume} {5}},\ \bibinfo {pages} {512} (\bibinfo {year} {2021})}\BibitemShut {NoStop}%
\bibitem [{\citenamefont {Dunjko}\ and\ \citenamefont {Briegel}(2018)}]{dunjko2018machine}%
  \BibitemOpen
  \bibfield  {author} {\bibinfo {author} {\bibfnamefont {V.}~\bibnamefont {Dunjko}}\ and\ \bibinfo {author} {\bibfnamefont {H.~J.}\ \bibnamefont {Briegel}},\ }\bibfield  {title} {{Machine learning \& artificial intelligence in the quantum domain: a review of recent progress},\ }\href@noop {} {\bibfield  {journal} {\bibinfo  {journal} {Reports on Progress in Physics}\ }\textbf {\bibinfo {volume} {81}},\ \bibinfo {pages} {074001} (\bibinfo {year} {2018})}\BibitemShut {NoStop}%
\bibitem [{\citenamefont {Lloyd}(1996)}]{lloyd1996universal}%
  \BibitemOpen
  \bibfield  {author} {\bibinfo {author} {\bibfnamefont {S.}~\bibnamefont {Lloyd}},\ }\bibfield  {title} {{Universal quantum simulators},\ }\href@noop {} {\bibfield  {journal} {\bibinfo  {journal} {Science}\ }\textbf {\bibinfo {volume} {273}},\ \bibinfo {pages} {1073} (\bibinfo {year} {1996})}\BibitemShut {NoStop}%
\bibitem [{\citenamefont {Abbas}\ et~al.(2021)\citenamefont {Abbas}, \citenamefont {Sutter}, \citenamefont {Zoufal}, \citenamefont {Lucchi}, \citenamefont {Figalli},\ and\ \citenamefont {Woerner}}]{abbas2021power}%
  \BibitemOpen
  \bibfield  {author} {\bibinfo {author} {\bibfnamefont {A.}~\bibnamefont {Abbas}}, \bibinfo {author} {\bibfnamefont {D.}~\bibnamefont {Sutter}}, \bibinfo {author} {\bibfnamefont {C.}~\bibnamefont {Zoufal}}, \bibinfo {author} {\bibfnamefont {A.}~\bibnamefont {Lucchi}}, \bibinfo {author} {\bibfnamefont {A.}~\bibnamefont {Figalli}},\ and\ \bibinfo {author} {\bibfnamefont {S.}~\bibnamefont {Woerner}},\ }\bibfield  {title} {{The power of quantum neural networks},\ }\href@noop {} {\bibfield  {journal} {\bibinfo  {journal} {Nature Computational Science}\ }\textbf {\bibinfo {volume} {1}},\ \bibinfo {pages} {403} (\bibinfo {year} {2021})}\BibitemShut {NoStop}%
\bibitem [{\citenamefont {Zoufal}(2021)}]{zoufal2021generative}%
  \BibitemOpen
  \bibfield  {author} {\bibinfo {author} {\bibfnamefont {C.}~\bibnamefont {Zoufal}},\ }\bibfield  {title} {{Generative Quantum Machine Learning},\ }\href@noop {} {\bibfield  {journal} {\bibinfo  {journal} {arXiv preprint arXiv:2111.12738}\ } (\bibinfo {year} {2021})}\BibitemShut {NoStop}%
\bibitem [{\citenamefont {Bartkiewicz}\ et~al.(2020)\citenamefont {Bartkiewicz}, \citenamefont {Gneiting}, \citenamefont {{\v{C}}ernoch}, \citenamefont {Jir{\'a}kov{\'a}}, \citenamefont {Lemr},\ and\ \citenamefont {Nori}}]{bartkiewicz2020experimental}%
  \BibitemOpen
  \bibfield  {author} {\bibinfo {author} {\bibfnamefont {K.}~\bibnamefont {Bartkiewicz}}, \bibinfo {author} {\bibfnamefont {C.}~\bibnamefont {Gneiting}}, \bibinfo {author} {\bibfnamefont {A.}~\bibnamefont {{\v{C}}ernoch}}, \bibinfo {author} {\bibfnamefont {K.}~\bibnamefont {Jir{\'a}kov{\'a}}}, \bibinfo {author} {\bibfnamefont {K.}~\bibnamefont {Lemr}},\ and\ \bibinfo {author} {\bibfnamefont {F.}~\bibnamefont {Nori}},\ }\bibfield  {title} {{Experimental kernel-based quantum machine learning in finite feature space},\ }\href@noop {} {\bibfield  {journal} {\bibinfo  {journal} {Scientific Reports}\ }\textbf {\bibinfo {volume} {10}},\ \bibinfo {pages} {1} (\bibinfo {year} {2020})}\BibitemShut {NoStop}%
\bibitem [{\citenamefont {Rebentrost}\ et~al.(2014)\citenamefont {Rebentrost}, \citenamefont {Mohseni},\ and\ \citenamefont {Lloyd}}]{rebentrost2014quantum}%
  \BibitemOpen
  \bibfield  {author} {\bibinfo {author} {\bibfnamefont {P.}~\bibnamefont {Rebentrost}}, \bibinfo {author} {\bibfnamefont {M.}~\bibnamefont {Mohseni}},\ and\ \bibinfo {author} {\bibfnamefont {S.}~\bibnamefont {Lloyd}},\ }\bibfield  {title} {{Quantum support vector machine for big data classification},\ }\href@noop {} {\bibfield  {journal} {\bibinfo  {journal} {Physical review letters}\ }\textbf {\bibinfo {volume} {113}},\ \bibinfo {pages} {130503} (\bibinfo {year} {2014})}\BibitemShut {NoStop}%
\bibitem [{\citenamefont {Li}\ et~al.(2015)\citenamefont {Li}, \citenamefont {Liu}, \citenamefont {Xu},\ and\ \citenamefont {Du}}]{li2015experimental}%
  \BibitemOpen
  \bibfield  {author} {\bibinfo {author} {\bibfnamefont {Z.}~\bibnamefont {Li}}, \bibinfo {author} {\bibfnamefont {X.}~\bibnamefont {Liu}}, \bibinfo {author} {\bibfnamefont {N.}~\bibnamefont {Xu}},\ and\ \bibinfo {author} {\bibfnamefont {J.}~\bibnamefont {Du}},\ }\bibfield  {title} {{Experimental realization of a quantum support vector machine},\ }\href@noop {} {\bibfield  {journal} {\bibinfo  {journal} {Physical review letters}\ }\textbf {\bibinfo {volume} {114}},\ \bibinfo {pages} {140504} (\bibinfo {year} {2015})}\BibitemShut {NoStop}%
\bibitem [{\citenamefont {Akhalwaya}\ et~al.(2022{\natexlab{a}})\citenamefont {Akhalwaya}, \citenamefont {Ubaru}, \citenamefont {Clarkson}, \citenamefont {Squillante}, \citenamefont {Jejjala}, \citenamefont {He}, \citenamefont {Naidoo}, \citenamefont {Kalantzis},\ and\ \citenamefont {Horesh}}]{akhalwaya2022exponential}%
  \BibitemOpen
  \bibfield  {author} {\bibinfo {author} {\bibfnamefont {I.~Y.}\ \bibnamefont {Akhalwaya}}, \bibinfo {author} {\bibfnamefont {S.}~\bibnamefont {Ubaru}}, \bibinfo {author} {\bibfnamefont {K.~L.}\ \bibnamefont {Clarkson}}, \bibinfo {author} {\bibfnamefont {M.~S.}\ \bibnamefont {Squillante}}, \bibinfo {author} {\bibfnamefont {V.}~\bibnamefont {Jejjala}}, \bibinfo {author} {\bibfnamefont {Y.-H.}\ \bibnamefont {He}}, \bibinfo {author} {\bibfnamefont {K.}~\bibnamefont {Naidoo}}, \bibinfo {author} {\bibfnamefont {V.}~\bibnamefont {Kalantzis}},\ and\ \bibinfo {author} {\bibfnamefont {L.}~\bibnamefont {Horesh}},\ }\bibfield  {title} {{Exponential advantage on noisy quantum computers},\ }\href@noop {} {\bibfield  {journal} {\bibinfo  {journal} {arXiv preprint arXiv:2209.09371}\ } (\bibinfo {year} {2022}{\natexlab{a}})}\BibitemShut {NoStop}%
\bibitem [{\citenamefont {Saini}\ et~al.(2020)\citenamefont {Saini}, \citenamefont {Khosla}, \citenamefont {Kaur},\ and\ \citenamefont {Singh}}]{saini2020quantum}%
  \BibitemOpen
  \bibfield  {author} {\bibinfo {author} {\bibfnamefont {S.}~\bibnamefont {Saini}}, \bibinfo {author} {\bibfnamefont {P.}~\bibnamefont {Khosla}}, \bibinfo {author} {\bibfnamefont {M.}~\bibnamefont {Kaur}},\ and\ \bibinfo {author} {\bibfnamefont {G.}~\bibnamefont {Singh}},\ }\bibfield  {title} {{Quantum driven machine learning},\ }\href@noop {} {\bibfield  {journal} {\bibinfo  {journal} {International Journal of Theoretical Physics}\ }\textbf {\bibinfo {volume} {59}},\ \bibinfo {pages} {4013} (\bibinfo {year} {2020})}\BibitemShut {NoStop}%
\bibitem [{\citenamefont {Acar}\ and\ \citenamefont {Yilmaz}(2021)}]{acar2021covid}%
  \BibitemOpen
  \bibfield  {author} {\bibinfo {author} {\bibfnamefont {E.}~\bibnamefont {Acar}}\ and\ \bibinfo {author} {\bibfnamefont {I.}~\bibnamefont {Yilmaz}},\ }\bibfield  {title} {{{COVID-19 detection on IBM quantum computer with classical-quantum transferlearning}},\ }\href@noop {} {\bibfield  {journal} {\bibinfo  {journal} {Turkish Journal of Electrical Engineering and Computer Sciences}\ }\textbf {\bibinfo {volume} {29}},\ \bibinfo {pages} {46} (\bibinfo {year} {2021})}\BibitemShut {NoStop}%
\bibitem [{\citenamefont {Hauke}\ et~al.(2020)\citenamefont {Hauke}, \citenamefont {Katzgraber}, \citenamefont {Lechner}, \citenamefont {Nishimori},\ and\ \citenamefont {Oliver}}]{hauke2020perspectives}%
  \BibitemOpen
  \bibfield  {author} {\bibinfo {author} {\bibfnamefont {P.}~\bibnamefont {Hauke}}, \bibinfo {author} {\bibfnamefont {H.~G.}\ \bibnamefont {Katzgraber}}, \bibinfo {author} {\bibfnamefont {W.}~\bibnamefont {Lechner}}, \bibinfo {author} {\bibfnamefont {H.}~\bibnamefont {Nishimori}},\ and\ \bibinfo {author} {\bibfnamefont {W.~D.}\ \bibnamefont {Oliver}},\ }\bibfield  {title} {{Perspectives of quantum annealing: Methods and implementations},\ }\href@noop {} {\bibfield  {journal} {\bibinfo  {journal} {Reports on Progress in Physics}\ }\textbf {\bibinfo {volume} {83}},\ \bibinfo {pages} {054401} (\bibinfo {year} {2020})}\BibitemShut {NoStop}%
\bibitem [{\citenamefont {Das}\ and\ \citenamefont {Chakrabarti}(2008)}]{das2008colloquium}%
  \BibitemOpen
  \bibfield  {author} {\bibinfo {author} {\bibfnamefont {A.}~\bibnamefont {Das}}\ and\ \bibinfo {author} {\bibfnamefont {B.~K.}\ \bibnamefont {Chakrabarti}},\ }\bibfield  {title} {{Colloquium: Quantum annealing and analog quantum computation},\ }\href@noop {} {\bibfield  {journal} {\bibinfo  {journal} {Reviews of Modern Physics}\ }\textbf {\bibinfo {volume} {80}},\ \bibinfo {pages} {1061} (\bibinfo {year} {2008})}\BibitemShut {NoStop}%
\bibitem [{\citenamefont {Finnila}\ et~al.(1994)\citenamefont {Finnila}, \citenamefont {Gomez}, \citenamefont {Sebenik}, \citenamefont {Stenson},\ and\ \citenamefont {Doll}}]{finnila1994quantum}%
  \BibitemOpen
  \bibfield  {author} {\bibinfo {author} {\bibfnamefont {A.~B.}\ \bibnamefont {Finnila}}, \bibinfo {author} {\bibfnamefont {M.~A.}\ \bibnamefont {Gomez}}, \bibinfo {author} {\bibfnamefont {C.}~\bibnamefont {Sebenik}}, \bibinfo {author} {\bibfnamefont {C.}~\bibnamefont {Stenson}},\ and\ \bibinfo {author} {\bibfnamefont {J.~D.}\ \bibnamefont {Doll}},\ }\bibfield  {title} {{Quantum annealing: A new method for minimizing multidimensional functions},\ }\href@noop {} {\bibfield  {journal} {\bibinfo  {journal} {Chemical physics letters}\ }\textbf {\bibinfo {volume} {219}},\ \bibinfo {pages} {343} (\bibinfo {year} {1994})}\BibitemShut {NoStop}%
\bibitem [{\citenamefont {Zoufal}\ et~al.(2019)\citenamefont {Zoufal}, \citenamefont {Lucchi},\ and\ \citenamefont {Woerner}}]{zoufal2019quantum}%
  \BibitemOpen
  \bibfield  {author} {\bibinfo {author} {\bibfnamefont {C.}~\bibnamefont {Zoufal}}, \bibinfo {author} {\bibfnamefont {A.}~\bibnamefont {Lucchi}},\ and\ \bibinfo {author} {\bibfnamefont {S.}~\bibnamefont {Woerner}},\ }\bibfield  {title} {{Quantum generative adversarial networks for learning and loading random distributions},\ }\href@noop {} {\bibfield  {journal} {\bibinfo  {journal} {npj Quantum Information}\ }\textbf {\bibinfo {volume} {5}},\ \bibinfo {pages} {1} (\bibinfo {year} {2019})}\BibitemShut {NoStop}%
\bibitem [{\citenamefont {Wiebe}\ et~al.(2012)\citenamefont {Wiebe}, \citenamefont {Braun},\ and\ \citenamefont {Lloyd}}]{wiebe2012quantum}%
  \BibitemOpen
  \bibfield  {author} {\bibinfo {author} {\bibfnamefont {N.}~\bibnamefont {Wiebe}}, \bibinfo {author} {\bibfnamefont {D.}~\bibnamefont {Braun}},\ and\ \bibinfo {author} {\bibfnamefont {S.}~\bibnamefont {Lloyd}},\ }\bibfield  {title} {{Quantum algorithm for data fitting},\ }\href@noop {} {\bibfield  {journal} {\bibinfo  {journal} {Physical review letters}\ }\textbf {\bibinfo {volume} {109}},\ \bibinfo {pages} {050505} (\bibinfo {year} {2012})}\BibitemShut {NoStop}%
\bibitem [{\citenamefont {Cerezo}\ et~al.(2021{\natexlab{a}})\citenamefont {Cerezo}, \citenamefont {Arrasmith}, \citenamefont {Babbush}, \citenamefont {Benjamin}, \citenamefont {Endo}, \citenamefont {Fujii}, \citenamefont {McClean}, \citenamefont {Mitarai}, \citenamefont {Yuan}, \citenamefont {Cincio} et~al.}]{cerezo2021variational}%
  \BibitemOpen
  \bibfield  {author} {\bibinfo {author} {\bibfnamefont {M.}~\bibnamefont {Cerezo}}, \bibinfo {author} {\bibfnamefont {A.}~\bibnamefont {Arrasmith}}, \bibinfo {author} {\bibfnamefont {R.}~\bibnamefont {Babbush}}, \bibinfo {author} {\bibfnamefont {S.~C.}\ \bibnamefont {Benjamin}}, \bibinfo {author} {\bibfnamefont {S.}~\bibnamefont {Endo}}, \bibinfo {author} {\bibfnamefont {K.}~\bibnamefont {Fujii}}, \bibinfo {author} {\bibfnamefont {J.~R.}\ \bibnamefont {McClean}}, \bibinfo {author} {\bibfnamefont {K.}~\bibnamefont {Mitarai}}, \bibinfo {author} {\bibfnamefont {X.}~\bibnamefont {Yuan}}, \bibinfo {author} {\bibfnamefont {L.}~\bibnamefont {Cincio}}, et~al.,\ }\bibfield  {title} {{Variational quantum algorithms},\ }\href@noop {} {\bibfield  {journal} {\bibinfo  {journal} {Nature Reviews Physics}\ }\textbf {\bibinfo {volume} {3}},\ \bibinfo {pages} {625} (\bibinfo {year} {2021}{\natexlab{a}})}\BibitemShut {NoStop}%
\bibitem [{\citenamefont {Beer}\ et~al.(2020)\citenamefont {Beer}, \citenamefont {Bondarenko}, \citenamefont {Farrelly}, \citenamefont {Osborne}, \citenamefont {Salzmann}, \citenamefont {Scheiermann},\ and\ \citenamefont {Wolf}}]{beer2020training}%
  \BibitemOpen
  \bibfield  {author} {\bibinfo {author} {\bibfnamefont {K.}~\bibnamefont {Beer}}, \bibinfo {author} {\bibfnamefont {D.}~\bibnamefont {Bondarenko}}, \bibinfo {author} {\bibfnamefont {T.}~\bibnamefont {Farrelly}}, \bibinfo {author} {\bibfnamefont {T.~J.}\ \bibnamefont {Osborne}}, \bibinfo {author} {\bibfnamefont {R.}~\bibnamefont {Salzmann}}, \bibinfo {author} {\bibfnamefont {D.}~\bibnamefont {Scheiermann}},\ and\ \bibinfo {author} {\bibfnamefont {R.}~\bibnamefont {Wolf}},\ }\bibfield  {title} {{Training deep quantum neural networks},\ }\href@noop {} {\bibfield  {journal} {\bibinfo  {journal} {Nature communications}\ }\textbf {\bibinfo {volume} {11}},\ \bibinfo {pages} {1} (\bibinfo {year} {2020})}\BibitemShut {NoStop}%
\bibitem [{\citenamefont {Schuld}\ et~al.(2017)\citenamefont {Schuld}, \citenamefont {Fingerhuth},\ and\ \citenamefont {Petruccione}}]{schuld2017implementing}%
  \BibitemOpen
  \bibfield  {author} {\bibinfo {author} {\bibfnamefont {M.}~\bibnamefont {Schuld}}, \bibinfo {author} {\bibfnamefont {M.}~\bibnamefont {Fingerhuth}},\ and\ \bibinfo {author} {\bibfnamefont {F.}~\bibnamefont {Petruccione}},\ }\bibfield  {title} {{Implementing a distance-based classifier with a quantum interference circuit},\ }\href@noop {} {\bibfield  {journal} {\bibinfo  {journal} {EPL (Europhysics Letters)}\ }\textbf {\bibinfo {volume} {119}},\ \bibinfo {pages} {60002} (\bibinfo {year} {2017})}\BibitemShut {NoStop}%
\bibitem [{\citenamefont {Killoran}\ et~al.(2019)\citenamefont {Killoran}, \citenamefont {Bromley}, \citenamefont {Arrazola}, \citenamefont {Schuld}, \citenamefont {Quesada},\ and\ \citenamefont {Lloyd}}]{killoran2019continuous}%
  \BibitemOpen
  \bibfield  {author} {\bibinfo {author} {\bibfnamefont {N.}~\bibnamefont {Killoran}}, \bibinfo {author} {\bibfnamefont {T.~R.}\ \bibnamefont {Bromley}}, \bibinfo {author} {\bibfnamefont {J.~M.}\ \bibnamefont {Arrazola}}, \bibinfo {author} {\bibfnamefont {M.}~\bibnamefont {Schuld}}, \bibinfo {author} {\bibfnamefont {N.}~\bibnamefont {Quesada}},\ and\ \bibinfo {author} {\bibfnamefont {S.}~\bibnamefont {Lloyd}},\ }\bibfield  {title} {{Continuous-variable quantum neural networks},\ }\href@noop {} {\bibfield  {journal} {\bibinfo  {journal} {Physical Review Research}\ }\textbf {\bibinfo {volume} {1}},\ \bibinfo {pages} {033063} (\bibinfo {year} {2019})}\BibitemShut {NoStop}%
\bibitem [{\citenamefont {Yano}\ et~al.(2021)\citenamefont {Yano}, \citenamefont {Suzuki}, \citenamefont {Itoh}, \citenamefont {Raymond},\ and\ \citenamefont {Yamamoto}}]{yano2020efficient}%
  \BibitemOpen
  \bibfield  {author} {\bibinfo {author} {\bibfnamefont {H.}~\bibnamefont {Yano}}, \bibinfo {author} {\bibfnamefont {Y.}~\bibnamefont {Suzuki}}, \bibinfo {author} {\bibfnamefont {K.~M.}\ \bibnamefont {Itoh}}, \bibinfo {author} {\bibfnamefont {R.}~\bibnamefont {Raymond}},\ and\ \bibinfo {author} {\bibfnamefont {N.}~\bibnamefont {Yamamoto}},\ }\bibfield  {title} {{Efficient Discrete Feature Encoding for Variational Quantum Classifier},\ }\href {https://doi.org/10.1109/TQE.2021.3103050} {\bibfield  {journal} {\bibinfo  {journal} {IEEE Transactions on Quantum Engineering}\ }\textbf {\bibinfo {volume} {2}},\ \bibinfo {pages} {1} (\bibinfo {year} {2021})}\BibitemShut {NoStop}%
\bibitem [{\citenamefont {Plesch}\ and\ \citenamefont {Brukner}(2011)}]{plesch2011quantum}%
  \BibitemOpen
  \bibfield  {author} {\bibinfo {author} {\bibfnamefont {M.}~\bibnamefont {Plesch}}\ and\ \bibinfo {author} {\bibfnamefont {{\v{C}}.}~\bibnamefont {Brukner}},\ }\bibfield  {title} {{Quantum-state preparation with universal gate decompositions},\ }\href@noop {} {\bibfield  {journal} {\bibinfo  {journal} {Physical Review A}\ }\textbf {\bibinfo {volume} {83}},\ \bibinfo {pages} {032302} (\bibinfo {year} {2011})}\BibitemShut {NoStop}%
\bibitem [{\citenamefont {Sanders}\ et~al.(2019)\citenamefont {Sanders}, \citenamefont {Low}, \citenamefont {Scherer},\ and\ \citenamefont {Berry}}]{sanders2019black}%
  \BibitemOpen
  \bibfield  {author} {\bibinfo {author} {\bibfnamefont {Y.~R.}\ \bibnamefont {Sanders}}, \bibinfo {author} {\bibfnamefont {G.~H.}\ \bibnamefont {Low}}, \bibinfo {author} {\bibfnamefont {A.}~\bibnamefont {Scherer}},\ and\ \bibinfo {author} {\bibfnamefont {D.~W.}\ \bibnamefont {Berry}},\ }\bibfield  {title} {{Black-box quantum state preparation without arithmetic},\ }\href@noop {} {\bibfield  {journal} {\bibinfo  {journal} {Physical review letters}\ }\textbf {\bibinfo {volume} {122}},\ \bibinfo {pages} {020502} (\bibinfo {year} {2019})}\BibitemShut {NoStop}%
\bibitem [{\citenamefont {Shende}\ et~al.(2005)\citenamefont {Shende}, \citenamefont {Bullock},\ and\ \citenamefont {Markov}}]{shende2005synthesis}%
  \BibitemOpen
  \bibfield  {author} {\bibinfo {author} {\bibfnamefont {V.~V.}\ \bibnamefont {Shende}}, \bibinfo {author} {\bibfnamefont {S.~S.}\ \bibnamefont {Bullock}},\ and\ \bibinfo {author} {\bibfnamefont {I.~L.}\ \bibnamefont {Markov}},\ }in\ \href@noop {} {\bibinfo {booktitle} {Proceedings of the 2005 Asia and South Pacific Design Automation Conference}}\ (\bibinfo {year} {2005})\ pp.\ \bibinfo {pages} {272--275}\BibitemShut {NoStop}%
\bibitem [{\citenamefont {Grover}(1996)}]{grover1996fast}%
  \BibitemOpen
  \bibfield  {author} {\bibinfo {author} {\bibfnamefont {L.~K.}\ \bibnamefont {Grover}},\ }in\ \href@noop {} {\bibinfo {booktitle} {Proceedings of the twenty-eighth annual ACM symposium on Theory of computing}}\ (\bibinfo {year} {1996})\ pp.\ \bibinfo {pages} {212--219}\BibitemShut {NoStop}%
\bibitem [{\citenamefont {Grover}(2000)}]{grover2000synthesis}%
  \BibitemOpen
  \bibfield  {author} {\bibinfo {author} {\bibfnamefont {L.~K.}\ \bibnamefont {Grover}},\ }\bibfield  {title} {{Synthesis of quantum superpositions by quantum computation},\ }\href@noop {} {\bibfield  {journal} {\bibinfo  {journal} {Physical review letters}\ }\textbf {\bibinfo {volume} {85}},\ \bibinfo {pages} {1334} (\bibinfo {year} {2000})}\BibitemShut {NoStop}%
\bibitem [{\citenamefont {Vazquez}\ and\ \citenamefont {Woerner}(2021)}]{vazquez2021efficient}%
  \BibitemOpen
  \bibfield  {author} {\bibinfo {author} {\bibfnamefont {A.~C.}\ \bibnamefont {Vazquez}}\ and\ \bibinfo {author} {\bibfnamefont {S.}~\bibnamefont {Woerner}},\ }\bibfield  {title} {{Efficient state preparation for quantum amplitude estimation},\ }\href@noop {} {\bibfield  {journal} {\bibinfo  {journal} {Physical Review Applied}\ }\textbf {\bibinfo {volume} {15}},\ \bibinfo {pages} {034027} (\bibinfo {year} {2021})}\BibitemShut {NoStop}%
\bibitem [{\citenamefont {Farhi}\ and\ \citenamefont {Neven}(2018)}]{basisencoding1}%
  \BibitemOpen
  \bibfield  {author} {\bibinfo {author} {\bibfnamefont {E.}~\bibnamefont {Farhi}}\ and\ \bibinfo {author} {\bibfnamefont {H.}~\bibnamefont {Neven}},\ }\bibfield  {title} {{Classification with quantum neural networks on near term processors},\ }\href@noop {} {\bibfield  {journal} {\bibinfo  {journal} {arXiv preprint arXiv:1802.06002}\ } (\bibinfo {year} {2018})}\BibitemShut {NoStop}%
\bibitem [{\citenamefont {Schuld}\ and\ \citenamefont {Petruccione}(2018{\natexlab{b}})}]{basisencoding2}%
  \BibitemOpen
  \bibfield  {author} {\bibinfo {author} {\bibfnamefont {M.}~\bibnamefont {Schuld}}\ and\ \bibinfo {author} {\bibfnamefont {F.}~\bibnamefont {Petruccione}},\ }\href@noop {} {\bibinfo {title} {Supervised learning with quantum computers}},\ Vol.~\bibinfo {volume} {17}\ (\bibinfo  {publisher} {Springer},\ \bibinfo {year} {2018})\BibitemShut {NoStop}%
\bibitem [{\citenamefont {Cao}\ and\ \citenamefont {Wang}(2021)}]{autoencoder3}%
  \BibitemOpen
  \bibfield  {author} {\bibinfo {author} {\bibfnamefont {C.}~\bibnamefont {Cao}}\ and\ \bibinfo {author} {\bibfnamefont {X.}~\bibnamefont {Wang}},\ }\bibfield  {title} {{Noise-assisted quantum autoencoder},\ }\href@noop {} {\bibfield  {journal} {\bibinfo  {journal} {Physical Review Applied}\ }\textbf {\bibinfo {volume} {15}},\ \bibinfo {pages} {054012} (\bibinfo {year} {2021})}\BibitemShut {NoStop}%
\bibitem [{\citenamefont {Pepper}\ et~al.(2019)\citenamefont {Pepper}, \citenamefont {Tischler},\ and\ \citenamefont {Pryde}}]{autoencoder2}%
  \BibitemOpen
  \bibfield  {author} {\bibinfo {author} {\bibfnamefont {A.}~\bibnamefont {Pepper}}, \bibinfo {author} {\bibfnamefont {N.}~\bibnamefont {Tischler}},\ and\ \bibinfo {author} {\bibfnamefont {G.~J.}\ \bibnamefont {Pryde}},\ }\bibfield  {title} {{Experimental realization of a quantum autoencoder: The compression of qutrits via machine learning},\ }\href@noop {} {\bibfield  {journal} {\bibinfo  {journal} {Physical review letters}\ }\textbf {\bibinfo {volume} {122}},\ \bibinfo {pages} {060501} (\bibinfo {year} {2019})}\BibitemShut {NoStop}%
\bibitem [{\citenamefont {De~Favereau}\ et~al.(2014)\citenamefont {De~Favereau}, \citenamefont {Delaere}, \citenamefont {Demin}, \citenamefont {Giammanco}, \citenamefont {Lemaitre}, \citenamefont {Mertens},\ and\ \citenamefont {Selvaggi}}]{delphes}%
  \BibitemOpen
  \bibfield  {author} {\bibinfo {author} {\bibfnamefont {J.}~\bibnamefont {De~Favereau}}, \bibinfo {author} {\bibfnamefont {C.}~\bibnamefont {Delaere}}, \bibinfo {author} {\bibfnamefont {P.}~\bibnamefont {Demin}}, \bibinfo {author} {\bibfnamefont {A.}~\bibnamefont {Giammanco}}, \bibinfo {author} {\bibfnamefont {V.}~\bibnamefont {Lemaitre}}, \bibinfo {author} {\bibfnamefont {A.}~\bibnamefont {Mertens}},\ and\ \bibinfo {author} {\bibfnamefont {M.}~\bibnamefont {Selvaggi}},\ }\bibfield  {title} {{DELPHES 3: a modular framework for fast simulation of a generic collider experiment},\ }\href@noop {} {\bibfield  {journal} {\bibinfo  {journal} {Journal of High Energy Physics}\ }\textbf {\bibinfo {volume} {2014}},\ \bibinfo {pages} {1} (\bibinfo {year} {2014})}\BibitemShut {NoStop}%
\bibitem [{\citenamefont {Chen}\ and\ \citenamefont {Guestrin}(2016)}]{xgboost}%
  \BibitemOpen
  \bibfield  {author} {\bibinfo {author} {\bibfnamefont {T.}~\bibnamefont {Chen}}\ and\ \bibinfo {author} {\bibfnamefont {C.}~\bibnamefont {Guestrin}},\ }in\ \href@noop {} {\bibinfo {booktitle} {Proceedings of the 22nd acm sigkdd international conference on knowledge discovery and data mining}}\ (\bibinfo {year} {2016})\ pp.\ \bibinfo {pages} {785--794}\BibitemShut {NoStop}%
\bibitem [{\citenamefont {Boser}\ et~al.(1992)\citenamefont {Boser}, \citenamefont {Guyon},\ and\ \citenamefont {Vapnik}}]{svm1}%
  \BibitemOpen
  \bibfield  {author} {\bibinfo {author} {\bibfnamefont {B.~E.}\ \bibnamefont {Boser}}, \bibinfo {author} {\bibfnamefont {I.~M.}\ \bibnamefont {Guyon}},\ and\ \bibinfo {author} {\bibfnamefont {V.~N.}\ \bibnamefont {Vapnik}},\ }in\ \href@noop {} {\bibinfo {booktitle} {Proceedings of the fifth annual workshop on Computational learning theory}}\ (\bibinfo {year} {1992})\ pp.\ \bibinfo {pages} {144--152}\BibitemShut {NoStop}%
\bibitem [{\citenamefont {Park}\ et~al.(2020)\citenamefont {Park}, \citenamefont {Blank},\ and\ \citenamefont {Petruccione}}]{park2020theory}%
  \BibitemOpen
  \bibfield  {author} {\bibinfo {author} {\bibfnamefont {D.~K.}\ \bibnamefont {Park}}, \bibinfo {author} {\bibfnamefont {C.}~\bibnamefont {Blank}},\ and\ \bibinfo {author} {\bibfnamefont {F.}~\bibnamefont {Petruccione}},\ }\bibfield  {title} {{The theory of the quantum kernel-based binary classifier},\ }\href@noop {} {\bibfield  {journal} {\bibinfo  {journal} {Physics Letters A}\ }\textbf {\bibinfo {volume} {384}},\ \bibinfo {pages} {126422} (\bibinfo {year} {2020})}\BibitemShut {NoStop}%
\bibitem [{\citenamefont {Byrd}\ et~al.(1995)\citenamefont {Byrd}, \citenamefont {Lu}, \citenamefont {Nocedal},\ and\ \citenamefont {Zhu}}]{byrd1995limited}%
  \BibitemOpen
  \bibfield  {author} {\bibinfo {author} {\bibfnamefont {R.~H.}\ \bibnamefont {Byrd}}, \bibinfo {author} {\bibfnamefont {P.}~\bibnamefont {Lu}}, \bibinfo {author} {\bibfnamefont {J.}~\bibnamefont {Nocedal}},\ and\ \bibinfo {author} {\bibfnamefont {C.}~\bibnamefont {Zhu}},\ }\bibfield  {title} {{A limited memory algorithm for bound constrained optimization},\ }\href@noop {} {\bibfield  {journal} {\bibinfo  {journal} {SIAM Journal on scientific computing}\ }\textbf {\bibinfo {volume} {16}},\ \bibinfo {pages} {1190} (\bibinfo {year} {1995})}\BibitemShut {NoStop}%
\bibitem [{\citenamefont {Kolodrubetz}\ et~al.(2017)\citenamefont {Kolodrubetz}, \citenamefont {Sels}, \citenamefont {Mehta},\ and\ \citenamefont {Polkovnikov}}]{kolodrubetz2017geometry}%
  \BibitemOpen
  \bibfield  {author} {\bibinfo {author} {\bibfnamefont {M.}~\bibnamefont {Kolodrubetz}}, \bibinfo {author} {\bibfnamefont {D.}~\bibnamefont {Sels}}, \bibinfo {author} {\bibfnamefont {P.}~\bibnamefont {Mehta}},\ and\ \bibinfo {author} {\bibfnamefont {A.}~\bibnamefont {Polkovnikov}},\ }\bibfield  {title} {{Geometry and non-adiabatic response in quantum and classical systems},\ }\href@noop {} {\bibfield  {journal} {\bibinfo  {journal} {Physics Reports}\ }\textbf {\bibinfo {volume} {697}},\ \bibinfo {pages} {1} (\bibinfo {year} {2017})}\BibitemShut {NoStop}%
\bibitem [{\citenamefont {Spall}(1998)}]{spall1998overview}%
  \BibitemOpen
  \bibfield  {author} {\bibinfo {author} {\bibfnamefont {J.~C.}\ \bibnamefont {Spall}},\ }\bibfield  {title} {{An overview of the simultaneous perturbation method for efficient optimization},\ }\href@noop {} {\bibfield  {journal} {\bibinfo  {journal} {Johns Hopkins apl technical digest}\ }\textbf {\bibinfo {volume} {19}},\ \bibinfo {pages} {482} (\bibinfo {year} {1998})}\BibitemShut {NoStop}%
\bibitem [{\citenamefont {Hansen}(2006)}]{hansen2006cma}%
  \BibitemOpen
  \bibfield  {author} {\bibinfo {author} {\bibfnamefont {N.}~\bibnamefont {Hansen}},\ }\bibfield  {title} {{The CMA evolution strategy: a comparing review},\ }\href@noop {} {\bibfield  {journal} {\bibinfo  {journal} {Towards a new evolutionary computation}\ ,\ \bibinfo {pages} {75}} (\bibinfo {year} {2006})}\BibitemShut {NoStop}%
\bibitem [{\citenamefont {Spall}(2000)}]{spall2}%
  \BibitemOpen
  \bibfield  {author} {\bibinfo {author} {\bibfnamefont {J.~C.}\ \bibnamefont {Spall}},\ }\bibfield  {title} {{Adaptive stochastic approximation by the simultaneous perturbation method},\ }\href@noop {} {\bibfield  {journal} {\bibinfo  {journal} {IEEE transactions on automatic control}\ }\textbf {\bibinfo {volume} {45}},\ \bibinfo {pages} {1839} (\bibinfo {year} {2000})}\BibitemShut {NoStop}%
\bibitem [{\citenamefont {Spall}(1997)}]{spall1}%
  \BibitemOpen
  \bibfield  {author} {\bibinfo {author} {\bibfnamefont {J.~C.}\ \bibnamefont {Spall}},\ }\bibfield  {title} {{A one-measurement form of simultaneous perturbation stochastic approximation},\ }\href@noop {} {\bibfield  {journal} {\bibinfo  {journal} {Automatica}\ }\textbf {\bibinfo {volume} {33}},\ \bibinfo {pages} {109} (\bibinfo {year} {1997})}\BibitemShut {NoStop}%
\bibitem [{\citenamefont {Sj{\"o}strand}\ et~al.(2006)\citenamefont {Sj{\"o}strand}, \citenamefont {Mrenna},\ and\ \citenamefont {Skands}}]{pythia}%
  \BibitemOpen
  \bibfield  {author} {\bibinfo {author} {\bibfnamefont {T.}~\bibnamefont {Sj{\"o}strand}}, \bibinfo {author} {\bibfnamefont {S.}~\bibnamefont {Mrenna}},\ and\ \bibinfo {author} {\bibfnamefont {P.}~\bibnamefont {Skands}},\ }\bibfield  {title} {{PYTHIA 6.4 physics and manual},\ }\href@noop {} {\bibfield  {journal} {\bibinfo  {journal} {Journal of High Energy Physics}\ }\textbf {\bibinfo {volume} {2006}},\ \bibinfo {pages} {026} (\bibinfo {year} {2006})}\BibitemShut {NoStop}%
\bibitem [{\citenamefont {Sharma}\ et~al.(2021)\citenamefont {Sharma}, \citenamefont {Gupta}, \citenamefont {Mehta},\ and\ \citenamefont {Lad}}]{sharma2021quantum}%
  \BibitemOpen
  \bibfield  {author} {\bibinfo {author} {\bibfnamefont {V.}~\bibnamefont {Sharma}}, \bibinfo {author} {\bibfnamefont {S.}~\bibnamefont {Gupta}}, \bibinfo {author} {\bibfnamefont {G.}~\bibnamefont {Mehta}},\ and\ \bibinfo {author} {\bibfnamefont {B.~K.}\ \bibnamefont {Lad}},\ }\bibfield  {title} {{A quantum-based diagnostics approach for additive manufacturing machine},\ }\href@noop {} {\bibfield  {journal} {\bibinfo  {journal} {IET Collaborative Intelligent Manufacturing}\ }\textbf {\bibinfo {volume} {3}},\ \bibinfo {pages} {184} (\bibinfo {year} {2021})}\BibitemShut {NoStop}%
\bibitem [{\citenamefont {Fawcett}(2006)}]{fawcett2006pattern}%
  \BibitemOpen
  \bibfield  {author} {\bibinfo {author} {\bibfnamefont {T.}~\bibnamefont {Fawcett}},\ }\href@noop {} {{Pattern Recognition Letters. Vol. 27}} (\bibinfo {year} {2006})\BibitemShut {NoStop}%
\bibitem [{\citenamefont {Nakaji}\ and\ \citenamefont {Yamamoto}(2021)}]{nakaji2021expressibility}%
  \BibitemOpen
  \bibfield  {author} {\bibinfo {author} {\bibfnamefont {K.}~\bibnamefont {Nakaji}}\ and\ \bibinfo {author} {\bibfnamefont {N.}~\bibnamefont {Yamamoto}},\ }\bibfield  {title} {{Expressibility of the alternating layered ansatz for quantum computation},\ }\href@noop {} {\bibfield  {journal} {\bibinfo  {journal} {Quantum}\ }\textbf {\bibinfo {volume} {5}},\ \bibinfo {pages} {434} (\bibinfo {year} {2021})}\BibitemShut {NoStop}%
\bibitem [{\citenamefont {Hand}\ and\ \citenamefont {Till}(2001)}]{hand2001simple}%
  \BibitemOpen
  \bibfield  {author} {\bibinfo {author} {\bibfnamefont {D.~J.}\ \bibnamefont {Hand}}\ and\ \bibinfo {author} {\bibfnamefont {R.~J.}\ \bibnamefont {Till}},\ }\bibfield  {title} {{{A simple generalisation of the area under the ROC curve for multiple class classification problems}},\ }\href@noop {} {\bibfield  {journal} {\bibinfo  {journal} {Machine learning}\ }\textbf {\bibinfo {volume} {45}},\ \bibinfo {pages} {171} (\bibinfo {year} {2001})}\BibitemShut {NoStop}%
\bibitem [{\citenamefont {Haug}\ et~al.(2021)\citenamefont {Haug}, \citenamefont {Self},\ and\ \citenamefont {Kim}}]{blockencoding2}%
  \BibitemOpen
  \bibfield  {author} {\bibinfo {author} {\bibfnamefont {T.}~\bibnamefont {Haug}}, \bibinfo {author} {\bibfnamefont {C.~N.}\ \bibnamefont {Self}},\ and\ \bibinfo {author} {\bibfnamefont {M.}~\bibnamefont {Kim}},\ }\bibfield  {title} {{Large-scale quantum machine learning},\ }\href@noop {} {\bibfield  {journal} {\bibinfo  {journal} {arXiv preprint arXiv:2108.01039}\ } (\bibinfo {year} {2021})}\BibitemShut {NoStop}%
\bibitem [{\citenamefont {Caro}\ et~al.(2021)\citenamefont {Caro}, \citenamefont {Gil-Fuster}, \citenamefont {Meyer}, \citenamefont {Eisert},\ and\ \citenamefont {Sweke}}]{blockencoding1}%
  \BibitemOpen
  \bibfield  {author} {\bibinfo {author} {\bibfnamefont {M.~C.}\ \bibnamefont {Caro}}, \bibinfo {author} {\bibfnamefont {E.}~\bibnamefont {Gil-Fuster}}, \bibinfo {author} {\bibfnamefont {J.~J.}\ \bibnamefont {Meyer}}, \bibinfo {author} {\bibfnamefont {J.}~\bibnamefont {Eisert}},\ and\ \bibinfo {author} {\bibfnamefont {R.}~\bibnamefont {Sweke}},\ }\bibfield  {title} {{Encoding-dependent generalization bounds for parametrized quantum circuits},\ }\href@noop {} {\bibfield  {journal} {\bibinfo  {journal} {Quantum}\ }\textbf {\bibinfo {volume} {5}},\ \bibinfo {pages} {582} (\bibinfo {year} {2021})}\BibitemShut {NoStop}%
\bibitem [{\citenamefont {Ren}\ et~al.(2022)\citenamefont {Ren}, \citenamefont {Li}, \citenamefont {Xu}, \citenamefont {Wang}, \citenamefont {Jiang}, \citenamefont {Jin}, \citenamefont {Zhu}, \citenamefont {Chen}, \citenamefont {Song}, \citenamefont {Zhang} et~al.}]{ren2022experimental}%
  \BibitemOpen
  \bibfield  {author} {\bibinfo {author} {\bibfnamefont {W.}~\bibnamefont {Ren}}, \bibinfo {author} {\bibfnamefont {W.}~\bibnamefont {Li}}, \bibinfo {author} {\bibfnamefont {S.}~\bibnamefont {Xu}}, \bibinfo {author} {\bibfnamefont {K.}~\bibnamefont {Wang}}, \bibinfo {author} {\bibfnamefont {W.}~\bibnamefont {Jiang}}, \bibinfo {author} {\bibfnamefont {F.}~\bibnamefont {Jin}}, \bibinfo {author} {\bibfnamefont {X.}~\bibnamefont {Zhu}}, \bibinfo {author} {\bibfnamefont {J.}~\bibnamefont {Chen}}, \bibinfo {author} {\bibfnamefont {Z.}~\bibnamefont {Song}}, \bibinfo {author} {\bibfnamefont {P.}~\bibnamefont {Zhang}}, et~al.,\ }\bibfield  {title} {{Experimental quantum adversarial learning with programmable superconducting qubits},\ }\href@noop {} {\bibfield  {journal} {\bibinfo  {journal} {Nature Computational Science}\ }\textbf {\bibinfo {volume} {2}},\ \bibinfo {pages} {711} (\bibinfo {year} {2022})}\BibitemShut {NoStop}%
\bibitem [{\citenamefont {Kermany}\ et~al.(2018)\citenamefont {Kermany}, \citenamefont {Goldbaum}, \citenamefont {Cai}, \citenamefont {Valentim}, \citenamefont {Liang}, \citenamefont {Baxter}, \citenamefont {McKeown}, \citenamefont {Yang}, \citenamefont {Wu}, \citenamefont {Yan} et~al.}]{PNEUMONIA_DS}%
  \BibitemOpen
  \bibfield  {author} {\bibinfo {author} {\bibfnamefont {D.~S.}\ \bibnamefont {Kermany}}, \bibinfo {author} {\bibfnamefont {M.}~\bibnamefont {Goldbaum}}, \bibinfo {author} {\bibfnamefont {W.}~\bibnamefont {Cai}}, \bibinfo {author} {\bibfnamefont {C.~C.}\ \bibnamefont {Valentim}}, \bibinfo {author} {\bibfnamefont {H.}~\bibnamefont {Liang}}, \bibinfo {author} {\bibfnamefont {S.~L.}\ \bibnamefont {Baxter}}, \bibinfo {author} {\bibfnamefont {A.}~\bibnamefont {McKeown}}, \bibinfo {author} {\bibfnamefont {G.}~\bibnamefont {Yang}}, \bibinfo {author} {\bibfnamefont {X.}~\bibnamefont {Wu}}, \bibinfo {author} {\bibfnamefont {F.}~\bibnamefont {Yan}}, et~al.,\ }\bibfield  {title} {{Identifying medical diagnoses and treatable diseases by image-based deep learning},\ }\href@noop {} {\bibfield  {journal} {\bibinfo  {journal} {Cell}\ }\textbf {\bibinfo {volume} {172}},\ \bibinfo {pages} {1122} (\bibinfo {year} {2018})}\BibitemShut {NoStop}%
\bibitem [{\citenamefont {Mathur}\ et~al.(2021)\citenamefont {Mathur}, \citenamefont {Landman}, \citenamefont {Li}, \citenamefont {Strahm}, \citenamefont {Kazdaghli}, \citenamefont {Prakash},\ and\ \citenamefont {Kerenidis}}]{mathur2021medical}%
  \BibitemOpen
  \bibfield  {author} {\bibinfo {author} {\bibfnamefont {N.}~\bibnamefont {Mathur}}, \bibinfo {author} {\bibfnamefont {J.}~\bibnamefont {Landman}}, \bibinfo {author} {\bibfnamefont {Y.~Y.}\ \bibnamefont {Li}}, \bibinfo {author} {\bibfnamefont {M.}~\bibnamefont {Strahm}}, \bibinfo {author} {\bibfnamefont {S.}~\bibnamefont {Kazdaghli}}, \bibinfo {author} {\bibfnamefont {A.}~\bibnamefont {Prakash}},\ and\ \bibinfo {author} {\bibfnamefont {I.}~\bibnamefont {Kerenidis}},\ }\bibfield  {title} {{Medical image classification via quantum neural networks},\ }\href@noop {} {\bibfield  {journal} {\bibinfo  {journal} {arXiv preprint arXiv:2109.01831}\ } (\bibinfo {year} {2021})}\BibitemShut {NoStop}%
\bibitem [{\citenamefont {Lu}\ et~al.(2020)\citenamefont {Lu}, \citenamefont {Duan},\ and\ \citenamefont {Deng}}]{lu2020quantum}%
  \BibitemOpen
  \bibfield  {author} {\bibinfo {author} {\bibfnamefont {S.}~\bibnamefont {Lu}}, \bibinfo {author} {\bibfnamefont {L.-M.}\ \bibnamefont {Duan}},\ and\ \bibinfo {author} {\bibfnamefont {D.-L.}\ \bibnamefont {Deng}},\ }\bibfield  {title} {{Quantum adversarial machine learning},\ }\href@noop {} {\bibfield  {journal} {\bibinfo  {journal} {Physical Review Research}\ }\textbf {\bibinfo {volume} {2}},\ \bibinfo {pages} {033212} (\bibinfo {year} {2020})}\BibitemShut {NoStop}%
\bibitem [{\citenamefont {Li}\ et~al.(2019{\natexlab{b}})\citenamefont {Li}, \citenamefont {Jia}, \citenamefont {Wen}, \citenamefont {Liu},\ and\ \citenamefont {Tao}}]{li2019orthogonal}%
  \BibitemOpen
  \bibfield  {author} {\bibinfo {author} {\bibfnamefont {S.}~\bibnamefont {Li}}, \bibinfo {author} {\bibfnamefont {K.}~\bibnamefont {Jia}}, \bibinfo {author} {\bibfnamefont {Y.}~\bibnamefont {Wen}}, \bibinfo {author} {\bibfnamefont {T.}~\bibnamefont {Liu}},\ and\ \bibinfo {author} {\bibfnamefont {D.}~\bibnamefont {Tao}},\ }\bibfield  {title} {{Orthogonal deep neural networks},\ }\href@noop {} {\bibfield  {journal} {\bibinfo  {journal} {IEEE transactions on pattern analysis and machine intelligence}\ }\textbf {\bibinfo {volume} {43}},\ \bibinfo {pages} {1352} (\bibinfo {year} {2019}{\natexlab{b}})}\BibitemShut {NoStop}%
\bibitem [{\citenamefont {Zhang}\ et~al.(2020)\citenamefont {Zhang}, \citenamefont {Hsieh}, \citenamefont {Liu},\ and\ \citenamefont {Tao}}]{zhang2020toward}%
  \BibitemOpen
  \bibfield  {author} {\bibinfo {author} {\bibfnamefont {K.}~\bibnamefont {Zhang}}, \bibinfo {author} {\bibfnamefont {M.-H.}\ \bibnamefont {Hsieh}}, \bibinfo {author} {\bibfnamefont {L.}~\bibnamefont {Liu}},\ and\ \bibinfo {author} {\bibfnamefont {D.}~\bibnamefont {Tao}},\ }\bibfield  {title} {{Toward trainability of quantum neural networks},\ }\href@noop {} {\bibfield  {journal} {\bibinfo  {journal} {arXiv preprint arXiv:2011.06258}\ } (\bibinfo {year} {2020})}\BibitemShut {NoStop}%
\bibitem [{\citenamefont {Pesah}\ et~al.(2021)\citenamefont {Pesah}, \citenamefont {Cerezo}, \citenamefont {Wang}, \citenamefont {Volkoff}, \citenamefont {Sornborger},\ and\ \citenamefont {Coles}}]{pesah2021absence}%
  \BibitemOpen
  \bibfield  {author} {\bibinfo {author} {\bibfnamefont {A.}~\bibnamefont {Pesah}}, \bibinfo {author} {\bibfnamefont {M.}~\bibnamefont {Cerezo}}, \bibinfo {author} {\bibfnamefont {S.}~\bibnamefont {Wang}}, \bibinfo {author} {\bibfnamefont {T.}~\bibnamefont {Volkoff}}, \bibinfo {author} {\bibfnamefont {A.~T.}\ \bibnamefont {Sornborger}},\ and\ \bibinfo {author} {\bibfnamefont {P.~J.}\ \bibnamefont {Coles}},\ }\bibfield  {title} {{Absence of barren plateaus in quantum convolutional neural networks},\ }\href@noop {} {\bibfield  {journal} {\bibinfo  {journal} {Physical Review X}\ }\textbf {\bibinfo {volume} {11}},\ \bibinfo {pages} {041011} (\bibinfo {year} {2021})}\BibitemShut {NoStop}%
\bibitem [{\citenamefont {Kerenidis}\ et~al.(2021)\citenamefont {Kerenidis}, \citenamefont {Landman},\ and\ \citenamefont {Mathur}}]{kerenidis2021classical}%
  \BibitemOpen
  \bibfield  {author} {\bibinfo {author} {\bibfnamefont {I.}~\bibnamefont {Kerenidis}}, \bibinfo {author} {\bibfnamefont {J.}~\bibnamefont {Landman}},\ and\ \bibinfo {author} {\bibfnamefont {N.}~\bibnamefont {Mathur}},\ }\bibfield  {title} {{Classical and quantum algorithms for orthogonal neural networks},\ }\href@noop {} {\bibfield  {journal} {\bibinfo  {journal} {arXiv preprint arXiv:2106.07198}\ } (\bibinfo {year} {2021})}\BibitemShut {NoStop}%
\bibitem [{\citenamefont {Lundberg}\ and\ \citenamefont {Lee}(2017)}]{SHAP}%
  \BibitemOpen
  \bibfield  {author} {\bibinfo {author} {\bibfnamefont {S.~M.}\ \bibnamefont {Lundberg}}\ and\ \bibinfo {author} {\bibfnamefont {S.-I.}\ \bibnamefont {Lee}},\ }\bibfield  {title} {{A unified approach to interpreting model predictions},\ }\href@noop {} {\bibfield  {journal} {\bibinfo  {journal} {Advances in neural information processing systems}\ }\textbf {\bibinfo {volume} {30}} (\bibinfo {year} {2017})}\BibitemShut {NoStop}%
\bibitem [{\citenamefont {Gianelle}\ et~al.(2022)\citenamefont {Gianelle}, \citenamefont {Koppenburg}, \citenamefont {Lucchesi}, \citenamefont {Nicotra}, \citenamefont {Rodrigues}, \citenamefont {Sestini}, \citenamefont {de~Vries},\ and\ \citenamefont {Zuliani}}]{gianelle2022quantum}%
  \BibitemOpen
  \bibfield  {author} {\bibinfo {author} {\bibfnamefont {A.}~\bibnamefont {Gianelle}}, \bibinfo {author} {\bibfnamefont {P.}~\bibnamefont {Koppenburg}}, \bibinfo {author} {\bibfnamefont {D.}~\bibnamefont {Lucchesi}}, \bibinfo {author} {\bibfnamefont {D.}~\bibnamefont {Nicotra}}, \bibinfo {author} {\bibfnamefont {E.}~\bibnamefont {Rodrigues}}, \bibinfo {author} {\bibfnamefont {L.}~\bibnamefont {Sestini}}, \bibinfo {author} {\bibfnamefont {J.}~\bibnamefont {de~Vries}},\ and\ \bibinfo {author} {\bibfnamefont {D.}~\bibnamefont {Zuliani}},\ }\bibfield  {title} {{Quantum Machine Learning for b-jet charge identification},\ }\href@noop {} {\bibfield  {journal} {\bibinfo  {journal} {Journal of High Energy Physics}\ }\textbf {\bibinfo {volume} {2022}},\ \bibinfo {pages} {1} (\bibinfo {year} {2022})}\BibitemShut {NoStop}%
\bibitem [{\citenamefont {Krunic}\ et~al.(2022)\citenamefont {Krunic}, \citenamefont {Fl{\"o}ther}, \citenamefont {Seegan}, \citenamefont {Earnest-Noble},\ and\ \citenamefont {Shehab}}]{krunic2022quantum}%
  \BibitemOpen
  \bibfield  {author} {\bibinfo {author} {\bibfnamefont {Z.}~\bibnamefont {Krunic}}, \bibinfo {author} {\bibfnamefont {F.~F.}\ \bibnamefont {Fl{\"o}ther}}, \bibinfo {author} {\bibfnamefont {G.}~\bibnamefont {Seegan}}, \bibinfo {author} {\bibfnamefont {N.~D.}\ \bibnamefont {Earnest-Noble}},\ and\ \bibinfo {author} {\bibfnamefont {O.}~\bibnamefont {Shehab}},\ }\bibfield  {title} {{Quantum kernels for real-world predictions based on electronic health records},\ }\href@noop {} {\bibfield  {journal} {\bibinfo  {journal} {IEEE Transactions on Quantum Engineering}\ }\textbf {\bibinfo {volume} {3}},\ \bibinfo {pages} {1} (\bibinfo {year} {2022})}\BibitemShut {NoStop}%
\bibitem [{\citenamefont {Lloyd}\ et~al.(2020)\citenamefont {Lloyd}, \citenamefont {Schuld}, \citenamefont {Ijaz}, \citenamefont {Izaac},\ and\ \citenamefont {Killoran}}]{lloyd2020quantum}%
  \BibitemOpen
  \bibfield  {author} {\bibinfo {author} {\bibfnamefont {S.}~\bibnamefont {Lloyd}}, \bibinfo {author} {\bibfnamefont {M.}~\bibnamefont {Schuld}}, \bibinfo {author} {\bibfnamefont {A.}~\bibnamefont {Ijaz}}, \bibinfo {author} {\bibfnamefont {J.}~\bibnamefont {Izaac}},\ and\ \bibinfo {author} {\bibfnamefont {N.}~\bibnamefont {Killoran}},\ }\bibfield  {title} {{Quantum embeddings for machine learning},\ }\href@noop {} {\bibfield  {journal} {\bibinfo  {journal} {arXiv preprint arXiv:2001.03622}\ } (\bibinfo {year} {2020})}\BibitemShut {NoStop}%
\bibitem [{\citenamefont {Brandao}\ et~al.(2016)\citenamefont {Brandao}, \citenamefont {Harrow},\ and\ \citenamefont {Horodecki}}]{brandao2016local}%
  \BibitemOpen
  \bibfield  {author} {\bibinfo {author} {\bibfnamefont {F.~G.}\ \bibnamefont {Brandao}}, \bibinfo {author} {\bibfnamefont {A.~W.}\ \bibnamefont {Harrow}},\ and\ \bibinfo {author} {\bibfnamefont {M.}~\bibnamefont {Horodecki}},\ }\bibfield  {title} {{Local random quantum circuits are approximate polynomial-designs},\ }\href@noop {} {\bibfield  {journal} {\bibinfo  {journal} {Communications in Mathematical Physics}\ }\textbf {\bibinfo {volume} {346}},\ \bibinfo {pages} {397} (\bibinfo {year} {2016})}\BibitemShut {NoStop}%
\bibitem [{\citenamefont {Harrigan}\ et~al.(2021)\citenamefont {Harrigan}, \citenamefont {Sung}, \citenamefont {Neeley}, \citenamefont {Satzinger}, \citenamefont {Arute}, \citenamefont {Arya}, \citenamefont {Atalaya}, \citenamefont {Bardin}, \citenamefont {Barends}, \citenamefont {Boixo} et~al.}]{harrigan2021quantum}%
  \BibitemOpen
  \bibfield  {author} {\bibinfo {author} {\bibfnamefont {M.~P.}\ \bibnamefont {Harrigan}}, \bibinfo {author} {\bibfnamefont {K.~J.}\ \bibnamefont {Sung}}, \bibinfo {author} {\bibfnamefont {M.}~\bibnamefont {Neeley}}, \bibinfo {author} {\bibfnamefont {K.~J.}\ \bibnamefont {Satzinger}}, \bibinfo {author} {\bibfnamefont {F.}~\bibnamefont {Arute}}, \bibinfo {author} {\bibfnamefont {K.}~\bibnamefont {Arya}}, \bibinfo {author} {\bibfnamefont {J.}~\bibnamefont {Atalaya}}, \bibinfo {author} {\bibfnamefont {J.~C.}\ \bibnamefont {Bardin}}, \bibinfo {author} {\bibfnamefont {R.}~\bibnamefont {Barends}}, \bibinfo {author} {\bibfnamefont {S.}~\bibnamefont {Boixo}}, et~al.,\ }\bibfield  {title} {{Quantum approximate optimization of non-planar graph problems on a planar superconducting processor},\ }\href@noop {} {\bibfield  {journal} {\bibinfo  {journal} {Nature Physics}\ }\textbf {\bibinfo {volume} {17}},\ \bibinfo {pages} {332} (\bibinfo {year} {2021})}\BibitemShut {NoStop}%
\bibitem [{\citenamefont {Arute}\ et~al.(2020)\citenamefont {Arute}, \citenamefont {Arya}, \citenamefont {Babbush}, \citenamefont {Bacon}, \citenamefont {Bardin}, \citenamefont {Barends}, \citenamefont {Boixo}, \citenamefont {Broughton}, \citenamefont {Buckley}, \citenamefont {Buell}, \citenamefont {Burkett}, \citenamefont {Bushnell}, \citenamefont {Chen}, \citenamefont {Chen}, \citenamefont {Chiaro}, \citenamefont {Collins}, \citenamefont {Courtney}, \citenamefont {Demura}, \citenamefont {Dunsworth}, \citenamefont {Farhi}, \citenamefont {Fowler}, \citenamefont {Foxen}, \citenamefont {Gidney}, \citenamefont {Giustina}, \citenamefont {Graff}, \citenamefont {Habegger}, \citenamefont {Harrigan}, \citenamefont {Ho}, \citenamefont {Hong}, \citenamefont {Huang}, \citenamefont {Huggins}, \citenamefont {Ioffe}, \citenamefont {Isakov}, \citenamefont {Jeffrey}, \citenamefont {Jiang}, \citenamefont {Jones}, \citenamefont {Kafri}, \citenamefont {Kechedzhi}, \citenamefont {Kelly}, \citenamefont {Kim}, \citenamefont
  {Klimov}, \citenamefont {Korotkov}, \citenamefont {Kostritsa}, \citenamefont {Landhuis}, \citenamefont {Laptev}, \citenamefont {Lindmark}, \citenamefont {Lucero}, \citenamefont {Martin}, \citenamefont {Martinis}, \citenamefont {McClean}, \citenamefont {McEwen}, \citenamefont {Megrant}, \citenamefont {Mi}, \citenamefont {Mohseni}, \citenamefont {Mruczkiewicz}, \citenamefont {Mutus}, \citenamefont {Naaman}, \citenamefont {Neeley}, \citenamefont {Neill}, \citenamefont {Neven}, \citenamefont {Niu}, \citenamefont {O'Brien}, \citenamefont {Ostby}, \citenamefont {Petukhov}, \citenamefont {Putterman}, \citenamefont {Quintana}, \citenamefont {Roushan}, \citenamefont {Rubin}, \citenamefont {Sank}, \citenamefont {Satzinger}, \citenamefont {Smelyanskiy}, \citenamefont {Strain}, \citenamefont {Sung}, \citenamefont {Szalay}, \citenamefont {Takeshita}, \citenamefont {Vainsencher}, \citenamefont {White}, \citenamefont {Wiebe}, \citenamefont {Yao}, \citenamefont {Yeh},\ and\ \citenamefont {Zalcman}}]{google2020hartree}%
  \BibitemOpen
  \bibfield  {author} {\bibinfo {author} {\bibfnamefont {F.~C.}\ \bibnamefont {Arute}}, \bibinfo {author} {\bibfnamefont {K.}~\bibnamefont {Arya}}, \bibinfo {author} {\bibfnamefont {R.}~\bibnamefont {Babbush}}, \bibinfo {author} {\bibfnamefont {D.}~\bibnamefont {Bacon}}, \bibinfo {author} {\bibfnamefont {J.}~\bibnamefont {Bardin}}, \bibinfo {author} {\bibfnamefont {R.}~\bibnamefont {Barends}}, \bibinfo {author} {\bibfnamefont {S.}~\bibnamefont {Boixo}}, \bibinfo {author} {\bibfnamefont {M.~B.}\ \bibnamefont {Broughton}}, \bibinfo {author} {\bibfnamefont {B.~B.}\ \bibnamefont {Buckley}}, \bibinfo {author} {\bibfnamefont {D.~A.}\ \bibnamefont {Buell}}, \bibinfo {author} {\bibfnamefont {B.}~\bibnamefont {Burkett}}, \bibinfo {author} {\bibfnamefont {N.}~\bibnamefont {Bushnell}}, \bibinfo {author} {\bibfnamefont {Y.}~\bibnamefont {Chen}}, \bibinfo {author} {\bibfnamefont {J.}~\bibnamefont {Chen}}, \bibinfo {author} {\bibfnamefont {B.}~\bibnamefont {Chiaro}}, \bibinfo {author} {\bibfnamefont {R.}~\bibnamefont
  {Collins}}, \bibinfo {author} {\bibfnamefont {W.}~\bibnamefont {Courtney}}, \bibinfo {author} {\bibfnamefont {S.}~\bibnamefont {Demura}}, \bibinfo {author} {\bibfnamefont {A.}~\bibnamefont {Dunsworth}}, \bibinfo {author} {\bibfnamefont {E.}~\bibnamefont {Farhi}}, \bibinfo {author} {\bibfnamefont {A.}~\bibnamefont {Fowler}}, \bibinfo {author} {\bibfnamefont {B.~R.}\ \bibnamefont {Foxen}}, \bibinfo {author} {\bibfnamefont {C.~M.}\ \bibnamefont {Gidney}}, \bibinfo {author} {\bibfnamefont {M.}~\bibnamefont {Giustina}}, \bibinfo {author} {\bibfnamefont {R.}~\bibnamefont {Graff}}, \bibinfo {author} {\bibfnamefont {S.}~\bibnamefont {Habegger}}, \bibinfo {author} {\bibfnamefont {M.~P.}\ \bibnamefont {Harrigan}}, \bibinfo {author} {\bibfnamefont {A.}~\bibnamefont {Ho}}, \bibinfo {author} {\bibfnamefont {S.}~\bibnamefont {Hong}}, \bibinfo {author} {\bibfnamefont {T.}~\bibnamefont {Huang}}, \bibinfo {author} {\bibfnamefont {W.~J.}\ \bibnamefont {Huggins}}, \bibinfo {author} {\bibfnamefont {L.}~\bibnamefont {Ioffe}},
  \bibinfo {author} {\bibfnamefont {S.}~\bibnamefont {Isakov}}, \bibinfo {author} {\bibfnamefont {E.}~\bibnamefont {Jeffrey}}, \bibinfo {author} {\bibfnamefont {Z.}~\bibnamefont {Jiang}}, \bibinfo {author} {\bibfnamefont {C.}~\bibnamefont {Jones}}, \bibinfo {author} {\bibfnamefont {D.}~\bibnamefont {Kafri}}, \bibinfo {author} {\bibfnamefont {K.}~\bibnamefont {Kechedzhi}}, \bibinfo {author} {\bibfnamefont {J.}~\bibnamefont {Kelly}}, \bibinfo {author} {\bibfnamefont {S.}~\bibnamefont {Kim}}, \bibinfo {author} {\bibfnamefont {P.}~\bibnamefont {Klimov}}, \bibinfo {author} {\bibfnamefont {A.}~\bibnamefont {Korotkov}}, \bibinfo {author} {\bibfnamefont {F.}~\bibnamefont {Kostritsa}}, \bibinfo {author} {\bibfnamefont {D.}~\bibnamefont {Landhuis}}, \bibinfo {author} {\bibfnamefont {P.}~\bibnamefont {Laptev}}, \bibinfo {author} {\bibfnamefont {M.}~\bibnamefont {Lindmark}}, \bibinfo {author} {\bibfnamefont {E.}~\bibnamefont {Lucero}}, \bibinfo {author} {\bibfnamefont {O.}~\bibnamefont {Martin}}, \bibinfo {author}
  {\bibfnamefont {J.}~\bibnamefont {Martinis}}, \bibinfo {author} {\bibfnamefont {J.~R.}\ \bibnamefont {McClean}}, \bibinfo {author} {\bibfnamefont {M.}~\bibnamefont {McEwen}}, \bibinfo {author} {\bibfnamefont {A.}~\bibnamefont {Megrant}}, \bibinfo {author} {\bibfnamefont {X.}~\bibnamefont {Mi}}, \bibinfo {author} {\bibfnamefont {M.}~\bibnamefont {Mohseni}}, \bibinfo {author} {\bibfnamefont {W.}~\bibnamefont {Mruczkiewicz}}, \bibinfo {author} {\bibfnamefont {J.}~\bibnamefont {Mutus}}, \bibinfo {author} {\bibfnamefont {O.}~\bibnamefont {Naaman}}, \bibinfo {author} {\bibfnamefont {M.}~\bibnamefont {Neeley}}, \bibinfo {author} {\bibfnamefont {C.}~\bibnamefont {Neill}}, \bibinfo {author} {\bibfnamefont {H.}~\bibnamefont {Neven}}, \bibinfo {author} {\bibfnamefont {M.~Y.}\ \bibnamefont {Niu}}, \bibinfo {author} {\bibfnamefont {T.~E.}\ \bibnamefont {O'Brien}}, \bibinfo {author} {\bibfnamefont {E.}~\bibnamefont {Ostby}}, \bibinfo {author} {\bibfnamefont {A.~G.}\ \bibnamefont {Petukhov}}, \bibinfo {author}
  {\bibfnamefont {H.}~\bibnamefont {Putterman}}, \bibinfo {author} {\bibfnamefont {C.}~\bibnamefont {Quintana}}, \bibinfo {author} {\bibfnamefont {P.}~\bibnamefont {Roushan}}, \bibinfo {author} {\bibfnamefont {N.}~\bibnamefont {Rubin}}, \bibinfo {author} {\bibfnamefont {D.}~\bibnamefont {Sank}}, \bibinfo {author} {\bibfnamefont {K.}~\bibnamefont {Satzinger}}, \bibinfo {author} {\bibfnamefont {V.}~\bibnamefont {Smelyanskiy}}, \bibinfo {author} {\bibfnamefont {D.}~\bibnamefont {Strain}}, \bibinfo {author} {\bibfnamefont {K.~J.}\ \bibnamefont {Sung}}, \bibinfo {author} {\bibfnamefont {M.}~\bibnamefont {Szalay}}, \bibinfo {author} {\bibfnamefont {T.~Y.}\ \bibnamefont {Takeshita}}, \bibinfo {author} {\bibfnamefont {A.}~\bibnamefont {Vainsencher}}, \bibinfo {author} {\bibfnamefont {T.}~\bibnamefont {White}}, \bibinfo {author} {\bibfnamefont {N.}~\bibnamefont {Wiebe}}, \bibinfo {author} {\bibfnamefont {J.}~\bibnamefont {Yao}}, \bibinfo {author} {\bibfnamefont {P.}~\bibnamefont {Yeh}},\ and\ \bibinfo {author}
  {\bibfnamefont {A.}~\bibnamefont {Zalcman}},\ }\bibfield  {title} {{Hartree-Fock on a Superconducting Qubit Quantum Computer},\ }\href {https://science.sciencemag.org/content/369/6507/1084} {\bibfield  {journal} {\bibinfo  {journal} {Science}\ }\textbf {\bibinfo {volume} {369}},\ \bibinfo {pages} {6507} (\bibinfo {year} {2020})}\BibitemShut {NoStop}%
\bibitem [{\citenamefont {Huang}\ et~al.(2021{\natexlab{d}})\citenamefont {Huang}, \citenamefont {Tan},\ and\ \citenamefont {Xu}}]{huang2021variational}%
  \BibitemOpen
  \bibfield  {author} {\bibinfo {author} {\bibfnamefont {R.}~\bibnamefont {Huang}}, \bibinfo {author} {\bibfnamefont {X.}~\bibnamefont {Tan}},\ and\ \bibinfo {author} {\bibfnamefont {Q.}~\bibnamefont {Xu}},\ }\bibfield  {title} {{Variational quantum tensor networks classifiers},\ }\href@noop {} {\bibfield  {journal} {\bibinfo  {journal} {Neurocomputing}\ }\textbf {\bibinfo {volume} {452}},\ \bibinfo {pages} {89} (\bibinfo {year} {2021}{\natexlab{d}})}\BibitemShut {NoStop}%
\bibitem [{\citenamefont {Johri}\ et~al.(2021)\citenamefont {Johri}, \citenamefont {Debnath}, \citenamefont {Mocherla}, \citenamefont {Singk}, \citenamefont {Prakash}, \citenamefont {Kim},\ and\ \citenamefont {Kerenidis}}]{johri2021nearest}%
  \BibitemOpen
  \bibfield  {author} {\bibinfo {author} {\bibfnamefont {S.}~\bibnamefont {Johri}}, \bibinfo {author} {\bibfnamefont {S.}~\bibnamefont {Debnath}}, \bibinfo {author} {\bibfnamefont {A.}~\bibnamefont {Mocherla}}, \bibinfo {author} {\bibfnamefont {A.}~\bibnamefont {Singk}}, \bibinfo {author} {\bibfnamefont {A.}~\bibnamefont {Prakash}}, \bibinfo {author} {\bibfnamefont {J.}~\bibnamefont {Kim}},\ and\ \bibinfo {author} {\bibfnamefont {I.}~\bibnamefont {Kerenidis}},\ }\bibfield  {title} {{Nearest centroid classification on a trapped ion quantum computer},\ }\href@noop {} {\bibfield  {journal} {\bibinfo  {journal} {npj Quantum Information}\ }\textbf {\bibinfo {volume} {7}},\ \bibinfo {pages} {1} (\bibinfo {year} {2021})}\BibitemShut {NoStop}%
\bibitem [{\citenamefont {Kwak}\ et~al.(2021)\citenamefont {Kwak}, \citenamefont {Yun}, \citenamefont {Jung},\ and\ \citenamefont {Kim}}]{kwak2021quantum}%
  \BibitemOpen
  \bibfield  {author} {\bibinfo {author} {\bibfnamefont {Y.}~\bibnamefont {Kwak}}, \bibinfo {author} {\bibfnamefont {W.~J.}\ \bibnamefont {Yun}}, \bibinfo {author} {\bibfnamefont {S.}~\bibnamefont {Jung}},\ and\ \bibinfo {author} {\bibfnamefont {J.}~\bibnamefont {Kim}},\ }in\ \href@noop {} {\bibinfo {booktitle} {2021 Twelfth International Conference on Ubiquitous and Future Networks (ICUFN)}}\ (\bibinfo {organization} {IEEE},\ \bibinfo {year} {2021})\ pp.\ \bibinfo {pages} {413--416}\BibitemShut {NoStop}%
\bibitem [{\citenamefont {Ramos-Calderer}\ et~al.(2021)\citenamefont {Ramos-Calderer}, \citenamefont {P{\'e}rez-Salinas}, \citenamefont {Garc{\'\i}a-Mart{\'\i}n}, \citenamefont {Bravo-Prieto}, \citenamefont {Cortada}, \citenamefont {Planaguma},\ and\ \citenamefont {Latorre}}]{ramos2021quantum}%
  \BibitemOpen
  \bibfield  {author} {\bibinfo {author} {\bibfnamefont {S.}~\bibnamefont {Ramos-Calderer}}, \bibinfo {author} {\bibfnamefont {A.}~\bibnamefont {P{\'e}rez-Salinas}}, \bibinfo {author} {\bibfnamefont {D.}~\bibnamefont {Garc{\'\i}a-Mart{\'\i}n}}, \bibinfo {author} {\bibfnamefont {C.}~\bibnamefont {Bravo-Prieto}}, \bibinfo {author} {\bibfnamefont {J.}~\bibnamefont {Cortada}}, \bibinfo {author} {\bibfnamefont {J.}~\bibnamefont {Planaguma}},\ and\ \bibinfo {author} {\bibfnamefont {J.~I.}\ \bibnamefont {Latorre}},\ }\bibfield  {title} {{Quantum unary approach to option pricing},\ }\href@noop {} {\bibfield  {journal} {\bibinfo  {journal} {Physical Review A}\ }\textbf {\bibinfo {volume} {103}},\ \bibinfo {pages} {032414} (\bibinfo {year} {2021})}\BibitemShut {NoStop}%
\bibitem [{\citenamefont {Kingma}\ and\ \citenamefont {Ba}(2014)}]{adam}%
  \BibitemOpen
  \bibfield  {author} {\bibinfo {author} {\bibfnamefont {D.~P.}\ \bibnamefont {Kingma}}\ and\ \bibinfo {author} {\bibfnamefont {J.}~\bibnamefont {Ba}},\ }\bibfield  {title} {{Adam: A method for stochastic optimization},\ }\href@noop {} {\bibfield  {journal} {\bibinfo  {journal} {arXiv preprint arXiv:1412.6980}\ } (\bibinfo {year} {2014})}\BibitemShut {NoStop}%
\bibitem [{\citenamefont {Araz}\ and\ \citenamefont {Spannowsky}(2021{\natexlab{a}})}]{araz_recons_2}%
  \BibitemOpen
  \bibfield  {author} {\bibinfo {author} {\bibfnamefont {J.~Y.}\ \bibnamefont {Araz}}\ and\ \bibinfo {author} {\bibfnamefont {M.}~\bibnamefont {Spannowsky}},\ }\bibfield  {title} {{Combine and conquer: event reconstruction with Bayesian Ensemble Neural Networks},\ }\href@noop {} {\bibfield  {journal} {\bibinfo  {journal} {Journal of High Energy Physics}\ }\textbf {\bibinfo {volume} {2021}},\ \bibinfo {pages} {1} (\bibinfo {year} {2021}{\natexlab{a}})}\BibitemShut {NoStop}%
\bibitem [{\citenamefont {Araz}\ and\ \citenamefont {Spannowsky}(2021{\natexlab{b}})}]{araz_recons1}%
  \BibitemOpen
  \bibfield  {author} {\bibinfo {author} {\bibfnamefont {J.~Y.}\ \bibnamefont {Araz}}\ and\ \bibinfo {author} {\bibfnamefont {M.}~\bibnamefont {Spannowsky}},\ }\bibfield  {title} {{Quantum-inspired event reconstruction with tensor networks: Matrix product states},\ }\href@noop {} {\bibfield  {journal} {\bibinfo  {journal} {Journal of High Energy Physics}\ }\textbf {\bibinfo {volume} {2021}},\ \bibinfo {pages} {1} (\bibinfo {year} {2021}{\natexlab{b}})}\BibitemShut {NoStop}%
\bibitem [{\citenamefont {Butterworth}\ et~al.(2016)\citenamefont {Butterworth}, \citenamefont {Carrazza}, \citenamefont {Cooper-Sarkar}, \citenamefont {De~Roeck}, \citenamefont {Feltesse}, \citenamefont {Forte}, \citenamefont {Gao}, \citenamefont {Glazov}, \citenamefont {Huston}, \citenamefont {Kassabov} et~al.}]{butterworth2016pdf4lhc}%
  \BibitemOpen
  \bibfield  {author} {\bibinfo {author} {\bibfnamefont {J.}~\bibnamefont {Butterworth}}, \bibinfo {author} {\bibfnamefont {S.}~\bibnamefont {Carrazza}}, \bibinfo {author} {\bibfnamefont {A.}~\bibnamefont {Cooper-Sarkar}}, \bibinfo {author} {\bibfnamefont {A.}~\bibnamefont {De~Roeck}}, \bibinfo {author} {\bibfnamefont {J.}~\bibnamefont {Feltesse}}, \bibinfo {author} {\bibfnamefont {S.}~\bibnamefont {Forte}}, \bibinfo {author} {\bibfnamefont {J.}~\bibnamefont {Gao}}, \bibinfo {author} {\bibfnamefont {S.}~\bibnamefont {Glazov}}, \bibinfo {author} {\bibfnamefont {J.}~\bibnamefont {Huston}}, \bibinfo {author} {\bibfnamefont {Z.}~\bibnamefont {Kassabov}}, et~al.,\ }\bibfield  {title} {{{PDF4LHC recommendations for LHC run II}},\ }\href@noop {} {\bibfield  {journal} {\bibinfo  {journal} {Journal of Physics G: Nuclear and Particle Physics}\ }\textbf {\bibinfo {volume} {43}},\ \bibinfo {pages} {023001} (\bibinfo {year} {2016})}\BibitemShut {NoStop}%
\bibitem [{\citenamefont {Placakyte}(2011)}]{placakyte2011parton}%
  \BibitemOpen
  \bibfield  {author} {\bibinfo {author} {\bibfnamefont {R.}~\bibnamefont {Placakyte}},\ }\bibfield  {title} {{Parton Distribution Functions},\ }\href@noop {} {\bibfield  {journal} {\bibinfo  {journal} {arXiv preprint arXiv:1111.5452}\ } (\bibinfo {year} {2011})}\BibitemShut {NoStop}%
\bibitem [{\citenamefont {Jolliffe}\ and\ \citenamefont {Cadima}(2016)}]{pca2}%
  \BibitemOpen
  \bibfield  {author} {\bibinfo {author} {\bibfnamefont {I.~T.}\ \bibnamefont {Jolliffe}}\ and\ \bibinfo {author} {\bibfnamefont {J.}~\bibnamefont {Cadima}},\ }\bibfield  {title} {{Principal component analysis: a review and recent developments},\ }\href@noop {} {\bibfield  {journal} {\bibinfo  {journal} {Philosophical Transactions of the Royal Society A: Mathematical, Physical and Engineering Sciences}\ }\textbf {\bibinfo {volume} {374}},\ \bibinfo {pages} {20150202} (\bibinfo {year} {2016})}\BibitemShut {NoStop}%
\bibitem [{\citenamefont {Kreplin}\ and\ \citenamefont {Roth}(2023)}]{kreplin2023reduction}%
  \BibitemOpen
  \bibfield  {author} {\bibinfo {author} {\bibfnamefont {D.}~\bibnamefont {Kreplin}}\ and\ \bibinfo {author} {\bibfnamefont {M.}~\bibnamefont {Roth}},\ }\bibfield  {title} {{Reduction of finite sampling noise in quantum neural networks},\ }\href@noop {} {\bibfield  {journal} {\bibinfo  {journal} {arXiv preprint arXiv:2306.01639}\ } (\bibinfo {year} {2023})}\BibitemShut {NoStop}%
\bibitem [{\citenamefont {Dua}\ and\ \citenamefont {Graff}(2017)}]{SUSY}%
  \BibitemOpen
  \bibfield  {author} {\bibinfo {author} {\bibfnamefont {D.}~\bibnamefont {Dua}}\ and\ \bibinfo {author} {\bibfnamefont {C.}~\bibnamefont {Graff}},\ }\href {http://archive.ics.uci.edu/ml} {{{UCI} Machine Learning Repository}} (\bibinfo {year} {2017})\BibitemShut {NoStop}%
\bibitem [{\citenamefont {Islam}\ et~al.(2022)\citenamefont {Islam}, \citenamefont {Ahmed}, \citenamefont {Barua},\ and\ \citenamefont {Begum}}]{islam2022systematic}%
  \BibitemOpen
  \bibfield  {author} {\bibinfo {author} {\bibfnamefont {M.~R.}\ \bibnamefont {Islam}}, \bibinfo {author} {\bibfnamefont {M.~U.}\ \bibnamefont {Ahmed}}, \bibinfo {author} {\bibfnamefont {S.}~\bibnamefont {Barua}},\ and\ \bibinfo {author} {\bibfnamefont {S.}~\bibnamefont {Begum}},\ }\bibfield  {title} {{A systematic review of explainable artificial intelligence in terms of different application domains and tasks},\ }\href@noop {} {\bibfield  {journal} {\bibinfo  {journal} {Applied Sciences}\ }\textbf {\bibinfo {volume} {12}},\ \bibinfo {pages} {1353} (\bibinfo {year} {2022})}\BibitemShut {NoStop}%
\bibitem [{\citenamefont {Saranya}\ and\ \citenamefont {Subhashini}(2023)}]{saranya2023systematic}%
  \BibitemOpen
  \bibfield  {author} {\bibinfo {author} {\bibfnamefont {A.}~\bibnamefont {Saranya}}\ and\ \bibinfo {author} {\bibfnamefont {R.}~\bibnamefont {Subhashini}},\ }\bibfield  {title} {{A systematic review of Explainable Artificial Intelligence models and applications: Recent developments and future trends},\ }\href@noop {} {\bibfield  {journal} {\bibinfo  {journal} {Decision Analytics Journal}\ ,\ \bibinfo {pages} {100230}} (\bibinfo {year} {2023})}\BibitemShut {NoStop}%
\bibitem [{\citenamefont {Steinm{\"u}ller}\ et~al.(2022)\citenamefont {Steinm{\"u}ller}, \citenamefont {Schulz}, \citenamefont {Graf},\ and\ \citenamefont {Herr}}]{steinmuller2022explainable}%
  \BibitemOpen
  \bibfield  {author} {\bibinfo {author} {\bibfnamefont {P.}~\bibnamefont {Steinm{\"u}ller}}, \bibinfo {author} {\bibfnamefont {T.}~\bibnamefont {Schulz}}, \bibinfo {author} {\bibfnamefont {F.}~\bibnamefont {Graf}},\ and\ \bibinfo {author} {\bibfnamefont {D.}~\bibnamefont {Herr}},\ }\bibfield  {title} {{e{X}plainable {AI} for Quantum Machine Learning},\ }\href@noop {} {\bibfield  {journal} {\bibinfo  {journal} {arXiv preprint arXiv:2211.01441}\ } (\bibinfo {year} {2022})}\BibitemShut {NoStop}%
\bibitem [{\citenamefont {Wack}\ et~al.(2021)\citenamefont {Wack}, \citenamefont {Paik}, \citenamefont {Javadi-Abhari}, \citenamefont {Jurcevic}, \citenamefont {Faro}, \citenamefont {Gambetta},\ and\ \citenamefont {Johnson}}]{wack2021quality}%
  \BibitemOpen
  \bibfield  {author} {\bibinfo {author} {\bibfnamefont {A.}~\bibnamefont {Wack}}, \bibinfo {author} {\bibfnamefont {H.}~\bibnamefont {Paik}}, \bibinfo {author} {\bibfnamefont {A.}~\bibnamefont {Javadi-Abhari}}, \bibinfo {author} {\bibfnamefont {P.}~\bibnamefont {Jurcevic}}, \bibinfo {author} {\bibfnamefont {I.}~\bibnamefont {Faro}}, \bibinfo {author} {\bibfnamefont {J.~M.}\ \bibnamefont {Gambetta}},\ and\ \bibinfo {author} {\bibfnamefont {B.~R.}\ \bibnamefont {Johnson}},\ }\bibfield  {title} {{Quality, speed, and scale: three key attributes to measure the performance of near-term quantum computers},\ }\href@noop {} {\bibfield  {journal} {\bibinfo  {journal} {arXiv preprint arXiv:2110.14108}\ } (\bibinfo {year} {2021})}\BibitemShut {NoStop}%
\bibitem [{\citenamefont {Wang}\ et~al.(2022)\citenamefont {Wang}, \citenamefont {Guo},\ and\ \citenamefont {Shan}}]{e24101467}%
  \BibitemOpen
  \bibfield  {author} {\bibinfo {author} {\bibfnamefont {J.}~\bibnamefont {Wang}}, \bibinfo {author} {\bibfnamefont {G.}~\bibnamefont {Guo}},\ and\ \bibinfo {author} {\bibfnamefont {Z.}~\bibnamefont {Shan}},\ }\bibfield  {title} {{SoK: Benchmarking the Performance of a Quantum Computer},\ }\bibfield  {journal} {\bibinfo  {journal} {Entropy}\ }\textbf {\bibinfo {volume} {24}},\ \href {https://doi.org/10.3390/e24101467} {10.3390/e24101467} (\bibinfo {year} {2022})\BibitemShut {NoStop}%
\bibitem [{Chatgpt()}]{chatgpt}%
  \BibitemOpen
  Chatgpt,\ \href@noop {} {{Chat{GPT}}},\ \bibinfo {howpublished} {\url{https://chat.openai.com/chat}} (\bibinfo {year} {2023}),\ \bibinfo {note} {accessed: June 5, 2023}\BibitemShut {NoStop}%
\bibitem [{IBMQuantum()}]{ibmq}%
  \BibitemOpen
  IBMQuantum,\ \href@noop {} {{{IBM Quantum}}},\ \bibinfo {howpublished} {\url{https://quantum-computing.ibm.com/}} (\bibinfo {year} {2019})\BibitemShut {NoStop}%
\bibitem [{\citenamefont {Jordan}\ and\ \citenamefont {Mitchell}(2015)}]{jordan2015machine}%
  \BibitemOpen
  \bibfield  {author} {\bibinfo {author} {\bibfnamefont {M.~I.}\ \bibnamefont {Jordan}}\ and\ \bibinfo {author} {\bibfnamefont {T.~M.}\ \bibnamefont {Mitchell}},\ }\bibfield  {title} {{Machine learning: Trends, perspectives, and prospects},\ }\href@noop {} {\bibfield  {journal} {\bibinfo  {journal} {Science}\ }\textbf {\bibinfo {volume} {349}},\ \bibinfo {pages} {255} (\bibinfo {year} {2015})}\BibitemShut {NoStop}%
\bibitem [{\citenamefont {Schuld}\ et~al.(2014)\citenamefont {Schuld}, \citenamefont {Sinayskiy},\ and\ \citenamefont {Petruccione}}]{schuld2014quest}%
  \BibitemOpen
  \bibfield  {author} {\bibinfo {author} {\bibfnamefont {M.}~\bibnamefont {Schuld}}, \bibinfo {author} {\bibfnamefont {I.}~\bibnamefont {Sinayskiy}},\ and\ \bibinfo {author} {\bibfnamefont {F.}~\bibnamefont {Petruccione}},\ }\bibfield  {title} {{The quest for a quantum neural network},\ }\href@noop {} {\bibfield  {journal} {\bibinfo  {journal} {Quantum Information Processing}\ }\textbf {\bibinfo {volume} {13}},\ \bibinfo {pages} {2567} (\bibinfo {year} {2014})}\BibitemShut {NoStop}%
\bibitem [{\citenamefont {Nielsen}\ and\ \citenamefont {Chuang}(2002)}]{nielsen2002quantum}%
  \BibitemOpen
  \bibfield  {author} {\bibinfo {author} {\bibfnamefont {M.~A.}\ \bibnamefont {Nielsen}}\ and\ \bibinfo {author} {\bibfnamefont {I.}~\bibnamefont {Chuang}},\ }\href@noop {} {{Quantum computation and quantum information}} (\bibinfo {year} {2002})\BibitemShut {NoStop}%
\bibitem [{\citenamefont {Shor}(1999)}]{shor1999polynomial}%
  \BibitemOpen
  \bibfield  {author} {\bibinfo {author} {\bibfnamefont {P.~W.}\ \bibnamefont {Shor}},\ }\bibfield  {title} {{Polynomial-time algorithms for prime factorization and discrete logarithms on a quantum computer},\ }\href@noop {} {\bibfield  {journal} {\bibinfo  {journal} {SIAM review}\ }\textbf {\bibinfo {volume} {41}},\ \bibinfo {pages} {303} (\bibinfo {year} {1999})}\BibitemShut {NoStop}%
\bibitem [{\citenamefont {Feynman}(2018)}]{feynman2018simulating}%
  \BibitemOpen
  \bibfield  {author} {\bibinfo {author} {\bibfnamefont {R.~P.}\ \bibnamefont {Feynman}},\ }in\ \href@noop {} {\bibinfo {booktitle} {Feynman and computation}}\ (\bibinfo  {publisher} {CRC Press},\ \bibinfo {year} {2018})\ pp.\ \bibinfo {pages} {133--153}\BibitemShut {NoStop}%
\bibitem [{\citenamefont {Allam~Jr}\ et~al.(2018)\citenamefont {Allam~Jr}, \citenamefont {Bahmanyar}, \citenamefont {Biswas}, \citenamefont {Dai}, \citenamefont {Galbany}, \citenamefont {Hlo{\v{z}}ek}, \citenamefont {Ishida}, \citenamefont {Jha}, \citenamefont {Jones}, \citenamefont {Kessler} et~al.}]{plasticc}%
  \BibitemOpen
  \bibfield  {author} {\bibinfo {author} {\bibfnamefont {T.}~\bibnamefont {Allam~Jr}}, \bibinfo {author} {\bibfnamefont {A.}~\bibnamefont {Bahmanyar}}, \bibinfo {author} {\bibfnamefont {R.}~\bibnamefont {Biswas}}, \bibinfo {author} {\bibfnamefont {M.}~\bibnamefont {Dai}}, \bibinfo {author} {\bibfnamefont {L.}~\bibnamefont {Galbany}}, \bibinfo {author} {\bibfnamefont {R.}~\bibnamefont {Hlo{\v{z}}ek}}, \bibinfo {author} {\bibfnamefont {E.~E.}\ \bibnamefont {Ishida}}, \bibinfo {author} {\bibfnamefont {S.~W.}\ \bibnamefont {Jha}}, \bibinfo {author} {\bibfnamefont {D.~O.}\ \bibnamefont {Jones}}, \bibinfo {author} {\bibfnamefont {R.}~\bibnamefont {Kessler}}, et~al.,\ }\bibfield  {title} {{The photometric lsst astronomical time-series classification challenge (plasticc): Data set},\ }\href@noop {} {\bibfield  {journal} {\bibinfo  {journal} {arXiv preprint arXiv:1810.00001}\ } (\bibinfo {year} {2018})}\BibitemShut {NoStop}%
\bibitem [{\citenamefont {Jr.}(2017)}]{titanic}%
  \BibitemOpen
  \bibfield  {author} {\bibinfo {author} {\bibfnamefont {F.~E.~H.}\ \bibnamefont {Jr.}},\ }\href@noop {} {{Titanic dataset}},\ \bibinfo {howpublished} {\url{https://www.openml.org/d/40945}} (\bibinfo {year} {2017})\BibitemShut {NoStop}%
\bibitem [{\citenamefont {Araz}\ and\ \citenamefont {Spannowsky}(2022)}]{araz2022classical}%
  \BibitemOpen
  \bibfield  {author} {\bibinfo {author} {\bibfnamefont {J.~Y.}\ \bibnamefont {Araz}}\ and\ \bibinfo {author} {\bibfnamefont {M.}~\bibnamefont {Spannowsky}},\ }\bibfield  {title} {{Classical versus quantum: Comparing tensor-network-based quantum circuits on {Large Hadron Collider data}},\ }\href@noop {} {\bibfield  {journal} {\bibinfo  {journal} {Physical Review A}\ }\textbf {\bibinfo {volume} {106}},\ \bibinfo {pages} {062423} (\bibinfo {year} {2022})}\BibitemShut {NoStop}%
\bibitem [{\citenamefont {Lloyd}\ et~al.(2014)\citenamefont {Lloyd}, \citenamefont {Mohseni},\ and\ \citenamefont {Rebentrost}}]{lloyd2014quantum}%
  \BibitemOpen
  \bibfield  {author} {\bibinfo {author} {\bibfnamefont {S.}~\bibnamefont {Lloyd}}, \bibinfo {author} {\bibfnamefont {M.}~\bibnamefont {Mohseni}},\ and\ \bibinfo {author} {\bibfnamefont {P.}~\bibnamefont {Rebentrost}},\ }\bibfield  {title} {{Quantum principal component analysis},\ }\href@noop {} {\bibfield  {journal} {\bibinfo  {journal} {Nature Physics}\ }\textbf {\bibinfo {volume} {10}},\ \bibinfo {pages} {631} (\bibinfo {year} {2014})}\BibitemShut {NoStop}%
\bibitem [{\citenamefont {Cerezo}\ et~al.(2021{\natexlab{b}})\citenamefont {Cerezo}, \citenamefont {Sone}, \citenamefont {Volkoff}, \citenamefont {Cincio},\ and\ \citenamefont {Coles}}]{cerezo2021cost}%
  \BibitemOpen
  \bibfield  {author} {\bibinfo {author} {\bibfnamefont {M.}~\bibnamefont {Cerezo}}, \bibinfo {author} {\bibfnamefont {A.}~\bibnamefont {Sone}}, \bibinfo {author} {\bibfnamefont {T.}~\bibnamefont {Volkoff}}, \bibinfo {author} {\bibfnamefont {L.}~\bibnamefont {Cincio}},\ and\ \bibinfo {author} {\bibfnamefont {P.~J.}\ \bibnamefont {Coles}},\ }\bibfield  {title} {{Cost function dependent barren plateaus in shallow parametrized quantum circuits},\ }\href@noop {} {\bibfield  {journal} {\bibinfo  {journal} {Nature communications}\ }\textbf {\bibinfo {volume} {12}},\ \bibinfo {pages} {1} (\bibinfo {year} {2021}{\natexlab{b}})}\BibitemShut {NoStop}%
\bibitem [{\citenamefont {Blance}\ and\ \citenamefont {Spannowsky}(2021)}]{blance2021quantum}%
  \BibitemOpen
  \bibfield  {author} {\bibinfo {author} {\bibfnamefont {A.}~\bibnamefont {Blance}}\ and\ \bibinfo {author} {\bibfnamefont {M.}~\bibnamefont {Spannowsky}},\ }\bibfield  {title} {{Quantum machine learning for particle physics using a variational quantum classifier},\ }\href@noop {} {\bibfield  {journal} {\bibinfo  {journal} {Journal of High Energy Physics}\ }\textbf {\bibinfo {volume} {2021}},\ \bibinfo {pages} {1} (\bibinfo {year} {2021})}\BibitemShut {NoStop}%
\bibitem [{\citenamefont {Arrasmith}\ et~al.(2020)\citenamefont {Arrasmith}, \citenamefont {Cincio}, \citenamefont {Somma},\ and\ \citenamefont {Coles}}]{arrasmith2020operator}%
  \BibitemOpen
  \bibfield  {author} {\bibinfo {author} {\bibfnamefont {A.}~\bibnamefont {Arrasmith}}, \bibinfo {author} {\bibfnamefont {L.}~\bibnamefont {Cincio}}, \bibinfo {author} {\bibfnamefont {R.~D.}\ \bibnamefont {Somma}},\ and\ \bibinfo {author} {\bibfnamefont {P.~J.}\ \bibnamefont {Coles}},\ }\bibfield  {title} {{Operator sampling for shot-frugal optimization in variational algorithms},\ }\href@noop {} {\bibfield  {journal} {\bibinfo  {journal} {arXiv preprint arXiv:2004.06252}\ } (\bibinfo {year} {2020})}\BibitemShut {NoStop}%
\bibitem [{\citenamefont {Blinov}\ et~al.(2021)\citenamefont {Blinov}, \citenamefont {Wu},\ and\ \citenamefont {Monroe}}]{blinov2021comparison}%
  \BibitemOpen
  \bibfield  {author} {\bibinfo {author} {\bibfnamefont {S.}~\bibnamefont {Blinov}}, \bibinfo {author} {\bibfnamefont {B.}~\bibnamefont {Wu}},\ and\ \bibinfo {author} {\bibfnamefont {C.}~\bibnamefont {Monroe}},\ }\bibfield  {title} {{Comparison of cloud-based ion trap and superconducting quantum computer architectures},\ }\href@noop {} {\bibfield  {journal} {\bibinfo  {journal} {AVS Quantum Science}\ }\textbf {\bibinfo {volume} {3}},\ \bibinfo {pages} {033801} (\bibinfo {year} {2021})}\BibitemShut {NoStop}%
\bibitem [{\citenamefont {Ostaszewski}\ et~al.(2021)\citenamefont {Ostaszewski}, \citenamefont {Grant},\ and\ \citenamefont {Benedetti}}]{ostaszewski2021structure}%
  \BibitemOpen
  \bibfield  {author} {\bibinfo {author} {\bibfnamefont {M.}~\bibnamefont {Ostaszewski}}, \bibinfo {author} {\bibfnamefont {E.}~\bibnamefont {Grant}},\ and\ \bibinfo {author} {\bibfnamefont {M.}~\bibnamefont {Benedetti}},\ }\bibfield  {title} {{Structure optimization for parameterized quantum circuits},\ }\href@noop {} {\bibfield  {journal} {\bibinfo  {journal} {Quantum}\ }\textbf {\bibinfo {volume} {5}},\ \bibinfo {pages} {391} (\bibinfo {year} {2021})}\BibitemShut {NoStop}%
\bibitem [{\citenamefont {Parrish}\ et~al.(2019)\citenamefont {Parrish}, \citenamefont {Iosue}, \citenamefont {Ozaeta},\ and\ \citenamefont {McMahon}}]{parrish2019jacobi}%
  \BibitemOpen
  \bibfield  {author} {\bibinfo {author} {\bibfnamefont {R.~M.}\ \bibnamefont {Parrish}}, \bibinfo {author} {\bibfnamefont {J.~T.}\ \bibnamefont {Iosue}}, \bibinfo {author} {\bibfnamefont {A.}~\bibnamefont {Ozaeta}},\ and\ \bibinfo {author} {\bibfnamefont {P.~L.}\ \bibnamefont {McMahon}},\ }\bibfield  {title} {{A Jacobi diagonalization and {A}nderson acceleration algorithm for variational quantum algorithm parameter optimization},\ }\href@noop {} {\bibfield  {journal} {\bibinfo  {journal} {arXiv preprint arXiv:1904.03206}\ } (\bibinfo {year} {2019})}\BibitemShut {NoStop}%
\bibitem [{\citenamefont {Chen}\ et~al.(2024)\citenamefont {Chen}, \citenamefont {Guang}, \citenamefont {Guo}, \citenamefont {Feng},\ and\ \citenamefont {Hou}}]{chen2024pure}%
  \BibitemOpen
  \bibfield  {author} {\bibinfo {author} {\bibfnamefont {R.}~\bibnamefont {Chen}}, \bibinfo {author} {\bibfnamefont {Z.}~\bibnamefont {Guang}}, \bibinfo {author} {\bibfnamefont {C.}~\bibnamefont {Guo}}, \bibinfo {author} {\bibfnamefont {G.}~\bibnamefont {Feng}},\ and\ \bibinfo {author} {\bibfnamefont {S.-Y.}\ \bibnamefont {Hou}},\ }\bibfield  {title} {{Pure quantum gradient descent algorithm and full quantum variational eigensolver},\ }\href@noop {} {\bibfield  {journal} {\bibinfo  {journal} {Frontiers of Physics}\ }\textbf {\bibinfo {volume} {19}},\ \bibinfo {pages} {21202} (\bibinfo {year} {2024})}\BibitemShut {NoStop}%
\bibitem [{\citenamefont {Mc~Keever}\ and\ \citenamefont {Lubasch}(2023)}]{mc2023classically}%
  \BibitemOpen
  \bibfield  {author} {\bibinfo {author} {\bibfnamefont {C.}~\bibnamefont {Mc~Keever}}\ and\ \bibinfo {author} {\bibfnamefont {M.}~\bibnamefont {Lubasch}},\ }\bibfield  {title} {{Classically optimized {H}amiltonian simulation},\ }\href@noop {} {\bibfield  {journal} {\bibinfo  {journal} {Physical Review Research}\ }\textbf {\bibinfo {volume} {5}},\ \bibinfo {pages} {023146} (\bibinfo {year} {2023})}\BibitemShut {NoStop}%
\bibitem [{\citenamefont {Nakanishi}\ et~al.(2020{\natexlab{a}})\citenamefont {Nakanishi}, \citenamefont {Fujii},\ and\ \citenamefont {Todo}}]{nakanishi2020sequential}%
  \BibitemOpen
  \bibfield  {author} {\bibinfo {author} {\bibfnamefont {K.~M.}\ \bibnamefont {Nakanishi}}, \bibinfo {author} {\bibfnamefont {K.}~\bibnamefont {Fujii}},\ and\ \bibinfo {author} {\bibfnamefont {S.}~\bibnamefont {Todo}},\ }\bibfield  {title} {{Sequential minimal optimization for quantum-classical hybrid algorithms},\ }\href@noop {} {\bibfield  {journal} {\bibinfo  {journal} {Physical Review Research}\ }\textbf {\bibinfo {volume} {2}},\ \bibinfo {pages} {043158} (\bibinfo {year} {2020}{\natexlab{a}})}\BibitemShut {NoStop}%
\bibitem [{\citenamefont {Vidal}\ and\ \citenamefont {Theis}(2018)}]{vidal2018calculus}%
  \BibitemOpen
  \bibfield  {author} {\bibinfo {author} {\bibfnamefont {J.~G.}\ \bibnamefont {Vidal}}\ and\ \bibinfo {author} {\bibfnamefont {D.~O.}\ \bibnamefont {Theis}},\ }\bibfield  {title} {{Calculus on parameterized quantum circuits},\ }\href@noop {} {\bibfield  {journal} {\bibinfo  {journal} {arXiv preprint arXiv:1812.06323}\ } (\bibinfo {year} {2018})}\BibitemShut {NoStop}%
\bibitem [{\citenamefont {Wu}\ and\ \citenamefont {Yoo}(2022)}]{wu2022challenges}%
  \BibitemOpen
  \bibfield  {author} {\bibinfo {author} {\bibfnamefont {S.~L.}\ \bibnamefont {Wu}}\ and\ \bibinfo {author} {\bibfnamefont {S.}~\bibnamefont {Yoo}},\ }\bibfield  {title} {{Challenges and opportunities in quantum machine learning for high-energy physics},\ }\href@noop {} {\bibfield  {journal} {\bibinfo  {journal} {Nature Reviews Physics}\ }\textbf {\bibinfo {volume} {4}},\ \bibinfo {pages} {143} (\bibinfo {year} {2022})}\BibitemShut {NoStop}%
\bibitem [{\citenamefont {Havl{\'\i}{\v{c}}ek}\ et~al.(2019)\citenamefont {Havl{\'\i}{\v{c}}ek}, \citenamefont {C{\'o}rcoles}, \citenamefont {Temme}, \citenamefont {Harrow}, \citenamefont {Kandala}, \citenamefont {Chow},\ and\ \citenamefont {Gambetta}}]{havlivcek2019supervised}%
  \BibitemOpen
  \bibfield  {author} {\bibinfo {author} {\bibfnamefont {V.}~\bibnamefont {Havl{\'\i}{\v{c}}ek}}, \bibinfo {author} {\bibfnamefont {A.~D.}\ \bibnamefont {C{\'o}rcoles}}, \bibinfo {author} {\bibfnamefont {K.}~\bibnamefont {Temme}}, \bibinfo {author} {\bibfnamefont {A.~W.}\ \bibnamefont {Harrow}}, \bibinfo {author} {\bibfnamefont {A.}~\bibnamefont {Kandala}}, \bibinfo {author} {\bibfnamefont {J.~M.}\ \bibnamefont {Chow}},\ and\ \bibinfo {author} {\bibfnamefont {J.~M.}\ \bibnamefont {Gambetta}},\ }\bibfield  {title} {{Supervised learning with quantum-enhanced feature spaces},\ }\href@noop {} {\bibfield  {journal} {\bibinfo  {journal} {Nature}\ }\textbf {\bibinfo {volume} {567}},\ \bibinfo {pages} {209} (\bibinfo {year} {2019})}\BibitemShut {NoStop}%
\bibitem [{\citenamefont {Ngairangbam}\ et~al.(2022)\citenamefont {Ngairangbam}, \citenamefont {Spannowsky},\ and\ \citenamefont {Takeuchi}}]{ngairangbam2022anomaly}%
  \BibitemOpen
  \bibfield  {author} {\bibinfo {author} {\bibfnamefont {V.~S.}\ \bibnamefont {Ngairangbam}}, \bibinfo {author} {\bibfnamefont {M.}~\bibnamefont {Spannowsky}},\ and\ \bibinfo {author} {\bibfnamefont {M.}~\bibnamefont {Takeuchi}},\ }\bibfield  {title} {{Anomaly detection in high-energy physics using a quantum autoencoder},\ }\href@noop {} {\bibfield  {journal} {\bibinfo  {journal} {Physical Review D}\ }\textbf {\bibinfo {volume} {105}},\ \bibinfo {pages} {095004} (\bibinfo {year} {2022})}\BibitemShut {NoStop}%
\bibitem [{\citenamefont {Blume-Kohout}(2010)}]{blume2010optimal}%
  \BibitemOpen
  \bibfield  {author} {\bibinfo {author} {\bibfnamefont {R.}~\bibnamefont {Blume-Kohout}},\ }\bibfield  {title} {{Optimal, reliable estimation of quantum states},\ }\href@noop {} {\bibfield  {journal} {\bibinfo  {journal} {New Journal of Physics}\ }\textbf {\bibinfo {volume} {12}},\ \bibinfo {pages} {043034} (\bibinfo {year} {2010})}\BibitemShut {NoStop}%
\bibitem [{\citenamefont {Carollo}\ et~al.(2019)\citenamefont {Carollo}, \citenamefont {Spagnolo}, \citenamefont {Dubkov},\ and\ \citenamefont {Valenti}}]{carollo2019quantumness}%
  \BibitemOpen
  \bibfield  {author} {\bibinfo {author} {\bibfnamefont {A.}~\bibnamefont {Carollo}}, \bibinfo {author} {\bibfnamefont {B.}~\bibnamefont {Spagnolo}}, \bibinfo {author} {\bibfnamefont {A.~A.}\ \bibnamefont {Dubkov}},\ and\ \bibinfo {author} {\bibfnamefont {D.}~\bibnamefont {Valenti}},\ }\bibfield  {title} {{On quantumness in multi-parameter quantum estimation},\ }\href@noop {} {\bibfield  {journal} {\bibinfo  {journal} {Journal of Statistical Mechanics: Theory and Experiment}\ }\textbf {\bibinfo {volume} {2019}},\ \bibinfo {pages} {094010} (\bibinfo {year} {2019})}\BibitemShut {NoStop}%
\bibitem [{\citenamefont {Cui}\ and\ \citenamefont {Fan}(2010)}]{cui2010correlations}%
  \BibitemOpen
  \bibfield  {author} {\bibinfo {author} {\bibfnamefont {J.}~\bibnamefont {Cui}}\ and\ \bibinfo {author} {\bibfnamefont {H.}~\bibnamefont {Fan}},\ }\bibfield  {title} {{Correlations in the Grover search},\ }\href@noop {} {\bibfield  {journal} {\bibinfo  {journal} {Journal of Physics A: Mathematical and Theoretical}\ }\textbf {\bibinfo {volume} {43}},\ \bibinfo {pages} {045305} (\bibinfo {year} {2010})}\BibitemShut {NoStop}%
\bibitem [{\citenamefont {Biron}\ et~al.(1999)\citenamefont {Biron}, \citenamefont {Biham}, \citenamefont {Biham}, \citenamefont {Grassl},\ and\ \citenamefont {Lidar}}]{biron1999generalized}%
  \BibitemOpen
  \bibfield  {author} {\bibinfo {author} {\bibfnamefont {D.}~\bibnamefont {Biron}}, \bibinfo {author} {\bibfnamefont {O.}~\bibnamefont {Biham}}, \bibinfo {author} {\bibfnamefont {E.}~\bibnamefont {Biham}}, \bibinfo {author} {\bibfnamefont {M.}~\bibnamefont {Grassl}},\ and\ \bibinfo {author} {\bibfnamefont {D.~A.}\ \bibnamefont {Lidar}},\ }\bibfield  {title} {{Generalized Grover search algorithm for arbitrary initial amplitude distribution},\ }\href@noop {} {\bibfield  {journal} {\bibinfo  {journal} {Lecture notes in computer science}\ ,\ \bibinfo {pages} {140}} (\bibinfo {year} {1999})}\BibitemShut {NoStop}%
\bibitem [{\citenamefont {Helstrom}(1969)}]{helstrom1969quantum}%
  \BibitemOpen
  \bibfield  {author} {\bibinfo {author} {\bibfnamefont {C.~W.}\ \bibnamefont {Helstrom}},\ }\bibfield  {title} {{Quantum detection and estimation theory},\ }\href@noop {} {\bibfield  {journal} {\bibinfo  {journal} {Journal of Statistical Physics}\ }\textbf {\bibinfo {volume} {1}},\ \bibinfo {pages} {231} (\bibinfo {year} {1969})}\BibitemShut {NoStop}%
\bibitem [{\citenamefont {D'Ariano}\ et~al.(1998)\citenamefont {D'Ariano}, \citenamefont {Macchiavello},\ and\ \citenamefont {Sacchi}}]{d1998general}%
  \BibitemOpen
  \bibfield  {author} {\bibinfo {author} {\bibfnamefont {G.}~\bibnamefont {D'Ariano}}, \bibinfo {author} {\bibfnamefont {C.}~\bibnamefont {Macchiavello}},\ and\ \bibinfo {author} {\bibfnamefont {M.}~\bibnamefont {Sacchi}},\ }\bibfield  {title} {{On the general problem of quantum phase estimation},\ }\href@noop {} {\bibfield  {journal} {\bibinfo  {journal} {Physics Letters A}\ }\textbf {\bibinfo {volume} {248}},\ \bibinfo {pages} {103} (\bibinfo {year} {1998})}\BibitemShut {NoStop}%
\bibitem [{\citenamefont {Dorner}\ et~al.(2009)\citenamefont {Dorner}, \citenamefont {Demkowicz-Dobrzanski}, \citenamefont {Smith}, \citenamefont {Lundeen}, \citenamefont {Wasilewski}, \citenamefont {Banaszek},\ and\ \citenamefont {Walmsley}}]{dorner2009optimal}%
  \BibitemOpen
  \bibfield  {author} {\bibinfo {author} {\bibfnamefont {U.}~\bibnamefont {Dorner}}, \bibinfo {author} {\bibfnamefont {R.}~\bibnamefont {Demkowicz-Dobrzanski}}, \bibinfo {author} {\bibfnamefont {B.~J.}\ \bibnamefont {Smith}}, \bibinfo {author} {\bibfnamefont {J.~S.}\ \bibnamefont {Lundeen}}, \bibinfo {author} {\bibfnamefont {W.}~\bibnamefont {Wasilewski}}, \bibinfo {author} {\bibfnamefont {K.}~\bibnamefont {Banaszek}},\ and\ \bibinfo {author} {\bibfnamefont {I.~A.}\ \bibnamefont {Walmsley}},\ }\bibfield  {title} {{Optimal quantum phase estimation},\ }\href@noop {} {\bibfield  {journal} {\bibinfo  {journal} {Physical review letters}\ }\textbf {\bibinfo {volume} {102}},\ \bibinfo {pages} {040403} (\bibinfo {year} {2009})}\BibitemShut {NoStop}%
\bibitem [{\citenamefont {Roszkowska}(2021)}]{roszkowska2021fintech}%
  \BibitemOpen
  \bibfield  {author} {\bibinfo {author} {\bibfnamefont {P.}~\bibnamefont {Roszkowska}},\ }\bibfield  {title} {{Fintech in financial reporting and audit for fraud prevention and safeguarding equity investments},\ }\href@noop {} {\bibfield  {journal} {\bibinfo  {journal} {Journal of Accounting \& Organizational Change}\ }\textbf {\bibinfo {volume} {17}},\ \bibinfo {pages} {164} (\bibinfo {year} {2021})}\BibitemShut {NoStop}%
\bibitem [{\citenamefont {Koshiyama}\ et~al.(2021)\citenamefont {Koshiyama}, \citenamefont {Kazim}, \citenamefont {Treleaven}, \citenamefont {Rai}, \citenamefont {Szpruch}, \citenamefont {Pavey}, \citenamefont {Ahamat}, \citenamefont {Leutner}, \citenamefont {Goebel}, \citenamefont {Knight} et~al.}]{koshiyama2021towards}%
  \BibitemOpen
  \bibfield  {author} {\bibinfo {author} {\bibfnamefont {A.}~\bibnamefont {Koshiyama}}, \bibinfo {author} {\bibfnamefont {E.}~\bibnamefont {Kazim}}, \bibinfo {author} {\bibfnamefont {P.}~\bibnamefont {Treleaven}}, \bibinfo {author} {\bibfnamefont {P.}~\bibnamefont {Rai}}, \bibinfo {author} {\bibfnamefont {L.}~\bibnamefont {Szpruch}}, \bibinfo {author} {\bibfnamefont {G.}~\bibnamefont {Pavey}}, \bibinfo {author} {\bibfnamefont {G.}~\bibnamefont {Ahamat}}, \bibinfo {author} {\bibfnamefont {F.}~\bibnamefont {Leutner}}, \bibinfo {author} {\bibfnamefont {R.}~\bibnamefont {Goebel}}, \bibinfo {author} {\bibfnamefont {A.}~\bibnamefont {Knight}}, et~al.,\ }\href@noop {} {{{Towards algorithm auditing: a survey on managing legal, ethical and technological risks of AI, ML and associated algorithms}}},\ \bibinfo {howpublished} {\url{https://papers.ssrn.com/sol3/papers.cfm?abstract_id=3778998}} (\bibinfo {year} {2021}),\ \bibinfo {note} {sSRN Working Paper}\BibitemShut {NoStop}%
\bibitem [{\citenamefont {Larocca}\ et~al.(2023)\citenamefont {Larocca}, \citenamefont {Ju}, \citenamefont {Garc{\'\i}a-Mart{\'\i}n}, \citenamefont {Coles},\ and\ \citenamefont {Cerezo}}]{larocca2023theory}%
  \BibitemOpen
  \bibfield  {author} {\bibinfo {author} {\bibfnamefont {M.}~\bibnamefont {Larocca}}, \bibinfo {author} {\bibfnamefont {N.}~\bibnamefont {Ju}}, \bibinfo {author} {\bibfnamefont {D.}~\bibnamefont {Garc{\'\i}a-Mart{\'\i}n}}, \bibinfo {author} {\bibfnamefont {P.~J.}\ \bibnamefont {Coles}},\ and\ \bibinfo {author} {\bibfnamefont {M.}~\bibnamefont {Cerezo}},\ }\bibfield  {title} {{Theory of overparametrization in quantum neural networks},\ }\href@noop {} {\bibfield  {journal} {\bibinfo  {journal} {Nature Computational Science}\ }\textbf {\bibinfo {volume} {3}},\ \bibinfo {pages} {542} (\bibinfo {year} {2023})}\BibitemShut {NoStop}%
\bibitem [{\citenamefont {Kiani}\ et~al.(2020)\citenamefont {Kiani}, \citenamefont {Lloyd},\ and\ \citenamefont {Maity}}]{kiani2020learning}%
  \BibitemOpen
  \bibfield  {author} {\bibinfo {author} {\bibfnamefont {B.~T.}\ \bibnamefont {Kiani}}, \bibinfo {author} {\bibfnamefont {S.}~\bibnamefont {Lloyd}},\ and\ \bibinfo {author} {\bibfnamefont {R.}~\bibnamefont {Maity}},\ }\bibfield  {title} {{Learning unitaries by gradient descent},\ }\href@noop {} {\bibfield  {journal} {\bibinfo  {journal} {arXiv preprint arXiv:2001.11897}\ } (\bibinfo {year} {2020})}\BibitemShut {NoStop}%
\bibitem [{\citenamefont {Wecker}\ et~al.(2015)\citenamefont {Wecker}, \citenamefont {Hastings},\ and\ \citenamefont {Troyer}}]{wecker2015progress}%
  \BibitemOpen
  \bibfield  {author} {\bibinfo {author} {\bibfnamefont {D.}~\bibnamefont {Wecker}}, \bibinfo {author} {\bibfnamefont {M.~B.}\ \bibnamefont {Hastings}},\ and\ \bibinfo {author} {\bibfnamefont {M.}~\bibnamefont {Troyer}},\ }\bibfield  {title} {{Progress towards practical quantum variational algorithms},\ }\href@noop {} {\bibfield  {journal} {\bibinfo  {journal} {Physical Review A}\ }\textbf {\bibinfo {volume} {92}},\ \bibinfo {pages} {042303} (\bibinfo {year} {2015})}\BibitemShut {NoStop}%
\bibitem [{\citenamefont {Hadfield}\ et~al.(2019)\citenamefont {Hadfield}, \citenamefont {Wang}, \citenamefont {O’gorman}, \citenamefont {Rieffel}, \citenamefont {Venturelli},\ and\ \citenamefont {Biswas}}]{hadfield2019quantum}%
  \BibitemOpen
  \bibfield  {author} {\bibinfo {author} {\bibfnamefont {S.}~\bibnamefont {Hadfield}}, \bibinfo {author} {\bibfnamefont {Z.}~\bibnamefont {Wang}}, \bibinfo {author} {\bibfnamefont {B.}~\bibnamefont {O’gorman}}, \bibinfo {author} {\bibfnamefont {E.~G.}\ \bibnamefont {Rieffel}}, \bibinfo {author} {\bibfnamefont {D.}~\bibnamefont {Venturelli}},\ and\ \bibinfo {author} {\bibfnamefont {R.}~\bibnamefont {Biswas}},\ }\bibfield  {title} {{From the quantum approximate optimization algorithm to a quantum alternating operator ansatz},\ }\href@noop {} {\bibfield  {journal} {\bibinfo  {journal} {Algorithms}\ }\textbf {\bibinfo {volume} {12}},\ \bibinfo {pages} {34} (\bibinfo {year} {2019})}\BibitemShut {NoStop}%
\bibitem [{\citenamefont {Wang}\ et~al.(2021{\natexlab{c}})\citenamefont {Wang}, \citenamefont {Ma}, \citenamefont {Hsieh},\ and\ \citenamefont {Yung}}]{wang2021quantum}%
  \BibitemOpen
  \bibfield  {author} {\bibinfo {author} {\bibfnamefont {X.}~\bibnamefont {Wang}}, \bibinfo {author} {\bibfnamefont {Y.}~\bibnamefont {Ma}}, \bibinfo {author} {\bibfnamefont {M.-H.}\ \bibnamefont {Hsieh}},\ and\ \bibinfo {author} {\bibfnamefont {M.-H.}\ \bibnamefont {Yung}},\ }\bibfield  {title} {{Quantum speedup in adaptive boosting of binary classification},\ }\href@noop {} {\bibfield  {journal} {\bibinfo  {journal} {Science China Physics, Mechanics \& Astronomy}\ }\textbf {\bibinfo {volume} {64}},\ \bibinfo {pages} {220311} (\bibinfo {year} {2021}{\natexlab{c}})}\BibitemShut {NoStop}%
\bibitem [{\citenamefont {Arunachalam}\ and\ \citenamefont {Maity}(2020)}]{arunachalam2020quantum}%
  \BibitemOpen
  \bibfield  {author} {\bibinfo {author} {\bibfnamefont {S.}~\bibnamefont {Arunachalam}}\ and\ \bibinfo {author} {\bibfnamefont {R.}~\bibnamefont {Maity}},\ }in\ \href@noop {} {\bibinfo {booktitle} {International Conference on Machine Learning}}\ (\bibinfo {organization} {PMLR},\ \bibinfo {year} {2020})\ pp.\ \bibinfo {pages} {377--387}\BibitemShut {NoStop}%
\bibitem [{\citenamefont {Izdebski}\ and\ \citenamefont {de~Wolf}(2020)}]{izdebski2020improved}%
  \BibitemOpen
  \bibfield  {author} {\bibinfo {author} {\bibfnamefont {A.}~\bibnamefont {Izdebski}}\ and\ \bibinfo {author} {\bibfnamefont {R.}~\bibnamefont {de~Wolf}},\ }\bibfield  {title} {{Improved quantum boosting},\ }\href@noop {} {\bibfield  {journal} {\bibinfo  {journal} {arXiv preprint arXiv:2009.08360}\ } (\bibinfo {year} {2020})}\BibitemShut {NoStop}%
\bibitem [{\citenamefont {Macaluso}\ et~al.(2020)\citenamefont {Macaluso}, \citenamefont {Clissa}, \citenamefont {Lodi},\ and\ \citenamefont {Sartori}}]{macaluso2020quantum}%
  \BibitemOpen
  \bibfield  {author} {\bibinfo {author} {\bibfnamefont {A.}~\bibnamefont {Macaluso}}, \bibinfo {author} {\bibfnamefont {L.}~\bibnamefont {Clissa}}, \bibinfo {author} {\bibfnamefont {S.}~\bibnamefont {Lodi}},\ and\ \bibinfo {author} {\bibfnamefont {C.}~\bibnamefont {Sartori}},\ }\bibfield  {title} {{Quantum ensemble for classification},\ }\href@noop {} {\bibfield  {journal} {\bibinfo  {journal} {arXiv preprint arXiv:2007.01028}\ } (\bibinfo {year} {2020})}\BibitemShut {NoStop}%
\bibitem [{\citenamefont {Endo}\ et~al.(2021)\citenamefont {Endo}, \citenamefont {Cai}, \citenamefont {Benjamin},\ and\ \citenamefont {Yuan}}]{endo2021hybrid}%
  \BibitemOpen
  \bibfield  {author} {\bibinfo {author} {\bibfnamefont {S.}~\bibnamefont {Endo}}, \bibinfo {author} {\bibfnamefont {Z.}~\bibnamefont {Cai}}, \bibinfo {author} {\bibfnamefont {S.~C.}\ \bibnamefont {Benjamin}},\ and\ \bibinfo {author} {\bibfnamefont {X.}~\bibnamefont {Yuan}},\ }\bibfield  {title} {{Hybrid quantum-classical algorithms and quantum error mitigation},\ }\href@noop {} {\bibfield  {journal} {\bibinfo  {journal} {Journal of the Physical Society of Japan}\ }\textbf {\bibinfo {volume} {90}},\ \bibinfo {pages} {032001} (\bibinfo {year} {2021})}\BibitemShut {NoStop}%
\bibitem [{\citenamefont {Wu}\ et~al.(2021{\natexlab{a}})\citenamefont {Wu}, \citenamefont {Chan}, \citenamefont {Guan}, \citenamefont {Sun}, \citenamefont {Wang}, \citenamefont {Zhou}, \citenamefont {Livny}, \citenamefont {Carminati}, \citenamefont {Di~Meglio}, \citenamefont {Li} et~al.}]{wu_vqc}%
  \BibitemOpen
  \bibfield  {author} {\bibinfo {author} {\bibfnamefont {S.~L.}\ \bibnamefont {Wu}}, \bibinfo {author} {\bibfnamefont {J.}~\bibnamefont {Chan}}, \bibinfo {author} {\bibfnamefont {W.}~\bibnamefont {Guan}}, \bibinfo {author} {\bibfnamefont {S.}~\bibnamefont {Sun}}, \bibinfo {author} {\bibfnamefont {A.}~\bibnamefont {Wang}}, \bibinfo {author} {\bibfnamefont {C.}~\bibnamefont {Zhou}}, \bibinfo {author} {\bibfnamefont {M.}~\bibnamefont {Livny}}, \bibinfo {author} {\bibfnamefont {F.}~\bibnamefont {Carminati}}, \bibinfo {author} {\bibfnamefont {A.}~\bibnamefont {Di~Meglio}}, \bibinfo {author} {\bibfnamefont {A.~C.}\ \bibnamefont {Li}}, et~al.,\ }\bibfield  {title} {{Application of quantum machine learning using the quantum variational classifier method to high energy physics analysis at the {LHC} on {IBM} quantum computer simulator and hardware with 10 qubits},\ }\href@noop {} {\bibfield  {journal} {\bibinfo  {journal} {Journal of Physics G: Nuclear and Particle Physics}\ }\textbf {\bibinfo {volume} {48}},\ \bibinfo
  {pages} {125003} (\bibinfo {year} {2021}{\natexlab{a}})}\BibitemShut {NoStop}%
\bibitem [{\citenamefont {Wu}\ et~al.(2021{\natexlab{b}})\citenamefont {Wu}, \citenamefont {Sun}, \citenamefont {Guan}, \citenamefont {Zhou}, \citenamefont {Chan}, \citenamefont {Cheng}, \citenamefont {Pham}, \citenamefont {Qian}, \citenamefont {Wang}, \citenamefont {Zhang} et~al.}]{wu_kernel}%
  \BibitemOpen
  \bibfield  {author} {\bibinfo {author} {\bibfnamefont {S.~L.}\ \bibnamefont {Wu}}, \bibinfo {author} {\bibfnamefont {S.}~\bibnamefont {Sun}}, \bibinfo {author} {\bibfnamefont {W.}~\bibnamefont {Guan}}, \bibinfo {author} {\bibfnamefont {C.}~\bibnamefont {Zhou}}, \bibinfo {author} {\bibfnamefont {J.}~\bibnamefont {Chan}}, \bibinfo {author} {\bibfnamefont {C.~L.}\ \bibnamefont {Cheng}}, \bibinfo {author} {\bibfnamefont {T.}~\bibnamefont {Pham}}, \bibinfo {author} {\bibfnamefont {Y.}~\bibnamefont {Qian}}, \bibinfo {author} {\bibfnamefont {A.~Z.}\ \bibnamefont {Wang}}, \bibinfo {author} {\bibfnamefont {R.}~\bibnamefont {Zhang}}, et~al.,\ }\bibfield  {title} {{Application of quantum machine learning using the quantum kernel algorithm on high energy physics analysis at the {LHC}},\ }\href@noop {} {\bibfield  {journal} {\bibinfo  {journal} {Physical Review Research}\ }\textbf {\bibinfo {volume} {3}},\ \bibinfo {pages} {033221} (\bibinfo {year} {2021}{\natexlab{b}})}\BibitemShut {NoStop}%
\bibitem [{\citenamefont {Or{\'u}s}(2019)}]{tnn1}%
  \BibitemOpen
  \bibfield  {author} {\bibinfo {author} {\bibfnamefont {R.}~\bibnamefont {Or{\'u}s}},\ }\bibfield  {title} {{Tensor networks for complex quantum systems},\ }\href@noop {} {\bibfield  {journal} {\bibinfo  {journal} {Nature Reviews Physics}\ }\textbf {\bibinfo {volume} {1}},\ \bibinfo {pages} {538} (\bibinfo {year} {2019})}\BibitemShut {NoStop}%
\bibitem [{\citenamefont {Verstraete}\ et~al.(2008)\citenamefont {Verstraete}, \citenamefont {Murg},\ and\ \citenamefont {Cirac}}]{tnn5}%
  \BibitemOpen
  \bibfield  {author} {\bibinfo {author} {\bibfnamefont {F.}~\bibnamefont {Verstraete}}, \bibinfo {author} {\bibfnamefont {V.}~\bibnamefont {Murg}},\ and\ \bibinfo {author} {\bibfnamefont {J.~I.}\ \bibnamefont {Cirac}},\ }\bibfield  {title} {{Matrix product states, projected entangled pair states, and variational renormalization group methods for quantum spin systems},\ }\href@noop {} {\bibfield  {journal} {\bibinfo  {journal} {Advances in physics}\ }\textbf {\bibinfo {volume} {57}},\ \bibinfo {pages} {143} (\bibinfo {year} {2008})}\BibitemShut {NoStop}%
\bibitem [{\citenamefont {Vidal}(2008)}]{tnn4}%
  \BibitemOpen
  \bibfield  {author} {\bibinfo {author} {\bibfnamefont {G.}~\bibnamefont {Vidal}},\ }\bibfield  {title} {{Class of quantum many-body states that can be efficiently simulated},\ }\href@noop {} {\bibfield  {journal} {\bibinfo  {journal} {Physical review letters}\ }\textbf {\bibinfo {volume} {101}},\ \bibinfo {pages} {110501} (\bibinfo {year} {2008})}\BibitemShut {NoStop}%
\bibitem [{\citenamefont {Thumwanit}\ et~al.(2021)\citenamefont {Thumwanit}, \citenamefont {Lortaraprasert}, \citenamefont {Yano},\ and\ \citenamefont {Raymond}}]{thumwanit2021trainable}%
  \BibitemOpen
  \bibfield  {author} {\bibinfo {author} {\bibfnamefont {N.}~\bibnamefont {Thumwanit}}, \bibinfo {author} {\bibfnamefont {C.}~\bibnamefont {Lortaraprasert}}, \bibinfo {author} {\bibfnamefont {H.}~\bibnamefont {Yano}},\ and\ \bibinfo {author} {\bibfnamefont {R.}~\bibnamefont {Raymond}},\ }\bibfield  {title} {{Trainable discrete feature embeddings for variational quantum classifier},\ }\href@noop {} {\bibfield  {journal} {\bibinfo  {journal} {arXiv preprint arXiv:2106.09415}\ } (\bibinfo {year} {2021})}\BibitemShut {NoStop}%
\bibitem [{\citenamefont {Grant}\ et~al.(2018)\citenamefont {Grant}, \citenamefont {Benedetti}, \citenamefont {Cao}, \citenamefont {Hallam}, \citenamefont {Lockhart}, \citenamefont {Stojevic}, \citenamefont {Green},\ and\ \citenamefont {Severini}}]{grant2018hierarchical}%
  \BibitemOpen
  \bibfield  {author} {\bibinfo {author} {\bibfnamefont {E.}~\bibnamefont {Grant}}, \bibinfo {author} {\bibfnamefont {M.}~\bibnamefont {Benedetti}}, \bibinfo {author} {\bibfnamefont {S.}~\bibnamefont {Cao}}, \bibinfo {author} {\bibfnamefont {A.}~\bibnamefont {Hallam}}, \bibinfo {author} {\bibfnamefont {J.}~\bibnamefont {Lockhart}}, \bibinfo {author} {\bibfnamefont {V.}~\bibnamefont {Stojevic}}, \bibinfo {author} {\bibfnamefont {A.~G.}\ \bibnamefont {Green}},\ and\ \bibinfo {author} {\bibfnamefont {S.}~\bibnamefont {Severini}},\ }\bibfield  {title} {{Hierarchical quantum classifiers},\ }\href@noop {} {\bibfield  {journal} {\bibinfo  {journal} {npj Quantum Information}\ }\textbf {\bibinfo {volume} {4}},\ \bibinfo {pages} {1} (\bibinfo {year} {2018})}\BibitemShut {NoStop}%
\bibitem [{\citenamefont {Shi}\ et~al.(2006)\citenamefont {Shi}, \citenamefont {Duan},\ and\ \citenamefont {Vidal}}]{tnn3}%
  \BibitemOpen
  \bibfield  {author} {\bibinfo {author} {\bibfnamefont {Y.-Y.}\ \bibnamefont {Shi}}, \bibinfo {author} {\bibfnamefont {L.-M.}\ \bibnamefont {Duan}},\ and\ \bibinfo {author} {\bibfnamefont {G.}~\bibnamefont {Vidal}},\ }\bibfield  {title} {{Classical simulation of quantum many-body systems with a tree tensor network},\ }\href@noop {} {\bibfield  {journal} {\bibinfo  {journal} {Physical review a}\ }\textbf {\bibinfo {volume} {74}},\ \bibinfo {pages} {022320} (\bibinfo {year} {2006})}\BibitemShut {NoStop}%
\bibitem [{\citenamefont {Perez-Garcia}\ et~al.(2008)\citenamefont {Perez-Garcia}, \citenamefont {Verstraete}, \citenamefont {Wolf},\ and\ \citenamefont {Cirac}}]{tnn2}%
  \BibitemOpen
  \bibfield  {author} {\bibinfo {author} {\bibfnamefont {D.}~\bibnamefont {Perez-Garcia}}, \bibinfo {author} {\bibfnamefont {F.}~\bibnamefont {Verstraete}}, \bibinfo {author} {\bibfnamefont {M.}~\bibnamefont {Wolf}},\ and\ \bibinfo {author} {\bibfnamefont {J.}~\bibnamefont {Cirac}},\ }\bibfield  {title} {{{Matrix Product State Representations Quantum Inf. Comput. 7, 401 (2007); L Tagliacozzo, T R. de Oliveira, S Iblisdir, and JI Latorre, Scaling of entanglement support for Matrix Product States}},\ }\href@noop {} {\bibfield  {journal} {\bibinfo  {journal} {Phys. Rev. B}\ }\textbf {\bibinfo {volume} {78}},\ \bibinfo {pages} {024410} (\bibinfo {year} {2008})}\BibitemShut {NoStop}%
\bibitem [{\citenamefont {Cappelletti}\ et~al.(2020)\citenamefont {Cappelletti}, \citenamefont {Erbanni},\ and\ \citenamefont {Keller}}]{cappelletti2020polyadic}%
  \BibitemOpen
  \bibfield  {author} {\bibinfo {author} {\bibfnamefont {W.}~\bibnamefont {Cappelletti}}, \bibinfo {author} {\bibfnamefont {R.}~\bibnamefont {Erbanni}},\ and\ \bibinfo {author} {\bibfnamefont {J.}~\bibnamefont {Keller}},\ }in\ \href@noop {} {\bibinfo {booktitle} {2020 IEEE International Conference on Quantum Computing and Engineering (QCE)}}\ (\bibinfo {organization} {IEEE},\ \bibinfo {year} {2020})\ pp.\ \bibinfo {pages} {22--29}\BibitemShut {NoStop}%
\bibitem [{\citenamefont {Peters}\ et~al.(2021)\citenamefont {Peters}, \citenamefont {Caldeira}, \citenamefont {Ho}, \citenamefont {Leichenauer}, \citenamefont {Mohseni}, \citenamefont {Neven}, \citenamefont {Spentzouris}, \citenamefont {Strain},\ and\ \citenamefont {Perdue}}]{peters2021machine}%
  \BibitemOpen
  \bibfield  {author} {\bibinfo {author} {\bibfnamefont {E.}~\bibnamefont {Peters}}, \bibinfo {author} {\bibfnamefont {J.}~\bibnamefont {Caldeira}}, \bibinfo {author} {\bibfnamefont {A.}~\bibnamefont {Ho}}, \bibinfo {author} {\bibfnamefont {S.}~\bibnamefont {Leichenauer}}, \bibinfo {author} {\bibfnamefont {M.}~\bibnamefont {Mohseni}}, \bibinfo {author} {\bibfnamefont {H.}~\bibnamefont {Neven}}, \bibinfo {author} {\bibfnamefont {P.}~\bibnamefont {Spentzouris}}, \bibinfo {author} {\bibfnamefont {D.}~\bibnamefont {Strain}},\ and\ \bibinfo {author} {\bibfnamefont {G.~N.}\ \bibnamefont {Perdue}},\ }\bibfield  {title} {{Machine learning of high dimensional data on a noisy quantum processor},\ }\href@noop {} {\bibfield  {journal} {\bibinfo  {journal} {npj Quantum Information}\ }\textbf {\bibinfo {volume} {7}},\ \bibinfo {pages} {1} (\bibinfo {year} {2021})}\BibitemShut {NoStop}%
\bibitem [{\citenamefont {Huang}\ et~al.(2022)\citenamefont {Huang}, \citenamefont {Broughton}, \citenamefont {Cotler}, \citenamefont {Chen}, \citenamefont {Li}, \citenamefont {Mohseni}, \citenamefont {Neven}, \citenamefont {Babbush}, \citenamefont {Kueng}, \citenamefont {Preskill} et~al.}]{huang2022quantum}%
  \BibitemOpen
  \bibfield  {author} {\bibinfo {author} {\bibfnamefont {H.-Y.}\ \bibnamefont {Huang}}, \bibinfo {author} {\bibfnamefont {M.}~\bibnamefont {Broughton}}, \bibinfo {author} {\bibfnamefont {J.}~\bibnamefont {Cotler}}, \bibinfo {author} {\bibfnamefont {S.}~\bibnamefont {Chen}}, \bibinfo {author} {\bibfnamefont {J.}~\bibnamefont {Li}}, \bibinfo {author} {\bibfnamefont {M.}~\bibnamefont {Mohseni}}, \bibinfo {author} {\bibfnamefont {H.}~\bibnamefont {Neven}}, \bibinfo {author} {\bibfnamefont {R.}~\bibnamefont {Babbush}}, \bibinfo {author} {\bibfnamefont {R.}~\bibnamefont {Kueng}}, \bibinfo {author} {\bibfnamefont {J.}~\bibnamefont {Preskill}}, et~al.,\ }\bibfield  {title} {{Quantum advantage in learning from experiments},\ }\href@noop {} {\bibfield  {journal} {\bibinfo  {journal} {Science}\ }\textbf {\bibinfo {volume} {376}},\ \bibinfo {pages} {1182} (\bibinfo {year} {2022})}\BibitemShut {NoStop}%
\bibitem [{\citenamefont {Aubry}\ and\ \citenamefont {Andr{\'e}}(1980)}]{aubry1980analyticity}%
  \BibitemOpen
  \bibfield  {author} {\bibinfo {author} {\bibfnamefont {S.}~\bibnamefont {Aubry}}\ and\ \bibinfo {author} {\bibfnamefont {G.}~\bibnamefont {Andr{\'e}}},\ }\bibfield  {title} {{Analyticity breaking and Anderson localization in incommensurate lattices},\ }\href@noop {} {\bibfield  {journal} {\bibinfo  {journal} {Ann. Israel Phys. Soc}\ }\textbf {\bibinfo {volume} {3}},\ \bibinfo {pages} {18} (\bibinfo {year} {1980})}\BibitemShut {NoStop}%
\bibitem [{\citenamefont {Li}\ et~al.(2021{\natexlab{b}})\citenamefont {Li}, \citenamefont {Alam}, \citenamefont {Congzhou}, \citenamefont {Wang}, \citenamefont {Dokholyan},\ and\ \citenamefont {Ghosh}}]{li2021drug}%
  \BibitemOpen
  \bibfield  {author} {\bibinfo {author} {\bibfnamefont {J.}~\bibnamefont {Li}}, \bibinfo {author} {\bibfnamefont {M.}~\bibnamefont {Alam}}, \bibinfo {author} {\bibfnamefont {M.~S.}\ \bibnamefont {Congzhou}}, \bibinfo {author} {\bibfnamefont {J.}~\bibnamefont {Wang}}, \bibinfo {author} {\bibfnamefont {N.~V.}\ \bibnamefont {Dokholyan}},\ and\ \bibinfo {author} {\bibfnamefont {S.}~\bibnamefont {Ghosh}},\ }in\ \href@noop {} {\bibinfo {booktitle} {2021 58th ACM/IEEE Design Automation Conference (DAC)}}\ (\bibinfo {organization} {IEEE},\ \bibinfo {year} {2021})\ pp.\ \bibinfo {pages} {1356--1359}\BibitemShut {NoStop}%
\bibitem [{\citenamefont {Cervera-Lierta}\ et~al.(2021)\citenamefont {Cervera-Lierta}, \citenamefont {Kottmann},\ and\ \citenamefont {Aspuru-Guzik}}]{cervera2021meta}%
  \BibitemOpen
  \bibfield  {author} {\bibinfo {author} {\bibfnamefont {A.}~\bibnamefont {Cervera-Lierta}}, \bibinfo {author} {\bibfnamefont {J.~S.}\ \bibnamefont {Kottmann}},\ and\ \bibinfo {author} {\bibfnamefont {A.}~\bibnamefont {Aspuru-Guzik}},\ }\bibfield  {title} {{Meta-variational quantum eigensolver: Learning energy profiles of parameterized hamiltonians for quantum simulation},\ }\href@noop {} {\bibfield  {journal} {\bibinfo  {journal} {PRX Quantum}\ }\textbf {\bibinfo {volume} {2}},\ \bibinfo {pages} {020329} (\bibinfo {year} {2021})}\BibitemShut {NoStop}%
\bibitem [{\citenamefont {Batra}\ et~al.(2021)\citenamefont {Batra}, \citenamefont {Zorn}, \citenamefont {Foil}, \citenamefont {Minerali}, \citenamefont {Gawriljuk}, \citenamefont {Lane},\ and\ \citenamefont {Ekins}}]{batra2021quantum}%
  \BibitemOpen
  \bibfield  {author} {\bibinfo {author} {\bibfnamefont {K.}~\bibnamefont {Batra}}, \bibinfo {author} {\bibfnamefont {K.~M.}\ \bibnamefont {Zorn}}, \bibinfo {author} {\bibfnamefont {D.~H.}\ \bibnamefont {Foil}}, \bibinfo {author} {\bibfnamefont {E.}~\bibnamefont {Minerali}}, \bibinfo {author} {\bibfnamefont {V.~O.}\ \bibnamefont {Gawriljuk}}, \bibinfo {author} {\bibfnamefont {T.~R.}\ \bibnamefont {Lane}},\ and\ \bibinfo {author} {\bibfnamefont {S.}~\bibnamefont {Ekins}},\ }\bibfield  {title} {{Quantum machine learning algorithms for drug discovery applications},\ }\href@noop {} {\bibfield  {journal} {\bibinfo  {journal} {Journal of chemical information and modeling}\ }\textbf {\bibinfo {volume} {61}},\ \bibinfo {pages} {2641} (\bibinfo {year} {2021})}\BibitemShut {NoStop}%
\bibitem [{\citenamefont {Easom-Mccaldin}\ et~al.(2021)\citenamefont {Easom-Mccaldin}, \citenamefont {Bouridane}, \citenamefont {Belatreche},\ and\ \citenamefont {Jiang}}]{easom2021depth}%
  \BibitemOpen
  \bibfield  {author} {\bibinfo {author} {\bibfnamefont {P.}~\bibnamefont {Easom-Mccaldin}}, \bibinfo {author} {\bibfnamefont {A.}~\bibnamefont {Bouridane}}, \bibinfo {author} {\bibfnamefont {A.}~\bibnamefont {Belatreche}},\ and\ \bibinfo {author} {\bibfnamefont {R.}~\bibnamefont {Jiang}},\ }\bibfield  {title} {{On Depth, Robustness and Performance Using the Data Re-Uploading Single-Qubit Classifier},\ }\href@noop {} {\bibfield  {journal} {\bibinfo  {journal} {IEEE Access}\ }\textbf {\bibinfo {volume} {9}},\ \bibinfo {pages} {65127} (\bibinfo {year} {2021})}\BibitemShut {NoStop}%
\bibitem [{\citenamefont {Uvarov}\ and\ \citenamefont {Biamonte}(2021)}]{uvarov2021barren}%
  \BibitemOpen
  \bibfield  {author} {\bibinfo {author} {\bibfnamefont {A.}~\bibnamefont {Uvarov}}\ and\ \bibinfo {author} {\bibfnamefont {J.~D.}\ \bibnamefont {Biamonte}},\ }\bibfield  {title} {{On barren plateaus and cost function locality in variational quantum algorithms},\ }\href@noop {} {\bibfield  {journal} {\bibinfo  {journal} {Journal of Physics A: Mathematical and Theoretical}\ }\textbf {\bibinfo {volume} {54}},\ \bibinfo {pages} {245301} (\bibinfo {year} {2021})}\BibitemShut {NoStop}%
\bibitem [{\citenamefont {Sweke}\ et~al.(2020{\natexlab{b}})\citenamefont {Sweke}, \citenamefont {Wilde}, \citenamefont {Meyer}, \citenamefont {Schuld}, \citenamefont {F{\"a}hrmann}, \citenamefont {Meynard-Piganeau},\ and\ \citenamefont {Eisert}}]{sweke2020stochastic}%
  \BibitemOpen
  \bibfield  {author} {\bibinfo {author} {\bibfnamefont {R.}~\bibnamefont {Sweke}}, \bibinfo {author} {\bibfnamefont {F.}~\bibnamefont {Wilde}}, \bibinfo {author} {\bibfnamefont {J.}~\bibnamefont {Meyer}}, \bibinfo {author} {\bibfnamefont {M.}~\bibnamefont {Schuld}}, \bibinfo {author} {\bibfnamefont {P.~K.}\ \bibnamefont {F{\"a}hrmann}}, \bibinfo {author} {\bibfnamefont {B.}~\bibnamefont {Meynard-Piganeau}},\ and\ \bibinfo {author} {\bibfnamefont {J.}~\bibnamefont {Eisert}},\ }\bibfield  {title} {{Stochastic gradient descent for hybrid quantum-classical optimization},\ }\href@noop {} {\bibfield  {journal} {\bibinfo  {journal} {Quantum}\ }\textbf {\bibinfo {volume} {4}},\ \bibinfo {pages} {314} (\bibinfo {year} {2020}{\natexlab{b}})}\BibitemShut {NoStop}%
\bibitem [{\citenamefont {Schuld}\ et~al.(2021)\citenamefont {Schuld}, \citenamefont {Sweke},\ and\ \citenamefont {Meyer}}]{schuld2021effect}%
  \BibitemOpen
  \bibfield  {author} {\bibinfo {author} {\bibfnamefont {M.}~\bibnamefont {Schuld}}, \bibinfo {author} {\bibfnamefont {R.}~\bibnamefont {Sweke}},\ and\ \bibinfo {author} {\bibfnamefont {J.~J.}\ \bibnamefont {Meyer}},\ }\bibfield  {title} {{Effect of data encoding on the expressive power of variational quantum-machine-learning models},\ }\href@noop {} {\bibfield  {journal} {\bibinfo  {journal} {Physical Review A}\ }\textbf {\bibinfo {volume} {103}},\ \bibinfo {pages} {032430} (\bibinfo {year} {2021})}\BibitemShut {NoStop}%
\bibitem [{\citenamefont {Sweke}\ et~al.(2021)\citenamefont {Sweke}, \citenamefont {Seifert}, \citenamefont {Hangleiter},\ and\ \citenamefont {Eisert}}]{sweke2021quantum}%
  \BibitemOpen
  \bibfield  {author} {\bibinfo {author} {\bibfnamefont {R.}~\bibnamefont {Sweke}}, \bibinfo {author} {\bibfnamefont {J.-P.}\ \bibnamefont {Seifert}}, \bibinfo {author} {\bibfnamefont {D.}~\bibnamefont {Hangleiter}},\ and\ \bibinfo {author} {\bibfnamefont {J.}~\bibnamefont {Eisert}},\ }\bibfield  {title} {{On the quantum versus classical learnability of discrete distributions},\ }\href@noop {} {\bibfield  {journal} {\bibinfo  {journal} {Quantum}\ }\textbf {\bibinfo {volume} {5}},\ \bibinfo {pages} {417} (\bibinfo {year} {2021})}\BibitemShut {NoStop}%
\bibitem [{\citenamefont {P{\'e}rez-Salinas}\ et~al.(2020)\citenamefont {P{\'e}rez-Salinas}, \citenamefont {Cervera-Lierta}, \citenamefont {Gil-Fuster},\ and\ \citenamefont {Latorre}}]{reupload}%
  \BibitemOpen
  \bibfield  {author} {\bibinfo {author} {\bibfnamefont {A.}~\bibnamefont {P{\'e}rez-Salinas}}, \bibinfo {author} {\bibfnamefont {A.}~\bibnamefont {Cervera-Lierta}}, \bibinfo {author} {\bibfnamefont {E.}~\bibnamefont {Gil-Fuster}},\ and\ \bibinfo {author} {\bibfnamefont {J.~I.}\ \bibnamefont {Latorre}},\ }\bibfield  {title} {{Data re-uploading for a universal quantum classifier},\ }\href@noop {} {\bibfield  {journal} {\bibinfo  {journal} {Quantum}\ }\textbf {\bibinfo {volume} {4}},\ \bibinfo {pages} {226} (\bibinfo {year} {2020})}\BibitemShut {NoStop}%
\bibitem [{\citenamefont {Muten}\ et~al.(2021)\citenamefont {Muten}, \citenamefont {Yusuf},\ and\ \citenamefont {Tomut}}]{muten2021modified}%
  \BibitemOpen
  \bibfield  {author} {\bibinfo {author} {\bibfnamefont {E.~R.}\ \bibnamefont {Muten}}, \bibinfo {author} {\bibfnamefont {T.~T.}\ \bibnamefont {Yusuf}},\ and\ \bibinfo {author} {\bibfnamefont {A.~V.}\ \bibnamefont {Tomut}},\ }in\ \href@noop {} {\bibinfo {booktitle} {2021 IEEE International Conference on Quantum Computing and Engineering (QCE)}}\ (\bibinfo {organization} {IEEE},\ \bibinfo {year} {2021})\ pp.\ \bibinfo {pages} {82--88}\BibitemShut {NoStop}%
\bibitem [{\citenamefont {Terashi}\ et~al.(2021{\natexlab{b}})\citenamefont {Terashi}, \citenamefont {Kaneda}, \citenamefont {Kishimoto}, \citenamefont {Saito}, \citenamefont {Sawada},\ and\ \citenamefont {Tanaka}}]{terashi2021event}%
  \BibitemOpen
  \bibfield  {author} {\bibinfo {author} {\bibfnamefont {K.}~\bibnamefont {Terashi}}, \bibinfo {author} {\bibfnamefont {M.}~\bibnamefont {Kaneda}}, \bibinfo {author} {\bibfnamefont {T.}~\bibnamefont {Kishimoto}}, \bibinfo {author} {\bibfnamefont {M.}~\bibnamefont {Saito}}, \bibinfo {author} {\bibfnamefont {R.}~\bibnamefont {Sawada}},\ and\ \bibinfo {author} {\bibfnamefont {J.}~\bibnamefont {Tanaka}},\ }\bibfield  {title} {{Event classification with quantum machine learning in high-energy physics},\ }\href@noop {} {\bibfield  {journal} {\bibinfo  {journal} {Computing and Software for Big Science}\ }\textbf {\bibinfo {volume} {5}},\ \bibinfo {pages} {1} (\bibinfo {year} {2021}{\natexlab{b}})}\BibitemShut {NoStop}%
\bibitem [{\citenamefont {Marrero}\ et~al.(2021)\citenamefont {Marrero}, \citenamefont {Kieferov{\'a}},\ and\ \citenamefont {Wiebe}}]{marrero2021entanglement}%
  \BibitemOpen
  \bibfield  {author} {\bibinfo {author} {\bibfnamefont {C.~O.}\ \bibnamefont {Marrero}}, \bibinfo {author} {\bibfnamefont {M.}~\bibnamefont {Kieferov{\'a}}},\ and\ \bibinfo {author} {\bibfnamefont {N.}~\bibnamefont {Wiebe}},\ }\bibfield  {title} {{Entanglement-induced barren plateaus},\ }\href@noop {} {\bibfield  {journal} {\bibinfo  {journal} {PRX Quantum}\ }\textbf {\bibinfo {volume} {2}},\ \bibinfo {pages} {040316} (\bibinfo {year} {2021})}\BibitemShut {NoStop}%
\bibitem [{\citenamefont {Hayashi}\ et~al.(2005)\citenamefont {Hayashi}, \citenamefont {Hashimoto},\ and\ \citenamefont {Horibe}}]{hayashi2005reexamination}%
  \BibitemOpen
  \bibfield  {author} {\bibinfo {author} {\bibfnamefont {A.}~\bibnamefont {Hayashi}}, \bibinfo {author} {\bibfnamefont {T.}~\bibnamefont {Hashimoto}},\ and\ \bibinfo {author} {\bibfnamefont {M.}~\bibnamefont {Horibe}},\ }\bibfield  {title} {{Reexamination of optimal quantum state estimation of pure states},\ }\href@noop {} {\bibfield  {journal} {\bibinfo  {journal} {Physical review A}\ }\textbf {\bibinfo {volume} {72}},\ \bibinfo {pages} {032325} (\bibinfo {year} {2005})}\BibitemShut {NoStop}%
\bibitem [{\citenamefont {Zhao}\ et~al.(2019)\citenamefont {Zhao}, \citenamefont {Fitzsimons},\ and\ \citenamefont {Fitzsimons}}]{zhao2019quantum}%
  \BibitemOpen
  \bibfield  {author} {\bibinfo {author} {\bibfnamefont {Z.}~\bibnamefont {Zhao}}, \bibinfo {author} {\bibfnamefont {J.~K.}\ \bibnamefont {Fitzsimons}},\ and\ \bibinfo {author} {\bibfnamefont {J.~F.}\ \bibnamefont {Fitzsimons}},\ }\bibfield  {title} {{Quantum-assisted Gaussian process regression},\ }\href@noop {} {\bibfield  {journal} {\bibinfo  {journal} {Physical Review A}\ }\textbf {\bibinfo {volume} {99}},\ \bibinfo {pages} {052331} (\bibinfo {year} {2019})}\BibitemShut {NoStop}%
\bibitem [{\citenamefont {Giovannetti}\ et~al.(2008{\natexlab{b}})\citenamefont {Giovannetti}, \citenamefont {Lloyd},\ and\ \citenamefont {Maccone}}]{giovannetti2008quantum}%
  \BibitemOpen
  \bibfield  {author} {\bibinfo {author} {\bibfnamefont {V.}~\bibnamefont {Giovannetti}}, \bibinfo {author} {\bibfnamefont {S.}~\bibnamefont {Lloyd}},\ and\ \bibinfo {author} {\bibfnamefont {L.}~\bibnamefont {Maccone}},\ }\bibfield  {title} {{Quantum random access memory},\ }\href@noop {} {\bibfield  {journal} {\bibinfo  {journal} {Physical review letters}\ }\textbf {\bibinfo {volume} {100}},\ \bibinfo {pages} {160501} (\bibinfo {year} {2008}{\natexlab{b}})}\BibitemShut {NoStop}%
\bibitem [{\citenamefont {Zaspel}\ et~al.(2018)\citenamefont {Zaspel}, \citenamefont {Huang}, \citenamefont {Harbrecht},\ and\ \citenamefont {von Lilienfeld}}]{zaspel2018boosting}%
  \BibitemOpen
  \bibfield  {author} {\bibinfo {author} {\bibfnamefont {P.}~\bibnamefont {Zaspel}}, \bibinfo {author} {\bibfnamefont {B.}~\bibnamefont {Huang}}, \bibinfo {author} {\bibfnamefont {H.}~\bibnamefont {Harbrecht}},\ and\ \bibinfo {author} {\bibfnamefont {O.~A.}\ \bibnamefont {von Lilienfeld}},\ }\bibfield  {title} {{Boosting quantum machine learning models with a multilevel combination technique: Pople diagrams revisited},\ }\href@noop {} {\bibfield  {journal} {\bibinfo  {journal} {Journal of chemical theory and computation}\ }\textbf {\bibinfo {volume} {15}},\ \bibinfo {pages} {1546} (\bibinfo {year} {2018})}\BibitemShut {NoStop}%
\bibitem [{\citenamefont {Von~Lilienfeld}(2013)}]{von2013first}%
  \BibitemOpen
  \bibfield  {author} {\bibinfo {author} {\bibfnamefont {O.~A.}\ \bibnamefont {Von~Lilienfeld}},\ }\bibfield  {title} {{First principles view on chemical compound space: Gaining rigorous atomistic control of molecular properties},\ }\href@noop {} {\bibfield  {journal} {\bibinfo  {journal} {International Journal of Quantum Chemistry}\ }\textbf {\bibinfo {volume} {113}},\ \bibinfo {pages} {1676} (\bibinfo {year} {2013})}\BibitemShut {NoStop}%
\bibitem [{\citenamefont {Rupp}\ et~al.(2018)\citenamefont {Rupp}, \citenamefont {Von~Lilienfeld},\ and\ \citenamefont {Burke}}]{rupp2018guest}%
  \BibitemOpen
  \bibfield  {author} {\bibinfo {author} {\bibfnamefont {M.}~\bibnamefont {Rupp}}, \bibinfo {author} {\bibfnamefont {O.~A.}\ \bibnamefont {Von~Lilienfeld}},\ and\ \bibinfo {author} {\bibfnamefont {K.}~\bibnamefont {Burke}},\ }\href@noop {} {{Guest editorial: Special topic on data-enabled theoretical chemistry}} (\bibinfo {year} {2018})\BibitemShut {NoStop}%
\bibitem [{\citenamefont {Akhalwaya}\ et~al.(2022{\natexlab{b}})\citenamefont {Akhalwaya}, \citenamefont {Ubaru}, \citenamefont {Clarkson}, \citenamefont {Squillante}, \citenamefont {Jejjala}, \citenamefont {He}, \citenamefont {Naidoo}, \citenamefont {Kalantzis},\ and\ \citenamefont {Horesh}}]{https://doi.org/10.48550/arxiv.2209.09371}%
  \BibitemOpen
  \bibfield  {author} {\bibinfo {author} {\bibfnamefont {I.~Y.}\ \bibnamefont {Akhalwaya}}, \bibinfo {author} {\bibfnamefont {S.}~\bibnamefont {Ubaru}}, \bibinfo {author} {\bibfnamefont {K.~L.}\ \bibnamefont {Clarkson}}, \bibinfo {author} {\bibfnamefont {M.~S.}\ \bibnamefont {Squillante}}, \bibinfo {author} {\bibfnamefont {V.}~\bibnamefont {Jejjala}}, \bibinfo {author} {\bibfnamefont {Y.-H.}\ \bibnamefont {He}}, \bibinfo {author} {\bibfnamefont {K.}~\bibnamefont {Naidoo}}, \bibinfo {author} {\bibfnamefont {V.}~\bibnamefont {Kalantzis}},\ and\ \bibinfo {author} {\bibfnamefont {L.}~\bibnamefont {Horesh}},\ }\href {https://doi.org/10.48550/ARXIV.2209.09371} {{Towards Quantum Advantage on Noisy Quantum Computers}} (\bibinfo {year} {2022}{\natexlab{b}})\BibitemShut {NoStop}%
\bibitem [{\citenamefont {Aaronson}(2015)}]{AaronsonFinePrint2015}%
  \BibitemOpen
  \bibfield  {author} {\bibinfo {author} {\bibfnamefont {S.}~\bibnamefont {Aaronson}},\ }\bibfield  {title} {{Read the fine print},\ }\href {https://doi.org/10.1038/nphys3272} {\bibfield  {journal} {\bibinfo  {journal} {Nature Physics}\ }\textbf {\bibinfo {volume} {11}},\ \bibinfo {pages} {291} (\bibinfo {year} {2015})}\BibitemShut {NoStop}%
\bibitem [{\citenamefont {Schuld}\ and\ \citenamefont {Killoran}(2022)}]{schuld2022quantum}%
  \BibitemOpen
  \bibfield  {author} {\bibinfo {author} {\bibfnamefont {M.}~\bibnamefont {Schuld}}\ and\ \bibinfo {author} {\bibfnamefont {N.}~\bibnamefont {Killoran}},\ }\bibfield  {title} {{Is quantum advantage the right goal for quantum machine learning?},\ }\href@noop {} {\bibfield  {journal} {\bibinfo  {journal} {Prx Quantum}\ }\textbf {\bibinfo {volume} {3}},\ \bibinfo {pages} {030101} (\bibinfo {year} {2022})}\BibitemShut {NoStop}%
\bibitem [{\citenamefont {Kerenidis}(2020)}]{KerenidisDataLoader2020}%
  \BibitemOpen
  \bibfield  {author} {\bibinfo {author} {\bibfnamefont {I.}~\bibnamefont {Kerenidis}},\ }\href@noop {} {{A method for loading classical data into quantum states for applications in machine learning and optimization}} (\bibinfo {year} {2020})\BibitemShut {NoStop}%
\bibitem [{\citenamefont {Nakanishi}\ et~al.(2020{\natexlab{b}})\citenamefont {Nakanishi}, \citenamefont {Fujii},\ and\ \citenamefont {Todo}}]{nakanishi2020}%
  \BibitemOpen
  \bibfield  {author} {\bibinfo {author} {\bibfnamefont {K.~M.}\ \bibnamefont {Nakanishi}}, \bibinfo {author} {\bibfnamefont {K.}~\bibnamefont {Fujii}},\ and\ \bibinfo {author} {\bibfnamefont {S.}~\bibnamefont {Todo}},\ }\bibfield  {title} {{Sequential minimal optimization for quantum-classical hybrid algorithms},\ }\href {https://doi.org/10.1103/physrevresearch.2.043158} {\bibfield  {journal} {\bibinfo  {journal} {Physical Review Research}\ }\textbf {\bibinfo {volume} {2}},\ \bibinfo {pages} {043158} (\bibinfo {year} {2020}{\natexlab{b}})}\BibitemShut {NoStop}%
\bibitem [{\citenamefont {Watanabe}\ et~al.(2021)\citenamefont {Watanabe}, \citenamefont {Raymond}, \citenamefont {Ohnishi}, \citenamefont {Kaminishi},\ and\ \citenamefont {Sugawara}}]{watanabe2021}%
  \BibitemOpen
  \bibfield  {author} {\bibinfo {author} {\bibfnamefont {H.~C.}\ \bibnamefont {Watanabe}}, \bibinfo {author} {\bibfnamefont {R.}~\bibnamefont {Raymond}}, \bibinfo {author} {\bibfnamefont {Y.-Y.}\ \bibnamefont {Ohnishi}}, \bibinfo {author} {\bibfnamefont {E.}~\bibnamefont {Kaminishi}},\ and\ \bibinfo {author} {\bibfnamefont {M.}~\bibnamefont {Sugawara}},\ }in\ \href {https://doi.org/https://doi.org/10.1109/QCE52317.2021.00026} {\bibinfo {booktitle} {2021 IEEE International Conference on Quantum Computing and Engineering (QCE)}}\ (\bibinfo {organization} {IEEE},\ \bibinfo {year} {2021})\ pp.\ \bibinfo {pages} {100--111}\BibitemShut {NoStop}%
\bibitem [{\citenamefont {Xia}\ et~al.(2023)\citenamefont {Xia}, \citenamefont {Zou}, \citenamefont {Qiu}, \citenamefont {Chen}, \citenamefont {Zhu}, \citenamefont {Li}, \citenamefont {Deng},\ and\ \citenamefont {Li}}]{xia2023configured}%
  \BibitemOpen
  \bibfield  {author} {\bibinfo {author} {\bibfnamefont {W.}~\bibnamefont {Xia}}, \bibinfo {author} {\bibfnamefont {J.}~\bibnamefont {Zou}}, \bibinfo {author} {\bibfnamefont {X.}~\bibnamefont {Qiu}}, \bibinfo {author} {\bibfnamefont {F.}~\bibnamefont {Chen}}, \bibinfo {author} {\bibfnamefont {B.}~\bibnamefont {Zhu}}, \bibinfo {author} {\bibfnamefont {C.}~\bibnamefont {Li}}, \bibinfo {author} {\bibfnamefont {D.-L.}\ \bibnamefont {Deng}},\ and\ \bibinfo {author} {\bibfnamefont {X.}~\bibnamefont {Li}},\ }\bibfield  {title} {{Configured quantum reservoir computing for multi-task machine learning},\ }\href@noop {} {\bibfield  {journal} {\bibinfo  {journal} {Science Bulletin}\ }\textbf {\bibinfo {volume} {68}},\ \bibinfo {pages} {2321} (\bibinfo {year} {2023})}\BibitemShut {NoStop}%
\bibitem [{\citenamefont {Huang}\ et~al.(2021{\natexlab{e}})\citenamefont {Huang}, \citenamefont {Wang}, \citenamefont {Song}, \citenamefont {Xu}, \citenamefont {Li}, \citenamefont {Wang}, \citenamefont {Guo}, \citenamefont {Song}, \citenamefont {Liu}, \citenamefont {Zheng} et~al.}]{huang2021quantum}%
  \BibitemOpen
  \bibfield  {author} {\bibinfo {author} {\bibfnamefont {K.}~\bibnamefont {Huang}}, \bibinfo {author} {\bibfnamefont {Z.-A.}\ \bibnamefont {Wang}}, \bibinfo {author} {\bibfnamefont {C.}~\bibnamefont {Song}}, \bibinfo {author} {\bibfnamefont {K.}~\bibnamefont {Xu}}, \bibinfo {author} {\bibfnamefont {H.}~\bibnamefont {Li}}, \bibinfo {author} {\bibfnamefont {Z.}~\bibnamefont {Wang}}, \bibinfo {author} {\bibfnamefont {Q.}~\bibnamefont {Guo}}, \bibinfo {author} {\bibfnamefont {Z.}~\bibnamefont {Song}}, \bibinfo {author} {\bibfnamefont {Z.-B.}\ \bibnamefont {Liu}}, \bibinfo {author} {\bibfnamefont {D.}~\bibnamefont {Zheng}}, et~al.,\ }\bibfield  {title} {{Quantum generative adversarial networks with multiple superconducting qubits},\ }\href@noop {} {\bibfield  {journal} {\bibinfo  {journal} {npj Quantum Information}\ }\textbf {\bibinfo {volume} {7}},\ \bibinfo {pages} {165} (\bibinfo {year} {2021}{\natexlab{e}})}\BibitemShut {NoStop}%
\bibitem [{\citenamefont {Avramouli}\ et~al.(2022)\citenamefont {Avramouli}, \citenamefont {Savvas}, \citenamefont {Vasilaki}, \citenamefont {Garani},\ and\ \citenamefont {Xenakis}}]{avramouli2022quantum}%
  \BibitemOpen
  \bibfield  {author} {\bibinfo {author} {\bibfnamefont {M.}~\bibnamefont {Avramouli}}, \bibinfo {author} {\bibfnamefont {I.}~\bibnamefont {Savvas}}, \bibinfo {author} {\bibfnamefont {A.}~\bibnamefont {Vasilaki}}, \bibinfo {author} {\bibfnamefont {G.}~\bibnamefont {Garani}},\ and\ \bibinfo {author} {\bibfnamefont {A.}~\bibnamefont {Xenakis}},\ }in\ \href@noop {} {\bibinfo {booktitle} {Proceedings of the 26th Pan-Hellenic Conference on Informatics}}\ (\bibinfo {year} {2022})\ pp.\ \bibinfo {pages} {394--401}\BibitemShut {NoStop}%
\bibitem [{\citenamefont {McArdle}\ et~al.(2020)\citenamefont {McArdle}, \citenamefont {Endo}, \citenamefont {Aspuru-Guzik}, \citenamefont {Benjamin},\ and\ \citenamefont {Yuan}}]{mcardle2020quantum}%
  \BibitemOpen
  \bibfield  {author} {\bibinfo {author} {\bibfnamefont {S.}~\bibnamefont {McArdle}}, \bibinfo {author} {\bibfnamefont {S.}~\bibnamefont {Endo}}, \bibinfo {author} {\bibfnamefont {A.}~\bibnamefont {Aspuru-Guzik}}, \bibinfo {author} {\bibfnamefont {S.~C.}\ \bibnamefont {Benjamin}},\ and\ \bibinfo {author} {\bibfnamefont {X.}~\bibnamefont {Yuan}},\ }\bibfield  {title} {{Quantum computational chemistry},\ }\href@noop {} {\bibfield  {journal} {\bibinfo  {journal} {Reviews of Modern Physics}\ }\textbf {\bibinfo {volume} {92}},\ \bibinfo {pages} {015003} (\bibinfo {year} {2020})}\BibitemShut {NoStop}%
\bibitem [{\citenamefont {Sajjan}\ et~al.(2022)\citenamefont {Sajjan}, \citenamefont {Li}, \citenamefont {Selvarajan}, \citenamefont {Sureshbabu}, \citenamefont {Kale}, \citenamefont {Gupta}, \citenamefont {Singh},\ and\ \citenamefont {Kais}}]{sajjan2022quantum}%
  \BibitemOpen
  \bibfield  {author} {\bibinfo {author} {\bibfnamefont {M.}~\bibnamefont {Sajjan}}, \bibinfo {author} {\bibfnamefont {J.}~\bibnamefont {Li}}, \bibinfo {author} {\bibfnamefont {R.}~\bibnamefont {Selvarajan}}, \bibinfo {author} {\bibfnamefont {S.~H.}\ \bibnamefont {Sureshbabu}}, \bibinfo {author} {\bibfnamefont {S.~S.}\ \bibnamefont {Kale}}, \bibinfo {author} {\bibfnamefont {R.}~\bibnamefont {Gupta}}, \bibinfo {author} {\bibfnamefont {V.}~\bibnamefont {Singh}},\ and\ \bibinfo {author} {\bibfnamefont {S.}~\bibnamefont {Kais}},\ }\bibfield  {title} {{Quantum machine learning for chemistry and physics},\ }\href@noop {} {\bibfield  {journal} {\bibinfo  {journal} {Chemical Society Reviews}\ } (\bibinfo {year} {2022})}\BibitemShut {NoStop}%
\bibitem [{\citenamefont {Wada}\ et~al.(2022)\citenamefont {Wada}, \citenamefont {Raymond}, \citenamefont {Sato},\ and\ \citenamefont {Watanabe}}]{WRSW_FQS_2022}%
  \BibitemOpen
  \bibfield  {author} {\bibinfo {author} {\bibfnamefont {K.}~\bibnamefont {Wada}}, \bibinfo {author} {\bibfnamefont {R.}~\bibnamefont {Raymond}}, \bibinfo {author} {\bibfnamefont {Y.}~\bibnamefont {Sato}},\ and\ \bibinfo {author} {\bibfnamefont {H.~C.}\ \bibnamefont {Watanabe}},\ }\href {https://doi.org/10.48550/ARXIV.2209.08535} {{Full optimization of a single-qubit gate on the generalized sequential quantum optimizer}} (\bibinfo {year} {2022})\BibitemShut {NoStop}%
\bibitem [{\citenamefont {Li}\ and\ \citenamefont {Deng}(2022)}]{li2022recent}%
  \BibitemOpen
  \bibfield  {author} {\bibinfo {author} {\bibfnamefont {W.}~\bibnamefont {Li}}\ and\ \bibinfo {author} {\bibfnamefont {D.-L.}\ \bibnamefont {Deng}},\ }\bibfield  {title} {{Recent advances for quantum classifiers},\ }\href@noop {} {\bibfield  {journal} {\bibinfo  {journal} {Science China Physics, Mechanics \& Astronomy}\ }\textbf {\bibinfo {volume} {65}},\ \bibinfo {pages} {220301} (\bibinfo {year} {2022})}\BibitemShut {NoStop}%
\bibitem [{\citenamefont {Noble}(2006)}]{noble2006support}%
  \BibitemOpen
  \bibfield  {author} {\bibinfo {author} {\bibfnamefont {W.~S.}\ \bibnamefont {Noble}},\ }\bibfield  {title} {{What is a support vector machine?},\ }\href@noop {} {\bibfield  {journal} {\bibinfo  {journal} {Nature biotechnology}\ }\textbf {\bibinfo {volume} {24}},\ \bibinfo {pages} {1565} (\bibinfo {year} {2006})}\BibitemShut {NoStop}%
\bibitem [{\citenamefont {Du}\ et~al.(2021)\citenamefont {Du}, \citenamefont {Qian},\ and\ \citenamefont {Tao}}]{QUDIO2021}%
  \BibitemOpen
  \bibfield  {author} {\bibinfo {author} {\bibfnamefont {Y.}~\bibnamefont {Du}}, \bibinfo {author} {\bibfnamefont {Y.}~\bibnamefont {Qian}},\ and\ \bibinfo {author} {\bibfnamefont {D.}~\bibnamefont {Tao}},\ }\href {https://doi.org/10.48550/ARXIV.2106.12819} {{Accelerating variational quantum algorithms with multiple quantum processors}} (\bibinfo {year} {2021})\BibitemShut {NoStop}%
\bibitem [{\citenamefont {Shen}\ et~al.(2024)\citenamefont {Shen}, \citenamefont {Jobst}, \citenamefont {Shishenina},\ and\ \citenamefont {Pollmann}}]{shen2024classification}%
  \BibitemOpen
  \bibfield  {author} {\bibinfo {author} {\bibfnamefont {K.}~\bibnamefont {Shen}}, \bibinfo {author} {\bibfnamefont {B.}~\bibnamefont {Jobst}}, \bibinfo {author} {\bibfnamefont {E.}~\bibnamefont {Shishenina}},\ and\ \bibinfo {author} {\bibfnamefont {F.}~\bibnamefont {Pollmann}},\ }\bibfield  {title} {{{Classification of the Fashion-MNIST Dataset on a Quantum Computer}},\ }\href@noop {} {\bibfield  {journal} {\bibinfo  {journal} {arXiv:2403.02405}\ } (\bibinfo {year} {2024})},\ \Eprint {https://arxiv.org/abs/2403.02405} {2403.02405 [quant-ph]} \BibitemShut {NoStop}%
\bibitem [{\citenamefont {Bowles}\ et~al.(2024)\citenamefont {Bowles}, \citenamefont {Ahmed},\ and\ \citenamefont {Schuld}}]{bowles2024better}%
  \BibitemOpen
  \bibfield  {author} {\bibinfo {author} {\bibfnamefont {J.}~\bibnamefont {Bowles}}, \bibinfo {author} {\bibfnamefont {S.}~\bibnamefont {Ahmed}},\ and\ \bibinfo {author} {\bibfnamefont {M.}~\bibnamefont {Schuld}},\ }\bibfield  {title} {{Better than classical? The subtle art of benchmarking quantum machine learning models},\ }\href@noop {} {\bibfield  {journal} {\bibinfo  {journal} {arXiv:2403.07059}\ } (\bibinfo {year} {2024})},\ \Eprint {https://arxiv.org/abs/2403.07059} {2403.07059 [quant-ph]} \BibitemShut {NoStop}%
\bibitem [{\citenamefont {Anagolum}\ et~al.(2024)\citenamefont {Anagolum}, \citenamefont {Alavisamani}, \citenamefont {Das}, \citenamefont {Qureshi}, \citenamefont {Kessler},\ and\ \citenamefont {Shi}}]{anagolum2024elivagar}%
  \BibitemOpen
  \bibfield  {author} {\bibinfo {author} {\bibfnamefont {S.}~\bibnamefont {Anagolum}}, \bibinfo {author} {\bibfnamefont {N.}~\bibnamefont {Alavisamani}}, \bibinfo {author} {\bibfnamefont {P.}~\bibnamefont {Das}}, \bibinfo {author} {\bibfnamefont {M.}~\bibnamefont {Qureshi}}, \bibinfo {author} {\bibfnamefont {E.}~\bibnamefont {Kessler}},\ and\ \bibinfo {author} {\bibfnamefont {Y.}~\bibnamefont {Shi}},\ }\bibfield  {title} {{{\'Eliv\'agar: Efficient Quantum Circuit Search for Classification}},\ }\href@noop {} {\bibfield  {journal} {\bibinfo  {journal} {arXiv:2401.09393}\ } (\bibinfo {year} {2024})},\ \Eprint {https://arxiv.org/abs/2401.09393} {2401.09393 [quant-ph]} \BibitemShut {NoStop}%
\bibitem [{\citenamefont {Koyasu}\ et~al.(2023)\citenamefont {Koyasu}, \citenamefont {Raymond},\ and\ \citenamefont {Imai}}]{kriieeeqce23}%
  \BibitemOpen
  \bibfield  {author} {\bibinfo {author} {\bibfnamefont {I.}~\bibnamefont {Koyasu}}, \bibinfo {author} {\bibfnamefont {R.}~\bibnamefont {Raymond}},\ and\ \bibinfo {author} {\bibfnamefont {H.}~\bibnamefont {Imai}},\ }in\ \href {https://doi.org/10.1109/QCE57702.2023.00059} {\bibinfo {booktitle} {2023 IEEE International Conference on Quantum Computing and Engineering (QCE)}},\ Vol.~\bibinfo {volume} {01}\ (\bibinfo {year} {2023})\ pp.\ \bibinfo {pages} {457--467}\BibitemShut {NoStop}%
\bibitem [{\citenamefont {Moradi}\ et~al.(2022)\citenamefont {Moradi}, \citenamefont {Brandner}, \citenamefont {Spielvogel}, \citenamefont {Krajnc}, \citenamefont {Hillmich}, \citenamefont {Wille}, \citenamefont {Drexler},\ and\ \citenamefont {Papp}}]{moradi2022clinical}%
  \BibitemOpen
  \bibfield  {author} {\bibinfo {author} {\bibfnamefont {S.}~\bibnamefont {Moradi}}, \bibinfo {author} {\bibfnamefont {C.}~\bibnamefont {Brandner}}, \bibinfo {author} {\bibfnamefont {C.}~\bibnamefont {Spielvogel}}, \bibinfo {author} {\bibfnamefont {D.}~\bibnamefont {Krajnc}}, \bibinfo {author} {\bibfnamefont {S.}~\bibnamefont {Hillmich}}, \bibinfo {author} {\bibfnamefont {R.}~\bibnamefont {Wille}}, \bibinfo {author} {\bibfnamefont {W.}~\bibnamefont {Drexler}},\ and\ \bibinfo {author} {\bibfnamefont {L.}~\bibnamefont {Papp}},\ }\bibfield  {title} {{Clinical data classification with noisy intermediate scale quantum computers},\ }\href@noop {} {\bibfield  {journal} {\bibinfo  {journal} {Scientific reports}\ }\textbf {\bibinfo {volume} {12}},\ \bibinfo {pages} {1851} (\bibinfo {year} {2022})}\BibitemShut {NoStop}%
\bibitem [{\citenamefont {Mardirosian}(2019)}]{mardirosian2019quantum}%
  \BibitemOpen
  \bibfield  {author} {\bibinfo {author} {\bibfnamefont {S.}~\bibnamefont {Mardirosian}},\ }\bibinfo {title} {Quantum-enhanced Supervised Learning with Variational Quantum Circuits},\ \href@noop {} {Ph.D. thesis},\ \bibinfo  {school} {PhD thesis. Leiden University} (\bibinfo {year} {2019})\BibitemShut {NoStop}%
\bibitem [{\citenamefont {Wang}\ et~al.(2024)\citenamefont {Wang}, \citenamefont {Baba-Yara},\ and\ \citenamefont {Chen}}]{wang2024justq}%
  \BibitemOpen
  \bibfield  {author} {\bibinfo {author} {\bibfnamefont {R.}~\bibnamefont {Wang}}, \bibinfo {author} {\bibfnamefont {F.}~\bibnamefont {Baba-Yara}},\ and\ \bibinfo {author} {\bibfnamefont {F.}~\bibnamefont {Chen}},\ }in\ \href@noop {} {\bibinfo {booktitle} {2024 29th Asia and South Pacific Design Automation Conference (ASP-DAC)}}\ (\bibinfo {organization} {IEEE},\ \bibinfo {year} {2024})\ pp.\ \bibinfo {pages} {121--126}\BibitemShut {NoStop}%
\bibitem [{\citenamefont {Innan}\ et~al.(2024)\citenamefont {Innan}, \citenamefont {Khan}, \citenamefont {Marchisio}, \citenamefont {Shafique},\ and\ \citenamefont {Bennai}}]{innan2024fedqnn}%
  \BibitemOpen
  \bibfield  {author} {\bibinfo {author} {\bibfnamefont {N.}~\bibnamefont {Innan}}, \bibinfo {author} {\bibfnamefont {M.~A.-Z.}\ \bibnamefont {Khan}}, \bibinfo {author} {\bibfnamefont {A.}~\bibnamefont {Marchisio}}, \bibinfo {author} {\bibfnamefont {M.}~\bibnamefont {Shafique}},\ and\ \bibinfo {author} {\bibfnamefont {M.}~\bibnamefont {Bennai}},\ }\bibfield  {title} {{{FedQNN: Federated Learning using Quantum Neural Networks}},\ }\href@noop {} {\bibfield  {journal} {\bibinfo  {journal} {arXiv preprint arXiv:2403.10861}\ } (\bibinfo {year} {2024})}\BibitemShut {NoStop}%
\bibitem [{\citenamefont {Peixoto}\ et~al.(2023)\citenamefont {Peixoto}, \citenamefont {Crispim~Rom{\~a}o}, \citenamefont {Oliveira},\ and\ \citenamefont {Ochoa}}]{peixoto2023fitting}%
  \BibitemOpen
  \bibfield  {author} {\bibinfo {author} {\bibfnamefont {M.~C.}\ \bibnamefont {Peixoto}}, \bibinfo {author} {\bibfnamefont {M.}~\bibnamefont {Crispim~Rom{\~a}o}}, \bibinfo {author} {\bibfnamefont {M.~G.~J.}\ \bibnamefont {Oliveira}},\ and\ \bibinfo {author} {\bibfnamefont {I.}~\bibnamefont {Ochoa}},\ }\bibfield  {title} {{Fitting a collider in a quantum computer: tackling the challenges of quantum machine learning for big datasets},\ }\href@noop {} {\bibfield  {journal} {\bibinfo  {journal} {Frontiers in Artificial Intelligence}\ }\textbf {\bibinfo {volume} {6}},\ \bibinfo {pages} {1268852} (\bibinfo {year} {2023})}\BibitemShut {NoStop}%
\bibitem [{\citenamefont {Cugini}\ et~al.(2023)\citenamefont {Cugini}, \citenamefont {Gerace}, \citenamefont {Govoni}, \citenamefont {Perego},\ and\ \citenamefont {Valsecchi}}]{cugini2023comparing}%
  \BibitemOpen
  \bibfield  {author} {\bibinfo {author} {\bibfnamefont {D.}~\bibnamefont {Cugini}}, \bibinfo {author} {\bibfnamefont {D.}~\bibnamefont {Gerace}}, \bibinfo {author} {\bibfnamefont {P.}~\bibnamefont {Govoni}}, \bibinfo {author} {\bibfnamefont {A.}~\bibnamefont {Perego}},\ and\ \bibinfo {author} {\bibfnamefont {D.}~\bibnamefont {Valsecchi}},\ }\bibfield  {title} {{Comparing quantum and classical machine learning for Vector Boson Scattering background reduction at the Large Hadron Collider},\ }\href@noop {} {\bibfield  {journal} {\bibinfo  {journal} {Quantum Machine Intelligence}\ }\textbf {\bibinfo {volume} {5}},\ \bibinfo {pages} {35} (\bibinfo {year} {2023})}\BibitemShut {NoStop}%
\bibitem [{\citenamefont {Nakayama}\ et~al.(2023)\citenamefont {Nakayama}, \citenamefont {Mitarai}, \citenamefont {Placidi}, \citenamefont {Sugimoto},\ and\ \citenamefont {Fujii}}]{nakayama2023vqe}%
  \BibitemOpen
  \bibfield  {author} {\bibinfo {author} {\bibfnamefont {A.}~\bibnamefont {Nakayama}}, \bibinfo {author} {\bibfnamefont {K.}~\bibnamefont {Mitarai}}, \bibinfo {author} {\bibfnamefont {L.}~\bibnamefont {Placidi}}, \bibinfo {author} {\bibfnamefont {T.}~\bibnamefont {Sugimoto}},\ and\ \bibinfo {author} {\bibfnamefont {K.}~\bibnamefont {Fujii}},\ }\bibfield  {title} {{{VQE}-generated quantum circuit dataset for machine learning},\ }\href@noop {} {\bibfield  {journal} {\bibinfo  {journal} {arXiv preprint arXiv:2302.09751}\ } (\bibinfo {year} {2023})}\BibitemShut {NoStop}%
\bibitem [{\citenamefont {Melo}\ et~al.(2023)\citenamefont {Melo}, \citenamefont {Earnest-Noble},\ and\ \citenamefont {Tacchino}}]{melo2023pulse}%
  \BibitemOpen
  \bibfield  {author} {\bibinfo {author} {\bibfnamefont {A.}~\bibnamefont {Melo}}, \bibinfo {author} {\bibfnamefont {N.}~\bibnamefont {Earnest-Noble}},\ and\ \bibinfo {author} {\bibfnamefont {F.}~\bibnamefont {Tacchino}},\ }\bibfield  {title} {{Pulse-efficient quantum machine learning},\ }\href@noop {} {\bibfield  {journal} {\bibinfo  {journal} {Quantum}\ }\textbf {\bibinfo {volume} {7}},\ \bibinfo {pages} {1130} (\bibinfo {year} {2023})}\BibitemShut {NoStop}%
\bibitem [{\citenamefont {Sim{\~o}es}\ et~al.(2023)\citenamefont {Sim{\~o}es}, \citenamefont {Huber}, \citenamefont {Meier}, \citenamefont {Smailov}, \citenamefont {F{\"u}chslin},\ and\ \citenamefont {Stockinger}}]{simoes2023experimental}%
  \BibitemOpen
  \bibfield  {author} {\bibinfo {author} {\bibfnamefont {R.~D.~M.}\ \bibnamefont {Sim{\~o}es}}, \bibinfo {author} {\bibfnamefont {P.}~\bibnamefont {Huber}}, \bibinfo {author} {\bibfnamefont {N.}~\bibnamefont {Meier}}, \bibinfo {author} {\bibfnamefont {N.}~\bibnamefont {Smailov}}, \bibinfo {author} {\bibfnamefont {R.~M.}\ \bibnamefont {F{\"u}chslin}},\ and\ \bibinfo {author} {\bibfnamefont {K.}~\bibnamefont {Stockinger}},\ }\bibfield  {title} {{Experimental evaluation of quantum machine learning algorithms},\ }\href@noop {} {\bibfield  {journal} {\bibinfo  {journal} {IEEE access}\ }\textbf {\bibinfo {volume} {11}},\ \bibinfo {pages} {6197} (\bibinfo {year} {2023})}\BibitemShut {NoStop}%
\bibitem [{\citenamefont {{scikit-learn developers}}(2022)}]{scikit-learn}%
  \BibitemOpen
  \bibfield  {author} {\bibinfo {author} {\bibnamefont {{scikit-learn developers}}},\ }\href {https://scikit-learn.org/stable/} {{scikit-learn: Machine Learning in Python}},\ \bibinfo {howpublished} {\url{https://scikit-learn.org/stable/modules/generated/sklearn.datasets.make_multilabel_classification.html?highlight=make_multilabel_classification}} (\bibinfo {year} {2022})\BibitemShut {NoStop}%
\bibitem [{\citenamefont {Vasques}\ et~al.(2023)\citenamefont {Vasques}, \citenamefont {Paik},\ and\ \citenamefont {Cif}}]{vasques2023application}%
  \BibitemOpen
  \bibfield  {author} {\bibinfo {author} {\bibfnamefont {X.}~\bibnamefont {Vasques}}, \bibinfo {author} {\bibfnamefont {H.}~\bibnamefont {Paik}},\ and\ \bibinfo {author} {\bibfnamefont {L.}~\bibnamefont {Cif}},\ }\bibfield  {title} {{Application of quantum machine learning using quantum kernel algorithms on multiclass neuron M-type classification},\ }\href@noop {} {\bibfield  {journal} {\bibinfo  {journal} {Scientific Reports}\ }\textbf {\bibinfo {volume} {13}},\ \bibinfo {pages} {11541} (\bibinfo {year} {2023})}\BibitemShut {NoStop}%
\bibitem [{\citenamefont {Bartkiewicz}\ et~al.(2023)\citenamefont {Bartkiewicz}, \citenamefont {Tulewicz}, \citenamefont {Roik},\ and\ \citenamefont {Lemr}}]{bartkiewicz2023synergic}%
  \BibitemOpen
  \bibfield  {author} {\bibinfo {author} {\bibfnamefont {K.}~\bibnamefont {Bartkiewicz}}, \bibinfo {author} {\bibfnamefont {P.}~\bibnamefont {Tulewicz}}, \bibinfo {author} {\bibfnamefont {J.}~\bibnamefont {Roik}},\ and\ \bibinfo {author} {\bibfnamefont {K.}~\bibnamefont {Lemr}},\ }\bibfield  {title} {{Synergic quantum generative machine learning},\ }\href@noop {} {\bibfield  {journal} {\bibinfo  {journal} {Scientific Reports}\ }\textbf {\bibinfo {volume} {13}},\ \bibinfo {pages} {12893} (\bibinfo {year} {2023})}\BibitemShut {NoStop}%
\end{thebibliography}%

\end{document}